\documentclass[twocolumn]{aastex63}
\usepackage{appendix}

\usepackage{amssymb}
\usepackage{amsmath}

\received{}
\revised{}
\accepted{}
\submitjournal{ApJ}

\shorttitle{SAGUARO}
\shortauthors{Paterson et al.}

\graphicspath{{./}{figures/}}

\begin{document}

\title{Searches after Gravitational Waves Using ARizona Observatories (SAGUARO): Observations and Analysis from Advanced LIGO/Virgo's Third Observing Run}

\correspondingauthor{K. Paterson}
\email{kerry.paterson@northwestern.edu}

\newcommand{\UA}{\affiliation{Steward Observatory, The University of Arizona, 933 North Cherry Avenue, Tucson, AZ 85721-0065, USA}}
\newcommand{\NU}{\affiliation{Center for Interdisciplinary Exploration and Research in Astrophysics and Department of Physics and Astronomy, \\ Northwestern University, 1800 Sherman Ave, Evanston, IL 60201, USA}}
\newcommand{\UCDavis}{\affiliation{Department of Physics, University of California, 1 Shields Avenue, Davis, CA 95616-5270, USA}}
\newcommand{\Padova}{\affiliation{Department of Physics and Astronomy Galileo Galilei, University of Padova, Vicolo dell'Osservatorio, 3, I-35122 Padova, Italy}}
\newcommand{\INAF}{\affiliation{INAF Osservatorio Astronomico di Padova, Vicolo dell'Osservatorio 5, I-35122 Padova, Italy}}
\newcommand{\INAFbol}{\affiliation{INAF - Osservatorio di Astrofisica e Scienza dello Spazio - Via Piero Gobetti 93/3, I-40129 Bologna, Italy}}
\newcommand{\LPL}{\affiliation{Lunar and Planetary Lab, Department of Planetary Sciences, University of Arizona, Tucson, AZ 85721, USA}}
\newcommand{\NOAO}{\affiliation{National Optical Astronomy Observatory, 950 North Cherry Avenue, Tucson, AZ 85719, USA}}
\newcommand{\PATAU}{\affiliation{The School of Physics and Astronomy, Tel Aviv University, Tel Aviv 69978, Israel}}
\newcommand{\TTU}{\affiliation{Department of Physics and Astronomy, Texas Tech University, Box 1051, Lubbock, TX 79409-1051, USA}}
\newcommand{\OSU}{\affiliation{Department  of  Astronomy,  The  Ohio  State University,  140  W.  18th  Ave.,  Columbus,  OH43210, USA}}
\newcommand{\LBT}{\affiliation{Large Binocular Telescope Observatory, 933 North Cherry Avenue, Tucson, AZ, USA}}
\newcommand{\RMC}{\affiliation{Department of Physics and Space Science Royal Military College of Canada P.O. Box 17000, Station Forces Kingston, ON K7K 7B4, Canada}}
\newcommand{\ASU}{\affiliation{School of Earth and Space Exploration, Arizona State University, Tempe, AZ 85287, USA}}
\newcommand{\MMT}{\affiliation{MMT Observatory, PO Box 210065, University of Arizona, Tucson, AZ 85721-0065, USA}}
\newcommand{\NAU}{\affiliation{Department of Astronomy and Planetary Science, Northern Arizona University, P.O. Box 6010, Flagstaff, AZ 86011, USA}}
\newcommand{\UAOptSci}{\affiliation{College of Optical Sciences, University of Arizona, 1630 E University Blvd, Tucson, AZ 85719, USA}}
\newcommand{\UNC}{\affiliation{Department of Physics and Astronomy, University of North Carolina at Chapel Hill, Chapel Hill, NC 27599, USA}}
\newcommand{\MSU}{\affiliation{Center for Data Intensive and Time Domain Astronomy, Department  of  Physics  and  Astronomy,  Michigan  State  University,East Lansing, MI 48824, USA}}
\newcommand{\UCSC}{\affiliation{Department of Astronomy and Astrophysics, University of California, Santa Cruz, CA 95064, USA}}
\newcommand{\STScI}{\affiliation{Space Telescope Science Institute, Baltimore, MD 21218, USA.}}
\newcommand{\Brandeis}{\affiliation{Department of Physics, Brandeis University, Waltham, MA 02453, USA}}
\newcommand{\LCO}{\affiliation{Las Cumbres Observatory, 6740 Cortona Drive, Suite 102, Goleta, CA 93117-5575, USA}}
\newcommand{\UToronto}{\affiliation{Department of Astronomy and Astrophysics, University of Toronto, 50 St. George Street, Toronto, Ontario, M5S 3H4 Canada}}
\newcommand{\NotreDame}{\affiliation{Department of Physics, University of Notre Dame, Notre Dame, IN 46556, USA}}
\newcommand{\UMN}{\affiliation{College of Science \& Engineering, Minnesota Institute for Astrophysics, University of Minnesota, 115 Union St. SE, Minneapolis, MN 55455, USA}}
\newcommand{\UT}{\affiliation{Department of Astronomy, University of Texas at Austin, Austin, TX 78712, USA}}
\newcommand{\JHU}{\affiliation{Department of Physics and Astronomy, The Johns Hopkins University, Baltimore, MD 21218, USA.}}
\newcommand{\VAT}{\affiliation{Vatican Observatory, 00120 Citt\`{a} del Vaticano, Vatican City State  }}
\newcommand{\HF}{\affiliation{Hubble Fellow}}
\newcommand{\Carnegie}{\affiliation{The Observatories of the Carnegie Institution for Science, 813 Santa Barbara St., Pasadena, CA 91101, USA}}
\author[0000-0001-8340-3486]{K.~Paterson}
\NU

\author[0000-0001-9589-3793]{M.~J. Lundquist}
\UA

\author[0000-0002-9267-6213]{J.~C. Rastinejad}
\NU

\author[0000-0002-7374-935X]{W.~Fong}
\NU

\author[0000-0003-4102-380X]{D.~J. Sand}
\UA

\author[0000-0003-0123-0062]{J.~E. Andrews}
\UA

\author[0000-0002-1546-9763]{R.~C. Amaro}
\UA

\author{O.~Eskandari}
\NU

\author{S.~Wyatt}
\UA

\author{P.~N. Daly}
\UA

\author{H.~Bradley}
\UA

\author{S.~Zhou-Wright}
\UA

\author[0000-0001-8818-0795]{S.~Valenti}
\UCDavis

\author[0000-0002-2898-6532]{S.~Yang}
\UCDavis\Padova\INAF

\author{E.~Christensen}
\LPL

\author[0000-0002-2575-2618]{A.~R. Gibbs}
\LPL

\author{F.~Shelly}
\LPL


\author[0000-0002-8826-3571]{C. Bilinski}
\UA

\author[0000-0002-8400-3705]{L. Chomiuk}
\MSU

\author[0000-0001-8104-3536]{A.~Corsi}
\TTU

\author{M.~R. Drout}
\UToronto\Carnegie

\author[0000-0002-2445-5275]{R.~J.~Foley}
\UCSC

\author[0000-0002-1315-9307]{P. Gabor}
\VAT

\author[0000-0003-4069-2817]{P. Garnavich}
\NotreDame

\author[0000-0001-9920-6057]{C.~J. Grier}
\UA

\author{E. Hamden}
\UA

\author[0000-0003-0000-0126]{H. Krantz}
\UA

\author[0000-0002-7157-500X]{E. Olszewski}
\UA

\author[0000-0002-8099-9023]{V. Paschalidis}
\UA

\author[0000-0002-5060-3673]{D. Reichart}
\UNC

\author[0000-0002-4410-5387]{A. Rest}
\STScI\JHU

\author[0000-0001-5510-2424]{N.~Smith}
\UA

\author[0000-0002-1468-9668]{J. Strader}
\MSU

\author[0000-0003-4580-3790]{D. Trilling}
\NAU

\author[0000-0003-0272-0418]{C. Veillet}
\LBT

\author[0000-0003-1892-2751]{R. M. Wagner}
\OSU\LBT

\author[0000-0001-6047-8469]{A. Zabludoff}
\UA

\begin{abstract}
With the conclusion of the third observing run for Advanced LIGO/Virgo (O3), we present a detailed analysis of both triggered and serendipitous observations of 17 gravitational wave (GW) events (7 triggered and 10 purely serendipitous) from the Searches After Gravitational-waves Using ARizona Observatories (SAGUARO) program. We searched a total of 4935 deg$^2$ down to a median 5$\sigma$ transient detection depth of 21.1 AB mag using the Mt Lemmon 1.5 m telescope, the discovery engine for SAGUARO. In addition to triggered events within 24~hours, our transient search encompassed a time interval following GW events of $<120$~hrs, providing observations on $\sim$ 1/2 of the events accessible to the Mt Lemmon 1.5 m telescope. We covered 2.1--86\% of the LVC total probability ($P_{\rm total}$) for individual events, with a median $P_{\rm total} \approx 8\%$ within $<120$~hours. Following improvements to our pipeline and the addition of serendipitous observations, we find a total of 7 new optical candidates across 5 GW events which we are unable to rule out after searching for additional information and comparing to kilonova models. Using both publicly available and our own late-time data, we investigated a total of 252 optical candidates for these 17 events, finding only 65\% were followed up in some capacity by the community. Of the total 252 candidates, we are able to rule out an additional 12 previously reported counterpart candidates. In light of these results, we discuss lessons learned from the SAGUARO GW counterpart search. We discuss how community coordination of observations and candidate follow-up, as well as the role of archival data, are crucial to improving the efficiency of follow-up efforts and preventing unnecessary duplication of effort with limited EM resources.
\end{abstract}


\section{Introduction}\label{sec:intro}

Since 2015, the Advanced Laser Interferometer Gravitational-Wave Observatory detectors (LIGO, \citealt{ligo09}), and later the Advanced Virgo Observatory \citep{,virgo15} detectors, have been detecting gravitational waves (GWs) produced by the inspirals of merging compact objects (neutron stars: NS and/or black holes: BH). During the first two observing runs (O1 and O2), the LIGO-Virgo Collaboration (LVC) announced 10 binary black hole (BBH) mergers and one binary neutron star (BNS) merger \citep{2019O1O2Catalog}. The single BNS merger, termed GW170817 \citep{lvc_gw170817}, also produced electromagnetic (EM) emission (AT2017gfo) from the $\gamma$-rays to the radio band, signaling the first GW-EM multi-messenger discovery \citep{LIGO_MMA}. AT2017gfo was discovered in the galaxy NGC~4993 \citep{Coulter17} at a distance of $40.7\pm1.4\pm1.9$ Mpc (random and systematic errors; \citealt{Cantiello2018}).

With the discovery of this EM counterpart, GW170817 was detected across the entire EM spectrum with a multitude of telescopes (c.f., \citealt{LIGO_MMA}). In-depth studies of the EM emission and its spectral and temporal evolution have provided a wealth of information about BNS mergers, such as direct evidence of the radioactive decay of r-process material demonstrating kilonovae as important sites for the production of heavy elements \citep{Arcavi+17,Chornock+17,Coulter+17,Cowperthwaite17,Drout+17,Diaz+17,Fong+17,Gall+2017,Hu+17,Kasliwal+17,Lipunov+17,McCully+17,Nicholl17,Pian+17,Pozanenko+17,Shappee+17,Smartt+17,Soares-Santos+17,Tanvir+17,Utsumi17,Valenti2017,villar17}. Through late-time follow-up and detailed modelling, the identification of a structured relativistic jet was also confirmed for GW170817 \citep{Haggard17,hallinan17,margutti17,margutti2018,Alexander2018,mooley18a,Fong19,ghirlanda19,hajela19,troja19}.  

The third observing run (O3) began on April 1, 2019 with a 1-month commissioning break in October 2019 and had an effective operational time of 10.9~months. Relative to the capabilities of O2, the increased sensitivities during the first 6 months of O3 (O3a) led to a median orientation-averaged range for BNS mergers of 108 Mpc, 135 Mpc and 45 Mpc for LIGO Hanford, LIGO Livingston, and Virgo, respectively; while the median 90\% localization during O3a was 860 deg$^2$ \citep{gwtc2}. During O3, a total of 56 non-retracted public alerts\footnote{https://gracedb.ligo.org} were released. Assuming an astrophysical origin for these 56 alerts, each GW candidate is assigned a probability of being the merger of two black holes (BBH), two neutron stars (BNS), a neutron star and a black hole (NSBH), or a more ambiguous case in which one of the components has a mass in the gap between NS and BH classifications (termed ``MassGap'', $3 < M/M_{\odot} < 5$; see \citealt{lvc_loc}). A fifth classification encompasses unmodeled short duration (t$<$1s) ``Burst'' signals, which do not fit the normal inspiral of compact objects and are of uncertain astrophysical origin.

This classification scheme resulted in 37 BBH, 7 BNS, 7 NSBH, 4 MassGap, and 1 burst event candidates that were detected during O3\footnote{Categorized by the highest non-terrestrial classification reported by the LVC for each event.}. For each event, we adopt the naming convention used by the LVC. Events have superevent names given by the letter ``S'' followed by the six digit UTC date and a lowercase incrementing letter. Events that are confirmed as real GW events by the LVC are renamed to ``GW'' followed by the six digit UTC date (e.g. GW190425). For O3a sources presented in the Gravitational-Wave Transient Catalog 2 (GWTC-2, \citealt{gwtc2}), we change the names to follow the naming convention adopted there. For events above the adopted False Alarm Rate (FAR) threshold, names include ``GW'' followed by the six digit UTC date and the six digit UTC time (e.g., GW190408\_181802); while sub-threshold events retain their superevent names. Previously published events (e.g. GW190425 and GW190521) keep their published names without the UTC time. These 56 non-retracted  events had 90\% localization regions that ranged in size from 18--24264 deg$^2$, depending largely on the number of detectors operational at the time. Other than the event detected by the burst pipeline (S200114f), which does not have an associated distance, the remaining 55 event candidates have inferred median distances of 108--5263~Mpc from the GW signals.

O3 comprised a number of firsts for the LVC, including the first detection of a candidate NSBH merger, GW190426\_152155 \citep{gcn24411}; the first detection of a candidate MassGap merger, S190814bv \citep{gcn25324}, which was later classified as a NSBH with a high mass-ratio, challenging current formation models and the predicted mass distribution of compact-object binaries (GW190814, \citealt{190814_paper}); and the first detection of a burst candidate, S200114f \citep{gcn26734}. In addition, the second BNS merger, GW190425 \citep{GW190425}, was significantly more massive than GW170817, as well as more massive than the Galactic NS binary population \citep{Safarzadeh2020}. Finally, O3 included the most massive BBH merger, GW190521, with BH masses of 91.4$^{+29.3}_{-17.5}$ $M_{\odot}$ and 66.8$^{+20.7}_{-20.7}$ $M_{\odot}$ (\citealt{lvc_gw190521,gwtc2}), having implications on various formation channels for these systems \citep{190521_imp_2020}.

Despite the wealth of new discoveries and GW event alerts, the large sky localizations coupled with the large median distances made it challenging for effective searches of associated UV, optical, or near-infrared (NIR) emission for a large number of events. Indeed, optical searches were dominated by instruments with wide fields-of-view (FOV), which largely adopted the approach of tiling the GW localization regions. Published examples include DECam/DES \citep{garcia20}, DECam/GROWTH \citep{Andreoni2019_190510,Andreoni20,Goldstein19}, ZTF/GROWTH \citep{coughlin19,Anand20,coughlin20,kasliwal20}, GRANDMA \citep{antier20}, GOTO \citep{gompertz2020}, ENGRAVE \citep{ackley2020} and CFHT \citep{vieira_cfht}; while other surveys like the MASTER-Network \citep{lipunov17}, Pan-STARRS \citep{chambers16}, ATLAS \citep{tonry18} and {\it Gaia} \citep{GaiaCo} have also contributed towards candidate searches. Meanwhile, telescopes with smaller FOV have opted for galaxy-targeted searches (Magellan; \citealt{gomez19}, MMT/SOAR; \citealt{Hosseinzadeh19}, J-GEM; \citealt{jgem}). The large majority of these surveys encompass triggered searches on GW events with a majority of the information distribution in real-time via The Gamma-ray Coordinates Network (GCN). Published papers on single events generally focus on the search for transient EM emission within a few days of the GW event and the in-depth analysis of the emission from candidates. Summary papers from groups such as ZTF/GROWTH \citep{Anand20,kasliwal20} have also explored the long term behaviour of their initially reported candidates, with efforts to follow up outstanding candidates through spectroscopic or photometric means.

Searches After Gravitational waves Using ARizona Observatories (SAGUARO) is a telescope network dedicated to the discovery and follow up of GW events\footnote{Although SAGUARO currently focuses on the follow-up of GW events, the methods presented in this paper could be applied to any transient of interest.}. In this paper, we present a summary of the SAGUARO campaign and analysis of our observations to find and study the optical counterparts to GW events that ran concurrent with LIGO/Virgo O3 \citep{Lundquist19}. In Section \ref{sec:overview} we provide an overview of SAGUARO, including our image subtraction and candidate vetting, summarize our coverage in O3, and discuss significant updates implemented during O3. We present a detailed description of our observations and each individual event in Section \ref{sec:ind}. In Section \ref{sec:discussion}, we discuss our findings. Lastly, we discuss the conclusions, lessons learned from O3, and future prospects in Section \ref{sec:conclusion}.

Unless otherwise stated, all magnitudes reported here are in \emph{Gaia} DR2 $G$-band and are converted to the AB system via $m_{AB}$ = $m_{Gaia}$ + 0.125 \citep{Maiz18}. We employ standard $\Lambda$CDM cosmology with H$_{0}$ = 69.6 km s$^{-1}$ Mpc$^{-1}$ and $\Omega_{M}$ = 0.286 \citep{Bennett+14}.

\begin{figure*}
\includegraphics[width=0.8\textwidth]{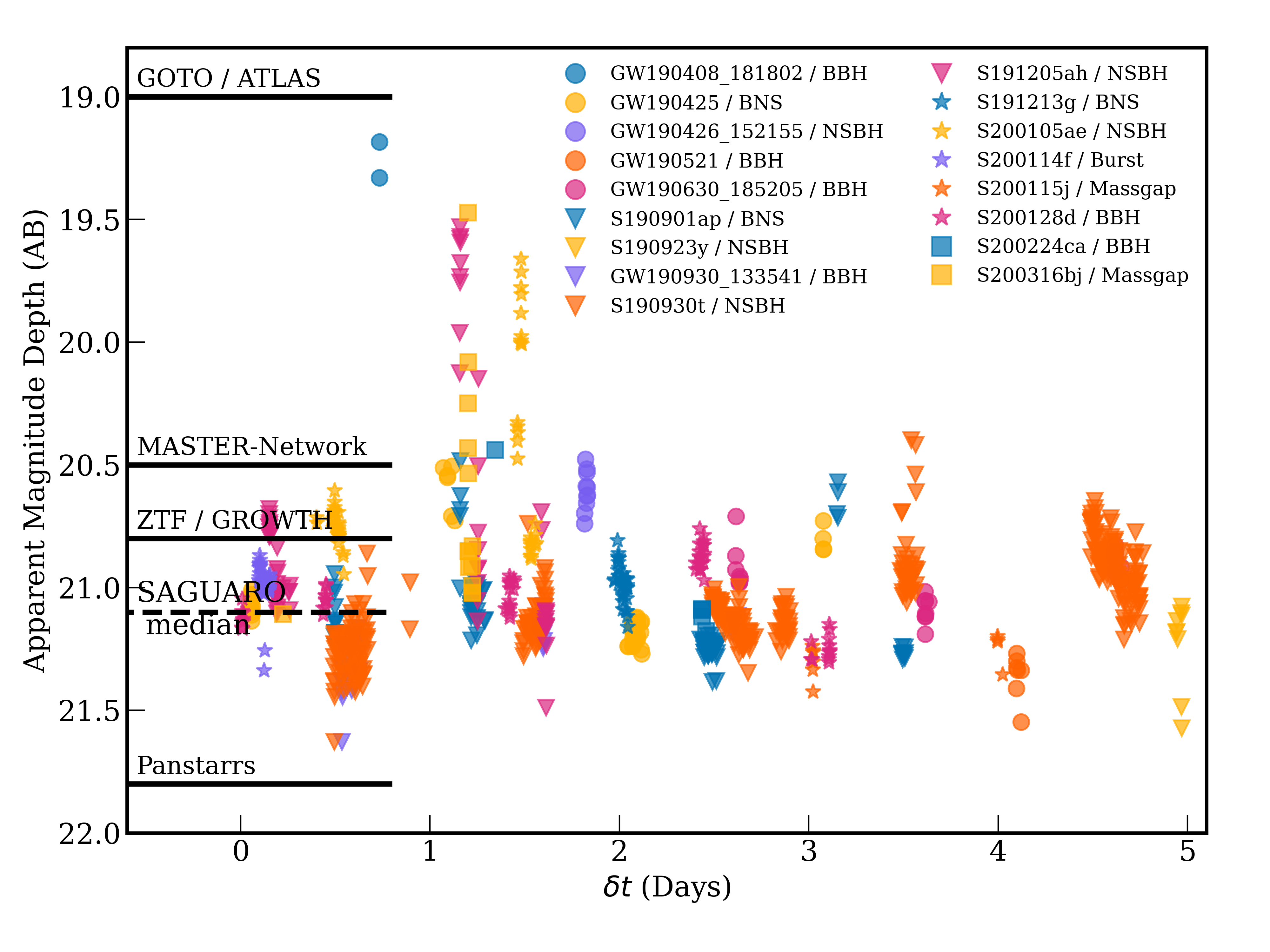}
\centering
\vspace{-0.2in}
\caption{The 5$\sigma$ transient depth of SAGUARO fields measured from the image subtraction process versus $\delta t$ (defined as the time after the GW event) for 17 events with triggered or serendipitous fields; all symbols represent upper limits. The reported depths (in various filters and sigma levels) of other GW follow-up surveys: GOTO \citep{gompertz2020}, ATLAS \citep{tonry18}, the MASTER-Network \citep{lipunov19}, ZTF/GROWTH \citep{kasliwal20}, and  Pan-STARRS \citep{chambers16}, are shown as black horizontal lines. We also note a 23.1~mag $i$-band median depth of the CFHT survey \citep{vieira_cfht}, a 23.0~mag $g$- or $r$-band median depth of the DECAM/GROWTH searches \citep{Anand20}, a 22.4~mag $i$-band median depth of DESGW (based on 2 events; \citealt{GCN27366, Morgan20}), and a 17.4~mag median limiting magnitude of BNS and NSBH events by the GRANDMA collaboration \citep{antier20}. The median 5$\sigma$ transient depth of SAGUARO fields, 21.1 mag, is deeper than many of the other GW follow-up surveys. \label{fig:Appmag}}
\end{figure*}

\begin{figure*}
\centering
\hspace{-2cm}
\includegraphics[width=0.8\textwidth]{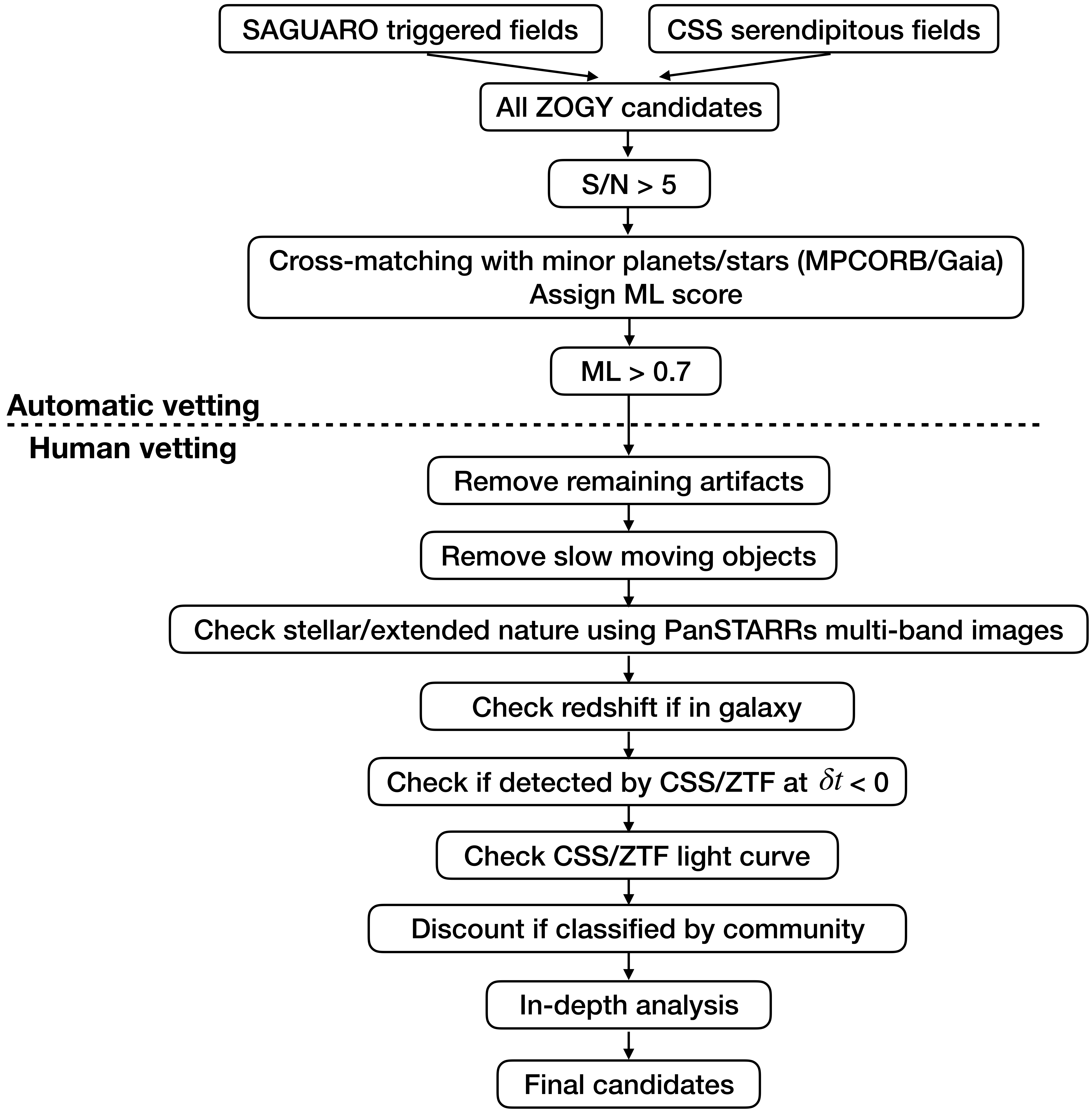}
\caption{Flow chart describing the automatic and human vetting of SAGUARO candidates during the in-depth candidate search presented here and described in Section \ref{sec:vetting}. All candidates from the image subtraction process using ZOGY \citep{ZOGY} with S/N $>$ 5 are cross-matched to various databases (such as the Minor Planet Center Orbit Database and {\it Gaia}) and assigned a machine learning (ML) score automatically after ingestion. Through a web interface, human vetting of candidates proceeds for all candidates with ML $>$ 0.7. This process includes removing any remaining artifacts or slow-moving objects, cross-checking with PanSTARRs images, checking the redshift of the candidate if in a galaxy, checking the CSS/ZTF light curve including the time of first detection, and community follow-up. A more in-depth study of the candidates is then performed. The details on the final candidates is summarized in Table \ref{tab:obs}. \label{fig:vetting}}
\end{figure*}

\section{Overview of SAGUARO} \label{sec:overview}
SAGUARO started operations in April 2019 and uses a network of telescopes\footnote{The facilities range from the 1.54 m Kuiper, 1.8 m VATT and 2.3 m Bok telescopes in Southern Arizona, to the 6.5 m MMT, the twin 6.5 m Magellan telescopes, the 2$\times$8.4 m Large Binocular Telescope, and the two Keck 10 m telescopes.}, primarily in Arizona. To facilitate the immediate follow-up of GW events, SAGUARO makes use of the Steward Observatory 1.5 m Mt. Lemmon telescope operated by the Catalina Sky Survey \citep[CSS;][]{CSS}. The telescope is equipped with a prime focus imager and a 10.5K$\times$10.5K CCD (0.77\arcsec\ per pixel), resulting in a 5 deg$^2$ FOV. It is operated with 2$\times$2 binning for an effective plate scale of 1.54\arcsec\ per pixel. CSS observes fixed fields on the sky, between declinations of $-$25~deg and +60~deg, while avoiding crowded regions in the Galactic plane (see Figure 1 of \citealt{Lundquist19}) and observes $\sim 24$ nights per month, avoiding the period around full moon. During normal operations, CSS searches for near-Earth objects (NEOs) and potentially hazardous asteroids (PHAs). To accomplish this, CSS obtains 4 $\times$ 30 s exposures per field, observing 12 fields over the span of $\sim$ 30 minutes (thus the typical survey speed is 120 deg$^2$ per hour; $\gtrsim$ 1000 deg$^2$ per night). This results in four images per field separated by $\sim$ 8 minutes, allowing for the identification of moving objects. SAGUARO observations that are triggered for GW counterpart searches (``triggered fields'') follow the same observing strategy. These fields are typically triggered in sets of 60, 120, or 180 deg$^2$ on the highest probability region of the GW localization that is observable. All images are taken without a filter and are calibrated to {\it Gaia} $G$-band using {\it Gaia} DR2 \citep{Gaiadr2}. For each 4$\times$30~s set of median-combined images, we show the 5$\sigma$ transient detection depth, defined as the magnitude of a source that can be detected through image subtraction with a significance (S/N) = 5, in Figure \ref{fig:Appmag}. The median 5$\sigma$ transient detection depth is $G\approx$ 21.1 mag. For astrometry, images are tied to {\it Gaia} DR2. The details of the follow-up capabilities, as well as an example of such follow-up, can be found in \citealt{Lundquist19}.

\subsection{Image subtraction and candidate vetting} \label{sec:vetting}
Deep reference images were constructed based on nearly three years of CSS data, with an average limiting magnitude of $G\approx 23.0$~mag. Data are run through a custom pipeline in real-time that performs image subtraction on each 4$\times$30 s median-combined image using ZOGY \citep{ZOGY}. Candidates with a S/N $>$ 5 are ingested into our database. PSF photometry is performed directly on the subtracted image, following the equations in \cite{ZOGY}, and calibrated to the CSS zero point calculated from {\it Gaia} DR2. A graphical summary of the entire candidate vetting process is shown in Figure \ref{fig:vetting}. ``Automatic vetting'' and simple ``human vetting'' were performed for triggered SAGUARO fields in real-time, with the results reported via GCN to the community. In this paper, we present a more detailed vetting of candidates in the context of the updates discussed in Section \ref{sec:updates}.

We make use of various catalogs and databases, including the Minor Planet Center Orbit Database\footnote{\url{http://www.minorplanetcenter.net/iau/MPCORB.html}} (MPCORB) to identify known moving objects, and the Transient Name Server\footnote{https://wis-tns.weizmann.ac.il/} (TNS) and the Zwicky Transient Facility (ZTF, \citealt{ZTF}) to identify known transients. We also cross-match candidates with the Galaxy List for the Advanced Detector Era (GLADE; \citealt{glade}) for possible associations to nearby cataloged galaxies, and the {\it Gaia} DR2 catalog \citep{Gaiadr2} to reduce contamination by stellar sources (see Section \ref{sec:updates}). Candidates are then assigned a machine learning (ML) score, based on the likelihood of being a real astrophysical source. A cut of ML score $>$ 0.7 is made before continuing (see Section \ref{sec:updates}). This process falls under our ``automatic vetting'' and is done in real-time without human intervention to prepare candidates for ``human vetting''.

We then move onto our human vetting through our web interface. We manually remove any remaining artifacts caused by satellite trails, bright stars or optical reflections. Slow moving objects which are not removed during the median stacking of the four images or found in the MPCORB were then rejected using our moving object calculator (see Section \ref{sec:mo}). Next, candidates coincident with point-like sources that we infer to be stellar with PanSTARRS1 multi-band images \citep{chambers16} that were not removed by the \emph{Gaia} cross-matching, were rejected as potential kilonova candidates. If candidates are clearly spatially coincident with a galaxy, the redshift of the host galaxy (if known) is compared to the distance inferred for the GW event reported by LIGO/Virgo. Candidates with detections prior to the GW event in either CSS or ZTF were ruled out as being related to the GW event. Candidates with light curves inconsistent with the rise or fall times of kilonovae (see Section \ref{sec:kilonova}) built from a combination of CSS and ZTF data were rejected. Additionally, candidates already ruled out by the community through spectroscopic or photometric means (reported either through GCNs or papers) were discounted. The remaining candidates are then analyzed in more detail to assess their association with the GW event and provide value-added information. First, individual images are checked to assess shifts in the source's shape and position, as well as the presence of the candidate in each image, to confirm the candidate as a real source. Next, a literature search for spatially coincident galaxies using a radius, $r < 4''$ (equating to spatial scales of $\sim$ 2--13 kpc for distances of 100--500 Mpc) is performed. If no galaxy is found within the literature, we search the Legacy Survey DR8 \citep{Dey2019} data for a detection or limits (based on nearby detections) on a faint coincident host. For candidates embedded within a galaxy, the position of the candidate and offset, $\delta r$, is reported relative to the position of the galaxy. For galaxies without measured redshifts, we check PanSTARRS DR2 \citep{PSDR2} for photometric redshifts ($z_{photo}$). The final candidates for all events presented here are summarized in Table \ref{tab:obs}, with the details discussed in-depth in the relevant individual event sections (Sections \ref{sec:S190901ap}, \ref{sec:S190930s}, \ref{sec:S190930t}, \ref{sec:S200105ae}, \ref{sec:S200224ca}).

In addition to vetting our own candidates, we make use of normal CSS operations to rule out candidates reported by the community that were not followed-up and are covered by CSS at a later date. For candidates within the CSS footprint, the positions of these candidates are put through the same automatic and human vetting procedures as described above to assess their viability as genuine kilonova.

\subsection{Coverage in O3} \label{sec:obs}
Starting alongside LIGO/Virgo's O3, SAGUARO had overlapping operations for 9.5 months with a duty cycle of 53.6\% for the 24~hr window following each event and 78.6\% for the 120~hr window after each event\footnote{The calculation of the duty cycle takes all observing constraints into account including the annual Arizona `monsoon' shutdown running mid-July to the end of August, the 1 month LIGO/Virgo shutdown between O3a and O3b, CSS moon constraints, and weather.}. Including all observing constraints, CSS was available to trigger on 27 non-retracted events with $\delta t < 24$~hrs (where $\delta t$ is defined as the time after the event) and on 38 events within $\delta t < 120$~hrs. Our trigger criteria focused on events which contained potential NS components (where the sum of the classifications containing a possible NS, $\sum_{\mathrm{(BNS,NSBH,MassGap})} >20\%$ at the time of trigger) for which CSS could cover part of the 90\% probability region when considering airmass and moon constraints described in \cite{Lundquist19}; but also included unusual events such as S200114f, the ``burst'' event. Two additional components of importance ``HasNS'' and ``HasRemnant'' accompany each event. These describe the probability that the event has a NS and the probability that there is matter outside the final compact object (calculated based on the masses and spins inferred), respectively. With these classifications, we would expect events with large HasRemnant values to produce EM emission. Taking our trigger criteria, the GW sky localization, and data quality under consideration, we present 7 triggered events during O3, including an initial trigger to test our system.

\section{SAGUARO System Updates} \label{sec:updates}
In the following section, we describe significant updates to the SAGUARO system for the latter part of O3 which build upon the infrastructure outlined in \citealt{Lundquist19}. First, we discuss the improvement of the machine learning algorithm, allowing for more accurate classifications between real and bogus candidates. Second, we include cross-matching with \emph{Gaia} proper motions measurements, to remove candidates clearly associated with stellar sources. Third, we introduce a more stream-lined approach to identify new, uncatalogued moving objects, with particular attention to slower moving objects that persist in the median images as candidates. Fourth, we implement infrastructure to identify and search through CSS fields which serendipitously overlap with observable GW events within $\delta t<5$~days, designated as ``serendipitous fields''. And lastly, we discuss the use of kilonova models to place constraints on the photometric evolution of candidates.

\subsection{Machine Learning}
Candidates are given a ML score between 0 (bogus) and 1 (real) based on a machine learning model to reduce the number of bogus candidates. Following the example of \citet{wright15}, this model was created using a random forest classifier that was trained on 10$\times$10 pixel substamps of a sample that included 3678 detections of known transients and 56062 bogus detections, including artifacts from bright stars and over-subtractions. To reduce overfitting in our machine learning model and improve generalization, we add Gaussian noise into our training sample data at the level of 1\% of the peak brightness. To improve rotation invariance, we add versions of each image in our training sample that have been rotated by 90, 180, and 270 degrees, effectively increasing our training set by four-fold.

We evaluated the ML scores on a test sample that included 1285 detections of known transients. By selecting a lower limit on the ML score of 0.70, such that we only assess candidates with ML $>$ 0.70, we recovered 94.3\% of these known transients with an average score of 0.87. Given this recovery rate of known transients, we use this score to limit the number of candidates that need human vetting.

\subsection{Gaia Matching}
To reduce the number of known variable stellar sources that produce residuals in subtracted images, all candidates are cross-matched against high proper motion objects in the {\it Gaia} DR2 catalog \citep{Gaiadr2} upon ingestion. If a candidate matches as a high proper motion source ($\mu > 3\sigma$) in the \emph{Gaia} catalog, it is classified in the candidate database as a stellar object and thus removed from our candidate vetting process. This reduces the number of candidates that need to be human vetted by $\sim$7.2\%.

\subsection{Moving Objects} \label{sec:mo}
Candidates are cross-matched against known moving objects from the MPCORB to remove these before human vetting. However, unknown moving objects, especially extremely slow-moving objects that move on the order of a pixel or less between each image and thus are not removed in the median image, can still persist as potential candidates. To help eliminate these slow-moving objects whose small shifts can be difficult to determine by eye, we implemented a moving object calculator. This calculator uses the position of the source in each of the four individual images, and determines the shift between the first and each of the subsequent images along with the standard deviation of these shifts. With this method, we are able to identify moving objects that move as little as $\sim$ 1 pixel between each of the individual images i.e. $\sim$1.6$''$/8 min.

\begin{figure*}
\centering
\hspace*{-1.8cm}
\includegraphics[width=1.18\textwidth]{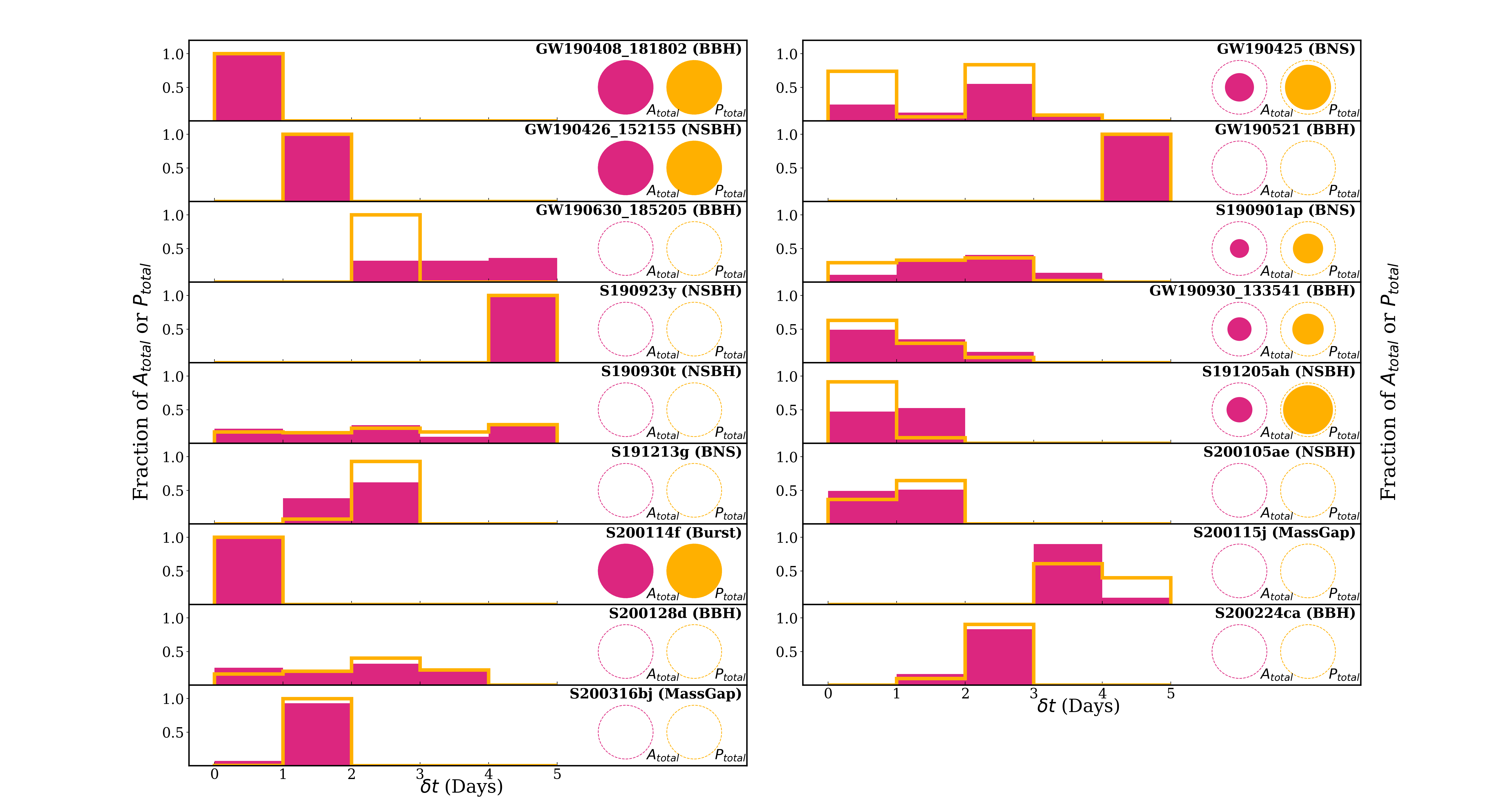}
\caption{Temporal and areal coverage of all GW events with significant follow-up from SAGUARO during O3. The histogram shows the fraction of the total area ($A_{\mathrm{total}}$, solid pink) and probability ($P_{\mathrm{total}}$, gold line) covered as a function of $\delta t$. The circles on the right show the area/probability covered by triggered fields (filled circle) normalized to the total covered area by including the serendipitous fields (open circle). As such, events with only a filled circle represent those without serendipitous coverage, while those with only the open circles show purely serendipitous events. \label{fig:comparison}}
\end{figure*}

\subsection{Serendipitous Fields}
\begin{figure*}
\centering
\includegraphics[width=0.8\textwidth]{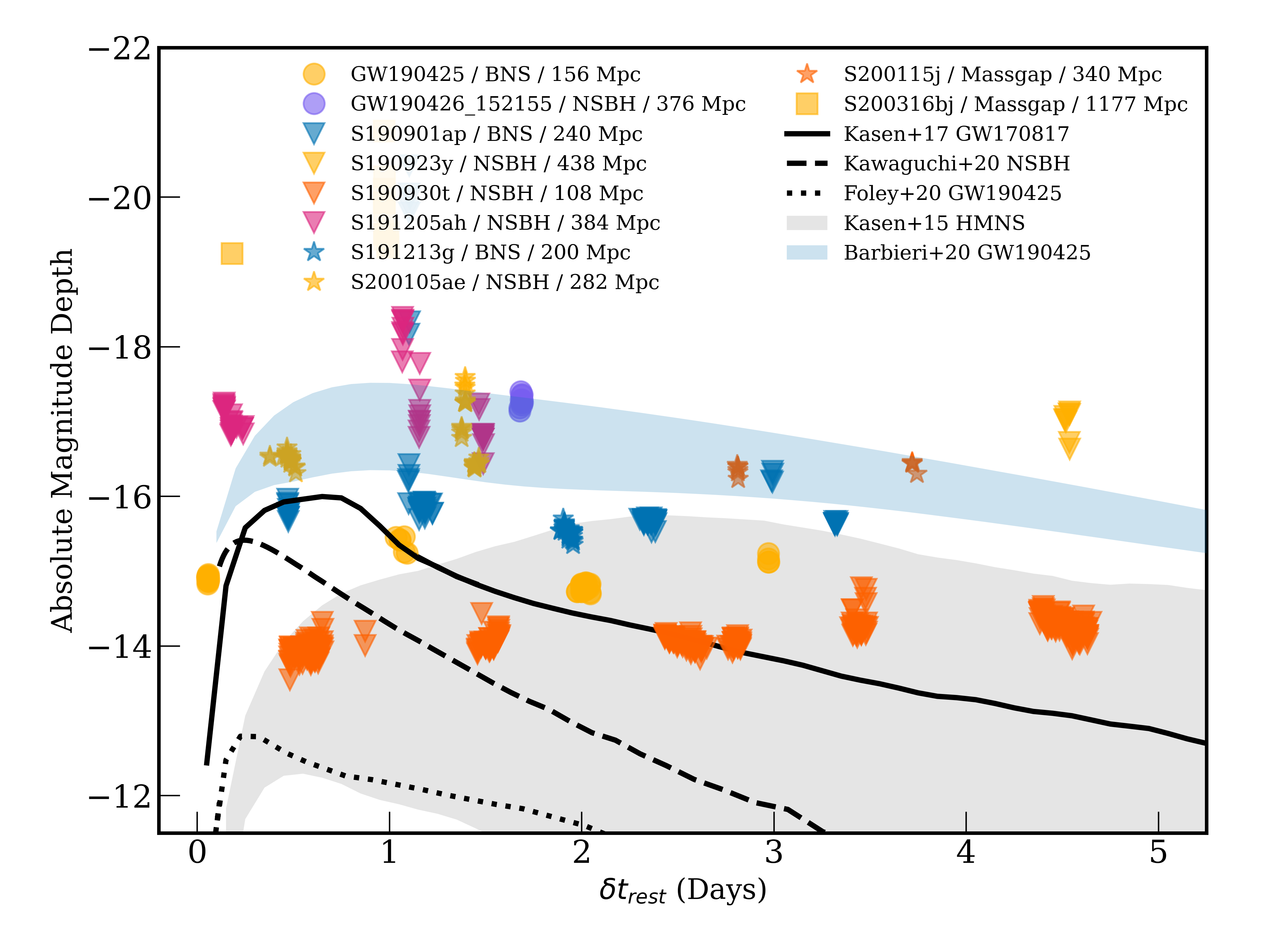}
\vspace{-0.2in}
\caption{The 5$\sigma$ transient depth in absolute magnitude space of individual SAGUARO fields with five kilonova models in the $r$-band. The solid black line represents a model for a GW170817/AT2017gfo-like kilonova \citep{kasen17}. The dashed black line shows a modeled kilonova from a NSBH merger viewed at $41<\theta<46$ degrees off-axis \citep{kawaguchi20}. The dotted line is a model for the unusually massive event GW190425 in the case of a BNS merger \citep{foley20}. The grey band models a slow-rising kilonova in the case of a NS remnant \citep{kasen15} for a prompt collapse (lower boundary), lifetime of 100 ms and an infinite lifetime (upper boundary). Finally, the blue band is a more optimistic model for the unusually massive event GW190425 in the case of a NSBH merger (upper boundary) and a BNS merger (lower boundary) \citep{barbieri20}. SAGUARO observations for BNS, NSBH and MassGap events are scaled to the median inferred distance reported by LIGO/Virgo. \label{fig:Absmag}}
\end{figure*}

While GW170817 peaked in the $r$-band at $\delta t\approx 0.65$~days and faded at a rate of $\sim$~1 magnitude per day  \citep[e.g.][]{villar17,kasen17,Nicholl17,Cowperthwaite17,Valenti2017}, kilonova models describing various progenitor and remnant scenarios predict a diversity in peak times and fading behaviors (e.g., \citealt{metzger14,kasen15,barbieri20}; see Section \ref{sec:kilonova} and Figure~\ref{fig:Absmag}). As CSS maps $\sim$ 1000 deg$^2$ of the sky each night in search of NEOs and PHAs, this provides the opportunity to analyze additional data outside our triggered area and time frame in search of optical counterparts. Thus motivated, we ingest incoming CSS survey data to search for fields which serendipitously fall within the 90\% contour of the localization of GW events within $\delta t <$ 5 days (i.e. $<$ 120 hr). We perform image subtraction between these fields and our deep reference images using ZOGY, and perform the same cross-matching and ingestion as described above. When only considering the triggered fields, SAGUARO covers between 15--180 deg$^2$ for each triggered event. The inclusion of these serendipitous fields results in an increased coverage of up to 2115 deg$^2$ for purely serendipitous events. For events with triggered fields, we see up to $\sim$ 9 times in additional area coverage (see Figure \ref{fig:comparison}). In total, we have serendipitous coverage up to $\sim$43\% of the LVC total probability ($P_{\mathrm{total}}$) for individual events. Thus, we present an additional 10 events with serendipitous fields, defined as those with coverage in the range of $\delta t < 120$~hr.

\subsection{Kilonova models} \label{sec:kilonova}
To place constraints on photometric evolution, for events classified as BNS, NSBH, or MassGap, we compare each candidate to a number of kilonova models which represent a large range of progenitor and remnant scenarios. To describe the data of GW170817/AT2017gfo, we choose the two-component model described in \cite{kasen17} which we smooth with a Savitsky-Golay filter. For NSBH events, we compare to a model by \cite{kawaguchi20} parameterized by an ejecta mass $M_{\rm ej} = 0.06 M_{\odot}$\footnote{This model divides the ejecta mass into dynamical ejecta ($M_{\rm dyn} = 0.02 M_{\odot}$) and post-merger ejecta mass ($M_{\rm pm} = 0.04 M_{\odot}$).}. We correct for the $\sim$0.2 mag artificial excess brightness (determined through private communication with the authors) due to the use of a restricted line list in calculating the ejecta opacity and smooth using a Savitsky-Golay filter. We also compare to 3 BNS models with resulting hyper-massive NS (HMNS) remnant lifetimes of 0~ms (prompt formation of a BH), 100~ms, and infinite (corresponding to a indefinitely stable remnant; \citealt{kasen15}). The overall effect of a long-lived NS remnant, as opposed to prompt collapse to a BH, is that the remnant radiates neutrinos into the post-merger disk, resulting in a bright and blue slow-rising kilonova \citep{Metzger17, perego14, metzger14}. Finally, to describe a massive merger event like GW190425 (with a chirp mass of $1.44^{+0.02}_{-0.02} M_{\odot}$; \citealt{GW190425}), we choose 2 models. The first model's ejecta masses are $\gtrsim 0.1 M_{\odot}$ with the DD2 equation of state, representing an optimistic transient brighter than AT2017gfo \citep{barbieri20}. The second assumes ejecta mass $M_{\rm ej} = 0.04 M_{\odot}$, velocity $v_{\rm ej} = 0.15c$, and lanthanide fraction $X_{\rm lan} = 10^{-2}$; and predicts a kilonova less luminous than AT2017gfo across all bands \citep{foley20}. We show these models in rest-frame time units ($\delta t_{rest}$) in Figure \ref{fig:Absmag}, along with the absolute magnitude of our 5$\sigma$ transient depth for individual fields. Even for the slowest rising model, given by an indefinitely stable remnant, the peak of emission occurs at $\delta t \sim$ 2.5 days. We can thus use the rise times to rule out candidates which continue to rise for $\delta t \gg 3$ days. Given the time evolution of all models, we can also use the fading time to rule out candidates which have detections on longer timescales, i.e. for candidates detected months after the GW event. In general, SAGUARO is able to search to depths that would allow the detection of kilonova emission out to $\sim$ 150--400 Mpc, assuming detection at peak brightness and depending on the model assumed (see Figure \ref{fig:Absmag}).

\begin{deluxetable*}{lccccccccccc}
\centering
\tabletypesize{\scriptsize}
\tablecolumns{15}
\tablewidth{0pc}
\tablecaption{Summary of SAGUARO Follow-up in O3
\label{tab:follow-up}}
\tablehead{
\colhead {Event}	 &
\colhead {Type$^\dagger$}	 &
\colhead {FAR}	 &
\colhead {Distance}	 &
\colhead {$\delta t_{\rm first}^\star$}	 &
\colhead {$\delta t$}	 &
\colhead {Fields}	 &
\colhead {5$\sigma$ Limit$^{\ddagger}$} &
\colhead {$A_{50\%}$} &		
\colhead {$A_{90\%}$} &
\colhead {$A_{\mathrm{total}}$} &
\colhead {$P_{\mathrm{total}}$} \\
\colhead {}		 &
\colhead {}		 &
\colhead {(Hz)}			 &
\colhead {(Mpc)}			 &
\colhead {(hr)}			 &
\colhead {(hr)} &
\colhead {observed}		 &
\colhead {(AB Mag)} &
\colhead {(deg$^2$)}	&
\colhead {(deg$^2$)}	&
\colhead {(deg$^2$)}	&
\colhead {(\%)}	\\
}
\startdata
GW190408\_181802 & BBH & 2.8e-18 & 1472.90 $\pm$ 357.88 & 17.6 & $<24$ & 3 & 19.2 & 5.7 & 12.9 & 15.0 & 13.73 \\
 \hline
 \hline
GW190425 & BNS & 4.5e-13 & 156.14 $\pm$ 41.37 & 1.4 & $<24$ & 12 & 21.1 & 60.0 & 60.0 & 60.0 & 3.68 \\
 & & & & & $24-48$ & 12 & 20.5 & 0.2 & 48.5 & 60.0 & 0.23\\
 & & & & & $48-72$ & 31 & 21.2 & 72.0 & 149.6 & 155.0 & 3.57\\
 & & & & & $72-96$ & 6 & 20.8 & 12.2 & 26.9 & 30.0 & 0.46\\
 \hline
 & & & & & & & 21.1 & 124.4 & 265.0 & 285.0 & 6.55 \\ 
 \hline
 \hline
GW190426\_152155 & NSBH & 1.9e-8 & 376.72 $\pm$ 100.44 & 43.6 & $24-48$ & 12 & 20.6 & 17.3 & 54.4 & 60.0 & 4.48 \\
 \hline
 \hline
GW190521 & BBH & 3.8e-9 & 3931.42 $\pm$ 953.03 & 98.3 & $96-120$ & 7 & 21.3 & 0.3 & 30.5 & 35.0 & 2.27 \\
 \hline
 \hline
GW190630\_185205 & BBH & 1.44e-13 & 925.67 $\pm$ 258.51 & 62.7 & $48-72$ & 8 & 21.0 & 25.3 & 38.2 & 40.0 & 9.07 \\
  & & & & & $72-96$ & 8 & 21.1 & 21.0 & 36.3 & 40.0 & 7.14\\
  & & & & & $96-120$ & 10 & 20.9 & 26.5 & 46.1 & 50.0 & 9.09\\
 \hline
 & & & & & & & 21.0 & 72.7 & 120.7 & 130.0 & 25.29 \\ 
 \hline
 \hline
S190901ap & BNS & 7.0e-9 & 240.87 $\pm$ 78.65 & 11.9 & $<24$ & 12 & 21.1 & 60.0 & 60.0 & 60.0 & 1.66\\
 & & & & & $24-48$ & 38 & 21.0 & 101.4& 190.0 & 190.0 & 1.87\\
 & & & & & $48-72$ & 44 & 21.2 & 119.6& 219.6 & 220.0 & 2.06\\
 & & & & & $72-96$ & 15 & 21.3 & 0.0& 69.4 & 75.0 & 0.19\\
 \hline
 & & & & & & & 21.2 & 281.0 & 539.0 & 545.0 & 5.79 \\
 \hline
 \hline
S190923y & NSBH & 4.78e-8 & 438.09 $\pm$ 132.95& 118.6 & $96-120$ & 9 & 21.2 & 28.3 & 43.1 & 45.0 & 4.05  \\
 \hline
 \hline
GW190930\_133541 & BBH & 3.0e-9 & 708.90 $\pm$ 190.66 & 12.79 & $<24$ & 32 & 21.3 & 72.1 & 153.0 & 160.0 & 8.33\\
 & & & & & $24-48$ & 24 & 21.2 & 0.0& 106.8 & 120.0 & 3.79\\
 & & & & & $48-72$ & 11 & 21.1 & 0.0& 43.2 & 55.0 & 1.00 \\
 \hline
 & & & & & & & 21.2 & 72.1 & 303.0 & 335.0 & 13.13 \\
 \hline
 \hline
S190930t & NSBH & 1.5e-8 & 108.49 $\pm$ 37.67 & 11.80 & $<24$ & 91 & 21.2 & 74.3 & 451.5 & 455.0 & 1.51 \\
 & & & & & $24-48$ & 61 & 21.2 & 219.2& 305.0 & 305.0 & 1.58 \\
 & & & & & $48-72$ & 115 & 21.1 & 207.3& 567.9 & 575.0 & 2.04 \\
 & & & & & $72-96$ & 41 & 21.0 & 205.0& 205.0 & 205.0 & 1.55 \\
 & & & & & $96-120$ & 117 & 20.9 & 263.7& 580.4 & 585.0 & 2.47 \\
 \hline
 & & & & & & & 21.1 & 964.4 & 2104.8 & 2120.0 & 9.13 \\
 \hline
 \hline
S191205ah & NSBH & 1.2e-8 & 384.97 $\pm$ 163.72 & 3.60 & $<24$ & 27 & 21.0 & 68.8 & 130.1 & 135.0 & 10.00 \\
 & & & & & $24-48$ & 30 & 20.9 & 4.6& 147.3 & 150.0 & 0.91 \\
 \hline
 & & & & & & & 20.9 & 73.4 & 277.4 & 285.0 & 10.91 \\
 \hline
 \hline
S191213g & BNS & 3.55e-8 & 200.86 $\pm$ 80.96 & 47.4 & $24-48$ & 8 & 20.9 & 2.5 & 39.6 & 40.0 & 0.65 \\
  & & & & & $48-72$ & 18 & 21.0 & 13.8 & 85.0 & 90.0 & 1.40 \\
  & & & & & $72-96$ & 2 & 21.0 & 0.0 & 9.2 & 10.0 & 0.05 \\
 \hline
 & & & & & & & 21.0 & 16.2 & 133.8 & 140.0 & 2.10 \\
 \hline
 \hline
S200105ae & NSBH & 7.67e-7 & 282.85 $\pm$ 73.78 & 20.7 & $<24$ & 23 & 19.64 & 6.5 & 104.3 & 115.0 & 0.90\\
& &  &  &  & $24-48$ & 30 & 20.4 & 55.0 & 139.1 & 150.0 & 1.97\\
 \hline
 & & & & & & & 20.7 & 61.5 & 240.9 & 260.0 & 2.84 \\
 \hline
 \hline
S200114f & IMBH & 1.2e-9 & n/a & 2.36 & $<24$ & 36 & 21.0 & 37.7 & 176.1 & 180.0 & 86.30 \\
 \hline
 \hline
S200115j & MassGap & 2.09e-11 & 332.22 $\pm$ 78.46 & 72.5 & $72-96$ & 9 & 21.3 & 0.0 & 37.5 & 45.0 &  2.07\\
& &  &  &  & $96-120$ & 2 & 21.4 & 0.0 & 9.9 & 10.0 & 0.57 \\
 \hline
 & & & & & & & 21.3 & 0.0 & 47.4 & 55.0 & 2.64 \\
 \hline
 \hline
S200128d & BBH & 1.65e-8 & 3701.59 $\pm$ 1264.51 & 0.16 & $<24$ & 16 & 21.1 & 4.1 & 72.3 & 80.0 & 1.93 \\
& &  &  &  & $24-48$ & 15 & 21.0 & 13.2 & 71.5 & 75.0 & 2.48 \\
& &  &  &  & $48-72$ & 20 & 20.9 & 42.0 & 92.6 & 100.0 & 4.66 \\
& &  &  &  & $72-96$ & 13 & 21.3 & 25.9 & 62.8 & 65.0 & 2.73  \\
 \hline
 & & & & & & & 21.0 & 85.2 & 299.1 & 320.0 & 11.80 \\
 \hline
 \hline
S200224ca & BBH & 1.61e-11 & 1575.00 $\pm$ 322.36 & 32.3 & $24-48$ & 1 & 20.4 & 0.0 & 3.9 & 5.0 &  1.18  \\
& &  &  &  & $48-72$ & 4 & 21.1 & 0.0 & 14.9 & 20.0 & 10.24  \\
 \hline
 & & & & & & & 21.1 & 0.0 & 18.8 & 25.0 & 11.42 \\
 \hline
 \hline
S200316bj & MassGap & 7.10e-11 & 1177.98 $\pm$ 283.01 & 5.4 & $<24$ & 2 & 21.1 & 0.0 & 5.1 & 10.0 & 0.24\\
& &  &  &  & $24-48$ & 18 & 20.9 & 22.1 & 88.9 & 90.0 & 73.00\\
 \hline
 & & & & & & & 20.9 & 22.1 & 93.9 & 100.0 & 73.24 \\
 \hline
 \hline
\enddata
\tablecomments{Magnitudes reported here are uncorrected for Galactic extinction and are reported in {\it Gaia} $G$-band.\\
The probability covered refers to the percent of the probability of the 50\%, 90\%, and total localizations that were covered by these observations.\\
$^\ddagger$ Median $5\sigma$ transient detection depth. \\
$^\star$ $\delta t$ of first field after the GW event. \\
$^\dagger$ Most likely classification based on GW probabilities \citep{Kapadia19}. \\
$^a$ Triggered events.
}
\end{deluxetable*}

\section{Individual events and Results} \label{sec:ind}
Here we (i) present a more detailed analysis of the search for candidate counterparts in triggered fields, given the improvements to our automatic vetting portion of the pipeline (Section \ref{sec:updates}) using updated localizations and (ii) extend this analysis to the serendipitous fields for all events that have SAGUARO coverage with $P_{\mathrm{total}}>2\%$.

For triggered fields, we vetted candidates in real-time to search for optical counterparts. Information on these fields were provided to the community through GCNs \citep{gcn25915,gcn26360,gcn24079,gcn24172,gcn26473,gcn26750,gcn26753} and uploaded to the Gravitational Wave Treasure Map\footnote{\url{http://treasuremap.space}} \citep{Wyatt20}. During O3, we reported observations of 6 new candidates found by SAGUARO in the triggered fields (SN2019wdb, \citealt{gcn26360}; SAGUARO20a, SAGUARO20b, SAGUARO20c, SAGUARO20d, and SAGUARO20e, \citealt{gcn26753}). We were also able to rule out 2 community-identified candidates: the first, AT2020vx, due to a previous CSS detection \citep{gcn26750}, and the second, SN2019ebq, due to spectroscopic classification \citep{Lundquist19}.

We exclude 4 GW events from our analysis that had serendipitous coverage $P_{\mathrm{total}}<2\%$. Three of these are candidate BBH mergers (for which bright EM counterparts are less likely), while the fourth is a candidate NSBH merger, S190910d \citep{gcn25723}, which has a $d$ = 632 $\pm$ 186 Mpc (such that our imaging does not reach sufficient depths to make any meaningful statements on NSBH kilonova emission). We also exclude an additional 2 events, 1 triggered (S191216ap; \citealt{gcn26454}) and 1 serendipitous (S191215w; \citealt{gcn26441}) from our analysis due to poor data quality from bad weather. 

Thus, in total, we present observations for 17~GW events, 7 triggered and 10 with only serendipitous coverage. In this section, we detail the GW discovery and relevant follow-up observations of these 17 events. For each, we summarize all relevant candidates reported by the community (focusing on actual candidates detected by the community versus survey limits), investigate whether they are still viable kilonova candidates with available information (checking GCNs, publicly available data uploaded to TNS, public ZTF data, peer-reviewed papers, and our own data), and present our own candidate search. We note SAGUARO follow-up of three of these events were also detailed in \citet{Lundquist19}; however, we include their details again here based on our updated system, results from additional serendipitous fields, as well as the follow-up of community candidates.

An overview of our observations, including the area covered of the 50\%, 90\%, and total localization ($A_{50\%}, A_{90\%}, A_{\mathrm{total}}$), as well as the percentage of the total probability ($P_{\mathrm{total}}$) covered for each event, is listed in Table~\ref{tab:follow-up}. The GW localization maps, along with our coverage, are displayed in Appendix \ref{appendix:A}. A summary of our final candidates is presented in Table \ref{tab:obs}. Hereafter, we use the common designations L1 (LIGO Livingston), H1 (LIGO Hanford), and V1 (Virgo). 

\subsection{GW190408\_181802} \label{sec:S190408an}
A candidate GW signal was identified on 2019-04-08 at 18:18:02.288 UTC using data from L1, H1, and V1 \citep{gcn24069}. The final classification for this event is $>$99\% BBH at a distance, $d$ = 1473 $\pm$ 358 Mpc. This event, as the first of the O3 alerts, served as a test for our system, and we triggered 3 fields. The initial follow-up for these triggered fields is discussed in \cite{Lundquist19}. No serendipitous fields were covered by CSS within our search interval of $\delta t < 120$~hr. The localization of the event, along with the CSS fields that fall within the 90\% probability region, is shown in Figure~\ref{fig:190408an_loc}.

A total of 11 candidates were initially reported by the community via GCN (MASTER-Network, PanSTARRS, \emph{Gaia}; \citealt{gcn24076,gcn24084,gcn24096,gcn24124,gcn24134}). Initial follow-up, either through photometry or spectroscopy, ruled out 4 of these as viable candidates \citep{gcn24090,gcn24092,gcn24097,gcn24106,gcn24154}. SN2019daj was found to have detections prior to the GW event \citep{gcn24078} and later classified as a Type Ia SN \citep{2019daj}. Later follow-up classified another candidate, SN2019dma, as a Type II SN \citep{AT2019dma}. Another candidate, AT2019dpg, was detected by routine ZTF survey operations out to $\sim$ 35 days after the GW event \citep{Nordin2019}. Having faded by $<1.5$ mag over a month, we consider it to be unrelated to the GW event based on photometric evolution. Searching public ZTF data, we find another candidate, AT2019deu, shows variability for over a year after the GW event \citep{ztfarchive}. We thus consider it to be unrelated to the GW event. Of the remaining four candidates, CSS observed the positions of AT2019doo, AT2019dmc, and AT2019dev. We found no detections down to a limit of 20.8, 21.0, and 21.2 mag at $\delta t$ = 74.7, 65.7, and 65.8 days, respectively. With these limits we are not able to rule out these as potential candidates. No other viable candidates were found within our data.

\subsection{GW190425} \label{sec:GW190425}
A candidate GW signal with classification $>$99\% BNS was identified on 2019-04-25 at 08:18:05.017 UTC using data from L1 and V1 \citep{gcn24168}. This event was later confirmed to be a real BNS merger with a total mass of $3.4_{-0.1}^{+0.3}~\mathrm{M_{\odot}}$ that occurred at $d = 159_{-71}^{+69}\ \mathrm{Mpc}$ \citep{GW190425}. We triggered 12 fields in the northern part of the localization with $\delta t < 24$~hr. The initial follow-up of these triggered fields are discussed in more detail in \cite{Lundquist19}. In total, we obtained additional serendipitous coverage of 49 fields (245 deg$^2$) at $\delta t < 96$~hr.  Including the triggered fields, the total probability coverage for this event was $P_{\mathrm{total}}$ = 6.55\% (Table~\ref{tab:follow-up}). The localization of the event, along with the CSS fields that fall within the 90\% region, is shown in Figure~\ref{fig:190425z_loc}.

A total of 54 candidate optical counterparts were reported by the community from ZTF, ATLAS, Pan-STARRS, \emph{Swift/UVOT}, and \emph{Gaia} (\citealt{gcn24191,gcn24197,gcn24210,gcn24262,gcn24296,gcn24311,gcn24345,gcn24354,gcn24362,gcn24366}). \citet{Hosseinzadeh19} compiled a list of all public follow-up searches that were reported for this event at the time, including these 54 candidates and an additional 15 pre-event transients reported by \citet{gcn24191}. Of these 54 candidates, 6 were excluded after spectroscopic classification \citep{gcn24200,gcn24204,gcn24205,gcn24215,gcn24221,gcn24321}, 8 had redshifts inconsistent with the inferred LIGO/Virgo distance \citep{gcn24269,gcn24311}, while an additional 9 were detected prior to the GW event \citep{gcn24302,gcn24223,gcn24356}. Five of these 54 candidates had marginal detections \citep{gcn24197}, with follow-up imaging failing to confirm these as real candidates (\citealt{gcn24202,gcn24224}). Later follow-up by \cite{coughlin19} ruled out an additional 4 candidates due to photometric evolution that was inconsistent with a kilonova. Routine survey operations by PanSTARRs resulted in photometric coverage of an additional candidate, AT2019eby, $\sim$ 1 month later after the GW event \citep{chambers16}. Having faded by $<0.4$ mag over a month, we consider it to be unrelated to the GW event based on photometric evolution. Of the remaining 21 candidates, the positions of 20 were covered by CSS survey operations over the subsequent months. Only one candidate, AT2019ech, was identified, just below our automatic vetting S/N threshold (with S/N $\approx$ 4.2) and 5$\sigma$ transient depth of 21.0 mag, $\sim$ 17 days later. Having faded by $<0.4$ mag, we are able to rule it out as a viable candidate. None of the other candidates were detected in our data. We find no new candidates within our serendipitous data.

\subsection{GW190426\_152155}
GW190426\_152155 was identified as a candidate GW event on 2019-04-26 15:21:55.337 UTC in data from H1, L1, and V1 \citep{gcn24237}. This event is classified as a MassGap, and is either a NSBH or BBH with masses of 5.7$^{+4.0}_{-2.3}$ and 1.5$^{+0.8}_{-0.5}~\mathrm{M_{\odot}}$ at $d = 380^{+190}_{-160}$ Mpc (\citealt{gcn24411,gwtc2}). We triggered 12 fields for this event in the interval $\delta t = 24-48$~hr (Table~\ref{tab:follow-up}). The initial follow up of these triggered fields are discussed in more detail in \cite{Lundquist19}. No additional fields within the 90\% localization were covered by CSS during normal operations over the subsequent 120 hr after the detection. The localization of the event, along with the CSS fields that fall within the 90\% region, is shown in Figure~\ref{fig:190426c_loc}.

A total of 30 optical candidate counterparts were reported by the community from LCOGT, DECam/GROWTH, ZTF, GRAWITA, and \emph{Gaia} (\citealt{gcn24249,gcn24268,gcn24283,gcn24331,gcn24340,gcn24344,gcn24355}). \citet{Hosseinzadeh19} compiled a list of all public follow-up searches that were reported for this event at the time. Of these candidates, 5 were spectroscopically classified and excluded as kilonova candidates (\citealt{gcn24317,gcn24359}), 3 were detected prior to the GW event (\citealt{gcn24349,gcn24251,gcn24357}), and 1 candidate had a redshift inconsistent with the inferred LIGO/Virgo distance \citep{gcn24331}. Later follow-up by \cite{kasliwal20} ruled out 10 candidates based on spectroscopic classification, photometric evolution, or association with a stellar source or artifact. CSS coverage of one of these sources, ZTF19aaslszp/AT2019snj, reveals a rising lightcurve $\sim$1 month after the GW event, consistent with the finding of \cite{kasliwal20} that it is unrelated to the GW event. Seven DECam/GROWTH candidates discussed in \cite{Goldstein19} were initially ruled out as they were outside the updated LALInference localization; however, GWTC-2 included a new localization that had these candidates within the 90\% region. Four of these candidates, DG19ftnb, DG19kqxe, DG19nmaf, and DG19zyaf, were covered by CSS $\sim1$ week after the event. None were detected to a limit of $\sim$21.0 mag. DG19zdwb was covered by CSS $\sim40$ days after the event and was not detected down to a limit of $\sim20.9$. The other 2 candidates, DG19ouub and DG19vkgf, were only covered by CSS operations $\sim278$ days after the GW event, with non-detections to a limit of 21.6 and 22.1 mag, respectively. With these limits we are not able to rule out any of these as potential candidates. None of the remaining four candidates were covered by CSS and no new candidates were found within our data. 

\subsection{GW190521}
GW190521 was identified as a candidate GW signal using data from H1, L1, and V1 on 2019-05-21 03:02:29.447 UTC \citep{gcn24621} and later confirmed to be a BBH merger of two unusually high mass components of 91.4$^{+29.3}_{-17.5}$ and 66.8$^{+20.7}_{-20.7}~\mathrm{M_{\odot}}$ (\citealt{lvc_gw190521,gwtc2}), at $d = 5.3^{+2.4}_{-2.6}$ Gpc \citep{gcn24640}. There are no triggered fields for this event, and our total serendipitous coverage is 7 fields (35 deg$^2$) in the interval $\delta t = 96-120$~hr, with $P_{\rm total}$ = 2.27\% (Table~\ref{tab:follow-up}). The localization of the event, along with the CSS fields that fall within the 90\% region, is shown in Figure~\ref{fig:S190521g}.

\cite{Graham2020} reported the detection of a potential counterpart, ZTF19abanrhr, which they found to be consistent with a kicked BBH merger inside the accretion disk of an AGN. \cite{Ashton2020}, however, found insufficient data to confidently associate this optical counterpart with the GW event. They determined the odds (defined as the ratio of the probabilities between the scenario where the AGN is caused by the BBH and a random coincidence) for this event to be 1--12, depending on the waveform model used, compared to the $10^6$ found for GW170817 and AT2017gfo. The location of ZTF19abanrhr and host, J124942.3+344929, is within the CSS footprint; however, we have no data for the period over the duration of the flare (which started at $\delta t \sim$ 50 days and had a a duration of $\sim$ 50 days). Our closest coverage was at $\delta t \sim 47$~days, 3 days before the approximated start of the flare, which can be used to help constraint that start of the flare and pre-flare activity. At this time, we find no detection of a transient or significant change in magnitude for the galaxy relative to the deep reference to a limit of 20.7 mag. Over the course of 53~epochs prior to the event, the source maintained a constant flux within $1\sigma$ of 19.00 $\pm$ 0.19 mag. Additionally, \cite{Podlesnyi2020} reported limits from a search for high-energy $\gamma$-rays in publicly available Fermi-LAT data in the 100--300 MeV range for this candidate. No other candidates were reported by the community, and a search of our serendipitous data found no candidates within the covered fields.

\subsection{GW190630\_185205}
A candidate GW signal was identified using data from L1 and V1 on 2019-06-30 at 18:52:05.180 UTC \citep{gcn24922}. The final classification for this event, GW190630\_185205, is a BBH merger with masses of 35.0$^{+6.9}_{-5.7}$ and 23.6$^{+5.2}_{-5.1}~\mathrm{M_{\odot}}$ at $d = 930^{+560}_{-400}$ Mpc \citep{gwtc2}. There are no triggered fields for this event. In total, we covered 26 fields serendipitously with $\delta t < 120$~hr, totaling 130 deg$^2$, with $P_{\rm total}$ = 25.29\% (Table \ref{tab:follow-up}). The localization of the event, along with the CSS fields that fall within the 90\% region, is shown in Figure~\ref{fig:S190630ag}. Only a single \emph{Gaia} candidate was reported within the localization \citep{gcn24977}; however no additional follow-up was conducted. This candidate was not covered by CSS. No new candidates were found within our serendipitous data.

\subsection{S190901ap} \label{sec:S190901ap}
A candidate GW signal was identified using data from L1 and V1 on 2019-09-01 at 23:31:01.838 UTC \citep{gcn25606}. The final classification for this event, S190901ap, is 86\% BNS at $d = 241 \pm 79$ Mpc, with a 14\% probability of being terrestrial in nature. We triggered 12 fields (60~deg$^{2}$) at $\delta t < 24$~hr. Additionally, we covered 97 fields serendipitously (485~deg$^{2}$) in the interval $\delta t = 24-96$~hr. In total, we observed 109 fields with $\delta t < 96$~hr, totaling 545 deg$^2$ with $P_{\rm total}$ = 5.79\% (Table \ref{tab:follow-up}). The localization of the event, along with the CSS fields that fall within the 90\% region, is shown in Figure~\ref{fig:S190901ap}.

Initially, 19 candidates were reported by the community via GCN (ZTF, \citealt{gcn25616,gcn25634,gcn25656}; MASTER-Network, \citealt{gcn25649}; GOTO, \citealt{gcn25654}; \emph{Gaia} \citealt{gcn25689}). Of these, 17 were subsequently ruled out as possible kilonova candidates by the community due to association with a galaxy at a distance outside the inferred LIGO/Virgo distance range, spectroscopic classification, or photometric evolution \citep{gcn25639,gcn25640,gcn25619,gcn25622,gcn25632,gcn25638,gcn25661, gcn25665,gcn25675,ZTF2020}. We find a detection $\sim$ 1 month later for ZTF19abvjnsm, initially ruled out by an inconsistent photo-$z$ \citep{gcn25616}, confirming it to be unrelated to the GW event due to inconsistent photometric evolution. Likewise, we also find a detection $\sim$ 1 month later for ZTF19abwsmmd/AT2019pnc, consistent with the findings of \cite{ZTF2020}, which ruled it out as a candidate due to a slow decline rate. Of the 2 candidates not followed up by the community, we find a detection $\sim$ 1 month later for Gaia19dzi/AT2019piw, showing inconsistent photometric evolution for a kilonova and thus ruling it out as a candidate.

During our search, we found a single new candidate, SAGUARO19k/AT2019aaid \citep{tnscands} at RA = 04h06m26.2s and Dec = $-$12d01m13s (see Table \ref{tab:obs}). Performing PSF photometry on the subtracted image, we obtain $m_{can} = 20.95 \pm 0.27$ mag at $\delta t_{rest} = 0.48$ days, making it consistent with an AT2017gfo-like kilonova when transforming to absolute magnitude space using the inferred distance from LIGO/Virgo. However, the candidate only has a single detection, with no pre-event data and no additional coverage of the field by CSS until three months after the event, by which time the source is no longer detected to a limit of 20.6 mag. Searching public ZTF data, we find non-detections to the limits of $r$ = 20.42 and 20.27 mag at $\delta t \sim$ 4.5 and 7.5 days respectively, and $g$ = 20.36 mag at $\delta t \sim$ 7.5 days. AT2019aaid is isolated with no cataloged galaxy at the position of the source and no coincident galaxy detected in our images. Searching the Legacy Survey data, we find no source at the position of this candidate with limits based on nearby sources of  $g > 25.56$, $r > 24.62$, and $z > 24.50$ mag. With no constraints on photometric evolution, we are not able to concretely conclude that this candidate is related or un-related to the GW event.

\subsection{S190923y}
S190923y was identified as a candidate GW signal using data from H1 and L1 on 2019-09-23 12:55:59.646 UTC \citep{gcn25814}. The final classification for this event is 68\% NSBH merger at $d = 438 \pm 133$ Mpc, and a 32\% probability of being terrestrial. There are no triggered fields for this event. In total, we covered 9 fields serendipitously in the interval $\delta t = 96-120$~hr, totaling 45 deg$^2$ with $P_{\rm total}$ = 4.05\% (Table \ref{tab:obs}). The localization of the event, along with the CSS fields that fall within the 90\% region, is shown in Figure~\ref{fig:S190923y}. A single candidate within the localization was reported by the MASTER-Network \citep{gcn25855}, but spectroscopic follow-up ruled it out due to its Galactic origin \citep{gcn25903}. Searching our serendipitous fields, we find no new candidates.

\subsection{GW190930\_133541} \label{sec:S190930s}
GW190930\_133541 was identified as a candidate GW event on 2019-09-30 13:35:41.247 UTC using data from L1 and H1 \citep{gcn25871}. The final classification for this event is a BBH with masses of 7.8 and 12.3 M$_{\odot}$ at $d = 709 \pm 79$ Mpc \citep{gwtc2}. We triggered 12 fields (60~deg$^{2}$) at $\delta t < 24$~hr. From CSS operations, we have an additional 55~fields (275~deg$^{2}$) of serendipitous coverage with $\delta t < 72$~hr. In total, we covered 67 fields with $\delta t < 72$~hr, totaling 335 deg$^2$ with $P_{\rm total}$ = 13.13\% (Table \ref{tab:follow-up}). The localization of the event, along with the CSS fields that fall within the 90\% region, is shown in Figure~\ref{fig:S190930s}.

Only two candidates were initially reported, both by the MASTER-Network \citep{gcn25897,gcn25900}. Both were reported as likely Galactic CVs, but only one was followed up (ATLAS, \citealt{gcn25922}). Searching through our data, we find a single new candidate, SAGUARO19n/AT2019aaig \citep{tnscands} at RA = 21h53m14.0s and Dec = $-$06d43m38s (see Table \ref{tab:obs}). Performing PSF photometry on the subtracted image, we obtain $m_{can} = 20.60 \pm 0.23$ mag at $\delta t_{rest} = 2.17$ days. As a BBH event however, we do not expect EM emission. Nevertheless, comparing AT2019aaig to the kilonova models discussed here, we find that AT2019aaig has a luminosity of 8.75$ \times 10^{42}$~erg~s$^{-1}$ at the median inferred distance of the GW event, making it $\sim$ 3 mag brighter, or 20 times more luminous, than the brightest model at $\delta t_{rest}$ = 2.17 days. Searching our data, we find a non-detection to the limit of 21.2 mag $\sim$ 12 days prior to the event and a non-detection to the limit of 20.9 mag seven months after the event. Searching the literature, we find AT2019aaig is coincident with the galaxy SDSS J215314.00-064338.3 \citep{SDSSDR12}. SDSS J215314.00-064338.3 has a $z_{photo}$ = 0.208 $\pm$ 0.030, within $3\sigma$ of the inferred LIGO/Virgo distance for this event. Searching PanSTARRS DR2, we find a $z_{photo}$ = 0.168 $\pm$ 0.041, in agreement with the SDSS $z_{photo}$, and in better agreement with the inferred LIGO/Virgo distance. We find an $\delta r \sim 0.5''$ between the position of AT2019aaig and the position of the galaxy from SDSS DR12.

\subsection{S190930t} \label{sec:S190930t}
S190930t was identified as a candidate GW event on 2019-09-30 14:34:07.685 UTC, nearly one hour after GW190930\_133541 was detected \cite{gcn25876}. S190930t was only identified in data from L1 and thus has a very large localization. The final classification for the event is 74$\%$ NSBH at $d = 108 \pm 38$ Mpc. As a result of the large sky localization for this event, no specific fields were triggered. In total, our serendipitous data covered 425 fields with $\delta t <120$~hr after the event, totaling 2125 deg$^2$ with $P_{\mathrm{total}}$ = 9.13\% (Table \ref{tab:follow-up}). The localization of the event, along with the CSS fields that fall within the 90\% region, is shown in Figure~\ref{fig:S190930t}.

Eleven candidates were reported by the community (MASTER-Network; \citealt{gcn25897,gcn25900}, ZTF; \citealt{gcn25899}, ATLAS; \citealt{gcn25922}, Swift/UVOT \citealt{gcn25901,gcn25964}), including the two candidates from the GW190930\_133541 localization (see Section \ref{sec:S190930s}). Six candidates were followed up in detail and ruled out as possible candidates due to spectroscopic classification (\citealt{gcn25921,2019rpr,2019rpp}), inconsistent photometric properties \citep{gcn25922} or associated distances outside the inferred LIGO/Virgo range \citep{gcn25964}. One of the remaining candidates, AT2019rpt, was observed several times by CSS after the GW event. Having faded by $\sim$2.5 mag over $\sim$136 days, we consider it to be unrelated to the GW event based on photometric evolution. None of the remaining 6 candidates were detected by CSS.

Searching through our serendipitous data, we find three new candidates (see Table \ref{tab:obs}). The first candidate, SAGUARO19l/AT2019aaie \citep{tnscands}, is at RA = 20h40m05.2s and Dec = $-$01h39m09s. Performing PSF photometry on the subtracted image, we obtain $m_{can} = 20.90 \pm 0.26$ mag at $\delta t_{rest} = 1.49$ days, making it consistent with the \cite{kawaguchi20} NSBH model discussed here when transforming to absolute magnitude space using the inferred LIGO/Virgo distance. Searching our data, we find a non-detection to the limit of 20.82 mag one month prior to the event and a non-detection to the limit of 20.96 mag eight months after the event. Searching the literature, we find AT2019aaie is coincident with a galaxy (SrcID: 472567686234) in the VISTA Hemisphere Survey DR4.1 (VHS, \citealt{McMahon2013}). Searching the Legacy Survey, we find the coincident galaxy (Brick: 3101m017, ObjID: 7531) with $g$ = 22.49, $r$ = 21.49, and $z$ = 20.88 mag. We find an $\delta r \sim 0.3''$ between the position of AT2019aaie and the position of the galaxy from the Legacy Survey. The second candidate, SAGUARO19m/AT2019aaif \citep{tnscands}, is at RA = 20h15m45.8s and Dec = $-$07d55m44s. Performing PSF photometry on the subtracted image, we obtain $m_{can} = 18.63 \pm 0.22$ mag at $\delta t_{rest} = 2.44$ days. Comparing the absolute magnitude of AT2019aaif to the kilonova models discussed here, we find that AT2019aaif overlaps with the \citet{kasen15} HMNS model when considering the error on the median inferred LIGO/Virgo distance. Searching our data, we find a non-detection to the limit of 20.97 mag three months prior to the event and a non-detection to the limit of 21.05 mag eight months after the event. AT2019aaif is isolated with no cataloged galaxy at the position of the source. No coincident galaxy is detected in our images, and it is not covered within the Legacy Survey footprint. The third candidate, SAGUARO19o/AT2019aaih \citep{tnscands}, is at RA = 23h22m03.4s and Dec = -23d15m11s. Performing PSF photometry on the subtracted image, we obtain $m_{can} = 20.28 \pm 0.31$ mag at $\delta t_{rest} = 4.49$ days. Comparing the absolute magnitude of AT2019aaih to the kilonova models discussed here, we find that AT2019aaih overlaps with the \cite{kasen15} HMNS model when considering the error on the median inferred LIGO/Virgo distance. Searching our data, we find only one other observation a year after the event, with a non-detection to the limit of 20.55 mag. AT2019aaih lies in the outskirts (with $\delta r = 7.9''$) of the GLADE galaxy PCG~802942, at $\sim$330 Mpc ($z_{photo}$ = 0.07 $\pm$ 0.03; \citealt{2MPZ}). This $z_{photo}$ lies within $3\sigma$ of the inferred LIGO/Virgo distance. Searching PanSTARRs DR2, however, we find a $z_{photo}$ = 0.113 $\pm$ 0.008, $>8\sigma$ from the inferred LIGO/Virgo distance. Given the disagreement between the $z_{photo}$ for the host, and the general uncertainties associated with $z_{photo}$ calculations, we do not rule AT2019aaih out as a potential candidate.

\subsection{S191205ah}
The GW candidate S191205ah was discovered using data from H1, L1, and V1 at 21:52:08.569 UTC by the {\tt gstlal} pipeline \citep{GCN26350}. The final classification is $93\%$ NSBH at $d = 385 \pm 164$ Mpc. We triggered 12 fields (60~deg$^{2}$) at $\delta t < 24$~hr and have an additional 45 serendipitous fields (225~deg$^{2}$) with $\delta t < 96$~hr. In total, we covered 57 fields with $\delta t < 96$~hr, totaling 285 deg$^2$ with $P_{\mathrm{total}}$ = 10.91\% (Table \ref{tab:follow-up}). The localization of the event, along with the CSS fields that fall within the 90\% region, is shown in Figure~\ref{fig:S191205ah}.

During the real-time analysis of our data, we reported a single SAGUARO candidate, SAGUARO19j \citep{gcn26360}. It was later found to have a previous detection in PanSTARRS \citep{gcn26364} and was subsequently classified as a SN \citep{yan20tns}. It was thus ruled out as being related to this GW event. An additional 13 candidates were reported by the community via GCN (ZTF; \citealt{gcn26416}, MASTER-Network \citealt{gcn26379,gcn26578}, \emph{Gaia} \citealt{gcn26397}). Of these, 9 were ruled out as possible kilonova candidates by the community due to spectroscopic classification or photometric evolution \citep{gcn26382,gcn26405,gcn26421,gcn26422,gcn26502,ZTF2020}. One of the remaining 4 candidates, AT2019wjr, was detected by ATLAS during routine survey operations two days after the initial detection. Still on the rise $\sim$ 7 days after the GW event, we consider it to be unrelated to the GW event based on photometric evolution. The remaining 3 candidates lie outside the CSS footprint and were thus not covered by us. Searching our data, we find no new additional candidates.

\subsection{S191213g}
A candidate GW signal was identified using data from H1, L1, and V1 on 2019-12-13 at 04:34:08.142 UTC \citep{gcn26402}. The final classification for this event, S191213g, is $76.8$\% BNS at $d = 201 \pm 81$ Mpc and a 23.2\% probability of being terrestrial. We did not trigger any fields for this event, but covered 28 fields serendipitously with $\delta t < 72$~hr, totalling 170 deg$^2$ with a $P_{\mathrm{total}}$ = 2.10\% (Table \ref{tab:follow-up}). The localization of the event, along with the CSS fields which fall within the 90\% region, is shown in Figure~\ref{fig:S191213g}.

In total, 21 candidates were reported in GCNs by ZTF, MASTER-Network, and Pan-STARRS \citep{andreoni19c,Stein19gcn, Denisenko19, mcbrien19gcn}. All candidates were subsequently ruled out and determined to be either SNe \citep{PerleyCopperwheat19, GCN26429, GCN26492, GCN26504}, due to AGN activity \citep{GCN26429}, stellar in nature \citep{GCN26492,Denisenko19}, or unrelated to S191213g \citep{coughlin20, andreoni19d}. Searching through our serendipitous fields, we find no new candidates. 

\subsection{S200105ae} \label{sec:S200105ae}
A sub-threshold candidate GW signal, with a high terrestrial probability (97.3\%), was identified in data from L1 and V1 on 2020-01-05 16:24:26.057 UTC \citep{gcn26640}. The event was not retracted, however, as the data suggested the probability of being astrophysical in nature was greater than calculated by the real-time processing due to its chirp structure \citep{gcn26640,gcn26657}. The offline parameter estimation classifies the event as a NSBH at $d = 283 \pm 74$ Mpc \citep{gcn26688} with a 90\% localization of 7373 deg$^2$. There are no triggered fields for this event. CSS serendipitously covered 53 fields with $\delta t < 48$~hr, totalling 265 deg$^2$ with $P_{\mathrm{total}}$ = 2.84\% (Table \ref{tab:follow-up}). The localization of the event, along with the CSS fields that fall within the 90\% region, is shown in Figure~\ref{fig:S200105ae}.

In total, 24 candidates were reported by ZTF \citep{GCN26673,GCN26810} and \emph{Gaia} \citep{GCN26686}. All candidates were ruled out through spectroscopic classification \citep{GCN26701,GCN26702,GCN26703,Anand2020}, slow photometric evolution, or identification as a stellar source or moving object \cite{Anand2020}. We detect 5 of the ZTF candidates in our data, all showing photometric evolution inconsistent with a kilonova, in agreement with \cite{Anand2020}. Searching our serendipitous fields, we find a single new candidate, SAGUARO20h/AT2020abgt \citep{tnscands}, at RA = 07h41m24.0s and Dec = 09d27m59s (see Table \ref{tab:obs}). Performing PSF photometry on the subtracted image, we obtain $m_{can} = 18.86 \pm 0.24$ mag at $\delta t_{rest} = 0.49$ days. Comparing the absolute magnitude of AT2020abgt to the kilonova models discussed here, we find that it is $\sim$3 mag brighter (a factor of 20 larger in luminosity) than the brightest model at $\delta t_{rest} = 0.49$ days, using the median inferred LIGO/Virgo distance. Given the uncertainty associated with NSBH models and the unknown parameters of the system, we do not rule out this candidate based purely on the above. Searching our data, we find a non-detection to the limit of 21.31 mag just over one month prior to the event and a non-detection to the limit of 21.07 mag one month after the event. AT2020abgt is isolated with no cataloged galaxy at the position of the source. No coincident galaxy is detected in our images, and it is not covered within the Legacy Survey footprint.

\subsection{S200114f} \label{sec:S200114f}
On 2020-01-14 02:08:18.23 UTC the coherent Wave Burst (cWB) pipeline \citep{Klimenko16} triggered on a candidate unmodelled burst signal, S200114f, using data from the L1, H1, and V1 \citep{lvcgcn26734}.  As a burst event, S200114f was detected without a template or prior knowledge of the waveform, with no constraint on the distance. The false alarm rate was 1 per 25.838 yr, and the 90\% localization covered 403 deg$^2$. We triggered 36 fields (180 deg$^2$) within $\delta t < 24$~hr, covering $P_{\mathrm{total}}$ = 86.30\% (Table \ref{tab:follow-up}). The localization of the event, along with the CSS fields that fall within the 90\% region, is shown in Figure~\ref{fig:200114f_globe_loc}.

In total, 33 optical candidates were reported by ZTF \citep{gcn26741,gcn26806} and SAGUARO \citep{gcn26753}. Ten of these were classified as SN and excluded (\citealt{gcn26764,gcn26830,gcn27073}), 2 (including one SAGUARO detection) were excluded for having previous detections (\citealt{gcn26750,gcn26806}), and 1 was classified as a QSO and excluded \citep{gcn27073}. Subsequent re-analysis of the 3 remaining original SAGUARO candidates suggest that 2 (SAGUARO20b and SAGUARO20d) have underlying point sources and are likely the result of stellar variability. Of the remaining 17 candidates, we find a detection of AT2020aco within our data prior to the event on 24 Nov 2019, thus ruling it out as being related to the GW event. Searching through our data, we find no new candidates for this event. A more in-depth discussion on the progenitor of this burst event, along with possible models and the efficiency of our search with regards to these ``burst'' events, will be presented in Lundquist et al., in prep.

\subsection{S200115j}
S200115j was identified as a candidate GW signal using data from H1, L1, and V1 on 2020-01-15 04:23:09.742 UTC \citep{gcn26759}. The final classification for this event is $>99\%$ MassGap at $d = 332 \pm 78$ Mpc. With a HasNS probability of $>99\%$, this would suggest that the event was a merger between one object of $3 < M/\mathrm{M}_{\odot} < 5 $ and a NS with $<3 \mathrm{M}_{\odot}$. The calculated HasRemnant for this event is $> 99\%$ \citep{gcn26807}, suggesting a high probability of EM emission. There are no triggered fields for this event, and our total serendipitous coverage is 11 fields (55 deg$^2$) in the interval $\delta t = 72-120$~hr, with $P_{\mathrm{total}}$ = 2.64\% (Table \ref{tab:follow-up}). The localization of the event, along with the CSS fields that fall within the 90\% region, is shown in Figure~\ref{fig:S200115j}. Three optical candidates were reported by the community, all from ZTF \citep{gcn26767}. Two of these candidates had inconsistent redshifts with the inferred LIGO/Virgo distance, while multiple non-detections in optical imaging follow-up of the third failed to confirm it as a source \citep{gcn26814,gcn26817,gcn26819,gcn26822}. Searching through our serendipitous fields, we find no new candidates. 

\subsection{S200128d}
A candidate GW signal, S200128d, was identified using data from H1 and L1 on 2020-01-28 02:20:11.903 UTC \citep{gcn26906}. The final classification for this event is 97\% BBH at $d = 3702 \pm 1265$ Mpc. There are no triggered fields for this event, and our total serendipitous coverage is 64 fields (320 deg$^2$) with $\delta t < 96$~hr and $P_{\mathrm{total}}$ = 11.80\% (Table \ref{tab:follow-up}). The localization of the event, along with the CSS fields that fall within the 90\% region, is shown in Figure~\ref{fig:S200128d}.

No counterparts were reported by the community. Searching our serendipitous fields, we find a single new candidate, SAGUARO20i/AT2020abgu \citep{tnscands}, at RA = 14h16m09.0s and Dec = $-$06d22m43s (see Table \ref{tab:obs}). Performing PSF photometry on the subtracted image, we obtain $m_{can} = 20.47 \pm 0.26$ mag at $\delta t_{rest} = 0.28$ days. As a BBH event however, we do not expect EM emission. Nevertheless, comparing the absolute magnitude of AT2020abgu to the kilonova models discussed here, we find that AT2020abgu is $\sim$ 6 mag brighter (a factor of 240 larger in luminosity) than the brightest model at $\delta t_{rest} = 0.28$ days, using the median inferred LIGO/Virgo distance. Searching our data, we find a non-detection to the limit of 20.72 mag just over one month prior to the event and a non-detection to the limit of 20.93 mag roughly three months after the event. Searching the Legacy Survey, we find AT2020abgu is coincident with a galaxy (Brick: 2140m065, Objid: 6389) with $g$ = 19.51, $r$ = 19.07, and $z$ = 18.73 mag. Searching PanSTARRS DR2, we find a $z_{photo}$ = 0.065 $\pm$ 0.039. This distance is vastly different from the inferred LIGO/Virgo distance (by $>3000$ Mpc), but, due to the large errors in the inferred LIGO/Virgo distance still falls within 3$\sigma$ of the median. From our images, AT2020abgu appears to be nuclear in nature with $\delta r \sim 1.7''$ between the position of AT2020abgu and the position of the galaxy from the Legacy Survey. The nature of both the merger (BBH and very far) and the candidate (extremely bright if assuming the median LIGO/Virgo distance, the host $z_{photo}$, and apparent nuclear position) implies that AT2020abgu is most likely due to AGN activity, or possibly a TDE, and not related to the GW event.

\subsection{S200224ca} \label{sec:S200224ca}
S200224ca was identified as a candidate GW signal using data from H1, L1, and V1 on 2020-02-24 22:22:34.406 UTC \citep{gcn27184}. The final classification for this event is $> 97\%$ BBH at $d = 1575 \pm 322$ Mpc \citep{gcn27262}. There are no triggered fields for this event, and our total serendipitous coverage is 5 fields (25 deg$^2$) in the interval $\delta t = 24-72$~hr, with $P_{\mathrm{total}}$ = 11.42\% (Table \ref{tab:follow-up}). The localization of the event, along with the CSS fields that fall within the 90\% region, is shown in Figure~\ref{fig:S200224ca}.

In total, 27 optical candidates were initially reported by the community (DESGW \citealt{gcn27227,GCN27366}, GRAWITA \citealt{gcn27230}). Initial follow-up ruled out 14 optical candidates due to photometric evolution consistent with a SN \citep{GCN27366} and thus too slow for a kilonova. While the DESGW candidates are generally too faint for the limiting mag of CSS, we are able to identify AT2020dlt and AT2020dlu below our automatic vetting S/N threshold (with S/N $\approx$ 2.2 and 2.9, and 5$\sigma$ transient depth of 21.6 mag and 21.1 mag, respectively) in images taken 6 days prior to the event. Thus, we are able to rule them out as being related to the GW event. Searching through our data, we find non-detections for the remaining 13 candidates, both prior to and after the event. No new candidates were found within our serendipitous data for this event.

\subsection{S200316bj} \label{sec:S200316bj}
A candidate GW signal, S200316bj, was identified using data from H1, L1, and V1 on 2020-03-16 21:57:56.157 UTC \citep{gcn27388}. The final classification for this event is $> 99\%$ MassGap at $d = 1178 \pm 283$ Mpc. With a HasNS probability of $<1\%$, this would suggest the event was a merger between one object of $3 < M/\mathrm{M}_{\odot} < 5 $ and a BH with $ > 5 \mathrm{M}_{\odot}$ \citep{gcn27419}. There are no triggered fields for this event, and our total serendipitous coverage is 20 fields (100 deg$^2$) at $\delta t < 48$~hr and $P_{\mathrm{total}}$ = 73.24\% (Table~\ref{tab:follow-up}). The localization of the event, along with the CSS fields that fall within the 90\% region, is shown in Figure~\ref{fig:S200316bj}. No candidates were reported by the community, and no candidates were found in our serendipitous data for this event.

\section{Discussion} \label{sec:discussion}
In total, we have presented observations of 17 GW events from O3 (see Table \ref{tab:follow-up}). We triggered observations (generally in sets of 60 deg$^2$) for 7 events: GW190408\_181802, GW190425, GW190426\_152155, S190901ap, GW190930\_133541, S191205ah, and S200114f. Our trigger criteria focused on events containing a potential NS or unusual events. Except for one event (GW190426\_152155), we were able to trigger fields within 24 hr of the GW alert. We find $\delta t$ = 1.4--43.6 hr for the first triggered field, with a median $\delta t$ = 11.9 hr across triggered events. Including fields from normal CSS operations that serendipitously overlap with the GW sky localization at $\delta t < 5$~days, we were able to increase our coverage of these triggered events, as well as include additional events with purely serendipitous coverage. We present observations for 10 additional events with serendipitous coverage that have $P_{\mathrm{total}} > 2\%$: GW190521, GW190630\_185205, S190923y, S190930t, S191213g, S200105ae, S200115j, S200128d, S200224ca, and S200316bj. Including triggered and serendipitous fields, we searched a total of 4755 deg$^2$, with $A_{\mathrm{total}}$ = 15--2120 deg$^2$ for these events. We covered up to $P_{\mathrm{total}}$ = 86\% for individual events, with a median $P_{\mathrm{total}} \approx 8\%$ within $\delta t < 5$~days of the GW event. The 5$\sigma$ transient depth of individual fields ranges from 17.1--21.6 mag, with a median of 21.1 mag. 

For a direct comparison to other surveys which reported on a number of events during O3, we compare the $P_{\mathrm{total}}$, the median depth of observations and the median time lag between the observations and the GW event. Our highest coverage (86\%) is comparable to the maximum covered by ZTF/GROWTH (88\%; \citealt{kasliwal20}), GOTO (95\%; \citealt{gompertz2020}), and GRANDMA (95\%; \citealt{antier20}). In terms of the median $P_{\mathrm{total}}$ covered across all events, we have a much lower number (8\% compared to $\sim$40\% from ZTF/GROWTH and GOTO) due to the number of low coverage events included in our analysis (2\% $<$ $P_{\mathrm{total}}$ $<$ 15\%). We highlight our coverage of the unusual burst event, S200114f, which has not been discussed in great detail by the community. In terms of the median depths of our observations, we are most comparable to (but deeper than) ZTF/GROWTH (see Figure~\ref{fig:Appmag}). Compared to the median trigger time for observations, we find comparable timescales for first trigger to GOTO and GRANDMA when looking at the triggering of a large number of events from O3, while ZTF/GROWTH usually had shorter delay times. Focusing on observations purely from O3a, DECAM/GROWTH \citep{Anand20} searched down to a deeper median depth compared to our observations (a median of $\approx$23 mag for 4 events) with coverage ranging from 8--98\%. In addition, there were a number of searches that targeted single (or two or three) events \citep{coughlin19,Anand2020,Thakur2020,vieira_cfht,ackley2020}. For these cases, we search to comparable depth for GW190425 as ZTF/GROWTH. For the NSBH events S200105ae and S200115j, we find similar search depths (with a slightly shallower median) as ZTF \citep{Anand2020}, who found that with search depths of $\approx$22 mag, one should be able to place strong constraints on the ejecta mass from these systems. For the BBH event GW190521, we also find comparable search depth (ranging from 21.2--22.3 for the respective filters) from DECam/GROWTH \citep{Andreoni2019_190510}. Although the NSBH event S190814bv \citep{gcn25324} was not covered by us, we can achieve deeper search depths in general than observations by the DDOTI \citep{Thakur2020}, while CFHT \citep{vieira_cfht}, ENGRAVE \citep{ackley2020}, and DECam/GROWTH \citep{Andreoni20} observations were deeper than our typical search depths. Likewise for the retracted NSBH event S190510g \citep{gcn24442}, DECam/GROWTH \citep{Andreoni2019_190510} reported 3$\sigma$ depths of 21.2--22.3 mag, while DECam/DES \citep{garcia20} reported 10$\sigma$ search depths of 20.58--21.72 mag, for $grz$ observations.

A number of updates were made to SAGUARO during O3, mostly improving our ability to classify and vet candidates (see Section \ref{sec:updates}). With these improvements, the availability of updated localizations, and the inclusion of serendipitous fields, we present a detailed analysis of these 17 events while taking advantage of later observations to rule out candidates based on their photometric evolution. To do this, we compare our candidates to a number of kilonova models: a GW170817/AT2017gfo-like kilonova \citep{kasen17}, a NSBH kilonova model \citep{kawaguchi20}, a HMNS remnant \citep{kasen15}, and a GW190425-like kilonova model \citep{barbieri20,foley20}.  Comparing the absolute magnitude of our observations to these models, SAGUARO is able to search to depths that would detect kilonova emission out to $\sim$ 150--400 Mpc (see Figure \ref{fig:Absmag}). This distance range assumes detection at peak brightness and depends on the model assumed. 

Of the 17 events presented here, 9 have BNS/NSBH/MassGap classifications from which we could expect some EM emission depending on the parameters of the system. Comparing these events directly to the models, SAGUARO should have been able to detect kilonova emission for S190901ap, S190930t, and S191213g for their respective models, as well as for GW190425, GW190426\_152155, S191205ah, S200105ae, and S200115j for the more optimistic models. Given the large localizations, however, it would have been extremely challenging to achieve both the necessary depths and areal coverage necessary to identify kilonovae.

After automatic and human vetting of our triggered and serendipitous fields (see Section \ref{sec:vetting}), we find a total of 7 viable candidates: AT2019aaid (S190901ap, BNS), AT2019aaig (GW190930\_133541, BBH), AT2019aaie, AT2019aaif and AT2019aaih (S190930t, NSBH), AT2020abgt (S200105ae, NSBH), and AT2020abgu (S200128d, BBH); from 5 GW events (see Table~\ref{tab:obs}). All candidates have a single detection within our data, along with limits on time scales that do not allow us to rule them out as optical counterparts. Their detected magnitudes range from 18.6 to 20.9 mag with $\delta t$ ranging from $\sim$0.5 to 4.5 days.

First, we compare the 5 candidates associated with BNS and NSBH mergers directly to the models and the luminosity of AT2017gfo at the median inferred LIGO/Virgo distance of each event. We also choose a range of models that represent the optical diversity of kilonovae, previously found in comparative studies between AT2017gfo and candidate kilonovae following short GRBs \citep{Ascenzi+19,Gompertz+18}. We find a single candidate, AT2019aaid (BNS), consistent with the $r$-band luminosity of AT2017gfo at $\delta t_{\rm rest} = 0.48$ days. As a potential candidate associated with an NSBH merger, AT2019aaie is consistent with the \cite{kawaguchi20} NSBH model discussed here at $\delta t_{rest} = 1.49$ days, while AT2020abgt is $\sim$20 times more luminous at $ \delta t_{rest} = 0.28$ days. Although AT2020abgt is more luminous than the NSBH model discussed here, we do not rule it out due to the uncertainty associated with NSBH models. The remaining candidates associated with the NSBH merger S190930t, AT2019aaif and AT2019aaih, only overlap with the optimistic HMNS model, where a stable remnant is expected to be a rare outcome \citep{MargalitMetzger2019}, when considering the error in the inferred LIGO/Virgo distance \citep{kasen15}. 

Likewise for our two BBH candidates (AT2019aaig and AT2020abgu), we do not expect EM emission, although there is much more uncertainty associated with the optical emission from BBH mergers (e.g. \citealt{Perna2018,Perna2019,Graham2020}). We note that both candidates are more luminous (by a factor of 20 and 240, respectively) than any model discussed here, and the limited predictions for any optical emission from BBH mergers overall predict fainter emission than BNS mergers and/or specific associations to AGN disks \citep{Stone2017,McKernan2018}. 

Four of our candidates have detectable galaxies within $4''$. Considering these galaxies as their hosts, we measure the offset between each candidate and the galaxy position. We find $\delta r = 0.3-7.9''$. Only 3 of these galaxies have redshifts (as $z_{photo}$), with 2 having multiple $z_{photo}$ values associated with them. Using each of the different photometric redshifts (Table~\ref{tab:obs}), this implies projected physical offsets of 1.4--16.3 kpc, well within the range of known short GRB offsets \citep{FongBerger13}. Overall, the precise nature of these seven candidates is inconclusive, although several demonstrate consistency with the properties and locations of kilonovae.

Looking at observations from the community for O3 as a whole, a total of 252 optical candidates were initially reported via GCN for the 17 GW events discussed here. Although much effort went into the search and discovery of candidates for these events, only 65\% were followed up in some capacity prior to this work, with only 25\% spectroscopically followed up.  Checking the status of community candidates not ruled out by initial (via GCN) or later (in published papers) follow up, we find 3 candidates (AT2019dpg, AT2019eby and AT2019wjr) with later detections that were automatically uploaded from various surveys to TNS. Using these data points, we were able to rule out these candidates as viable kilonova counterparts due to their inconsistent photometric evolution. Searching through public data from ZTF, we were able to find another candidate, AT2019deu, which displayed photometric behaviour inconsistent with a kilonova. From our own data, we were able to rule out an additional 8 candidates: AT2019ech, AT2019piw, AT2019rpt, SAGUARO20b, SAGUARO20d, AT2020aco, AT2020dlt and AT2020dlu. While some of these candidates were ruled out due to inconsistent photometric evolution based on later observations, 3 of these had detections prior to the GW event and thus could have been ruled out at the time of initial discovery. These 12 candidates make up $\sim$5\% of the total number of counterpart candidates associated with the 17 GW events in this paper.

\section{Conclusions and future prospects} \label{sec:conclusion}
In this paper, we presented a detailed analysis of both triggered and serendipitous observations of 17 events (7 triggered and 10 purely serendipitous) from O3. The serendipitous coverage of events provided both an increase in coverage for $\delta t < 5$~days, as well as information about long-term light curve evolution for both community and SAGUARO candidates. From our data, we report 7 viable kilonova candidates from 5 different GW events. Although each candidate only has a single detection, making it difficult to concretely tie them to a kilonova model, we cannot completely rule them out due to lack of information and the uncertainty surrounding the predicted kilonova light curves. Searching through publicly available data, we also found several outstanding candidates that could be ruled out due to later photometric observations.

The lack of follow-up for such a large number of candidates highlights the need for community coordination during candidate follow-up and vetting (although the number of candidates may decrease in the future as localizations improve; \citealt{2018LivingRev}). This is strengthened by the fact that many candidates were followed up, classified, and ruled out by multiple different groups, while others were not revisited at all. During their wide-field plus galaxy targeted search of GW190814, \cite{Thakur2020} also noted that the lack of follow-up from the community left $\sim$25\% of candidates without a classification (see also, \citealt{coughlin19}). They found brighter candidates were generally reported earlier, with follow-up efforts for classification concentrated within $\delta t < 4$~days with often duplicated efforts. Projects such as the Gravitational Wave Treasure Map \citep{Wyatt20}, which promote coordination of observations, will be increasingly important as we move towards more GW detections in future runs. 
A more comprehensive effort, tracking down and classifying remaining counterpart candidates for all of the GW events of O3, would certainly bear fruit and shed light on follow-up strategies going forward. To this end, having a candidate database, where candidate status and planned observations could be tracked, would be greatly beneficial. Real-time information from the LVC about event parameters such as the chirp mass, mass ratio, and inclination of the system, as well as prompt updates regarding classification and sky localizations would also greatly improve the efficiency of EM counterpart searches.

Looking to future runs, O4 is scheduled to start no earlier than mid-2022, with significant upgrades to LIGO and Virgo and the addition of KAGRA\footnote{Further into the future (O5 and beyond), will also see the addition of LIGO-India.}. The upgrades to LIGO/Virgo/KAGRA are expected to increase the BNS detection range for O4 to 160--190, 90--120, and 25--130 Mpc, respectively \citep{2018LivingRev}. Given the value added by the serendipitous observations provided by CSS and current kilonova model predictions, we plan to extend our SAGUARO real-time candidate searching to include serendipitous fields with $\delta t < 5$~days for O4 and beyond. Another planned improvement is the real-time comparison of candidates to kilonova models during the vetting process to provide value-added information about the likelihood of a candidate being associated with an event. Comparing candidates to a grid of kilonova models will also provide limits on model parameters such as ejecta mass, electron fraction, energy, density, and viewing angle. \cite{Almualla2020} presented an in-depth analysis of serendipitous kilonova detection using wide-field surveys, showing how cadence and the choice of filters could be used to optimize serendipitous kilonova detection. Previous studies by \citet{Cowperthwaite2018} also looked at how observations could be tailored to detect kilonova emission in the context of contaminating sources. Improvements to SAGUARO, such as the addition of other discovery telescopes, would provide resources to increase the likelihood of detecting kilonova emission from GW triggers, as well as independently from GW detections.

Although the planned increase in sensitivity will allow the detection of mergers out to greater distances, EM follow-up is greatly restricted by the depth of current telescope surveys (see Figure \ref{fig:Appmag}). With SAGUARO's current discovery depth, we are able to probe the peak brightness of kilonova out to $\sim$ 150--400 Mpc (see Figure \ref{fig:Absmag}). For surveys such as CFHT, which have deeper observations, the discovery space could go out to $\sim$ 1000 Mpc for unusually massive events like GW190425 (assuming the optimistic case); while shallower surveys that probe $<$ 19 mag can only reach $\sim$ 150 Mpc for the same scenario. The issue of survey depth is not easily solved and is generally a trade off with FOV. Given the large localizations seen in O3, and thus the fairly low $P_{\mathrm{total}}$ for many events, a trade off with FOV is often not feasible. The addition of KAGRA, however, will provide localizations which are a factor $\sim$ 1.4 better  \citep{2018LivingRev} and presumably increase the efficiency of EM follow-up searches.

\acknowledgments

We gratefully acknowledge Kyohei Kawaguchi and Claudio Barbieri for generously sharing their models with us. SAGUARO is supported by the National Science Foundation (NSF) under Award Nos.\ AST-1909358 and AST-1908972. Time domain research by D.J.S. is supported by NSF grants AST-1821987, 1813466, \& 1908972, and by the Heising-Simons Foundation under grant \#2020-1864.  Research by K.P., J.C.R., and W.F. is also supported by NSF Award No.\ AST-1814782. The UCSC team is supported in part by NASA grant NNG17PX03C, NSF grant AST-1815935, the Gordon \& Betty Moore Foundation, the Heising-Simons Foundation, and by a fellowship from the David and Lucile Packard Foundation to R.J.F. J.S. acknowledges support from the Packard Foundation. V.P. acknowledges support from NSF grant PHY-1912619. A.C. acknowledges support from NSF Award No. 1907975.
The operation of the facilities of Steward Observatory are supported in part by the state of Arizona.
This research has made use of data and/or services provided by the International Astronomical Union's Minor Planet Center.

\facilities{SO:1.5m}

\software{
astropy \citep{2013A&A...558A..33A,astropy},   
The IDL Astronomy User's Library \citep{IDLforever}, SCAMP \citep{scamp,scamp2}, SWarp \citep{swarp}, IRAF \citep{iraf1,iraf2},
          SExtractor \citep{Bertin1996}, ZOGY (\url{https://github.com/pmvreeswijk/ZOGY})
          }

\clearpage

\bibliography{biblio}

\begin{thebibliography}{}
\expandafter\ifx\csname natexlab\endcsname\relax\def\natexlab#1{#1}\fi
\providecommand{\url}[1]{\href{#1}{#1}}
\providecommand{\dodoi}[1]{doi:~\href{http://doi.org/#1}{\nolinkurl{#1}}}
\providecommand{\doeprint}[1]{\href{http://ascl.net/#1}{\nolinkurl{http://ascl.net/#1}}}
\providecommand{\doarXiv}[1]{\href{https://arxiv.org/abs/#1}{\nolinkurl{https://arxiv.org/abs/#1}}}

\bibitem[{{Abbott} {et~al.}(2009){Abbott}, {Abbott}, {Adhikari}, {Ajith},
  {Allen}, {Allen}, {Amin}, {Anderson}, {Anderson}, {Arain}, {Araya}, {Armand
  ula}, {Armor}, {Aso}, {Aston}, {Aufmuth}, {Aulbert}, {Babak}, {Baker},
  {Ballmer}, {Barker}, {Barker}, {Barr}, {Barriga}, {Barsotti}, {Barton},
  {Bartos}, {Bassiri}, {Bastarrika}, {Behnke}, {Benacquista}, {Betzwieser},
  {Beyersdorf}, {Bilenko}, {Billingsley}, {Biswas}, {Black}, {Blackburn},
  {Blackburn}, {Blair}, {Bland}, {Bodiya}, {Bogue}, {Bork}, {Boschi}, {Bose},
  {Brady}, {Braginsky}, {Brau}, {Bridges}, {Brinkmann}, {Brooks}, {Brown},
  {Brummit}, {Brunet}, {Bullington}, {Buonanno}, {Burmeister}, {Byer},
  {Cadonati}, {Camp}, {Cannizzo}, {Cannon}, {Cao}, {Cardenas}, {Caride},
  {Castaldi}, {Caudill}, {Cavagli{\`a}}, {Cepeda}, {Chalermsongsak}, \&
  {Chalkley}}]{ligo09}
{Abbott}, B.~P., {Abbott}, R., {Adhikari}, R., {et~al.} 2009, Reports on
  Progress in Physics, 72, 076901, \dodoi{10.1088/0034-4885/72/7/076901}

\bibitem[{{Abbott} {et~al.}(2016){Abbott}, {Abbott}, {Abbott}, {Abernathy},
  {Acernese}, {Ackley}, {Adams}, {Adams}, {Addesso}, {Adhikari}, \&
  et~al.}]{lvc_loc}
{Abbott}, B.~P., {Abbott}, R., {Abbott}, T.~D., {et~al.} 2016, Living Reviews
  in Relativity, 19, \dodoi{10.1007/lrr-2016-1}

\bibitem[{{Abbott} {et~al.}(2017{\natexlab{a}}){Abbott}, {Abbott}, {Abbott},
  {Acernese}, {Ackley}, {Adams}, {Adams}, {Addesso}, {Adhikari}, {Adya}, \&
  et~al.}]{lvc_gw170817}
---. 2017{\natexlab{a}}, Physical Review Letters, 119, 161101,
  \dodoi{10.1103/PhysRevLett.119.161101}

\bibitem[{{Abbott} {et~al.}(2017{\natexlab{b}}){Abbott}, {Abbott}, {Abbott},
  {Acernese}, {Ackley}, {Adams}, {Adams}, {Addesso}, {Adhikari}, {Adya},
  {Affeldt}, {Afrough}, {Agarwal}, {Agathos}, {Agatsuma}, {Aggarwal}, {Aguiar},
  {Aiello}, {Ain}, {Ajith}, {Allen}, {Allen}, {Allocca}, {Altin}, {Amato},
  {Ananyeva}, {Anderson}, {Anderson}, {Angelova}, {Antier}, {Appert}, {Arai},
  {Araya}, {Areeda}, {Arnaud}, {Arun}, {Ascenzi}, {Ashton}, {Ast}, {Aston},
  {Astone}, {Atallah}, {Aufmuth}, {Aulbert}, {AultONeal}, {Austin},
  {Avila-Alvarez}, {Babak}, {Bacon}, {Bader}, {Bae}, {Baker}, {Baldaccini},
  {Ballardin}, {Ballmer}, {Banagiri}, {Barayoga}, {Barclay}, {Barish},
  {Barker}, {Barkett}, {Barone}, {Barr}, {Barsotti}, {Barsuglia}, {Barta},
  {Barthelmy}, {Bartlett}, {Bartos}, {Bassiri}, {Basti}, {Batch}, {Bawaj},
  {Bayley}, {Bazzan}, {B{\'e}csy}, {Beer}, {Bejger}, {Belahcene}, {Bell},
  {Berger}, {Bergmann}, {Bero}, {Berry}, {Bersanetti}, {Bertolini},
  {Betzwieser}, {Bhagwat}, {Bhandare}, {Bilenko}, {Billingsley}, {Billman},
  {Birch}, {Birney}, {Birnholtz}, {Biscans}, {Biscoveanu}, {Bisht}, {Bitossi},
  \& {Biwer}}]{LIGO_MMA}
---. 2017{\natexlab{b}}, \apjl, 848, L12, \dodoi{10.3847/2041-8213/aa91c9}

\bibitem[{{Abbott} {et~al.}(2018){Abbott}, {Abbott}, {Abbott}, {Abernathy},
  {Acernese}, {Ackley}, {Adams}, {Adams}, {Addesso}, {Adhikari}, {Adya},
  {Affeldt}, {Agathos}, {Agatsuma}, {Aggarwal}, {Aguiar}, {Aiello}, {Ain},
  {Ajith}, {Akutsu}, {Allen}, {Allocca}, {Altin}, {Ananyeva}, {Anderson},
  {Anderson}, {Ando}, {Appert}, {Arai}, {Araya}, {Araya}, {Areeda}, {Arnaud},
  {Arun}, {Asada}, {Ascenzi}, {Ashton}, {Aso}, {Ast}, {Aston}, {Astone},
  {Atsuta}, {Aufmuth}, {Aulbert}, {Avila-Alvarez}, {Awai}, {Babak}, {Bacon},
  {Bader}, {Baiotti}, {Baker}, {Baldaccini}, {Ballardin}, {Ballmer},
  {Barayoga}, {Barclay}, {Barish}, {Barker}, {Barone}, {Barr}, {Barsotti},
  {Barsuglia}, {Barta}, {Bartlett}, {Barton}, {Bartos}, {Bassiri}, {Basti},
  {Batch}, {Baune}, {Bavigadda}, {Bazzan}, {B{\'e}csy}, {Beer}, {Bejger},
  {Belahcene}, {Belgin}, {Bell}, {Berger}, \& {Bergmann}}]{2018LivingRev}
---. 2018, Living Reviews in Relativity, 21, 3,
  \dodoi{10.1007/s41114-018-0012-9}

\bibitem[{{Abbott} {et~al.}(2019){Abbott}, {Abbott}, {Abbott}, {Abraham},
  {Acernese}, {Ackley}, {Adams}, {Adhikari}, {Adya}, {Affeldt}, {Agathos},
  {Agatsuma}, {Aggarwal}, {Aguiar}, {Aiello}, {Ain}, {Ajith}, {Allen},
  {Allocca}, {Aloy}, {Altin}, {Amato}, {Ananyeva}, {Anderson}, {Anderson},
  {Angelova}, {Antier}, {Appert}, {Arai}, {LIGO Scientific Collaboration}, \&
  {Virgo Collaboration}}]{2019O1O2Catalog}
---. 2019, Physical Review X, 9, 031040, \dodoi{10.1103/PhysRevX.9.031040}

\bibitem[{{Abbott} {et~al.}(2020{\natexlab{a}}){Abbott}, {Abbott}, {Abbott},
  {Abraham}, {Acernese}, {Ackley}, {Adams}, {Adhikari}, {Adya}, {Affeldt},
  {Agathos}, {Agatsuma}, {Aggarwal}, {Aguiar}, {Aiello}, {Ain}, {Ajith},
  {Allen}, {Allocca}, {Aloy}, {Altin}, {Amato}, {Anand}, {Ananyeva},
  {Anderson}, {Anderson}, {Angelova}, {Antier}, {Appert}, {Arai}, {Araya},
  {Areeda}, {Ar{\`e}ne}, {Arnaud}, {Aronson}, {Arun}, {Ascenzi}, {Ashton},
  {Aston}, {Astone}, {Aubin}, {Aufmuth}, {AultONeal}, {Austin}, {Avendano},
  {Avila-Alvarez}, {Babak}, {Bacon}, {Badaracco}, {Bader}, {Bae}, \&
  {Baird}}]{GW190425}
---. 2020{\natexlab{a}}, \apjl, 892, L3, \dodoi{10.3847/2041-8213/ab75f5}

\bibitem[{{Abbott} {et~al.}(2020{\natexlab{b}}){Abbott}, {Abbott}, {Abraham},
  {Acernese}, {Ackley}, {Adams}, {Adams}, {Adhikari}, {Adya}, {Affeldt},
  {Agathos}, {Agatsuma}, {Aggarwal}, {Aguiar}, {Aiello}, {Ain}, {Ajith},
  {Akcay}, {Allen}, {Allocca}, {Altin}, {Amato}, {Anand}, {Ananyeva},
  {Anderson}, {Anderson}, {Angelova}, {Ansoldi}, {Antelis}, {Antier}, {Appert},
  {Arai}, {Araya}, {Areeda}, {Ar{\`e}ne}, {Arnaud}, {Aronson}, {Arun}, {Asali},
  {Ascenzi}, {Ashton}, {Aston}, {Astone}, {Aubin}, {Aufmuth}, {AultONeal},
  {Austin}, {Avendano}, {Babak}, {Badaracco}, {Bader}, {Bae}, {Baer}, \&
  {Bagnasco}}]{gwtc2}
{Abbott}, R., {Abbott}, T.~D., {Abraham}, S., {et~al.} 2020{\natexlab{b}},
  arXiv e-prints, arXiv:2010.14527.
\newblock \doarXiv{2010.14527}

\bibitem[{{Abbott} {et~al.}(2020{\natexlab{c}}){Abbott}, {Abbott}, {Abraham},
  {Acernese}, {Ackley}, {Adams}, {Adhikari}, {Adya}, {Affeldt}, {Agathos},
  {Agatsuma}, {Aggarwal}, {Aguiar}, {Aich}, {Aiello}, {Ain}, {Ajith}, {Akcay},
  {Allen}, {Allocca}, {Altin}, {Amato}, {Anand}, {Ananyeva}, {Anderson},
  {Anderson}, {Angelova}, {Ansoldi}, {Antier}, {Appert}, {Arai}, {Araya},
  {Areeda}, {Ar{\`e}ne}, {Arnaud}, {Aronson}, {Arun}, {Asali}, {Ascenzi},
  {Ashton}, {Aston}, {Astone}, {Aubin}, {Aufmuth}, {AultONeal}, {Austin},
  {Avendano}, {Babak}, {Bacon}, {Badaracco}, {Bader}, {Bae}, {Baer}, {Baird},
  {Baldaccini}, {Ballardin}, {Ballmer}, {Bals}, {Balsamo}, {Baltus},
  {Banagiri}, {Bankar}, {Bankar}, {Barayoga}, {Barbieri}, {Barish}, {Barker},
  {Barkett}, {Barneo}, {Barone}, {Barr}, {Barsotti}, {Barsuglia}, {Barta},
  {Bartlett}, {Bartos}, {Bassiri}, {Basti}, {Bawaj}, {Bayley}, {Bazzan},
  {B{\'e}csy}, {Bejger}, {Belahcene}, {Bell}, {Beniwal}, {Benjamin}, {Benkel},
  {Bentley}, {Bergamin}, {Berger}, {Bergmann}, {Bernuzzi}, {Berry},
  {Bersanetti}, {Bertolini}, {Betzwieser}, {Bhand are}, {Bhandari}, {Bidler},
  {Biggs}, {Bilenko}, {Billingsley}, {Birney}, {Birnholtz}, {Biscans},
  {Bischi}, {Biscoveanu}, {Bisht}, {Bissenbayeva}, {Bitossi}, {Bizouard},
  {Blackburn}, {Blackman}, {Blair}, {Blair}, {Blair}, {Bobba}, {Bode}, \&
  {Boer}}]{190814_paper}
---. 2020{\natexlab{c}}, \apjl, 896, L44, \dodoi{10.3847/2041-8213/ab960f}

\bibitem[{{Abbott} {et~al.}(2020{\natexlab{d}}){Abbott}, {Abbott}, {Abraham},
  {Acernese}, {Ackley}, {Adams}, {Adhikari}, {Adya}, {Affeldt}, {Agathos},
  {Agatsuma}, {Aggarwal}, {Aguiar}, {Aich}, {Aiello}, {Ain}, {Ajith}, {Akcay},
  {Allen}, {Allocca}, {Altin}, {Amato}, {Anand}, {Ananyeva}, {Anderson},
  {Anderson}, {Angelova}, {Ansoldi}, {Antier}, {Appert}, {Arai}, {Araya},
  {Areeda}, {Ar{\`e}ne}, {Arnaud}, {Aronson}, {Arun}, {Asali}, {Ascenzi},
  {Ashton}, {Aston}, {Astone}, {Aubin}, {Aufmuth}, {AultONeal}, {Austin},
  {Avendano}, {Babak}, {Bacon}, {Badaracco}, {Bader}, {Bae}, {Baer}, {Baird},
  {Baldaccini}, {Ballardin}, {Ballmer}, {Bals}, {Balsamo}, {Baltus},
  {Banagiri}, {Bankar}, {Bankar}, {Barayoga}, {Barbieri}, {Barish}, {Barker},
  {Barkett}, {Barneo}, {Barone}, {Barr}, {Barsotti}, {Barsuglia}, {Barta},
  {Bartlett}, {Bartos}, {Bassiri}, {Basti}, {Bawaj}, {Bayley}, {Bazzan},
  {B{\'e}csy}, {Bejger}, {Belahcene}, {Bell}, {Beniwal}, {Benjamin}, {Bentley},
  {Bergamin}, {Berger}, {Bergmann}, {Bernuzzi}, {Berry}, {Bersanetti},
  {Bertolini}, {Betzwieser}, {Bhandare}, {Bhandari}, {Bidler}, {Biggs},
  {Bilenko}, {Billingsley}, {Birney}, {Birnholtz}, {Biscans}, {Bischi},
  {Biscoveanu}, {Bisht}, {Bissenbayeva}, {Bitossi}, {Bizouard}, {Blackburn},
  {Blackman}, {Blair}, {Blair}, {Blair}, {Bobba}, {Bode}, {Boer}, {Boetzel},
  {Bogaert}, {Bondu}, {Bonilla}, {Bonnand}, {Booker}, {Boom}, {Bork}, {Boschi},
  {Bose}, {Bossilkov}, {Bosveld}, {Bouffanais}, {Bozzi}, {Bradaschia}, {Brady},
  {Bramley}, {Branchesi}, {Brau}, {Breschi}, {Briant}, {Briggs}, {Brighenti},
  {Brillet}, {Brinkmann}, {Brockill}, {Brooks}, {Brooks}, {Brown}, {Brunett},
  {Bruno}, {Bruntz}, {Buikema}, {Bulik}, {Bulten}, {Buonanno}, {Buscicchio},
  {Buskulic}, {Byer}, {Cabero}, {Cadonati}, {Cagnoli}, {Cahillane}, {Bustillo},
  {Callaghan}, {Callister}, {Calloni}, {Camp}, {Canepa}, {Cannon}, {Cao},
  {Cao}, {Carapella}, {Carbognani}, {Caride}, {Carney}, {Carullo}, {Diaz},
  {Casentini}, {Casta{\~n}eda}, {Caudill}, {Cavagli{\`a}}, {Cavalier},
  {Cavalieri}, {Cella}, {Cerd{\'a}-Dur{\'a}n}, {Cesarini}, {Chaibi},
  {Chakravarti}, {Chan}, {Chan}, {Chao}, {Charlton}, {Chase},
  {Chassande-Mottin}, {Chatterjee}, {Chaturvedi}, {Chatziioannou}, {Chen},
  {Chen}, {Chen}, {Cheng}, {Cheong}, {Chia}, {Chiadini}, {Chierici},
  {Chincarini}, {Chiummo}, {Cho}, {Cho}, {Cho}, {Christensen}, {Chu}, {Chua},
  {Chung}, {Chung}, {Ciani}, {Ciecielag}, {Cie{\'s}lar}, {Ciobanu}, {Ciolfi},
  {Cipriano}, {Cirone}, {Clara}, {Clark}, {Clearwater}, {Clesse}, {Cleva},
  {Coccia}, {Cohadon}, {Cohen}, {Colleoni}, {Collette}, {Collins}, {Colpi},
  {Constancio}, {Conti}, {Cooper}, {Corban}, {Corbitt}, {Cordero-Carri{\'o}n},
  {Corezzi}, {Corley}, {Cornish}, {Corre}, {Corsi}, {Cortese}, {Costa},
  {Cotesta}, {Coughlin}, {Coughlin}, {Coulon}, {Countryman}, {Couvares},
  {Covas}, {Coward}, {Cowart}, {Coyne}, {Coyne}, {Creighton}, {Creighton},
  {Cripe}, {Croquette}, {Crowder}, {Cudell}, {Cullen}, {Cumming}, {Cummings},
  {Cunningham}, {Cuoco}, {Curylo}, {Canton}, {D{\'a}lya}, {Dana},
  {Daneshgaran-Bajastani}, {D'Angelo}, {Danilishin}, {D'Antonio}, {Danzmann},
  {Darsow-Fromm}, {Dasgupta}, {Datrier}, {Dattilo}, {Dave}, {Davier}, {Davies},
  {Davis}, {Daw}, {DeBra}, {Deenadayalan}, {Degallaix}, {De Laurentis},
  {Del{\'e}glise}, {Delfavero}, {De Lillo}, {Del Pozzo}, {DeMarchi},
  {D'Emilio}, {Demos}, {Dent}, {De Pietri}, {De Rosa}, {De Rossi}, {DeSalvo},
  {de Varona}, {Dhurandhar}, {D{\'\i}az}, {Diaz-Ortiz}, {Dietrich}, {Di Fiore},
  {Di Fronzo}, {Di Giorgio}, {Di Giovanni}, {Di Giovanni}, {Di Girolamo}, {Di
  Lieto}, {Ding}, {Di Pace}, {Di Palma}, {Di Renzo}, {Divakarla}, {Dmitriev},
  {Doctor}, {Donovan}, {Dooley}, {Doravari}, {Dorrington}, {Downes}, {Drago},
  {Driggers}, {Du}, {Ducoin}, {Dupej}, {Durante}, {D'Urso}, {Dwyer}, {Easter},
  {Eddolls}, {Edelman}, {Edo}, {Edy}, {Effler}, {Ehrens}, {Eichholz},
  {Eikenberry}, {Eisenmann}, {Eisenstein}, {Ejlli}, {Errico}, {Essick},
  {Estelles}, {Estevez}, {Etienne}, {Etzel}, {Evans}, {Evans}, {Ewing},
  {Fafone}, {Fairhurst}, {Fan}, {Farinon}, {Farr}, {Farr}, {Fauchon-Jones},
  {Favata}, {Fays}, {Fazio}, {Feicht}, {Fejer}, {Feng}, {Fenyvesi}, {Ferguson},
  {Fernandez-Galiana}, {Ferrante}, {Ferreira}, {Ferreira}, {Fidecaro}, {Fiori},
  {Fiorucci}, {Fishbach}, {Fisher}, {Fittipaldi}, {Fitz-Axen}, {Fiumara},
  {Flaminio}, {Floden}, {Flynn}, {Fong}, {Font}, {Forsyth}, {Fournier},
  {Frasca}, {Frasconi}, {Frei}, {Freise}, {Frey}, {Frey}, {Fritschel},
  {Frolov}, {Fronz{\`e}}, {Fulda}, {Fyffe}, {Gabbard}, {Gadre}, {Gaebel},
  {Gair}, {Galaudage}, {Ganapathy}, {Gaonkar}, {Garc{\'\i}a-Quir{\'o}s},
  {Garufi}, {Gateley}, {Gaudio}, {Gayathri}, {Gemme}, {Genin}, {Gennai},
  {George}, {George}, {Gergely}, {Ghonge}, {Ghosh}, {Ghosh}, {Ghosh},
  {Giacomazzo}, {Giaime}, {Giardina}, {Gibson}, {Gier}, {Gill}, {Glanzer},
  {Gniesmer}, {Godwin}, {Goetz}, {Goetz}, {Gohlke}, {Goncharov},
  {Gonz{\'a}lez}, {Gopakumar}, {Gossan}, {Gosselin}, {Gouaty}, {Grace},
  {Grado}, {Granata}, {Grant}, {Gras}, {Grassia}, {Gray}, {Gray}, {Greco},
  {Green}, {Green}, {Gretarsson}, {Griggs}, {Grignani}, {Grimaldi}, {Grimm},
  {Grote}, {Grunewald}, {Gruning}, {Guidi}, {Guimaraes}, {Guix{\'e}}, {Gulati},
  {Guo}, {Gupta}, {Gupta}, {Gupta}, {Gustafson}, {Gustafson}, {Haegel},
  {Halim}, {Hall}, {Hamilton}, {Hammond}, {Haney}, {Hanke}, {Hanks}, {Hanna},
  {Hannam}, {Hannuksela}, {Hansen}, {Hanson}, {Harder}, {Hardwick}, {Haris},
  {Harms}, {Harry}, {Harry}, {Hasskew}, {Haster}, {Haughian}, {Hayes}, {Healy},
  {Heidmann}, {Heintze}, {Heinze}, {Heitmann}, {Hellman}, {Hello}, {Hemming},
  {Hendry}, {Heng}, {Hennes}, {Hennig}, {Heurs}, {Hild}, {Hinderer}, {Hoback},
  {Hochheim}, {Hofgard}, {Hofman}, {Holgado}, {Holland}, {Holt}, {Holz},
  {Hopkins}, {Horst}, {Hough}, {Howell}, {Hoy}, {Huang}, {H{\"u}bner},
  {Huerta}, {Huet}, {Hughey}, {Hui}, {Husa}, {Huttner}, {Huxford},
  {Huynh-Dinh}, {Idzkowski}, {Iess}, {Inchauspe}, {Ingram}, {Intini}, {Isac},
  {Isi}, {Iyer}, {Jacqmin}, {Jadhav}, {Jadhav}, {James}, {Jani}, {Janthalur},
  {Jaranowski}, {Jariwala}, {Jaume}, {Jenkins}, {Jiang}, {Johns},
  {Johnson-McDaniel}, {Jones}, {Jones}, {Jones}, {Jones}, {Jones}, {Jonker},
  {Ju}, {Junker}, {Kalaghatgi}, {Kalogera}, {Kamai}, {Kandhasamy}, {Kang},
  {Kanner}, {Kapadia}, {Karki}, {Kashyap}, {Kasprzack}, {Kastaun},
  {Katsanevas}, {Katsavounidis}, {Katzman}, {Kaufer}, {Kawabe},
  {K{\'e}f{\'e}lian}, {Keitel}, {Keivani}, {Kennedy}, {Key}, {Khadka},
  {Khalili}, {Khan}, {Khan}, {Khan}, {Khazanov}, {Khetan}, {Khursheed},
  {Kijbunchoo}, {Kim}, {Kim}, {Kim}, {Kim}, {Kim}, {Kim}, {Kim}, {Kimball},
  {King}, {Kinley-Hanlon}, {Kirchhoff}, {Kissel}, {Kleybolte}, {Klimenko},
  {Knowles}, {Knyazev}, {Koch}, {Koehlenbeck}, {Koekoek}, {Koley},
  {Kondrashov}, {Kontos}, {Koper}, {Korobko}, {Korth}, {Kovalam}, {Kozak},
  {Kringel}, {Krishnendu}, {Kr{\'o}lak}, {Krupinski}, {Kuehn}, {Kumar},
  {Kumar}, {Kumar}, {Kumar}, {Kumar}, {Kuo}, {Kutynia}, {Lackey}, {Laghi},
  {Lalande}, {Lam}, {Lamberts}, {Landry}, {Lane}, {Lang}, {Lange}, {Lantz},
  {Lanza}, {La Rosa}, {Lartaux-Vollard}, {Lasky}, {Laxen}, {Lazzarini},
  {Lazzaro}, {Leaci}, {Leavey}, {Lecoeuche}, {Lee}, {Lee}, {Lee}, {Lee}, {Lee},
  {Lehmann}, {Leroy}, {Letendre}, {Levin}, {Li}, {Li}, {li}, {Li}, {Li},
  {Linde}, {Linker}, {Linley}, {Littenberg}, {Liu}, {Liu},
  {Llorens-Monteagudo}, {Lo}, {Lockwood}, {London}, {Longo}, {Lorenzini},
  {Loriette}, {Lormand}, {Losurdo}, {Lough}, {Lousto}, {Lovelace}, {L{\"u}ck},
  {Lumaca}, {Lundgren}, {Ma}, {Macas}, {Macfoy}, {MacInnis}, {Macleod},
  {MacMillan}, {Macquet}, {Hernandez}, {Maga{\~n}a-Sandoval}, {Magee},
  {Majorana}, {Maksimovic}, {Malik}, {Man}, {Mandic}, {Mangano}, {Mansell},
  {Manske}, {Mantovani}, {Mapelli}, {Marchesoni}, {Marion}, {M{\'a}rka},
  {M{\'a}rka}, {Markakis}, {Markosyan}, {Markowitz}, {Maros}, {Marquina},
  {Marsat}, {Martelli}, {Martin}, {Martin}, {Martinez}, {Martynov},
  {Masalehdan}, {Mason}, {Massera}, {Masserot}, {Massinger}, {Masso-Reid},
  {Mastrogiovanni}, {Matas}, {Matichard}, {Mavalvala}, {Maynard}, {McCann},
  {McCarthy}, {McClelland}, {McCormick}, {McCuller}, {McGuire}, {McIsaac},
  {McIver}, {McManus}, {McRae}, {McWilliams}, {Meacher}, {Meadors}, {Mehmet},
  {Mehta}, {Villa}, {Melatos}, {Mendell}, {Mercer}, {Mereni}, {Merfeld},
  {Merilh}, {Merritt}, {Merzougui}, {Meshkov}, {Messenger}, {Messick},
  {Metzdorff}, {Meyers}, {Meylahn}, {Mhaske}, {Miani}, {Miao}, {Michaloliakos},
  {Michel}, {Middleton}, {Milano}, {Miller}, {Millhouse}, {Mills}, {Milotti},
  {Milovich-Goff}, {Minazzoli}, {Minenkov}, {Mishkin}, {Mishra}, {Mistry},
  {Mitra}, {Mitrofanov}, {Mitselmakher}, {Mittleman}, {Mo}, {Mogushi},
  {Mohapatra}, {Mohite}, {Molina-Ruiz}, {Mondin}, {Montani}, {Moore}, {Moraru},
  {Morawski}, {Moreno}, {Morisaki}, {Mours}, {Mow-Lowry}, {Mozzon},
  {Muciaccia}, {Mukherjee}, {Mukherjee}, {Mukherjee}, {Mukherjee}, {Mukund},
  {Mullavey}, {Munch}, {Mu{\~n}iz}, {Murray}, {Nagar}, {Nardecchia},
  {Naticchioni}, {Nayak}, {Neil}, {Neilson}, {Nelemans}, {Nelson}, {Nery},
  {Neunzert}, {Ng}, {Ng}, {Nguyen}, {Nguyen}, {Nichols}, {Nichols}, {Nissanke},
  {Nocera}, {Noh}, {North}, {Nothard}, {Nuttall}, {Oberling}, {O'Brien},
  {Oganesyan}, {Ogin}, {Oh}, {Oh}, {Ohme}, {Ohta}, {Okada}, {Oliver},
  {Olivetto}, {Oppermann}, {Oram}, {O'Reilly}, {Ormiston}, {Ortega},
  {O'Shaughnessy}, {Ossokine}, {Osthelder}, {Ottaway}, {Overmier}, {Owen},
  {Pace}, {Pagano}, {Page}, {Pagliaroli}, {Pai}, {Pai}, {Palamos}, {Palashov},
  {Palomba}, {Pan}, {Panda}, {Pang}, {Pankow}, {Pannarale}, {Pant}, {Paoletti},
  {Paoli}, {Parida}, {Parker}, {Pascucci}, {Pasqualetti}, {Passaquieti},
  {Passuello}, {Patricelli}, {Payne}, {Pearlstone}, {Pechsiri}, {Pedersen},
  {Pedraza}, {Pele}, {Penn}, {Perego}, {Perez}, {P{\'e}rigois}, {Perreca},
  {Perri{\`e}s}, {Petermann}, {Pfeiffer}, {Phelps}, {Phukon}, {Piccinni},
  {Pichot}, {Piendibene}, {Piergiovanni}, {Pierro}, {Pillant}, {Pinard},
  {Pinto}, {Piotrzkowski}, {Pirello}, {Pitkin}, {Plastino}, {Poggiani}, {Pong},
  {Ponrathnam}, {Popolizio}, {Porter}, {Powell}, {Prajapati}, {Prasai},
  {Prasanna}, {Pratten}, {Prestegard}, {Principe}, {Prodi}, {Prokhorov},
  {Punturo}, {Puppo}, {P{\"u}rrer}, {Qi}, {Quetschke}, {Quinonez}, {Raab},
  {Raaijmakers}, {Radkins}, {Radulesco}, {Raffai}, {Rafferty}, {Raja}, {Rajan},
  {Rajbhandari}, {Rakhmanov}, {Ramirez}, {Ramos-Buades}, {Rana}, {Rao},
  {Rapagnani}, {Raymond}, {Razzano}, {Read}, {Regimbau}, {Rei}, {Reid},
  {Reitze}, {Rettegno}, {Ricci}, {Richardson}, {Richardson}, {Ricker},
  {Riemenschneider}, {Riles}, {Rizzo}, {Robertson}, {Robinet}, {Rocchi},
  {Rodriguez-Soto}, {Rolland}, {Rollins}, {Roma}, {Romanelli}, {Romano},
  {Romel}, {Romero-Shaw}, {Romie}, {Rose}, {Rose}, {Rose}, {Rosi{\'n}ska},
  {Rosofsky}, {Ross}, {Rowan}, {Rowlinson}, {Roy}, {Roy}, {Roy}, {Ruggi},
  {Rutins}, {Ryan}, {Sachdev}, {Sadecki}, {Sakellariadou}, {Salafia},
  {Salconi}, {Saleem}, {Samajdar}, {Sanchez}, {Sanchez}, {Sanchis-Gual},
  {Sanders}, {Santiago}, {Santos}, {Sarin}, {Sassolas}, {Sathyaprakash},
  {Sauter}, {Savage}, {Savant}, {Sawant}, {Sayah}, {Schaetzl}, {Schale},
  {Scheel}, {Scheuer}, {Schmidt}, {Schnabel}, {Schofield}, {Sch{\"o}nbeck},
  {Schreiber}, {Schulte}, {Schutz}, {Schwarm}, {Schwartz}, {Scott}, {Scott},
  {Seidel}, {Sellers}, {Sengupta}, {Sennett}, {Sentenac}, {Sequino}, {Sergeev},
  {Setyawati}, {Shaddock}, {Shaffer}, {Shahriar}, {Sharifi}, {Sharma},
  {Sharma}, {Shawhan}, {Shen}, {Shikauchi}, {Shink}, {Shoemaker}, {Shoemaker},
  {Shukla}, {ShyamSundar}, {Siellez}, {Sieniawska}, {Sigg}, {Singer}, {Singh},
  {Singh}, {Singha}, {Singhal}, {Sintes}, {Sipala}, {Skliris}, {Slagmolen},
  {Slaven-Blair}, {Smetana}, {Smith}, {Smith}, {Somala}, {Son}, {Soni},
  {Sorazu}, {Sordini}, {Sorrentino}, {Souradeep}, {Sowell}, {Spencer}, {Spera},
  {Srivastava}, {Srivastava}, {Staats}, {Stachie}, {Standke}, {Steer},
  {Steinke}, {Steinlechner}, {Steinlechner}, {Steinmeyer}, {Stevenson},
  {Stocks}, {Stops}, {Stover}, {Strain}, {Stratta}, {Strunk}, {Sturani},
  {Stuver}, {Sudhagar}, {Sudhir}, {Summerscales}, {Sun}, {Sunil}, {Sur},
  {Suresh}, {Sutton}, {Swinkels}, {Szczepa{\'n}czyk}, {Tacca}, {Tait},
  {Talbot}, {Tanasijczuk}, {Tanner}, {Tao}, {T{\'a}pai}, {Tapia}, {San Martin},
  {Tasson}, {Taylor}, {Tenorio}, {Terkowski}, {Thirugnanasambandam}, {Thomas},
  {Thomas}, {Thompson}, {Thondapu}, {Thorne}, {Thrane}, {Tinsman}, {Saravanan},
  {Tiwari}, {Tiwari}, {Tiwari}, {Toland}, {Tonelli}, {Tornasi},
  {Torres-Forn{\'e}}, {Torrie}, {Tosta e Melo}, {T{\"o}yr{\"a}}, {Trail},
  {Travasso}, {Traylor}, {Tringali}, {Tripathee}, {Trovato}, {Trudeau},
  {Tsang}, {Tse}, {Tso}, {Tsukada}, {Tsuna}, {Tsutsui}, {Turconi}, {Ubhi},
  {Udall}, {Ueno}, {Ugolini}, {Unnikrishnan}, {Urban}, {Usman}, {U0tina},
  {Vahlbruch}, {Vajente}, {Valdes}, {Valentini}, {van Bakel}, {van Beuzekom},
  {van den Brand}, {Van Den Broeck}, {Vander-Hyde}, {van der Schaaf}, {Van
  Heijningen}, {van Veggel}, {Vardaro}, {Varma}, {Vass}, {Vas{\'u}th},
  {Vecchio}, {Vedovato}, {Veitch}, {Veitch}, {Venkateswara}, {Venugopalan},
  {Verkindt}, {Veske}, {Vetrano}, {Vicer{\'e}}, {Viets}, {Vinciguerra}, {Vine},
  {Vinet}, {Vitale}, {Vivanco}, {Vo}, {Vocca}, {Vorvick}, {Vyatchanin}, {Wade},
  {Wade}, {Wade}, {Walet}, {Walker}, {Wallace}, {Wallace}, {Walsh}, {Wang},
  {Wang}, {Wang}, {Ward}, {Warden}, {Warner}, {Was}, {Watchi}, {Weaver}, {Wei},
  {Weinert}, {Weinstein}, {Weiss}, {Wellmann}, {Wen}, {We{\ss}els},
  {Westhouse}, {Wette}, {Whelan}, {Whiting}, {Whittle}, {Wilken}, {Williams},
  {Willis}, {Willke}, {Winkler}, {Wipf}, {Wittel}, {Woan}, {Woehler},
  {Wofford}, {Wong}, {Wright}, {Wu}, {Wysocki}, {Xiao}, {Yamamoto}, {Yang},
  {Yang}, {Yang}, {Yap}, {Yazback}, {Yeeles}, {Yu}, {Yu}, {Yuen},
  {Zadro{\.z}ny}, {Zadro{\.z}ny}, {Zanolin}, {Zelenova}, {Zendri}, {Zevin},
  {Zhang}, {Zhang}, {Zhang}, {Zhao}, {Zhao}, {Zhou}, {Zhou}, {Zhu},
  {Zimmerman}, {Zlochower}, {Zucker}, {Zweizig}, {LIGO Scientific
  Collaboration}, \& {Virgo Collaboration}}]{190521_imp_2020}
---. 2020{\natexlab{d}}, \apjl, 900, L13, \dodoi{10.3847/2041-8213/aba493}

\bibitem[{{Acernese} {et~al.}(2015){Acernese}, {Agathos}, {Agatsuma}, {Aisa},
  {Allemandou}, {Allocca}, {Amarni}, {Astone}, {Balestri}, {Ballardin},
  {Barone}, {Baronick}, {Barsuglia}, {Basti}, {Basti}, {Bauer}, {Bavigadda},
  {Bejger}, {Beker}, {Belczynski}, {Bersanetti}, {Bertolini}, {Bitossi},
  {Bizouard}, {Bloemen}, {Blom}, {Boer}, {Bogaert}, {Bondi}, {Bondu},
  {Bonelli}, {Bonnand}, {Boschi}, {Bosi}, {Bouedo}, {Bradaschia}, {Branchesi},
  {Briant}, {Brillet}, {Brisson}, {Bulik}, {Bulten}, {Buskulic}, {Buy},
  {Cagnoli}, {Calloni}, {Campeggi}, {Canuel}, {Carbognani}, {Cavalier},
  {Cavalieri}, {Cella}, {Cesarini}, {Chassande-Mottin}, \&
  {Chincarini}}]{virgo15}
{Acernese}, F., {Agathos}, M., {Agatsuma}, K., {et~al.} 2015, Classical and
  Quantum Gravity, 32, 024001, \dodoi{10.1088/0264-9381/32/2/024001}

\bibitem[{{Ackley} {et~al.}(2019){Ackley}, {Mata-Sanchez}, {Mong}, {Cutter},
  {Ulaczyk}, {Lyman}, {Steeghs}, {Ramsay}, {Galloway}, {Makrygianni},
  {Kennedy}, {Obradovic}, {Dyer}, {Dhillon}, {O'Brien}, {Pollacco}, {Thrane},
  {Poshyachinda}, {Palle}, {Wiersema}, {Marsh}, {West}, {Gompertz}, {Stanway},
  {Casey}, {Brown}, {Rol}, {Mullaney}, {Littlefair}, {Daw}, {Maund},
  {Starling}, {Eyles}, {Tooke}, {Sawangwit}, {Mkrtichian}, {Awiphan},
  {Aukkaravittayapun}, {Irawati}, {Breton}, {Heikkila}, {Kotak}, \&
  {Nuttall}}]{gcn25654}
{Ackley}, K., {Mata-Sanchez}, D., {Mong}, Y.~L., {et~al.} 2019, GRB Coordinates
  Network, 25654, 1

\bibitem[{{Ackley} {et~al.}(2020){Ackley}, {Amati}, {Barbieri}, {Bauer},
  {Benetti}, {Bernardini}, {Bhirombhakdi}, {Botticella}, {Branchesi},
  {Brocato}, {Bruun}, {Bulla}, {Campana}, {Cappellaro}, {Castro-Tirado},
  {Chambers}, {Chaty}, {Chen}, {Ciolfi}, {Coleiro}, {Copperwheat}, {Covino},
  {Cutter}, {D'Ammando}, {D'Avanzo}, {De Cesare}, {D'Elia}, {Della Valle},
  {Denneau}, {De Pasquale}, {Dhillon}, {Dyer}, {Elias-Rosa}, {Evans},
  {Eyles-Ferris}, {Fiore}, {Fraser}, {Fruchter}, {Fynbo}, {Galbany}, {Gall},
  {Galloway}, {Getman}, {Ghirlanda}, {Gillanders}, {Gomboc}, {Gompertz},
  {Gonz{\'a}lez-Fern{\'a}ndez}, {Gonz{\'a}lez-Gait{\'a}n}, {Grado}, {Greco},
  {Gromadzki}, {Groot}, {Guti{\'e}rrez}, {Heikkil{\"a}}, {Heintz}, {Hjorth},
  {Hu}, {Huber}, {Inserra}, {Izzo}, {Japelj}, {Jerkstrand}, {Jin}, {Jonker},
  {Kankare}, {Kann}, {Kennedy}, {Kim}, {Klose}, {Kool}, {Kotak},
  {Kuncarayakti}, {Lamb}, {Leloudas}, {Levan}, {Longo}, {Lowe}, {Lyman},
  {Magnier}, {Maguire}, {Maiorano}, {Mandel}, {Mapelli}, {Mattila}, {McBrien},
  {Melandri}, {Micha{\l}owski}, {Milvang-Jensen}, {Moran}, {Nicastro},
  {Nicholl}, {Nicuesa Guelbenzu}, {Nuttal}, {Oates}, {O'Brien}, {Onori},
  {Palazzi}, {Patricelli}, {Perego}, {Torres}, {Perley}, {Pian}, {Pignata},
  {Piranomonte}, {Poshyachinda}, {Possenti}, {Pumo}, {Quirola-V{\'a}squez},
  {Ragosta}, {Ramsay}, {Rau}, {Rest}, {Reynolds}, {Rosetti}, {Rossi},
  {Rosswog}, {Sabha}, {Sagu{\'e}s Carracedo}, {Salafia}, {Salmon},
  {Salvaterra}, {Savaglio}, {Sbordone}, {Schady}, {Schipani}, {Schultz},
  {Schweyer}, {Smartt}, {Smith}, {Smith}, {Sollerman}, {Srivastav}, {Stanway},
  {Starling}, {Steeghs}, {Stratta}, {Stubbs}, {Tanvir}, {Testa}, {Thrane},
  {Tonry}, {Turatto}, {Ulaczyk}, {van der Horst}, {Vergani}, {Walton},
  {Watson}, {Wiersema}, {Wiik}, {Wyrzykowski}, {Yang}, {Yi}, \&
  {Young}}]{ackley2020}
{Ackley}, K., {Amati}, L., {Barbieri}, C., {et~al.} 2020, \aap, 643, A113,
  \dodoi{10.1051/0004-6361/202037669}

\bibitem[{{Ahumada}(2020)}]{GCN26810}
{Ahumada}, T. 2020, GRB Coordinates Network, 26810, 1

\bibitem[{{Ahumada} {et~al.}(2020{\natexlab{a}}){Ahumada}, {Ztf}, \& {Growth
  Collaborations}}]{gcn26817}
{Ahumada}, T., {Ztf}, T., \& {Growth Collaborations}. 2020{\natexlab{a}}, GRB
  Coordinates Network, 26817, 1

\bibitem[{{Ahumada} {et~al.}(2020{\natexlab{b}}){Ahumada}, {Ztf}, \& {Growth
  Collaborations}}]{gcn26822}
---. 2020{\natexlab{b}}, GRB Coordinates Network, 26822, 1

\bibitem[{{Alam} {et~al.}(2015){Alam}, {Albareti}, {Allende Prieto}, {Anders},
  {Anderson}, {Anderton}, {Andrews}, {Armengaud}, {Aubourg}, {Bailey}, {Basu},
  {Bautista}, {Beaton}, {Beers}, {Bender}, {Berlind}, {Beutler}, {Bhardwaj},
  {Bird}, {Bizyaev}, {Blake}, {Blanton}, {Blomqvist}, {Bochanski}, {Bolton},
  {Bovy}, {Shelden Bradley}, {Brandt}, {Brauer}, {Brinkmann}, {Brown},
  {Brownstein}, {Burden}, {Burtin}, {Busca}, {Cai}, {Capozzi}, {Carnero
  Rosell}, {Carr}, {Carrera}, {Chambers}, {Chaplin}, {Chen}, {Chiappini},
  {Chojnowski}, {Chuang}, {Clerc}, {Comparat}, {Covey}, {Croft}, {Cuesta},
  {Cunha}, {da Costa}, {Da Rio}, {Davenport}, {Dawson}, {De Lee}, {Delubac},
  {Deshpande}, {Dhital}, {Dutra-Ferreira}, {Dwelly}, {Ealet}, {Ebelke},
  {Edmondson}, {Eisenstein}, {Ellsworth}, {Elsworth}, {Epstein}, {Eracleous},
  {Escoffier}, {Esposito}, {Evans}, {Fan}, {Fern{\'a}ndez-Alvar}, {Feuillet},
  {Filiz Ak}, {Finley}, {Finoguenov}, {Flaherty}, {Fleming}, {Font-Ribera},
  {Foster}, {Frinchaboy}, {Galbraith-Frew}, {Garc{\'\i}a},
  {Garc{\'\i}a-Hern{\'a}ndez}, {Garc{\'\i}a P{\'e}rez}, {Gaulme}, {Ge},
  {G{\'e}nova-Santos}, {Georgakakis}, {Ghezzi}, {Gillespie}, {Girardi},
  {Goddard}, {Gontcho}, {Gonz{\'a}lez Hern{\'a}ndez}, {Grebel}, {Green},
  {Grieb}, {Grieves}, {Gunn}, {Guo}, {Harding}, {Hasselquist}, {Hawley},
  {Hayden}, {Hearty}, {Hekker}, {Ho}, {Hogg}, {Holley-Bockelmann}, {Holtzman},
  {Honscheid}, {Huber}, {Huehnerhoff}, {Ivans}, {Jiang}, {Johnson},
  {Kinemuchi}, {Kirkby}, {Kitaura}, {Klaene}, {Knapp}, {Kneib}, {Koenig},
  {Lam}, {Lan}, {Lang}, {Laurent}, {Le Goff}, {Leauthaud}, {Lee}, {Lee},
  {Licquia}, {Liu}, {Long}, {L{\'o}pez-Corredoira}, {Lorenzo-Oliveira},
  {Lucatello}, {Lundgren}, {Lupton}, {Mack}, {Mahadevan}, {Maia}, {Majewski},
  {Malanushenko}, {Malanushenko}, {Manchado}, {Manera}, {Mao}, {Maraston},
  {Marchwinski}, {Margala}, {Martell}, {Martig}, {Masters}, {Mathur},
  {McBride}, {McGehee}, {McGreer}, {McMahon}, {M{\'e}nard}, {Menzel},
  {Merloni}, {M{\'e}sz{\'a}ros}, {Miller}, {Miralda-Escud{\'e}}, {Miyatake},
  {Montero-Dorta}, {More}, {Morganson}, {Morice-Atkinson}, {Morrison},
  {Mosser}, {Muna}, {Myers}, {Nand ra}, {Newman}, {Neyrinck}, {Nguyen},
  {Nichol}, {Nidever}, {Noterdaeme}, {Nuza}, {O'Connell}, {O'Connell},
  {O'Connell}, {Ogando}, {Olmstead}, {Oravetz}, {Oravetz}, {Osumi}, {Owen},
  {Padgett}, {Padmanabhan}, {Paegert}, {Palanque-Delabrouille}, {Pan},
  {Parejko}, {P{\^a}ris}, {Park}, {Pattarakijwanich}, {Pellejero-Ibanez},
  {Pepper}, {Percival}, {P{\'e}rez-Fournon}, {Ṕrez-Ra`fols}, {Petitjean},
  {Pieri}, {Pinsonneault}, {Porto de Mello}, {Prada}, {Prakash},
  {Price-Whelan}, {Protopapas}, {Raddick}, {Rahman}, {Reid}, {Rich}, {Rix},
  {Robin}, {Rockosi}, {Rodrigues}, {Rodr{\'\i}guez-Torres}, {Roe}, {Ross},
  {Ross}, {Rossi}, {Ruan}, {Rubi{\~n}o-Mart{\'\i}n}, {Rykoff},
  {Salazar-Albornoz}, {Salvato}, {Samushia}, {S{\'a}nchez}, {Santiago},
  {Sayres}, {Schiavon}, {Schlegel}, {Schmidt}, {Schneider}, {Schultheis},
  {Schwope}, {Sc{\'o}ccola}, {Scott}, {Sellgren}, {Seo}, {Serenelli}, {Shane},
  {Shen}, {Shetrone}, {Shu}, {Silva Aguirre}, {Sivarani}, {Skrutskie},
  {Slosar}, {Smith}, {Sobreira}, {Souto}, {Stassun}, {Steinmetz}, {Stello},
  {Strauss}, {Streblyanska}, {Suzuki}, {Swanson}, {Tan}, {Tayar}, {Terrien},
  {Thakar}, {Thomas}, {Thomas}, {Thompson}, {Tinker}, {Tojeiro}, {Troup},
  {Vargas-Maga{\~n}a}, {Vazquez}, {Verde}, {Viel}, {Vogt}, {Wake}, {Wang},
  {Weaver}, {Weinberg}, {Weiner}, {White}, {Wilson}, {Wisniewski},
  {Wood-Vasey}, {Ye`che}, {York}, {Zakamska}, {Zamora}, {Zasowski}, {Zehavi},
  {Zhao}, {Zheng}, {Zhou}, {Zhou}, {Zou}, \& {Zhu}}]{SDSSDR12}
{Alam}, S., {Albareti}, F.~D., {Allende Prieto}, C., {et~al.} 2015, \apjs, 219,
  12, \dodoi{10.1088/0067-0049/219/1/12}

\bibitem[{{Alexander} {et~al.}(2018){Alexander}, {Margutti}, {Blanchard},
  {Fong}, {Berger}, {Hajela}, {Eftekhari}, {Chornock}, {Cowperthwaite},
  {Giannios}, {Guidorzi}, {Kathirgamaraju}, {MacFadyen}, {Metzger}, {Nicholl},
  {Sironi}, {Villar}, {Williams}, {Xie}, \& {Zrake}}]{Alexander2018}
{Alexander}, K.~D., {Margutti}, R., {Blanchard}, P.~K., {et~al.} 2018, \apjl,
  863, L18, \dodoi{10.3847/2041-8213/aad637}

\bibitem[{{Almualla} {et~al.}(2020){Almualla}, {Anand}, {Coughlin}, {Dietrich},
  {Guessoum}, {Sagu{\'e}s Carracedo}, {Ahumada}, {Andreoni}, {Antier}, {Bellm},
  {Bulla}, \& {Singer}}]{Almualla2020}
{Almualla}, M., {Anand}, S., {Coughlin}, M.~W., {et~al.} 2020, arXiv e-prints,
  arXiv:2011.10421.
\newblock \doarXiv{2011.10421}

\bibitem[{{Anand} {et~al.}(2020{\natexlab{a}}){Anand}, {Zwicky Transient
  Facility (Ztf) Collaboration}, \& {Global Relay Of Observatories Watching
  Transients Happen (Growth) Collaboration}}]{gcn26767}
{Anand}, S., {Zwicky Transient Facility (Ztf) Collaboration}, \& {Global Relay
  Of Observatories Watching Transients Happen (Growth) Collaboration}.
  2020{\natexlab{a}}, GRB Coordinates Network, 26767, 1

\bibitem[{{Anand} {et~al.}(2019){Anand}, {Kasliwal}, {Coughlin}, {Ahumada},
  {Perley}, {Bhalerao}, {Goldstein}, {Singer}, {Andreoni}, {Duev}, {Cenko},
  {Bellm}, {de}, {Ho}, {Kumar}, {Waratkar}, {Copperwheat}, {Cunningham},
  {Ghosh}, {Goobar}, {Kaplan}, {Sollerman}, {Bloom}, {Bulla}, {Kawai}, {Yatsu},
  {Murata}, {Hanayama}, {Horiuchi}, {Anupama}, \& {Helou}}]{gcn24311}
{Anand}, S., {Kasliwal}, M.~M., {Coughlin}, M.~W., {et~al.} 2019, GRB
  Coordinates Network, 24311, 1

\bibitem[{{Anand} {et~al.}(2020{\natexlab{b}}){Anand}, {Andreoni}, {Goldstein},
  {Kasliwal}, {Ahumada}, {Barnes}, {Bloom}, {Bulla}, {Cenko}, {Cooke},
  {Coughlin}, {Nugent}, \& {Singer}}]{Anand20}
{Anand}, S., {Andreoni}, I., {Goldstein}, D.~A., {et~al.} 2020{\natexlab{b}},
  arXiv e-prints, arXiv:2003.05516.
\newblock \doarXiv{2003.05516}

\bibitem[{{Anand} {et~al.}(2020{\natexlab{c}}){Anand}, {Coughlin}, {Kasliwal},
  {Bulla}, {Ahumada}, {Sagu{\'e}s Carracedo}, {Almualla}, {Andreoni}, {Stein},
  {Foucart}, {Singer}, {Sollerman}, {Bellm}, {Bolin}, {Caballero-Garc{\'\i}a},
  {Castro-Tirado}, {Cenko}, {De}, {Dekany}, {Duev}, {Feeney}, {Fremling},
  {Goldstein}, {Golkhou}, {Graham}, {Guessoum}, {Hankins}, {Hu}, {Kong},
  {Kool}, {Kulkarni}, {Kumar}, {Laher}, {Masci}, {Mr{\'o}z}, {Nissanke},
  {Porter}, {Reusch}, {Riddle}, {Rosnet}, {Rusholme}, {Serabyn},
  {S{\'a}nchez-Ram{\'\i}rez}, {Rigault}, {Shupe}, {Smith}, {Soumagnac},
  {Walters}, \& {Valeev}}]{Anand2020}
{Anand}, S., {Coughlin}, M.~W., {Kasliwal}, M.~M., {et~al.} 2020{\natexlab{c}},
  Nature Astronomy, \dodoi{10.1038/s41550-020-1183-3}

\bibitem[{{Andreoni}(2020{\natexlab{a}})}]{gcn26741}
{Andreoni}, I. 2020{\natexlab{a}}, GRB Coordinates Network, 26741, 1

\bibitem[{{Andreoni}(2020{\natexlab{b}})}]{gcn26806}
---. 2020{\natexlab{b}}, GRB Coordinates Network, 26806, 1

\bibitem[{{Andreoni} {et~al.}(2019{\natexlab{a}}){Andreoni}, {Anand}, \&
  {Kasliwal}}]{gcn24349}
{Andreoni}, I., {Anand}, S., \& {Kasliwal}, M. 2019{\natexlab{a}}, GRB
  Coordinates Network, 24349, 1

\bibitem[{{Andreoni} {et~al.}(2019{\natexlab{b}}){Andreoni}, {Anand},
  {Kasliwal}, {Coughlin}, \& {Global Relay of Transients Watching Observatories
  Happen Collaboration}}]{andreoni19d}
{Andreoni}, I., {Anand}, S., {Kasliwal}, M.~M., {Coughlin}, M.~M., \& {Global
  Relay of Transients Watching Observatories Happen Collaboration}.
  2019{\natexlab{b}}, GRB Coordinates Network, 26432, 1

\bibitem[{{Andreoni} \& {Bellm}(2019{\natexlab{a}})}]{gcn24356}
{Andreoni}, I., \& {Bellm}, E. 2019{\natexlab{a}}, GRB Coordinates Network,
  24356, 1

\bibitem[{{Andreoni} \& {Bellm}(2019{\natexlab{b}})}]{gcn24357}
---. 2019{\natexlab{b}}, GRB Coordinates Network, 24357, 1

\bibitem[{{Andreoni} {et~al.}(2019{\natexlab{c}}){Andreoni}, {Cenko}, {Masci},
  \& {Graham}}]{gcn24302}
{Andreoni}, I., {Cenko}, S.~B., {Masci}, F., \& {Graham}, M.
  2019{\natexlab{c}}, GRB Coordinates Network, 24302, 1

\bibitem[{{Andreoni} {et~al.}(2019{\natexlab{d}}){Andreoni}, {Goldstein},
  {Anand}, {Coughlin}, {Singer}, {Ahumada}, {Medford}, {Kool}, {Webb}, {Bulla},
  {Bloom}, {Kasliwal}, {Nugent}, {Bagdasaryan}, {Barnes}, {Cook}, {Cooke},
  {Duev}, {Fremling}, {Gatkine}, {Golkhou}, {Kong}, {Mahabal},
  {Mart{\'\i}nez-Palomera}, {Tao}, \& {Zhang}}]{Andreoni2019_190510}
{Andreoni}, I., {Goldstein}, D.~A., {Anand}, S., {et~al.} 2019{\natexlab{d}},
  \apjl, 881, L16, \dodoi{10.3847/2041-8213/ab3399}

\bibitem[{{Andreoni} {et~al.}(2019{\natexlab{e}}){Andreoni}, {Goldstein},
  {Coughlin}, {Kasliwal}, {Nugent}, {Zhang}, {Palomera}, {Anand}, {Bloom},
  {Cenko}, {Cooke}, \& {Singer}}]{gcn24268}
{Andreoni}, I., {Goldstein}, D.~A., {Coughlin}, M., {et~al.}
  2019{\natexlab{e}}, GRB Coordinates Network, 24268, 1

\bibitem[{{Andreoni} {et~al.}(2019{\natexlab{f}}){Andreoni}, {Anand},
  {Coughlin}, {De}, {Kasliwal}, {Duev}, {Bellm}, {Stein}, {Reusch}, {Cenko},
  {Graham}, {Zwicky Transient Facility Collaboration}, \& {Global Relay of
  Transients Watching Observatories Happen Collaboration}}]{gcn26416}
{Andreoni}, I., {Anand}, S., {Coughlin}, M.~W., {et~al.} 2019{\natexlab{f}},
  GRB Coordinates Network, 26416, 1

\bibitem[{{Andreoni} {et~al.}(2019{\natexlab{g}}){Andreoni}, {Anand}, {Bellm},
  {Kool}, {Perley}, {Singer}, {Coughlin}, {Ahumada}, {Kumar}, \&
  {Kasliwal}}]{andreoni19c}
{Andreoni}, I., {Anand}, S., {Bellm}, E., {et~al.} 2019{\natexlab{g}}, GRB
  Coordinates Network, 26424, 1

\bibitem[{{Andreoni} {et~al.}(2020){Andreoni}, {Goldstein}, {Kasliwal},
  {Nugent}, {Zhou}, {Newman}, {Bulla}, {Foucart}, {Hotokezaka}, {Nakar},
  {Nissanke}, {Raaijmakers}, {Bloom}, {De}, {Jencson}, {Ward}, {Ahumada},
  {Anand}, {Buckley}, {Caballero-Garc{\'\i}a}, {Castro-Tirado}, \&
  {Copperwheat}}]{Andreoni20}
{Andreoni}, I., {Goldstein}, D.~A., {Kasliwal}, M.~M., {et~al.} 2020, \apj,
  890, 131, \dodoi{10.3847/1538-4357/ab6a1b}

\bibitem[{{Antier} {et~al.}(2020){Antier}, {Agayeva}, {Almualla}, {Awiphan},
  {Baransky}, {Barynova}, {Beradze}, {Bla{\v{z}}ek}, {Bo{\"e}r}, {Burkhonov},
  {Christensen}, {Coleiro}, {Corre}, {Coughlin}, {Crisp}, {Dietrich}, {Ducoin},
  {Duverne}, {Marchal-Duval}, {Gendre}, {Gokuldass}, {Eggenstein}, {Eymar},
  {Hello}, {Howell}, {Ismailov}, {Kann}, {Karpov}, {Klotz}, {Kochiashvili},
  {Lachaud}, {Leroy}, {Lin}, {Li}, {Ma{\v{s}}ek}, {Mo}, {Menard}, {Morris},
  {Noysena}, {Orange}, {Prouza}, {Rattanamala}, {Sadibekova}, {Saint-Gelais},
  {Serrau}, {Simon}, {Stachie}, {Th{\"o}ne}, {Tillayev}, {Turpin}, {de Ugarte
  Postigo}, {Vasylenko}, {Vidadi}, {Was}, {Wang}, {Zhang}, {Zhang}, \&
  {Zhang}}]{antier20}
{Antier}, S., {Agayeva}, S., {Almualla}, M., {et~al.} 2020, \mnras,
  \dodoi{10.1093/mnras/staa1846}

\bibitem[{{Arcavi} {et~al.}(2019{\natexlab{a}}){Arcavi}, {McCully},
  {Hiramatsu}, {Howell}, {Burke}, \& {Pellegrino}}]{gcn24249}
{Arcavi}, I., {McCully}, C., {Hiramatsu}, D., {et~al.} 2019{\natexlab{a}}, GRB
  Coordinates Network, 24249, 1

\bibitem[{{Arcavi} {et~al.}(2019{\natexlab{b}}){Arcavi}, {McCully},
  {Hiramatsu}, {Howell}, {Burke}, \& {Pellegrino}}]{gcn24251}
---. 2019{\natexlab{b}}, GRB Coordinates Network, 24251, 1

\bibitem[{{Arcavi} {et~al.}(2017){Arcavi}, {Hosseinzadeh}, {Howell}, {McCully},
  {Poznanski}, {Kasen}, {Barnes}, {Zaltzman}, {Vasylyev}, {Maoz}, \&
  {Valenti}}]{Arcavi+17}
{Arcavi}, I., {Hosseinzadeh}, G., {Howell}, D.~A., {et~al.} 2017, \nat, 551,
  64, \dodoi{10.1038/nature24291}

\bibitem[{{Ascenzi} {et~al.}(2019){Ascenzi}, {Coughlin}, {Dietrich}, {Foley},
  {Ramirez-Ruiz}, {Piranomonte}, {Mockler}, {Murguia-Berthier}, {Fryer},
  {Lloyd-Ronning}, \& {Rosswog}}]{Ascenzi+19}
{Ascenzi}, S., {Coughlin}, M.~W., {Dietrich}, T., {et~al.} 2019, \mnras, 486,
  672, \dodoi{10.1093/mnras/stz891}

\bibitem[{{Ashton} {et~al.}(2020){Ashton}, {Ackley}, {Maga{\~n}a Hernand ez},
  \& {Piotrzkowski}}]{Ashton2020}
{Ashton}, G., {Ackley}, K., {Maga{\~n}a Hernand ez}, I., \& {Piotrzkowski}, B.
  2020, arXiv e-prints, arXiv:2009.12346.
\newblock \doarXiv{2009.12346}

\bibitem[{{Astropy Collaboration} {et~al.}(2013){Astropy Collaboration},
  {Robitaille}, {Tollerud}, {Greenfield}, {Droettboom}, {Bray}, {Aldcroft},
  {Davis}, {Ginsburg}, {Price-Whelan}, {Kerzendorf}, {Conley}, {Crighton},
  {Barbary}, {Muna}, {Ferguson}, {Grollier}, {Parikh}, {Nair}, {Unther},
  {Deil}, {Woillez}, {Conseil}, {Kramer}, {Turner}, {Singer}, {Fox}, {Weaver},
  {Zabalza}, {Edwards}, {Azalee Bostroem}, {Burke}, {Casey}, {Crawford},
  {Dencheva}, {Ely}, {Jenness}, {Labrie}, {Lim}, {Pierfederici}, {Pontzen},
  {Ptak}, {Refsdal}, {Servillat}, \& {Streicher}}]{2013A&A...558A..33A}
{Astropy Collaboration}, {Robitaille}, T.~P., {Tollerud}, E.~J., {et~al.} 2013,
  \aap, 558, A33, \dodoi{10.1051/0004-6361/201322068}

\bibitem[{{Barbieri} {et~al.}(2020){Barbieri}, {Salafia}, {Colpi}, {Ghirlanda},
  \& {Perego}}]{barbieri20}
{Barbieri}, C., {Salafia}, O.~S., {Colpi}, M., {Ghirlanda}, G., \& {Perego}, A.
  2020, arXiv e-prints, arXiv:2002.09395.
\newblock \doarXiv{2002.09395}

\bibitem[{{Bennett} {et~al.}(2014){Bennett}, {Larson}, {Weiland}, \&
  {Hinshaw}}]{Bennett+14}
{Bennett}, C.~L., {Larson}, D., {Weiland}, J.~L., \& {Hinshaw}, G. 2014, \apj,
  794, 135, \dodoi{10.1088/0004-637X/794/2/135}

\bibitem[{{Bertin}(2006)}]{scamp}
{Bertin}, E. 2006, in Astronomical Society of the Pacific Conference Series,
  Vol. 351, Astronomical Data Analysis Software and Systems XV, ed.
  C.~{Gabriel}, C.~{Arviset}, D.~{Ponz}, \& S.~{Enrique}, 112

\bibitem[{{Bertin}(2010{\natexlab{a}})}]{scamp2}
{Bertin}, E. 2010{\natexlab{a}}, {SCAMP: Automatic Astrometric and Photometric
  Calibration}.
\newblock \doeprint{1010.063}

\bibitem[{{Bertin}(2010{\natexlab{b}})}]{swarp}
---. 2010{\natexlab{b}}, {SWarp: Resampling and Co-adding FITS Images
  Together}.
\newblock \doeprint{1010.068}

\bibitem[{{Bertin} \& {Arnouts}(1996)}]{Bertin1996}
{Bertin}, E., \& {Arnouts}, S. 1996, \aaps, 117, 393,
  \dodoi{10.1051/aas:1996164}

\bibitem[{{Bilicki} {et~al.}(2014){Bilicki}, {Jarrett}, {Peacock}, {Cluver}, \&
  {Steward}}]{2MPZ}
{Bilicki}, M., {Jarrett}, T.~H., {Peacock}, J.~A., {Cluver}, M.~E., \&
  {Steward}, L. 2014, \apjs, 210, 9, \dodoi{10.1088/0067-0049/210/1/9}

\bibitem[{{Breeveld} {et~al.}(2019){Breeveld}, {Kuin}, {Marshall}, {Brown},
  {Oates}, {Siegel}, {de Pasquale}, {Barthelmy}, {Lien}, {Palmer}, {Sakamoto},
  {Beardmore}, {Burrows}, {Campana}, {Cenko}, {Cusumano}, {D'Ai}, {D'Avanzo},
  {D'Elia}, {Emery}, {Evans}, {Giommi}, {Gronwall}, {Hartmann}, {Kennea},
  {Krimm}, {Meland ri}, {Mingo}, {Nousek}, {O'Brien}, {Osborne}, {Pagani},
  {Page}, {Perri}, {Racusin}, {Sbarufatti}, {Tagliaferri}, {Tohuvavohu}, \&
  {Troja}}]{gcn24296}
{Breeveld}, A.~A., {Kuin}, N.~P.~M., {Marshall}, F.~E., {et~al.} 2019, GRB
  Coordinates Network, 24296, 1

\bibitem[{{Brennan} {et~al.}(2019){Brennan}, {Killestein}, {Fraser}, {Jonker},
  {Maguire}, {Perez Torres}, \& {GW@WHT Collaboration}}]{GCN26429}
{Brennan}, S., {Killestein}, T., {Fraser}, M., {et~al.} 2019, GRB Coordinates
  Network, 26429, 1

\bibitem[{{Buckley} {et~al.}(2019{\natexlab{a}}){Buckley}, {Ciroi}, {Orio},
  {Jah}, \& {Mikolajewska}}]{gcn25903}
{Buckley}, D., {Ciroi}, S., {Orio}, M., {Jah}, S., \& {Mikolajewska}, J.
  2019{\natexlab{a}}, GRB Coordinates Network, 25903, 1

\bibitem[{{Buckley} {et~al.}(2019{\natexlab{b}}){Buckley}, {Jha}, {Cooke}, \&
  {Mogotsi}}]{gcn24205}
{Buckley}, D.~A.~H., {Jha}, S.~W., {Cooke}, J., \& {Mogotsi}, M.
  2019{\natexlab{b}}, GRB Coordinates Network, 24205, 1

\bibitem[{{Burdge} {et~al.}(2019{\natexlab{a}}){Burdge}, {Kasliwal}, {Perley},
  \& {Growth Collaboration}}]{gcn25638}
{Burdge}, K., {Kasliwal}, M.~M., {Perley}, D.~A., \& {Growth Collaboration}.
  2019{\natexlab{a}}, GRB Coordinates Network, 25638, 1

\bibitem[{{Burdge} {et~al.}(2019{\natexlab{b}}){Burdge}, {Perley}, {Kasliwal},
  \& {Growth Collaboration}}]{gcn25639}
{Burdge}, K., {Perley}, D.~A., {Kasliwal}, M.~M., \& {Growth Collaboration}.
  2019{\natexlab{b}}, GRB Coordinates Network, 25639, 1

\bibitem[{{Burke} {et~al.}(2019){Burke}, {Arcavi}, {Howell}, {Hiramatsu},
  {McCully}, \& {Valenti}}]{2019daj}
{Burke}, J., {Arcavi}, I., {Howell}, D.~A., {et~al.} 2019, Transient Name
  Server Classification Report, 2019-596, 1

\bibitem[{{Cantiello} {et~al.}(2018){Cantiello}, {Jensen}, {Blakeslee},
  {Berger}, {Levan}, {Tanvir}, {Raimondo}, {Brocato}, {Alexander}, {Blanchard},
  {Branchesi}, {Cano}, {Chornock}, {Covino}, {Cowperthwaite}, {D'Avanzo},
  {Eftekhari}, {Fong}, {Fruchter}, {Grado}, {Hjorth}, {Holz}, {Lyman},
  {Mandel}, {Margutti}, {Nicholl}, {Villar}, \& {Williams}}]{Cantiello2018}
{Cantiello}, M., {Jensen}, J.~B., {Blakeslee}, J.~P., {et~al.} 2018, \apjl,
  854, L31, \dodoi{10.3847/2041-8213/aaad64}

\bibitem[{{Cartier} {et~al.}(2020){Cartier}, {Olivares}, {Rodr{\'\i}guez},
  {Meza-Retamal}, {Quirola}, {Antilen}, {Allam}, {Butner}, {Tucker},
  {Soares-Santos}, {Makler}, {Bom}, {Santana-Silva}, {Martinez-Vazquez},
  {Garcia}, {Herner}, {Annis}, {Palmese}, {Sherman}, {Morgan}, {Bachmann},
  {Davis}, \& {Desgw Team}}]{gcn26830}
{Cartier}, R., {Olivares}, F., {Rodr{\'\i}guez}, {\'O}., {et~al.} 2020, GRB
  Coordinates Network, 26830, 1

\bibitem[{{Castro-Tirado} \& {Font}(2020)}]{GCN26703}
{Castro-Tirado}, A.~J., \& {Font}, J. 2020, GRB Coordinates Network, 26703, 1

\bibitem[{{Castro-Tirado} {et~al.}(2019){Castro-Tirado}, {Hu}, {Valeev},
  {Sokolov}, {Fernandez-Garcia}, {Carrasco}, {Castellon}, {Caballero-Garcia},
  \& {Geier}}]{GCN26492}
{Castro-Tirado}, A.~J., {Hu}, Y.~D., {Valeev}, A.~F., {et~al.} 2019, GRB
  Coordinates Network, 26492, 1

\bibitem[{{Chambers} {et~al.}(2016){Chambers}, {Magnier}, {Metcalfe},
  {Flewelling}, {Huber}, {Waters}, {Denneau}, {Draper}, {Farrow}, {Finkbeiner},
  {Holmberg}, {Koppenhoefer}, {Price}, {Rest}, {Saglia}, {Schlafly}, {Smartt},
  {Sweeney}, {Wainscoat}, {Burgett}, {Chastel}, {Grav}, {Heasley}, {Hodapp},
  {Jedicke}, {Kaiser}, {Kudritzki}, {Luppino}, {Lupton}, {Monet}, {Morgan},
  {Onaka}, {Shiao}, {Stubbs}, {Tonry}, {White}, {Ba{\~n}ados}, {Bell},
  {Bender}, {Bernard}, {Boegner}, {Boffi}, {Botticella}, {Calamida},
  {Casertano}, {Chen}, {Chen}, {Cole}, {Deacon}, {Frenk}, {Fitzsimmons},
  {Gezari}, {Gibbs}, {Goessl}, {Goggia}, {Gourgue}, {Goldman}, {Grant},
  {Grebel}, {Hambly}, {Hasinger}, {Heavens}, {Heckman}, {Henderson}, {Henning},
  {Holman}, {Hopp}, {Ip}, {Isani}, {Jackson}, {Keyes}, {Koekemoer}, {Kotak},
  {Le}, {Liska}, {Long}, {Lucey}, {Liu}, {Martin}, {Masci}, {McLean}, {Mindel},
  {Misra}, {Morganson}, {Murphy}, {Obaika}, {Narayan}, {Nieto-Santisteban},
  {Norberg}, {Peacock}, {Pier}, {Postman}, {Primak}, {Rae}, {Rai}, {Riess},
  {Riffeser}, {Rix}, {R{\"o}ser}, {Russel}, {Rutz}, {Schilbach}, {Schultz},
  {Scolnic}, {Strolger}, {Szalay}, {Seitz}, {Small}, {Smith}, {Soderblom},
  {Taylor}, {Thomson}, {Taylor}, {Thakar}, {Thiel}, {Thilker}, {Unger},
  {Urata}, {Valenti}, {Wagner}, {Walder}, {Walter}, {Watters}, {Werner},
  {Wood-Vasey}, \& {Wyse}}]{chambers16}
{Chambers}, K.~C., {Magnier}, E.~A., {Metcalfe}, N., {et~al.} 2016, arXiv
  e-prints, arXiv:1612.05560.
\newblock \doarXiv{1612.05560}

\bibitem[{{Chen} {et~al.}(2019){Chen}, {Schweyer}, {Rossi}, {Heintz},
  {Gromadzki}, {Bolmer}, \& {Schady}}]{gcn24097}
{Chen}, T.~W., {Schweyer}, T., {Rossi}, A., {et~al.} 2019, GRB Coordinates
  Network, 24097, 1

\bibitem[{{Chornock} {et~al.}(2017){Chornock}, {Berger}, {Kasen},
  {Cowperthwaite}, {Nicholl}, {Villar}, {Alexander}, {Blanchard}, {Eftekhari},
  {Fong}, {Margutti}, {Williams}, {Annis}, {Brout}, {Brown}, {Chen}, {Drout},
  {Farr}, {Foley}, {Frieman}, {Fryer}, {Herner}, {Holz}, {Kessler}, {Matheson},
  {Metzger}, {Quataert}, {Rest}, {Sako}, {Scolnic}, {Smith}, \&
  {Soares-Santos}}]{Chornock+17}
{Chornock}, R., {Berger}, E., {Kasen}, D., {et~al.} 2017, \apjl, 848, L19,
  \dodoi{10.3847/2041-8213/aa905c}

\bibitem[{{Christensen} {et~al.}(2018){Christensen}, {Africano}, {Farneth},
  {Fuls}, {Gibbs}, {Grauer}, {Groeller}, {Johnson}, {Kowalski}, {Larson},
  {Leonard}, {Seaman}, \& {Shelly}}]{CSS}
{Christensen}, E., {Africano}, B., {Farneth}, G., {et~al.} 2018, in
  AAS/Division for Planetary Sciences Meeting Abstracts, 310.10

\bibitem[{{Coughlin} {et~al.}(2019{\natexlab{a}}){Coughlin}, {Anand}, \&
  {Ahumada}}]{gcn24223}
{Coughlin}, M.~W., {Anand}, S., \& {Ahumada}, T. 2019{\natexlab{a}}, GRB
  Coordinates Network, 24223, 1

\bibitem[{{Coughlin} {et~al.}(2019{\natexlab{b}}){Coughlin}, {Ahumada},
  {Anand}, {De}, {Hankins}, {Kasliwal}, {Singer}, {Bellm}, {Andreoni}, {Cenko},
  {Cooke}, {Copperwheat}, {Dugas}, {Jencson}, {Perley}, {Yu}, {Bhalerao},
  {Kumar}, {Bloom}, {Anupama}, {Ashley}, {Bagdasaryan}, {Biswas}, {Buckley},
  {Burdge}, {Cook}, {Cromer}, {Cunningham}, {D'A{\`\i}}, {Dekany}, {Delacroix},
  {Dichiara}, {Duev}, {Dutta}, {Feeney}, {Frederick}, {Gatkine}, {Ghosh},
  {Goldstein}, {Golkhou}, {Goobar}, {Graham}, {Hanayama}, {Horiuchi}, {Hung},
  {Jha}, {Kong}, {Giomi}, {Kaplan}, {Karambelkar}, {Kowalski}, {Kulkarni},
  {Kupfer}, {Masci}, {Mazzali}, {Moore}, {Mogotsi}, {Neill}, {Ngeow},
  {Mart{\'\i}nez-Palomera}, {La Parola}, {Pavana}, {Ofek}, {Patil}, {Riddle},
  {Rigault}, {Rusholme}, {Serabyn}, {Shupe}, {Sharma}, {Singh}, {Sollerman},
  {Soon}, {Staats}, {Taggart}, {Tan}, {Travouillon}, {Troja}, {Waratkar}, \&
  {Yatsu}}]{coughlin19}
{Coughlin}, M.~W., {Ahumada}, T., {Anand}, S., {et~al.} 2019{\natexlab{b}},
  \apjl, 885, L19, \dodoi{10.3847/2041-8213/ab4ad8}

\bibitem[{{Coughlin} {et~al.}(2019{\natexlab{c}}){Coughlin}, {Kasliwal},
  {Perley}, {Goobar}, {Singer}, {Anand}, {Ahumada}, {Andreoni}, {Bellm}, {de},
  {Biswas}, {Nissanke}, {Duev}, {Cenko}, {Goldstein}, {Ho}, {Bhalerao},
  {Kumar}, {Karambelkar}, {Deshmukh}, {Saraogi}, {Anupama}, {Copperwheat},
  {Sollerman}, {Bloom}, {Bulla}, {Graham}, {Yan}, {Fremling}, {Gatkine}, \&
  {Miller}}]{gcn24283}
{Coughlin}, M.~W., {Kasliwal}, M.~M., {Perley}, D.~A., {et~al.}
  2019{\natexlab{c}}, GRB Coordinates Network, 24283, 1

\bibitem[{{Coughlin} {et~al.}(2020){Coughlin}, {Dietrich}, {Antier},
  {Almualla}, {Anand}, {Bulla}, {Foucart}, {Guessoum}, {Hotokezaka}, {Kumar},
  {Raaijmakers}, \& {Nissanke}}]{coughlin20}
{Coughlin}, M.~W., {Dietrich}, T., {Antier}, S., {et~al.} 2020, \mnras, 497,
  1181, \dodoi{10.1093/mnras/staa1925}

\bibitem[{{Coulter} {et~al.}(2017{\natexlab{a}}){Coulter}, {Foley},
  {Kilpatrick}, {Drout}, {Piro}, {Shappee}, {Siebert}, {Simon}, {Ulloa},
  {Kasen}, {Madore}, {Murguia-Berthier}, {Pan}, {Prochaska}, {Ramirez-Ruiz},
  {Rest}, \& {Rojas-Bravo}}]{Coulter17}
{Coulter}, D.~A., {Foley}, R.~J., {Kilpatrick}, C.~D., {et~al.}
  2017{\natexlab{a}}, Science, 358, 1556, \dodoi{10.1126/science.aap9811}

\bibitem[{{Coulter} {et~al.}(2017{\natexlab{b}}){Coulter}, {Foley},
  {Kilpatrick}, {Drout}, {Piro}, {Shappee}, {Siebert}, {Simon}, {Ulloa},
  {Kasen}, {Madore}, {Murguia-Berthier}, {Pan}, {Prochaska}, {Ramirez-Ruiz},
  {Rest}, \& {Rojas-Bravo}}]{Coulter+17}
---. 2017{\natexlab{b}}, Science, 358, 1556, \dodoi{10.1126/science.aap9811}

\bibitem[{{Cowperthwaite} {et~al.}(2017){Cowperthwaite}, {Berger}, {Villar},
  {Metzger}, {Nicholl}, {Chornock}, {Blanchard}, {Fong}, {Margutti},
  {Soares-Santos}, {Alexander}, {Allam}, {Annis}, {Brout}, {Brown}, {Butler},
  {Chen}, {Diehl}, {Doctor}, {Drout}, {Eftekhari}, {Farr}, {Finley}, {Foley},
  {Frieman}, {Fryer}, {Garc{\'{\i}}a-Bellido}, {Gill}, {Guillochon}, {Herner},
  {Holz}, {Kasen}, {Kessler}, {Marriner}, {Matheson}, {Neilsen}, {Quataert},
  {Palmese}, {Rest}, {Sako}, {Scolnic}, {Smith}, {Tucker}, {Williams},
  {Balbinot}, {Carlin}, {Cook}, {Durret}, {Li}, {Lopes}, {Louren{\c c}o},
  {Marshall}, {Medina}, {Muir}, {Mu{\~n}oz}, {Sauseda}, {Schlegel}, {Secco},
  {Vivas}, {Wester}, {Zenteno}, {Zhang}, {Abbott}, {Banerji}, {Bechtol},
  {Benoit-L{\'e}vy}, {Bertin}, {Buckley-Geer}, {Burke}, {Capozzi}, {Carnero
  Rosell}, {Carrasco Kind}, {Castander}, {Crocce}, {Cunha}, {D'Andrea}, {da
  Costa}, {Davis}, {DePoy}, {Desai}, {Dietrich}, {Drlica-Wagner}, {Eifler},
  {Evrard}, {Fernandez}, {Flaugher}, {Fosalba}, {Gaztanaga}, {Gerdes},
  {Giannantonio}, {Goldstein}, {Gruen}, {Gruendl}, {Gutierrez}, {Honscheid},
  {Jain}, {James}, {Jeltema}, {Johnson}, {Johnson}, {Kent}, {Krause}, {Kron},
  {Kuehn}, {Nuropatkin}, {Lahav}, {Lima}, {Lin}, {Maia}, {March}, {Martini},
  {McMahon}, {Menanteau}, {Miller}, {Miquel}, {Mohr}, {Neilsen}, {Nichol},
  {Ogando}, {Plazas}, {Roe}, {Romer}, {Roodman}, {Rykoff}, {Sanchez},
  {Scarpine}, {Schindler}, {Schubnell}, {Sevilla-Noarbe}, {Smith}, {Smith},
  {Sobreira}, {Suchyta}, {Swanson}, {Tarle}, {Thomas}, {Thomas}, {Troxel},
  {Vikram}, {Walker}, {Wechsler}, {Weller}, {Yanny}, \&
  {Zuntz}}]{Cowperthwaite17}
{Cowperthwaite}, P.~S., {Berger}, E., {Villar}, V.~A., {et~al.} 2017, \apjl,
  848, L17, \dodoi{10.3847/2041-8213/aa8fc7}

\bibitem[{{Cowperthwaite} {et~al.}(2018){Cowperthwaite}, {Berger}, {Rest},
  {Chornock}, {Scolnic}, {Williams}, {Fong}, {Drout}, {Foley}, {Margutti},
  {Lunnan}, {Metzger}, \& {Quataert}}]{Cowperthwaite2018}
{Cowperthwaite}, P.~S., {Berger}, E., {Rest}, A., {et~al.} 2018, \apj, 858, 18,
  \dodoi{10.3847/1538-4357/aabad9}

\bibitem[{{Dahiwale} \& {Fremling}(2020)}]{2019rpp}
{Dahiwale}, A., \& {Fremling}, C. 2020, Transient Name Server Classification
  Report, 2020-1935, 1

\bibitem[{{Dahiwale} {et~al.}(2019){Dahiwale}, {Fremling}, \&
  {Dugas}}]{2019rpr}
{Dahiwale}, A., {Fremling}, C., \& {Dugas}, A. 2019, Transient Name Server
  Classification Report, 2019-2026, 1

\bibitem[{{D{\'a}lya} {et~al.}(2018){D{\'a}lya}, {Galg{\'o}czi}, {Dobos},
  {Frei}, {Heng}, {Macas}, {Messenger}, {Raffai}, \& {de Souza}}]{glade}
{D{\'a}lya}, G., {Galg{\'o}czi}, G., {Dobos}, L., {et~al.} 2018, \mnras, 479,
  2374, \dodoi{10.1093/mnras/sty1703}

\bibitem[{{de} {et~al.}(2020){de}, {Ztf}, \& {Growth
  Collaborations}}]{gcn26814}
{de}, K., {Ztf}, T., \& {Growth Collaborations}. 2020, GRB Coordinates Network,
  26814, 1

\bibitem[{{Denisenko}(2019)}]{Denisenko19}
{Denisenko}, D. 2019, GRB Coordinates Network, 26470, 1

\bibitem[{{Dey} {et~al.}(2019){Dey}, {Schlegel}, {Lang}, {Blum}, {Burleigh},
  {Fan}, {Findlay}, {Finkbeiner}, {Herrera}, {Juneau}, {Landriau}, {Levi},
  {McGreer}, {Meisner}, {Myers}, {Moustakas}, {Nugent}, {Patej}, {Schlafly},
  {Walker}, {Valdes}, {Weaver}, {Y{\`e}che}, {Zou}, {Zhou}, {Abareshi},
  {Abbott}, {Abolfathi}, {Aguilera}, {Alam}, {Allen}, {Alvarez}, {Annis},
  {Ansarinejad}, {Aubert}, {Beechert}, {Bell}, {BenZvi}, {Beutler}, {Bielby},
  {Bolton}, {Brice{\~n}o}, {Buckley-Geer}, {Butler}, {Calamida}, {Carlberg},
  {Carter}, {Casas}, {Castander}, {Choi}, {Comparat}, {Cukanovaite}, {Delubac},
  {DeVries}, {Dey}, {Dhungana}, {Dickinson}, {Ding}, {Donaldson}, {Duan},
  {Duckworth}, {Eftekharzadeh}, {Eisenstein}, {Etourneau}, {Fagrelius},
  {Farihi}, {Fitzpatrick}, {Font-Ribera}, {Fulmer}, {G{\"a}nsicke},
  {Gaztanaga}, {George}, {Gerdes}, {Gontcho}, {Gorgoni}, {Green}, {Guy},
  {Harmer}, {Hernand ez}, {Honscheid}, {Huang}, {James}, {Jannuzi}, {Jiang},
  {Joyce}, {Karcher}, {Karkar}, {Kehoe}, {Kneib}, {Kueter-Young}, {Lan},
  {Lauer}, {Le Guillou}, {Le Van Suu}, {Lee}, {Lesser}, {Perreault Levasseur},
  {Li}, {Mann}, {Marshall}, {Mart{\'\i}nez-V{\'a}zquez}, {Martini}, {du Mas des
  Bourboux}, {McManus}, {Meier}, {M{\'e}nard}, {Metcalfe},
  {Mu{\~n}oz-Guti{\'e}rrez}, {Najita}, {Napier}, {Narayan}, {Newman}, {Nie},
  {Nord}, {Norman}, {Olsen}, {Paat}, {Palanque-Delabrouille}, {Peng},
  {Poppett}, {Poremba}, {Prakash}, {Rabinowitz}, {Raichoor}, {Rezaie},
  {Robertson}, {Roe}, {Ross}, {Ross}, {Rudnick}, {Safonova}, {Saha},
  {S{\'a}nchez}, {Savary}, {Schweiker}, {Scott}, {Seo}, {Shan}, {Silva},
  {Slepian}, {Soto}, {Sprayberry}, {Staten}, {Stillman}, {Stupak}, {Summers},
  {Sien Tie}, {Tirado}, {Vargas-Maga{\~n}a}, {Vivas}, {Wechsler}, {Williams},
  {Yang}, {Yang}, {Yapici}, {Zaritsky}, {Zenteno}, {Zhang}, {Zhang}, {Zhou}, \&
  {Zhou}}]{Dey2019}
{Dey}, A., {Schlegel}, D.~J., {Lang}, D., {et~al.} 2019, \aj, 157, 168,
  \dodoi{10.3847/1538-3881/ab089d}

\bibitem[{{D{\'\i}az} {et~al.}(2017){D{\'\i}az}, {Macri}, {Garcia Lambas},
  {Mendes de Oliveira}, {Nilo Castell{\'o}n}, {Ribeiro}, {S{\'a}nchez},
  {Schoenell}, {Abramo}, {Akras}, {Alcaniz}, {Artola}, {Beroiz}, {Bonoli},
  {Cabral}, {Camuccio}, {Castillo}, {Chavushyan}, {Coelho}, {Colazo},
  {Costa-Duarte}, {Cuevas Larenas}, {DePoy}, {Dom{\'\i}nguez Romero},
  {Dultzin}, {Fern{\'a}ndez}, {Garc{\'\i}a}, {Girardini}, {Gon{\c{c}}alves},
  {Gon{\c{c}}alves}, {Gurovich}, {Jim{\'e}nez-Teja}, {Kanaan}, {Lares}, {Lopes
  de Oliveira}, {L{\'o}pez-Cruz}, {Marshall}, {Melia}, {Molino}, {Padilla},
  {Pe{\~n}uela}, {Placco}, {Qui{\~n}ones}, {Ram{\'\i}rez Rivera}, {Renzi},
  {Riguccini}, {R{\'\i}os-L{\'o}pez}, {Rodriguez}, {Sampedro}, {Schneiter},
  {Sodr{\'e}}, {Starck}, {Torres-Flores}, {Tornatore}, \&
  {Zadro{\.z}ny}}]{Diaz+17}
{D{\'\i}az}, M.~C., {Macri}, L.~M., {Garcia Lambas}, D., {et~al.} 2017, \apjl,
  848, L29, \dodoi{10.3847/2041-8213/aa9060}

\bibitem[{{Drout} {et~al.}(2017){Drout}, {Piro}, {Shappee}, {Kilpatrick},
  {Simon}, {Contreras}, {Coulter}, {Foley}, {Siebert}, {Morrell}, {Boutsia},
  {Di Mille}, {Holoien}, {Kasen}, {Kollmeier}, {Madore}, {Monson},
  {Murguia-Berthier}, {Pan}, {Prochaska}, {Ramirez-Ruiz}, {Rest}, {Adams},
  {Alatalo}, {Ba{\~n}ados}, {Baughman}, {Beers}, {Bernstein}, {Bitsakis},
  {Campillay}, {Hansen}, {Higgs}, {Ji}, {Maravelias}, {Marshall}, {Moni Bidin},
  {Prieto}, {Rasmussen}, {Rojas-Bravo}, {Strom}, {Ulloa},
  {Vargas-Gonz{\'a}lez}, {Wan}, \& {Whitten}}]{Drout+17}
{Drout}, M.~R., {Piro}, A.~L., {Shappee}, B.~J., {et~al.} 2017, Science, 358,
  1570, \dodoi{10.1126/science.aaq0049}

\bibitem[{{Flewelling}(2018)}]{PSDR2}
{Flewelling}, H. 2018, in American Astronomical Society Meeting Abstracts, Vol.
  231, American Astronomical Society Meeting Abstracts \#231, 436.01

\bibitem[{{Foley} {et~al.}(2020){Foley}, {Coulter}, {Kilpatrick}, {Piro},
  {Ramirez-Ruiz}, \& {Schwab}}]{foley20}
{Foley}, R.~J., {Coulter}, D.~A., {Kilpatrick}, C.~D., {et~al.} 2020, \mnras,
  494, 190, \dodoi{10.1093/mnras/staa725}

\bibitem[{{Fong} \& {Berger}(2013)}]{FongBerger13}
{Fong}, W., \& {Berger}, E. 2013, \apj, 776, 18,
  \dodoi{10.1088/0004-637X/776/1/18}

\bibitem[{{Fong} {et~al.}(2017){Fong}, {Berger}, {Blanchard}, {Margutti},
  {Cowperthwaite}, {Chornock}, {Alexander}, {Metzger}, {Villar}, {Nicholl},
  {Eftekhari}, {Williams}, {Annis}, {Brout}, {Brown}, {Chen}, {Doctor},
  {Diehl}, {Holz}, {Rest}, {Sako}, \& {Soares-Santos}}]{Fong+17}
{Fong}, W., {Berger}, E., {Blanchard}, P.~K., {et~al.} 2017, \apjl, 848, L23,
  \dodoi{10.3847/2041-8213/aa9018}

\bibitem[{{Fong} {et~al.}(2019){Fong}, {Blanchard}, {Alexander}, {Strader},
  {Margutti}, {Hajela}, {Villar}, {Wu}, {Ye}, {Berger}, {Chornock},
  {Coppejans}, {Cowperthwaite}, {Eftekhari}, {Giannios}, {Guidorzi},
  {Kathirgamaraju}, {Laskar}, {Macfadyen}, {Metzger}, {Nicholl}, {Paterson},
  {Terreran}, {Sand}, {Sironi}, {Williams}, {Xie}, \& {Zrake}}]{Fong19}
{Fong}, W., {Blanchard}, P.~K., {Alexander}, K.~D., {et~al.} 2019, \apjl, 883,
  L1, \dodoi{10.3847/2041-8213/ab3d9e}

\bibitem[{{Gaia Collaboration} {et~al.}(2016){Gaia Collaboration}, {Prusti},
  {de Bruijne}, {Brown}, {Vallenari}, {Babusiaux}, {Bailer-Jones}, {Bastian},
  {Biermann}, {Evans}, {Eyer}, {Jansen}, {Jordi}, {Klioner}, {Lammers},
  {Lindegren}, {Luri}, {Mignard}, {Milligan}, {Panem}, {Poinsignon},
  {Pourbaix}, {Randich}, {Sarri}, {Sartoretti}, {Siddiqui}, {Soubiran},
  {Valette}, {van Leeuwen}, {Walton}, {Aerts}, {Arenou}, {Cropper}, {Drimmel},
  {H{\o}g}, {Katz}, {Lattanzi}, {O'Mullane}, {Grebel}, {Holland}, {Huc},
  {Passot}, {Bramante}, {Cacciari}, {Casta{\~n}eda}, {Chaoul}, {Cheek}, {De
  Angeli}, {Fabricius}, {Guerra}, {Hern{\'a}ndez}, {Jean-Antoine-Piccolo},
  {Masana}, {Messineo}, {Mowlavi}, {Nienartowicz}, {Ord{\'o}{\~n}ez-Blanco},
  {Panuzzo}, {Portell}, {Richards}, {Riello}, {Seabroke}, {Tanga},
  {Th{\'e}venin}, {Torra}, {Els}, {Gracia-Abril}, {Comoretto},
  {Garcia-Reinaldos}, {Lock}, {Mercier}, {Altmann}, {Andrae}, {Astraatmadja},
  {Bellas-Velidis}, {Benson}, {Berthier}, {Blomme}, {Busso}, {Carry},
  {Cellino}, {Clementini}, {Cowell}, {Creevey}, {Cuypers}, {Davidson}, {De
  Ridder}, {de Torres}, {Delchambre}, {Dell'Oro}, {Ducourant}, {Fr{\'e}mat},
  {Garc{\'\i}a-Torres}, {Gosset}, {Halbwachs}, {Hambly}, {Harrison}, {Hauser},
  {Hestroffer}, {Hodgkin}, {Huckle}, {Hutton}, {Jasniewicz}, {Jordan},
  {Kontizas}, {Korn}, {Lanzafame}, {Manteiga}, {Moitinho}, {Muinonen},
  {Osinde}, {Pancino}, {Pauwels}, {Petit}, {Recio-Blanco}, {Robin}, {Sarro},
  {Siopis}, {Smith}, {Smith}, {Sozzetti}, {Thuillot}, {van Reeven}, {Viala},
  {Abbas}, {Abreu Aramburu}, {Accart}, {Aguado}, {Allan}, {Allasia},
  {Altavilla}, {{\'A}lvarez}, {Alves}, {Anderson}, {Andrei}, {Anglada Varela},
  {Antiche}, {Antoja}, {Ant{\'o}n}, {Arcay}, {Atzei}, {Ayache}, {Bach},
  {Baker}, {Balaguer-N{\'u}{\~n}ez}, {Barache}, {Barata}, {Barbier}, {Barblan},
  {Baroni}, {Barrado y Navascu{\'e}s}, {Barros}, {Barstow}, {Becciani},
  {Bellazzini}, {Bellei}, {Bello Garc{\'\i}a}, {Belokurov}, {Bendjoya},
  {Berihuete}, {Bianchi}, {Bienaym{\'e}}, {Billebaud}, {Blagorodnova},
  {Blanco-Cuaresma}, {Boch}, {Bombrun}, {Borrachero}, {Bouquillon}, {Bourda},
  {Bouy}, {Bragaglia}, {Breddels}, {Brouillet}, {Br{\"u}semeister},
  {Bucciarelli}, {Budnik}, {Burgess}, {Burgon}, {Burlacu}, {Busonero}, {Buzzi},
  {Caffau}, {Cambras}, {Campbell}, {Cancelliere}, {Cantat-Gaudin}, {Carlucci},
  {Carrasco}, {Castellani}, {Charlot}, {Charnas}, {Charvet}, {Chassat},
  {Chiavassa}, {Clotet}, {Cocozza}, {Collins}, {Collins}, {Costigan}, {Crifo},
  {Cross}, {Crosta}, {Crowley}, {Dafonte}, {Damerdji}, {Dapergolas}, {David},
  {David}, {De Cat}, {de Felice}, {de Laverny}, {De Luise}, {De March}, {de
  Martino}, {de Souza}, {Debosscher}, {del Pozo}, {Delbo}, {Delgado},
  {Delgado}, {di Marco}, {Di Matteo}, {Diakite}, {Distefano}, {Dolding}, {Dos
  Anjos}, {Drazinos}, {Dur{\'a}n}, {Dzigan}, {Ecale}, {Edvardsson}, {Enke},
  {Erdmann}, {Escolar}, {Espina}, {Evans}, {Eynard Bontemps}, {Fabre},
  {Fabrizio}, {Faigler}, {Falc{\~a}o}, {Farr{\`a}s Casas}, {Faye}, {Federici},
  {Fedorets}, {Fern{\'a}ndez-Hern{\'a}ndez}, {Fernique}, {Fienga}, {Figueras},
  {Filippi}, {Findeisen}, {Fonti}, {Fouesneau}, {Fraile}, {Fraser}, {Fuchs},
  {Furnell}, {Gai}, {Galleti}, {Galluccio}, {Garabato}, {Garc{\'\i}a-Sedano},
  {Gar{\'e}}, {Garofalo}, {Garralda}, {Gavras}, {Gerssen}, {Geyer}, {Gilmore},
  {Girona}, {Giuffrida}, {Gomes}, {Gonz{\'a}lez-Marcos},
  {Gonz{\'a}lez-N{\'u}{\~n}ez}, {Gonz{\'a}lez-Vidal}, {Granvik}, {Guerrier},
  {Guillout}, {Guiraud}, {G{\'u}rpide}, {Guti{\'e}rrez-S{\'a}nchez}, {Guy},
  {Haigron}, {Hatzidimitriou}, {Haywood}, {Heiter}, {Helmi}, {Hobbs},
  {Hofmann}, {Holl}, {Holland}, {Hunt}, {Hypki}, {Icardi}, {Irwin}, {Jevardat
  de Fombelle}, {Jofr{\'e}}, {Jonker}, {Jorissen}, {Julbe}, {Karampelas},
  {Kochoska}, {Kohley}, {Kolenberg}, {Kontizas}, {Koposov}, {Kordopatis},
  {Koubsky}, {Kowalczyk}, {Krone-Martins}, {Kudryashova}, {Kull}, {Bachchan},
  {Lacoste-Seris}, {Lanza}, {Lavigne}, {Le Poncin-Lafitte}, {Lebreton},
  {Lebzelter}, {Leccia}, {Leclerc}, {Lecoeur-Taibi}, {Lemaitre}, {Lenhardt},
  {Leroux}, {Liao}, {Licata}, {Lindstr{\o}m}, {Lister}, {Livanou}, {Lobel},
  {L{\"o}ffler}, {L{\'o}pez}, {Lopez-Lozano}, {Lorenz}, {Loureiro},
  {MacDonald}, {Magalh{\~a}es Fernandes}, {Managau}, {Mann}, {Mantelet},
  {Marchal}, {Marchant}, {Marconi}, {Marie}, {Marinoni}, {Marrese},
  {Marschalk{\'o}}, {Marshall}, {Mart{\'\i}n-Fleitas}, {Martino}, {Mary},
  {Matijevi{\v{c}}}, {Mazeh}, {McMillan}, {Messina}, {Mestre}, {Michalik},
  {Millar}, {Miranda}, {Molina}, {Molinaro}, {Molinaro}, {Moln{\'a}r},
  {Moniez}, {Montegriffo}, {Monteiro}, {Mor}, {Mora}, {Morbidelli}, {Morel},
  {Morgenthaler}, {Morley}, {Morris}, {Mulone}, {Muraveva}, {Musella},
  {Narbonne}, {Nelemans}, {Nicastro}, {Noval}, {Ord{\'e}novic},
  {Ordieres-Mer{\'e}}, {Osborne}, {Pagani}, {Pagano}, {Pailler}, {Palacin},
  {Palaversa}, {Parsons}, {Paulsen}, {Pecoraro}, {Pedrosa}, {Pentik{\"a}inen},
  {Pereira}, {Pichon}, {Piersimoni}, {Pineau}, {Plachy}, {Plum}, {Poujoulet},
  {Pr{\v{s}}a}, {Pulone}, {Ragaini}, {Rago}, {Rambaux}, {Ramos-Lerate},
  {Ranalli}, {Rauw}, {Read}, {Regibo}, {Renk}, {Reyl{\'e}}, {Ribeiro},
  {Rimoldini}, {Ripepi}, {Riva}, {Rixon}, {Roelens}, {Romero-G{\'o}mez},
  {Rowell}, {Royer}, {Rudolph}, {Ruiz-Dern}, {Sadowski}, {Sagrist{\`a}
  Sell{\'e}s}, {Sahlmann}, {Salgado}, {Salguero}, {Sarasso}, {Savietto},
  {Schnorhk}, {Schultheis}, {Sciacca}, {Segol}, {Segovia}, {Segransan},
  {Serpell}, {Shih}, {Smareglia}, {Smart}, {Smith}, {Solano}, {Solitro},
  {Sordo}, {Soria Nieto}, {Souchay}, {Spagna}, {Spoto}, {Stampa}, {Steele},
  {Steidelm{\"u}ller}, {Stephenson}, {Stoev}, {Suess}, {S{\"u}veges}, {Surdej},
  {Szabados}, {Szegedi-Elek}, {Tapiador}, {Taris}, {Tauran}, {Taylor},
  {Teixeira}, {Terrett}, {Tingley}, {Trager}, {Turon}, {Ulla}, {Utrilla},
  {Valentini}, {van Elteren}, {Van Hemelryck}, {van Leeuwen}, {Varadi},
  {Vecchiato}, {Veljanoski}, {Via}, {Vicente}, {Vogt}, {Voss}, {Votruba},
  {Voutsinas}, {Walmsley}, {Weiler}, {Weingrill}, {Werner}, {Wevers},
  {Whitehead}, {Wyrzykowski}, {Yoldas}, {{\v{Z}}erjal}, {Zucker}, {Zurbach},
  {Zwitter}, {Alecu}, {Allen}, {Allende Prieto}, {Amorim},
  {Anglada-Escud{\'e}}, {Arsenijevic}, {Azaz}, {Balm}, {Beck}, {Bernstein},
  {Bigot}, {Bijaoui}, {Blasco}, {Bonfigli}, {Bono}, {Boudreault}, {Bressan},
  {Brown}, {Brunet}, {Bunclark}, {Buonanno}, {Butkevich}, {Carret}, {Carrion},
  {Chemin}, {Ch{\'e}reau}, {Corcione}, {Darmigny}, {de Boer}, {de Teodoro}, {de
  Zeeuw}, {Delle Luche}, {Domingues}, {Dubath}, {Fodor}, {Fr{\'e}zouls},
  {Fries}, {Fustes}, {Fyfe}, {Gallardo}, {Gallegos}, {Gardiol}, {Gebran},
  {Gomboc}, {G{\'o}mez}, {Grux}, {Gueguen}, {Heyrovsky}, {Hoar}, {Iannicola},
  {Isasi Parache}, {Janotto}, {Joliet}, {Jonckheere}, {Keil}, {Kim},
  {Klagyivik}, {Klar}, {Knude}, {Kochukhov}, {Kolka}, {Kos}, {Kutka}, {Lainey},
  {LeBouquin}, {Liu}, {Loreggia}, {Makarov}, {Marseille}, {Martayan},
  {Martinez-Rubi}, {Massart}, {Meynadier}, {Mignot}, {Munari}, {Nguyen},
  {Nordlander}, {Ocvirk}, {O'Flaherty}, {Olias Sanz}, {Ortiz}, {Osorio},
  {Oszkiewicz}, {Ouzounis}, {Palmer}, {Park}, {Pasquato}, {Peltzer}, {Peralta},
  {P{\'e}turaud}, {Pieniluoma}, {Pigozzi}, {Poels}, {Prat}, {Prod'homme},
  {Raison}, {Rebordao}, {Risquez}, {Rocca-Volmerange}, {Rosen}, {Ruiz-Fuertes},
  {Russo}, {Sembay}, {Serraller Vizcaino}, {Short}, {Siebert}, {Silva},
  {Sinachopoulos}, {Slezak}, {Soffel}, {Sosnowska}, {Strai{\v{z}}ys}, {ter
  Linden}, {Terrell}, {Theil}, {Tiede}, {Troisi}, {Tsalmantza}, {Tur},
  {Vaccari}, {Vachier}, {Valles}, {Van Hamme}, {Veltz}, {Virtanen}, {Wallut},
  {Wichmann}, {Wilkinson}, {Ziaeepour}, \& {Zschocke}}]{GaiaCo}
{Gaia Collaboration}, {Prusti}, T., {de Bruijne}, J.~H.~J., {et~al.} 2016,
  \aap, 595, A1, \dodoi{10.1051/0004-6361/201629272}

\bibitem[{{Gaia Collaboration} {et~al.}(2018){Gaia Collaboration}, {Brown},
  {Vallenari}, {Prusti}, {de Bruijne}, {Babusiaux}, {Bailer-Jones}, {Biermann},
  {Evans}, {Eyer}, \& et~al.}]{Gaiadr2}
{Gaia Collaboration}, {Brown}, A.~G.~A., {Vallenari}, A., {et~al.} 2018, \aap,
  616, A1, \dodoi{10.1051/0004-6361/201833051}

\bibitem[{Gall {et~al.}(2017)Gall, Hjorth, Rosswog, Tanvir, \&
  Levan}]{Gall+2017}
Gall, C., Hjorth, J., Rosswog, S., Tanvir, N.~R., \& Levan, A.~J. 2017, The
  Astrophysical Journal, 849, L19, \dodoi{10.3847/2041-8213/aa93f9}

\bibitem[{{Garcia} {et~al.}(2020){Garcia}, {Morgan}, {Herner}, {Palmese},
  {Soares-Santos}, {Annis}, {Brout}, {Vivas}, {Drlica-Wagner}, {Santana-Silva},
  {Tucker}, {Allam}, {Wiesner}, {Garc{\'\i}a-Bellido}, {Gill}, {Sako},
  {Kessler}, {Davis}, {Scolnic}, {Casares}, {Chen}, {Conselice}, {Cooke},
  {Doctor}, {Foley}, {Horvath}, {Howell}, {Kilpatrick}, {Lidman}, {Olivares
  E.}, {Paz-Chinch{\'o}n}, {Pineda-G.}, {Quirola-V{\'a}squez}, {Rest},
  {Sherman}, {Abbott}, {Aguena}, {Avila}, {Bertin}, {Bhargava}, {Brooks},
  {Burke}, {Carnero Rosell}, {Carrasco Kind}, {Carretero}, {Costanzi}, {da
  Costa}, {Desai}, {Diehl}, {Dietrich}, {Doel}, {Everett}, {Flaugher},
  {Fosalba}, {Friedel}, {Frieman}, {Gaztanaga}, {Gerdes}, {Gruen}, {Gruendl},
  {Gschwend}, {Gutierrez}, {Hinton}, {Hollowood}, {Honscheid}, {James},
  {Kuehn}, {Kuropatkin}, {Lahav}, {Lima}, {Maia}, {March}, {Marshall},
  {Menanteau}, {Miquel}, {Ogando}, {Plazas}, {Romer}, {Roodman}, {Sanchez},
  {Scarpine}, {Schubnell}, {Serrano}, {Sevilla-Noarbe}, {Smith}, {Suchyta},
  {Swanson}, {Tarle}, {Thomas}, {Varga}, {Walker}, {Weller}, \& {DES
  Collaboration}}]{garcia20}
{Garcia}, A., {Morgan}, R., {Herner}, K., {et~al.} 2020, \apj, 903, 75,
  \dodoi{10.3847/1538-4357/abb823}

\bibitem[{{Ghirlanda} {et~al.}(2019){Ghirlanda}, {Salafia}, {Paragi},
  {Giroletti}, {Yang}, {Marcote}, {Blanchard}, {Agudo}, {An}, {Bernardini},
  {Beswick}, {Branchesi}, {Campana}, {Casadio}, {Chassand e-Mottin}, {Colpi},
  {Covino}, {D'Avanzo}, {D'Elia}, {Frey}, {Gawronski}, {Ghisellini}, {Gurvits},
  {Jonker}, {van Langevelde}, {Melandri}, {Moldon}, {Nava}, {Perego},
  {Perez-Torres}, {Reynolds}, {Salvaterra}, {Tagliaferri}, {Venturi},
  {Vergani}, \& {Zhang}}]{ghirlanda19}
{Ghirlanda}, G., {Salafia}, O.~S., {Paragi}, Z., {et~al.} 2019, Science, 363,
  968, \dodoi{10.1126/science.aau8815}

\bibitem[{{Goldstein} {et~al.}(2019){Goldstein}, {Andreoni}, {Nugent},
  {Kasliwal}, {Coughlin}, {Anand}, {Bloom}, {Mart{\'\i}nez-Palomera}, {Zhang},
  {Ahumada}, {Bagdasaryan}, {Cooke}, {De}, {Duev}, {Fremling}, {Gatkine},
  {Graham}, {Ofek}, {Singer}, \& {Yan}}]{Goldstein19}
{Goldstein}, D.~A., {Andreoni}, I., {Nugent}, P.~E., {et~al.} 2019, arXiv
  e-prints, arXiv:1905.06980.
\newblock \doarXiv{1905.06980}

\bibitem[{{Gomez} {et~al.}(2019){Gomez}, {Hosseinzadeh}, {Cowperthwaite},
  {Villar}, {Berger}, {Gardner}, {Alexand er}, {Blanchard}, {Chornock},
  {Drout}, {Eftekhari}, {Fong}, {Gill}, {Margutti}, {Nicholl}, {Paterson}, \&
  {Williams}}]{gomez19}
{Gomez}, S., {Hosseinzadeh}, G., {Cowperthwaite}, P.~S., {et~al.} 2019, \apjl,
  884, L55, \dodoi{10.3847/2041-8213/ab4ad5}

\bibitem[{{Gompertz} {et~al.}(2018){Gompertz}, {Levan}, {Tanvir}, {Hjorth},
  {Covino}, {Evans}, {Fruchter}, {Gonz{\'a}lez-Fern{\'a}ndez}, {Jin}, {Lyman},
  {Oates}, {O'Brien}, \& {Wiersema}}]{Gompertz+18}
{Gompertz}, B.~P., {Levan}, A.~J., {Tanvir}, N.~R., {et~al.} 2018, \apj, 860,
  62, \dodoi{10.3847/1538-4357/aac206}

\bibitem[{Gompertz {et~al.}(2020)Gompertz, Cutter, Steeghs, Galloway, Lyman,
  Ulaczyk, Dyer, Ackley, Dhillon, O'Brien, Ramsay, Poshyachinda, Kotak,
  Nuttall, Breton, Pallé, Pollacco, Thrane, Aukkaravittayapun, Awiphan, Brown,
  Burhanudin, Chote, Chrimes, Daw, Duffy, Eyles-Ferris, Heikkilä, Irawati,
  Kennedy, Killestein, Levan, Littlefair, Makrygianni, Marsh, Sánchez,
  Mattila, Maund, McCormac, Mkrtichian, Mong, Mullaney, Müller, Obradovic,
  Rol, Sawangwit, Stanway, Starling, Strøm, Tooke, West, \&
  Wiersema}]{gompertz2020}
Gompertz, B.~P., Cutter, R., Steeghs, D., {et~al.} 2020, Searching for
  Electromagnetic Counterparts to Gravitational-wave Merger Events with the
  Prototype Gravitational-wave Optical Transient Observer (GOTO-4).
\newblock \doarXiv{2004.00025}

\bibitem[{{Grado} {et~al.}(2020){Grado}, {Cappellaro}, {Brocato}, {Carini},
  {Covino}, {D'Avanzo}, {Getman}, {Nicastro}, {Rossi}, \& {GRAWITA
  Collaboration}}]{gcn27230}
{Grado}, A., {Cappellaro}, E., {Brocato}, E., {et~al.} 2020, GRB Coordinates
  Network, 27230, 1

\bibitem[{{Graham} {et~al.}(2019){Graham}, {Kulkarni}, {Bellm}, {Adams},
  {Barbarino}, {Blagorodnova}, {Bodewits}, {Bolin}, {Brady}, {Cenko}, {Chang},
  {Coughlin}, {De}, {Eadie}, {Farnham}, {Feindt}, {Franckowiak}, {Fremling},
  {Gal-yam}, {Gezari}, {Ghosh}, {Goldstein}, {Golkhou}, {Goobar}, {Ho},
  {Huppenkothen}, {Ivezic}, {Jones}, {Juric}, {Kaplan}, {Kasliwal}, {Kelley},
  {Kupfer}, {Lee}, {Lin}, {Lunnan}, {Mahabal}, {Miller}, {Ngeow}, {Nugent},
  {Ofek}, {Prince}, {Rauch}, {van Roestel}, {Schulze}, {Singer}, {Sollerman},
  {Taddia}, {Yan}, {Ye}, {Yu}, {Andreoni}, {Barlow}, {Bauer}, {Beck},
  {Belicki}, {Biswas}, {Brinnel}, {Brooke}, {Bue}, {Bulla}, {Burdge},
  {Burruss}, {Connolly}, {Cromer}, {Cunningham}, {Dekany}, {Delacroix},
  {Desai}, {Duev}, {Hacopians}, {Hale}, {Helou}, {Henning}, {Hover},
  {Hillenbrand}, {Howell}, {Hung}, {Imel}, {Ip}, {Jackson}, {Kaspi}, {Kaye},
  {Kowalski}, {Kramer}, {Kuhn}, {Landry}, {Laher}, {Mao}, {Masci}, {Monkewitz},
  {Murphy}, {Nordin}, {Patterson}, {Penprase}, {Porter}, {Rebbapragada},
  {Reiley}, {Riddle}, {Rigault}, {Rodriguez}, {Rusholme}, {van Santen},
  {Shupe}, {Smith}, {Soumagnac}, {Stein}, {Surace}, {Szkody}, {Terek}, {van
  Sistine}, {van Velzen}, {Vestrand}, {Walters}, {Ward}, {Zhang}, \&
  {Zolkower}}]{ZTF}
{Graham}, M.~J., {Kulkarni}, S.~R., {Bellm}, E.~C., {et~al.} 2019, arXiv
  e-prints.
\newblock \doarXiv{1902.01945}

\bibitem[{{Graham} {et~al.}(2020){Graham}, {Ford}, {McKernan}, {Ross}, {Stern},
  {Burdge}, {Coughlin}, {Djorgovski}, {Drake}, {Duev}, {Kasliwal}, {Mahabal},
  {van Velzen}, {Belecki}, {Bellm}, {Burruss}, {Cenko}, {Cunningham}, {Helou},
  {Kulkarni}, {Masci}, {Prince}, {Reiley}, {Rodriguez}, {Rusholme}, {Smith}, \&
  {Soumagnac}}]{Graham2020}
{Graham}, M.~J., {Ford}, K.~E.~S., {McKernan}, B., {et~al.} 2020, \prl, 124,
  251102, \dodoi{10.1103/PhysRevLett.124.251102}

\bibitem[{{Haggard} {et~al.}(2017){Haggard}, {Nynka}, {Ruan}, {Kalogera},
  {Cenko}, {Evans}, \& {Kennea}}]{Haggard17}
{Haggard}, D., {Nynka}, M., {Ruan}, J.~J., {et~al.} 2017, \apjl, 848, L25,
  \dodoi{10.3847/2041-8213/aa8ede}

\bibitem[{{Hajela} {et~al.}(2019){Hajela}, {Margutti}, {Alexander},
  {Kathirgamaraju}, {Baldeschi}, {Guidorzi}, {Giannios}, {Fong}, {Wu},
  {MacFadyen}, {Paggi}, {Berger}, {Blanchard}, {Chornock}, {Coppejans},
  {Cowperthwaite}, {Eftekhari}, {Gomez}, {Hosseinzadeh}, {Laskar}, {Metzger},
  {Nicholl}, {Paterson}, {Radice}, {Sironi}, {Terreran}, {Villar}, {Williams},
  {Xie}, \& {Zrake}}]{hajela19}
{Hajela}, A., {Margutti}, R., {Alexander}, K.~D., {et~al.} 2019, \apjl, 886,
  L17, \dodoi{10.3847/2041-8213/ab5226}

\bibitem[{{Hallinan} {et~al.}(2017){Hallinan}, {Corsi}, {Mooley}, {Hotokezaka},
  {Nakar}, {Kasliwal}, {Kaplan}, {Frail}, {Myers}, {Murphy}, {De}, {Dobie},
  {Allison}, {Bannister}, {Bhalerao}, {Chandra}, {Clarke}, {Giacintucci}, {Ho},
  {Horesh}, {Kassim}, {Kulkarni}, {Lenc}, {Lockman}, {Lynch}, {Nichols},
  {Nissanke}, {Palliyaguru}, {Peters}, {Piran}, {Rana}, {Sadler}, \&
  {Singer}}]{hallinan17}
{Hallinan}, G., {Corsi}, A., {Mooley}, K.~P., {et~al.} 2017, Science, 358,
  1579, \dodoi{10.1126/science.aap9855}

\bibitem[{{Hosseinzadeh} {et~al.}(2019){Hosseinzadeh}, {Cowperthwaite},
  {Gomez}, {Villar}, {Nicholl}, {Margutti}, {Berger}, {Chornock}, {Paterson},
  {Fong}, {Short}, {Alexander}, {Blanchard}, {Braga}, {Cartier}, {Coppejans},
  {Eftekhari}, {Laskar}, {Patton}, {Pelisoli}, {Reichart}, {Terreran}, \&
  {Williams}}]{Hosseinzadeh19}
{Hosseinzadeh}, G., {Cowperthwaite}, P.~S., {Gomez}, S., {et~al.} 2019, arXiv
  e-prints, arXiv:1905.02186.
\newblock \doarXiv{1905.02186}

\bibitem[{{Hu} {et~al.}(2017){Hu}, {Wu}, {Andreoni}, {Ashley}, {Cooke}, {Cui},
  {Du}, {Dai}, {Gu}, {Hu}, {Lu}, {Li}, {Li}, {Liang}, {Liu}, {Ma}, {Shang},
  {Sun}, {Suntzeff}, {Tao}, {Udden}, {Wang}, {Wang}, {Wen}, {Xiao}, {Su},
  {Yang}, {Yang}, {Yuan}, {Zhou}, {Zhang}, {Zhou}, \& {Zhu}}]{Hu+17}
{Hu}, L., {Wu}, X., {Andreoni}, I., {et~al.} 2017, Science Bulletin, 62, 1433,
  \dodoi{10.1016/j.scib.2017.10.006}

\bibitem[{{Hu} {et~al.}(2020){Hu}, {Caballero Garcia}, {Font}, \& {a larger
  collaboration}}]{GCN26701}
{Hu}, Y.~D., {Caballero Garcia}, M.~D., {Font}, J., \& {a larger
  collaboration}. 2020, GRB Coordinates Network, 26701, 1

\bibitem[{{Hu} {et~al.}(2019{\natexlab{a}}){Hu}, {Geier}, \& {a larger
  collaboration}}]{gcn26502}
{Hu}, Y.~D., {Geier}, S., \& {a larger collaboration}. 2019{\natexlab{a}}, GRB
  Coordinates Network, 26502, 1

\bibitem[{{Hu} {et~al.}(2019{\natexlab{b}}){Hu}, {Valeev}, {Castro-Tirado},
  {Fernand ez-Garcia}, {Sokolov}, {Carrasco}, {Castellon}, \&
  {Scarpa}}]{gcn26405}
{Hu}, Y.~D., {Valeev}, A.~F., {Castro-Tirado}, A.~J., {et~al.}
  2019{\natexlab{b}}, GRB Coordinates Network, 26405, 1

\bibitem[{{Hu} {et~al.}(2019{\natexlab{c}}){Hu}, {Valeev}, {Castro-Tirado},
  {Fernand ez-Garcia}, {Sokolov}, {Carrasco}, {Castellon}, \&
  {Scarpa}}]{gcn26422}
---. 2019{\natexlab{c}}, GRB Coordinates Network, 26422, 1

\bibitem[{{Hu} {et~al.}(2019{\natexlab{d}}){Hu}, {Castro-Tirado}, {Valeev},
  {Sokolov}, {Sanchez-Ramirez}, {Li}, {Ayala}, {Fernandez-Garcia}, {Aceituno},
  {Carrasco}, {Castellon}, {Perez}, {Caballero-Garcia}, {Pandey}, {Garcia}, \&
  {Geier}}]{gcn24359}
{Hu}, Y.~D., {Castro-Tirado}, A.~J., {Valeev}, A.~F., {et~al.}
  2019{\natexlab{d}}, GRB Coordinates Network, 24359, 1

\bibitem[{{Izzo} {et~al.}(2019{\natexlab{a}}){Izzo}, {Carini}, {Yang},
  {Tomasella}, {Cappellaro}, {Greco}, {Martone}, {Benetti}, {Botticella},
  {Branchesi}, {Campana}, {Covino}, {D'Avanzo}, {D'Elia}, {Valle}, {Melandri},
  {Piranomonte}, {Rossi}, \& {Brocato}}]{gcn24340}
{Izzo}, L., {Carini}, R., {Yang}, S., {et~al.} 2019{\natexlab{a}}, GRB
  Coordinates Network, 24340, 1

\bibitem[{{Izzo} {et~al.}(2019{\natexlab{b}}){Izzo}, {Leloudas}, {Bruun},
  {Heintz}, {Milvang-Jensen}, {Hjorth}, {Fynbo}, {Malesani}, {Piranomonte},
  {Pursimo}, {Martikainen}, \& {et al.}}]{gcn25675}
{Izzo}, L., {Leloudas}, G., {Bruun}, S.~H., {et~al.} 2019{\natexlab{b}}, GRB
  Coordinates Network, 25675, 1

\bibitem[{{Jonker} {et~al.}(2019{\natexlab{a}}){Jonker}, {Mata-Sanchez},
  {Fraser}, {Gil}, {Maguire}, {Levan}, {Padilla Torres}, {Steeghs}, {Smartt},
  {Castro}, \& {Valladares}}]{gcn24221}
{Jonker}, P., {Mata-Sanchez}, D., {Fraser}, M., {et~al.} 2019{\natexlab{a}},
  GRB Coordinates Network, 24221, 1

\bibitem[{{Jonker} {et~al.}(2019{\natexlab{b}}){Jonker}, {Fraser}, {Torres},
  {Rodriguez-Gil}, {Maguire}, \& {Kostrzewa-Rutkowska}}]{gcn24154}
{Jonker}, P.~G., {Fraser}, M., {Torres}, M., {et~al.} 2019{\natexlab{b}}, GRB
  Coordinates Network, 24154, 1

\bibitem[{{Kankare} {et~al.}(2019){Kankare}, {Lundqvist}, {Kotak}, {Mattila},
  {Torres}, {Heikkil{\"a}}, {Kuncarayakti}, {Reynolds}, {Moran}, {Steeghs},
  {Lyman}, {Pursimo}, {Martikainen}, \& {GOTO NOT Collaboration}}]{gcn25665}
{Kankare}, E., {Lundqvist}, P., {Kotak}, R., {et~al.} 2019, GRB Coordinates
  Network, 25665, 1

\bibitem[{{Kapadia} {et~al.}(2019){Kapadia}, {Caudill}, {Creighton}, {Farr},
  {Mendell}, {Weinstein}, {Cannon}, {Fong}, {Godwin}, {Lo}, {Magee}, {Meacher},
  {Messick}, {Mohite}, {Mukherjee}, \& {Sachdev}}]{Kapadia19}
{Kapadia}, S.~J., {Caudill}, S., {Creighton}, J.~D.~E., {et~al.} 2019, arXiv
  e-prints.
\newblock \doarXiv{1903.06881}

\bibitem[{{Karambelkar} {et~al.}(2019){Karambelkar}, {De}, {Van Roestel}, \&
  {Kasliwal}}]{gcn25921}
{Karambelkar}, V., {De}, K., {Van Roestel}, J., \& {Kasliwal}, M.~M. 2019, GRB
  Coordinates Network, 25921, 1

\bibitem[{{Kasen} {et~al.}(2015){Kasen}, {Fern{\'a}ndez}, \&
  {Metzger}}]{kasen15}
{Kasen}, D., {Fern{\'a}ndez}, R., \& {Metzger}, B.~D. 2015, MNRAS, 450, 1777,
  \dodoi{10.1093/mnras/stv721}

\bibitem[{{Kasen} {et~al.}(2017){Kasen}, {Metzger}, {Barnes}, {Quataert}, \&
  {Ramirez-Ruiz}}]{kasen17}
{Kasen}, D., {Metzger}, B., {Barnes}, J., {Quataert}, E., \& {Ramirez-Ruiz}, E.
  2017, Nature, 551, 80, \dodoi{10.1038/nature24453}

\bibitem[{{Kasliwal} {et~al.}(2017){Kasliwal}, {Nakar}, {Singer}, {Kaplan},
  {Cook}, {Van Sistine}, {Lau}, {Fremling}, {Gottlieb}, {Jencson}, {Adams},
  {Feindt}, {Hotokezaka}, {Ghosh}, {Perley}, {Yu}, {Piran}, {Allison},
  {Anupama}, {Balasubramanian}, {Bannister}, {Bally}, {Barnes}, {Barway},
  {Bellm}, {Bhalerao}, {Bhattacharya}, {Blagorodnova}, {Bloom}, {Brady},
  {Cannella}, {Chatterjee}, {Cenko}, {Cobb}, {Copperwheat}, {Corsi}, {De},
  {Dobie}, {Emery}, {Evans}, {Fox}, {Frail}, {Frohmaier}, {Goobar}, {Hallinan},
  {Harrison}, {Helou}, {Hinderer}, {Ho}, {Horesh}, {Ip}, {Itoh}, {Kasen},
  {Kim}, {Kuin}, {Kupfer}, {Lynch}, {Madsen}, {Mazzali}, {Miller}, {Mooley},
  {Murphy}, {Ngeow}, {Nichols}, {Nissanke}, {Nugent}, {Ofek}, {Qi}, {Quimby},
  {Rosswog}, {Rusu}, {Sadler}, {Schmidt}, {Sollerman}, {Steele}, {Williamson},
  {Xu}, {Yan}, {Yatsu}, {Zhang}, \& {Zhao}}]{Kasliwal+17}
{Kasliwal}, M.~M., {Nakar}, E., {Singer}, L.~P., {et~al.} 2017, Science, 358,
  1559, \dodoi{10.1126/science.aap9455}

\bibitem[{{Kasliwal} {et~al.}(2019){Kasliwal}, {Coughlin}, {Bellm}, {Singer},
  {de}, {Andreoni}, {Duev}, {Anand }, {Ahumada}, {Cenko}, {Goldstein}, {Ho},
  {Perley}, {Bhalerao}, {Kumar}, {Sharma}, {Goobar}, {Kaplan}, {Sollerman},
  {Bloom}, {Bulla}, {Kawai}, {Yatsu}, {Murata}, {Hanayama}, {Horiuchi},
  {Anupama}, {Rigault}, {Barbarino}, {Biswas}, {Cook}, \& {Helou}}]{gcn24191}
{Kasliwal}, M.~M., {Coughlin}, M.~W., {Bellm}, E.~C., {et~al.} 2019, GRB
  Coordinates Network, 24191, 1

\bibitem[{{Kasliwal} {et~al.}(2020{\natexlab{a}}){Kasliwal}, {Anand},
  {Ahumada}, {Stein}, {Sagues Carracedo}, {Andreoni}, {Coughlin}, {Singer},
  {Kool}, {De}, {Kumar}, {AlMualla}, {Yao}, {Bulla}, {Dobie}, {Reusch},
  {Perley}, {Cenko}, {Bhalerao}, {Kaplan}, {Sollerman}, {Goobar},
  {Copperwheat}, {Bellm}, {Anupama}, {Corsi}, {Nissanke}, {Agudo},
  {Bagdasaryan}, {Barway}, {Belicki}, {Bloom}, {Bolin}, {Buckley}, {Burdge},
  {Burruss}, {Caballero-Garc{\i}a}, {Cannella}, {Castro-Tirado}, {Cook},
  {Cooke}, {Cunningham}, {Dahiwale}, {Deshmukh}, {Dichiara}, {Duev}, {Dutta},
  {Feeney}, {Franckowiak}, {Frederick}, {Fremling}, {Gal-Yam}, {Gatkine},
  {Ghosh}, {Goldstein}, {Golkhou}, {Graham}, {Graham}, {Hankins}, {Helou},
  {Hu}, {Ip}, {Jaodand}, {Karambelkar}, {Kong}, {Kowalski}, {Khandagale},
  {Kulkarni}, {Kumar}, {Laher}, {Li}, {Mahabal}, {Masci}, {Miller}, {Mogotsi},
  {Mohite}, {Mooley}, {Mroz}, {Newman}, {Ngeow}, {Oates}, {Patil}, {Pandey},
  {Pavana}, {Pian}, {Riddle}, {Sanchez-Ram{\i}rez}, {Sharma}, {Singh}, {Smith},
  {Soumagnac}, {Taggart}, {Tan}, {Tzanidakis}, {Troja}, {Valeev}, {Walters},
  {Waratkar}, {Webb}, {Yu}, {Zhang}, {Zhou}, \& {Zolkower}}]{kasliwal20}
{Kasliwal}, M.~M., {Anand}, S., {Ahumada}, T., {et~al.} 2020{\natexlab{a}},
  arXiv e-prints, arXiv:2006.11306.
\newblock \doarXiv{2006.11306}

\bibitem[{{Kasliwal} {et~al.}(2020{\natexlab{b}}){Kasliwal}, {Anand},
  {Ahumada}, {Stein}, {Sagues Carracedo}, {Andreoni}, {Coughlin}, {Singer},
  {Kool}, {De}, {Kumar}, {AlMualla}, {Yao}, {Bulla}, {Dobie}, {Reusch},
  {Perley}, {Cenko}, {Bhalerao}, {Kaplan}, {Sollerman}, {Goobar},
  {Copperwheat}, {Bellm}, {Anupama}, {Corsi}, {Nissanke}, {Agudo},
  {Bagdasaryan}, {Barway}, {Belicki}, {Bloom}, {Bolin}, {Buckley}, {Burdge},
  {Burruss}, {Caballero-Garc{\i}a}, {Cannella}, {Castro-Tirado}, {Cook},
  {Cooke}, {Cunningham}, {Dahiwale}, {Deshmukh}, {Dichiara}, {Duev}, {Dutta},
  {Feeney}, {Franckowiak}, {Frederick}, {Fremling}, {Gal-Yam}, {Gatkine},
  {Ghosh}, {Goldstein}, {Golkhou}, {Graham}, {Graham}, {Hankins}, {Helou},
  {Hu}, {Ip}, {Jaodand}, {Karambelkar}, {Kong}, {Kowalski}, {Khandagale},
  {Kulkarni}, {Kumar}, {Laher}, {Li}, {Mahabal}, {Masci}, {Miller}, {Mogotsi},
  {Mohite}, {Mooley}, {Mroz}, {Newman}, {Ngeow}, {Oates}, {Patil}, {Pandey},
  {Pavana}, {Pian}, {Riddle}, {Sanchez-Ram{\i}rez}, {Sharma}, {Singh}, {Smith},
  {Soumagnac}, {Taggart}, {Tan}, {Tzanidakis}, {Troja}, {Valeev}, {Walters},
  {Waratkar}, {Webb}, {Yu}, {Zhang}, {Zhou}, \& {Zolkower}}]{ZTF2020}
---. 2020{\natexlab{b}}, arXiv e-prints, arXiv:2006.11306.
\newblock \doarXiv{2006.11306}

\bibitem[{{Kawaguchi} {et~al.}(2020){Kawaguchi}, {Shibata}, \&
  {Tanaka}}]{kawaguchi20}
{Kawaguchi}, K., {Shibata}, M., \& {Tanaka}, M. 2020, \apj, 893, 153,
  \dodoi{10.3847/1538-4357/ab8309}

\bibitem[{{Klimenko} {et~al.}(2016){Klimenko}, {Vedovato}, {Drago}, {Salemi},
  {Tiwari}, {Prodi}, {Lazzaro}, {Ackley}, {Tiwari}, {Da Silva}, \&
  {Mitselmakher}}]{Klimenko16}
{Klimenko}, S., {Vedovato}, G., {Drago}, M., {et~al.} 2016, \prd, 93, 042004,
  \dodoi{10.1103/PhysRevD.93.042004}

\bibitem[{{Kool} {et~al.}(2019){Kool}, {Stein}, {Sharma}, {Karambelkar},
  {Kasliwal}, {Perley}, {Brinnel}, {Nordin}, {Anand}, {Coughlin}, {Singer},
  {Andreoni}, {Waratkar}, {Kumar}, {Khandagale}, {Deshmukh}, {Bhalerao},
  {Anupama}, {Dobie}, {Cenko}, {Ahmuda}, {Bellm}, {Kong}, {Franckowiak},
  {Gatkine}, {Ztf Collaboration}, \& {Growth Collaboration}}]{gcn25616}
{Kool}, E., {Stein}, R., {Sharma}, Y., {et~al.} 2019, GRB Coordinates Network,
  25616, 1

\bibitem[{{Kostrzewa-Rutkowska} \& {Gaia Alerts Team}(2020)}]{GCN26686}
{Kostrzewa-Rutkowska}, Z., \& {Gaia Alerts Team}. 2020, GRB Coordinates
  Network, 26686, 1

\bibitem[{{Kostrzewa-Rutkowska}
  {et~al.}(2019{\natexlab{a}}){Kostrzewa-Rutkowska}, {Hodgkin}, {Delgado},
  {Harrison}, {van Leeuwen}, {Rixon}, {Yoldas}, {Eappachen}, \&
  {Jonker}}]{gcn24124}
{Kostrzewa-Rutkowska}, Z., {Hodgkin}, S., {Delgado}, A., {et~al.}
  2019{\natexlab{a}}, GRB Coordinates Network, 24124, 1

\bibitem[{{Kostrzewa-Rutkowska}
  {et~al.}(2019{\natexlab{b}}){Kostrzewa-Rutkowska}, {Hodgkin}, {Delgado},
  {Harrison}, {van Leeuwen}, {Rixon}, {Yoldas}, {Eappachen}, \&
  {Jonker}}]{gcn24134}
---. 2019{\natexlab{b}}, GRB Coordinates Network, 24134, 1

\bibitem[{{Kostrzewa-Rutkowska}
  {et~al.}(2019{\natexlab{c}}){Kostrzewa-Rutkowska}, {Hodgkin}, {Delgado},
  {Harrison}, {van Leeuwen}, {Rixon}, {Yoldas}, {Eappachen}, \&
  {Jonker}}]{gcn24106}
---. 2019{\natexlab{c}}, GRB Coordinates Network, 24106, 1

\bibitem[{{Kostrzewa-Rutkowska}
  {et~al.}(2019{\natexlab{d}}){Kostrzewa-Rutkowska}, {Hodgkin}, {Delgado},
  {Harrison}, {van Leeuwen}, {Rixon}, {Yoldas}, {Eappachen}, \&
  {Jonker}}]{gcn24345}
---. 2019{\natexlab{d}}, GRB Coordinates Network, 24345, 1

\bibitem[{{Kostrzewa-Rutkowska}
  {et~al.}(2019{\natexlab{e}}){Kostrzewa-Rutkowska}, {Hodgkin}, {Delgado},
  {Harrison}, {van Leeuwen}, {Rixon}, {Yoldas}, {Eappachen}, \&
  {Jonker}}]{gcn24354}
---. 2019{\natexlab{e}}, GRB Coordinates Network, 24354, 1

\bibitem[{{Kostrzewa-Rutkowska}
  {et~al.}(2019{\natexlab{f}}){Kostrzewa-Rutkowska}, {Hodgkin}, {Delgado},
  {Harrison}, {van Leeuwen}, {Rixon}, {Yoldas}, {Eappachen}, \&
  {Jonker}}]{gcn24362}
---. 2019{\natexlab{f}}, GRB Coordinates Network, 24362, 1

\bibitem[{{Kostrzewa-Rutkowska}
  {et~al.}(2019{\natexlab{g}}){Kostrzewa-Rutkowska}, {Hodgkin}, {Delgado},
  {Harrison}, {van Leeuwen}, {Rixon}, {Yoldas}, {Eappachen}, \&
  {Jonker}}]{gcn24366}
---. 2019{\natexlab{g}}, GRB Coordinates Network, 24366, 1

\bibitem[{{Kostrzewa-Rutkowska}
  {et~al.}(2019{\natexlab{h}}){Kostrzewa-Rutkowska}, {Hodgkin}, {Delgado},
  {Harrison}, {van Leeuwen}, {Rixon}, {Yoldas}, {Eappachen}, \&
  {Jonker}}]{gcn24344}
---. 2019{\natexlab{h}}, GRB Coordinates Network, 24344, 1

\bibitem[{{Kostrzewa-Rutkowska}
  {et~al.}(2019{\natexlab{i}}){Kostrzewa-Rutkowska}, {Hodgkin}, {Delgado},
  {Harrison}, {van Leeuwen}, {Rixon}, {Yoldas}, {Eappachen}, \&
  {Jonker}}]{gcn24355}
---. 2019{\natexlab{i}}, GRB Coordinates Network, 24355, 1

\bibitem[{{Kostrzewa-Rutkowska}
  {et~al.}(2019{\natexlab{j}}){Kostrzewa-Rutkowska}, {Eappachen}, {Hodgkin},
  {Delgado}, {Harrison}, {van Leeuwen}, {Rixon}, {Yoldas}, {Jonker}, \& {Gaia
  Alerts Team}}]{gcn24977}
{Kostrzewa-Rutkowska}, Z., {Eappachen}, D., {Hodgkin}, S., {et~al.}
  2019{\natexlab{j}}, GRB Coordinates Network, 24977, 1

\bibitem[{{Kostrzewa-Rutkowska}
  {et~al.}(2019{\natexlab{k}}){Kostrzewa-Rutkowska}, {Hodgkin}, {Delgado},
  {Harrison}, {van Leeuwen}, {Rixon}, {Yoldas}, {Eappachen}, {Jonker}, \& {Gaia
  Alerts Team}}]{gcn25689}
{Kostrzewa-Rutkowska}, Z., {Hodgkin}, S., {Delgado}, A., {et~al.}
  2019{\natexlab{k}}, GRB Coordinates Network, 25689, 1

\bibitem[{{Kostrzewa-Rutkowska}
  {et~al.}(2019{\natexlab{l}}){Kostrzewa-Rutkowska}, {Eappachen}, {Hodgkin},
  {Delgado}, {Harrison}, {van Leeuwen}, {Rixon}, {Yoldas}, {Jonker}, \& {Gaia
  Alerts Team}}]{gcn26397}
{Kostrzewa-Rutkowska}, Z., {Eappachen}, D., {Hodgkin}, S., {et~al.}
  2019{\natexlab{l}}, GRB Coordinates Network, 26397, 1

\bibitem[{{Kumar} {et~al.}(2019){Kumar}, {Dutta}, {Waratkar}, {Deshmukh},
  {Khandagale}, {Singh}, {Kool}, {Bhalerao}, {Anupama}, {Stanzin}, \& {Growth
  Collaboration}}]{gcn25632}
{Kumar}, H., {Dutta}, A., {Waratkar}, G., {et~al.} 2019, GRB Coordinates
  Network, 25632, 1

\bibitem[{{Landsman}(1993)}]{IDLforever}
{Landsman}, W.~B. 1993, in Astronomical Society of the Pacific Conference
  Series, Vol.~52, Astronomical Data Analysis Software and Systems II, ed.
  R.~J. {Hanisch}, R.~J.~V. {Brissenden}, \& J.~{Barnes}, 246

\bibitem[{{Ligo Scientific Collaboration} \& {VIRGO
  Collaboration}(2019)}]{gcn24411}
{Ligo Scientific Collaboration}, \& {VIRGO Collaboration}. 2019, GRB
  Coordinates Network, Circular Service, No.~24411, \#1 (2019/April-0), 24411

\bibitem[{{LIGO Scientific Collaboration} \& {Virgo
  Collaboration}(2019{\natexlab{a}})}]{gcn25324}
{LIGO Scientific Collaboration}, \& {Virgo Collaboration}. 2019{\natexlab{a}},
  GRB Coordinates Network, 25324, 1

\bibitem[{{LIGO Scientific Collaboration} \& {Virgo
  Collaboration}(2019{\natexlab{b}})}]{gcn25723}
---. 2019{\natexlab{b}}, GRB Coordinates Network, 25723, 1

\bibitem[{{LIGO Scientific Collaboration} \& {Virgo
  Collaboration}(2019{\natexlab{c}})}]{gcn26454}
---. 2019{\natexlab{c}}, GRB Coordinates Network, 26454, 1

\bibitem[{{LIGO Scientific Collaboration} \& {Virgo
  Collaboration}(2019{\natexlab{d}})}]{gcn26441}
---. 2019{\natexlab{d}}, GRB Coordinates Network, 26441, 1

\bibitem[{{Ligo Scientific Collaboration} \& {VIRGO
  Collaboration}(2019{\natexlab{a}})}]{gcn24069}
{Ligo Scientific Collaboration}, \& {VIRGO Collaboration}. 2019{\natexlab{a}},
  GRB Coordinates Network, Circular Service, No.~24069, \#1 (2019/April-0),
  24069

\bibitem[{{Ligo Scientific Collaboration} \& {VIRGO
  Collaboration}(2019{\natexlab{b}})}]{gcn24168}
---. 2019{\natexlab{b}}, GRB Coordinates Network, Circular Service, No.~24168,
  \#1 (2019/April-0), 24168

\bibitem[{{Ligo Scientific Collaboration} \& {VIRGO
  Collaboration}(2019{\natexlab{c}})}]{gcn24237}
---. 2019{\natexlab{c}}, GRB Coordinates Network, Circular Service, No.~24237,
  \#1 (2019/April-0), 24237

\bibitem[{{Ligo Scientific Collaboration} \& {VIRGO
  Collaboration}(2019{\natexlab{d}})}]{gcn24621}
---. 2019{\natexlab{d}}, GRB Coordinates Network, 24621, 1

\bibitem[{{Ligo Scientific Collaboration} \& {VIRGO
  Collaboration}(2019{\natexlab{e}})}]{gcn24640}
---. 2019{\natexlab{e}}, GRB Coordinates Network, 24640, 1

\bibitem[{{Ligo Scientific Collaboration} \& {VIRGO
  Collaboration}(2019{\natexlab{f}})}]{gcn24922}
---. 2019{\natexlab{f}}, GRB Coordinates Network, 24922, 1

\bibitem[{{LIGO Scientific Collaboration} \& {Virgo
  Collaboration}(2019{\natexlab{a}})}]{gcn25606}
{LIGO Scientific Collaboration}, \& {Virgo Collaboration}. 2019{\natexlab{a}},
  GRB Coordinates Network, 25606, 1

\bibitem[{{LIGO Scientific Collaboration} \& {Virgo
  Collaboration}(2019{\natexlab{b}})}]{gcn25814}
---. 2019{\natexlab{b}}, GRB Coordinates Network, 25814, 1

\bibitem[{{LIGO Scientific Collaboration} \& {Virgo
  Collaboration}(2019{\natexlab{c}})}]{gcn25871}
---. 2019{\natexlab{c}}, GRB Coordinates Network, 25871, 1

\bibitem[{{LIGO Scientific Collaboration} \& {Virgo
  Collaboration}(2019{\natexlab{d}})}]{gcn25876}
---. 2019{\natexlab{d}}, GRB Coordinates Network, 25876, 1

\bibitem[{{LIGO Scientific Collaboration} \& {Virgo
  Collaboration}(2019{\natexlab{e}})}]{GCN26350}
---. 2019{\natexlab{e}}, GRB Coordinates Network, 26350, 1

\bibitem[{{LIGO Scientific Collaboration} \& {Virgo
  Collaboration}(2019{\natexlab{f}})}]{gcn26402}
---. 2019{\natexlab{f}}, GRB Coordinates Network, 26402, 1

\bibitem[{{Ligo Scientific Collaboration} \& {VIRGO
  Collaboration}(2019)}]{gcn24442}
{Ligo Scientific Collaboration}, \& {VIRGO Collaboration}. 2019, GRB
  Coordinates Network, 24442, 1

\bibitem[{{LIGO Scientific Collaboration} \& {Virgo
  Collaboration}(2020{\natexlab{a}})}]{gcn26734}
{LIGO Scientific Collaboration}, \& {Virgo Collaboration}. 2020{\natexlab{a}},
  GRB Coordinates Network, 26734, 1

\bibitem[{{LIGO Scientific Collaboration} \& {Virgo
  Collaboration}(2020{\natexlab{b}})}]{gcn26640}
---. 2020{\natexlab{b}}, GRB Coordinates Network, 26640, 1

\bibitem[{{LIGO Scientific Collaboration} \& {Virgo
  Collaboration}(2020{\natexlab{c}})}]{gcn26657}
---. 2020{\natexlab{c}}, GRB Coordinates Network, 26657, 1

\bibitem[{{LIGO Scientific Collaboration} \& {Virgo
  Collaboration}(2020{\natexlab{d}})}]{gcn26688}
---. 2020{\natexlab{d}}, GRB Coordinates Network, 26688, 1

\bibitem[{{LIGO Scientific Collaboration} \& {Virgo
  Collaboration}(2020{\natexlab{e}})}]{lvcgcn26734}
---. 2020{\natexlab{e}}, GRB Coordinates Network, 26734, 1

\bibitem[{{LIGO Scientific Collaboration} \& {Virgo
  Collaboration}(2020{\natexlab{f}})}]{gcn26759}
---. 2020{\natexlab{f}}, GRB Coordinates Network, 26759, 1

\bibitem[{{LIGO Scientific Collaboration} \& {Virgo
  Collaboration}(2020{\natexlab{g}})}]{gcn26807}
---. 2020{\natexlab{g}}, GRB Coordinates Network, 26807, 1

\bibitem[{{LIGO Scientific Collaboration} \& {Virgo
  Collaboration}(2020{\natexlab{h}})}]{gcn26906}
---. 2020{\natexlab{h}}, GRB Coordinates Network, 26906, 1

\bibitem[{{LIGO Scientific Collaboration} \& {Virgo
  Collaboration}(2020{\natexlab{i}})}]{gcn27184}
---. 2020{\natexlab{i}}, GRB Coordinates Network, 27184, 1

\bibitem[{{LIGO Scientific Collaboration} \& {Virgo
  Collaboration}(2020{\natexlab{j}})}]{gcn27262}
---. 2020{\natexlab{j}}, GRB Coordinates Network, 27262, 1

\bibitem[{{LIGO Scientific Collaboration} \& {Virgo
  Collaboration}(2020{\natexlab{k}})}]{gcn27388}
---. 2020{\natexlab{k}}, GRB Coordinates Network, 27388, 1

\bibitem[{{LIGO Scientific Collaboration} \& {Virgo
  Collaboration}(2020{\natexlab{l}})}]{gcn27419}
---. 2020{\natexlab{l}}, GRB Coordinates Network, 27419, 1

\bibitem[{{Lipunov} {et~al.}(2019{\natexlab{a}}){Lipunov}, {Pogrosheva},
  {Gorbovskoy}, {Tyurina}, {Kornilov}, {Vlasenko}, {Vladimirov}, {Zimnukhov},
  {Kuznetsov}, {Balanutsa}, {Gorbunov}, {Chasovnikov}, {Grinshpun}, {Balakin},
  {Podesta}, {Lopez}, {Francile}, {Podesta}, {Levato}, {Rebolo}, {Serra},
  {Lodieu}, {Israelian}, {Suarez-Andres}, {Buckley}, {Gress}, {Budnev},
  {Ershova}, {Yurkov}, {Gabovich}, {Sergienko}, {Kobcev}, {Tlatov}, {Senik},
  {Parhomenko}, \& {Dormidontov}}]{gcn24076}
{Lipunov}, V., {Pogrosheva}, T., {Gorbovskoy}, E., {et~al.} 2019{\natexlab{a}},
  GRB Coordinates Network, 24076, 1

\bibitem[{{Lipunov} {et~al.}(2019{\natexlab{b}}){Lipunov}, {Pogrosheva},
  {Gorbovskoy}, {Tyurina}, {Kornilov}, {Vlasenko}, {Vladimirov}, {Zimnukhov},
  {Kuznetsov}, {Balanutsa}, {Gorbunov}, {Balakin}, {Chasovnikov}, {Grinshpun},
  {Zhirkov}, {Pozdnyakov}, {Topolev}, {Pogrosheva}, {Shumkov}, {Podesta},
  {Lopez}, {Francile}, {Podesta}, {Levato}, {Rebolo}, {Serra}, {Lodieu},
  {Israelian}, {Suarez-Andres}, {Buckley}, {Gress}, {Budnev}, {Ershova},
  {Yurkov}, {Gabovich}, {Sergienko}, {Kobcev}, {Tlatov}, {Senik}, {Parhomenko},
  \& {Dormidontov}}]{gcn24084}
---. 2019{\natexlab{b}}, GRB Coordinates Network, 24084, 1

\bibitem[{{Lipunov} {et~al.}(2019{\natexlab{c}}){Lipunov}, {Pogrosheva},
  {Gorbovskoy}, {Balakin}, {Kornilov}, {Tyurina}, {Balanutsa}, {Kuznetsov},
  {Vladimirov}, {Vlasenko}, {Gorbunov}, {Zimnukhov}, {Senik}, {Pogrosheva},
  {Kuvshinov}, {Chasovnikov}, {Topolev}, {Posdnyakov}, {Rebolo}, {Serra},
  {Yurkov}, {Gabovich}, {Sergienko}, {Tlatov}, {Dormidontov}, {Podesta},
  {Lopez}, {Podesta}, {Francile}, {Levato}, {Buckley}, {Gres}, {Budnev}, \&
  {Ershova}}]{gcn25649}
---. 2019{\natexlab{c}}, GRB Coordinates Network, 25649, 1

\bibitem[{{Lipunov} {et~al.}(2019{\natexlab{d}}){Lipunov}, {Gorbovskoy},
  {Kornilov}, {Tiurina}, {Balanutsa}, {Kuznetsov}, {Vladimirov}, {Vlasenko},
  {Gorbunov}, {Zimnukhov}, {Senik}, {Pogrosheva}, {Pozdnyakov}, {Chasovnikov},
  {Topolev}, {Balakin}, {Kuvshinov}, {Podesta}, {Lopez}, {Podesta}, {Francile},
  {Levato}, {Rebolo}, {Serra}, {Buckley}, {Gres}, {Budnev}, {Tlatov},
  {Dormidontov}, {Yurkov}, {Gabovich}, \& {Sergienko}}]{gcn25855}
{Lipunov}, V., {Gorbovskoy}, E., {Kornilov}, V., {et~al.} 2019{\natexlab{d}},
  GRB Coordinates Network, 25855, 1

\bibitem[{{Lipunov} {et~al.}(2019{\natexlab{e}}){Lipunov}, {Gorbovskoy},
  {Kornilov}, {Tiurina}, {Balakin}, {Balanutsa}, {Kuznetsov}, {Vladimirov},
  {Vlasenko}, {Gorbunov}, {Zimnukhov}, {Senik}, {Pogrosheva}, {Pozdnyakov},
  {Chasovnikov}, {Topolev}, {Kuvshinov}, {Podesta}, {Lopez}, {Podesta},
  {Francile}, {Levato}, {Rebolo}, {Serra}, {Buckley}, {Gres}, {Budnev},
  {Tlatov}, {Dormidontov}, {Yurkov}, {Gabovich}, \& {Sergienko}}]{gcn25897}
---. 2019{\natexlab{e}}, GRB Coordinates Network, 25897, 1

\bibitem[{{Lipunov} {et~al.}(2019{\natexlab{f}}){Lipunov}, {Gorbovskoy},
  {Kornilov}, {Tiurina}, {Balakin}, {Balanutsa}, {Kuznetsov}, {Vladimirov},
  {Vlasenko}, {Gorbunov}, {Zimnukhov}, {Senik}, {Pogrosheva}, {Pozdnyakov},
  {Chasovnikov}, {Topolev}, {Kuvshinov}, {Podesta}, {Lopez}, {Podesta},
  {Francile}, {Levato}, {Rebolo}, {Serra}, {Buckley}, {Gres}, {Budnev},
  {Tlatov}, {Dormidontov}, {Yurkov}, {Gabovich}, \& {Sergienko}}]{gcn25900}
---. 2019{\natexlab{f}}, GRB Coordinates Network, 25900, 1

\bibitem[{{Lipunov} {et~al.}(2019{\natexlab{g}}){Lipunov}, {Gorbovskoy},
  {Kornilov}, {Tyurina}, {Balanutsa}, {Kuznetsov}, {Balakin}, {Vladimirov},
  {Vlasenko}, {Gorbunov}, {Zimnukhov}, {Senik}, {Pogrosheva}, {Kuvshinov},
  {Podesta}, {Lopez}, {Podesta}, {Francile}, {Levato}, {Rebolo}, {Serra},
  {Buckley}, {Gres}, {Budnev}, {Ershova}, {Tlatov}, {Dormidontov}, {Yurkov},
  {Gabovich}, \& {Sergienko}}]{gcn26379}
---. 2019{\natexlab{g}}, GRB Coordinates Network, 26379, 1

\bibitem[{{Lipunov} {et~al.}(2019{\natexlab{h}}){Lipunov}, {Gorbovskoy},
  {Kornilov}, {Tyurina}, {Balanutsa}, {Kuznetsov}, {Vladimirov}, {Vlasenko},
  {Gorbunov}, {Zimnukhov}, {Senik}, {Chasovnikov}, {Zhirkov}, {Pozdnyakov},
  {Balakin}, {Pogrosheva}, {Rebolo}, {Serra}, {Buckley}, {Podesta}, {Lopez},
  {Podesta}, \& {Francile}}]{gcn26578}
---. 2019{\natexlab{h}}, GRB Coordinates Network, 26578, 1

\bibitem[{{Lipunov} {et~al.}(2017{\natexlab{a}}){Lipunov}, {Gorbovskoy},
  {Kornilov}, {.~Tyurina}, {Balanutsa}, {Kuznetsov}, {Vlasenko}, {Kuvshinov},
  {Gorbunov}, {Buckley}, {Krylov}, {Podesta}, {Lopez}, {Podesta}, {Levato},
  {Saffe}, {Mallamachi}, {Potter}, {Budnev}, {Gress}, {Ishmuhametova},
  {Vladimirov}, {Zimnukhov}, {Yurkov}, {Sergienko}, {Gabovich}, {Rebolo},
  {Serra-Ricart}, {Israelyan}, {Chazov}, {Wang}, {Tlatov}, \&
  {Panchenko}}]{Lipunov+17}
{Lipunov}, V.~M., {Gorbovskoy}, E., {Kornilov}, V.~G., {et~al.}
  2017{\natexlab{a}}, \apjl, 850, L1, \dodoi{10.3847/2041-8213/aa92c0}

\bibitem[{{Lipunov} {et~al.}(2017{\natexlab{b}}){Lipunov}, {Gorbovskoy},
  {Kornilov}, {.~Tyurina}, {Balanutsa}, {Kuznetsov}, {Vlasenko}, {Kuvshinov},
  {Gorbunov}, {Buckley}, {Krylov}, {Podesta}, {Lopez}, {Podesta}, {Levato},
  {Saffe}, {Mallamachi}, {Potter}, {Budnev}, {Gress}, {Ishmuhametova},
  {Vladimirov}, {Zimnukhov}, {Yurkov}, {Sergienko}, {Gabovich}, {Rebolo},
  {Serra-Ricart}, {Israelyan}, {Chazov}, {Wang}, {Tlatov}, \&
  {Panchenko}}]{lipunov17}
---. 2017{\natexlab{b}}, \apjl, 850, L1, \dodoi{10.3847/2041-8213/aa92c0}

\bibitem[{{Lipunov} {et~al.}(2019{\natexlab{i}}){Lipunov}, {Vladimirov},
  {Gorbovskoi}, {Kuznetsov}, {Zimnukhov}, {Balanutsa}, {Kornilov}, {Tyurina},
  {Gress}, {Vlasenko}, {Gabovich}, {Yurkov}, {Kuvshinov}, \&
  {Senik}}]{lipunov19}
{Lipunov}, V.~M., {Vladimirov}, V.~V., {Gorbovskoi}, E.~S., {et~al.}
  2019{\natexlab{i}}, Astronomy Reports, 63, 293,
  \dodoi{10.1134/S1063772919040073}

\bibitem[{{Lundquist} {et~al.}(2019{\natexlab{a}}){Lundquist}, {Paterson},
  {Fong}, {Sand}, {Andrews}, {Shivaei}, {Daly}, {Valenti}, {Yang},
  {Christensen}, {Gibbs}, {Shelly}, {Wyatt}, {Eskandari}, {Kuhn}, {Amaro},
  {Arcavi}, {Behroozi}, {Butler}, {Chomiuk}, {Corsi}, {Drout}, {Egami}, {Fan},
  {Foley}, {Frye}, {Gabor}, {Green}, {Grier}, {Guzman}, {Hamden}, {Howell},
  {Jannuzi}, {Kelly}, {Milne}, {Moe}, {Nugent}, {Olszewski}, {Palazzi},
  {Paschalidis}, {Psaltis}, {Reichart}, {Rest}, {Rossi}, {Schroeder}, {Smith},
  {Smith}, {Spekkens}, {Strader}, {Stark}, {Trilling}, {Veillet}, {Wagner},
  {Weiner}, {Wheeler}, {Williams}, \& {Zabludoff}}]{Lundquist19}
{Lundquist}, M.~J., {Paterson}, K., {Fong}, W., {et~al.} 2019{\natexlab{a}},
  \apjl, 881, L26, \dodoi{10.3847/2041-8213/ab32f2}

\bibitem[{{Lundquist} {et~al.}(2019{\natexlab{b}}){Lundquist}, {Paterson},
  {Sand}, {Valenti}, {Yang}, {Wyatt}, {Christensen}, {Gibbs}, \&
  {Shelly}}]{gcn24079}
{Lundquist}, M.~J., {Paterson}, K., {Sand}, D.~J., {et~al.} 2019{\natexlab{b}},
  GRB Coordinates Network, 24079, 1

\bibitem[{{Lundquist} {et~al.}(2019{\natexlab{c}}){Lundquist}, {Paterson},
  {Sand}, {Valenti}, {Yang}, {Wyatt}, {Christensen}, {Gibbs}, {Shelly}, \&
  {Andrews}}]{gcn24172}
---. 2019{\natexlab{c}}, GRB Coordinates Network, 24172, 1

\bibitem[{{Lundquist} {et~al.}(2019{\natexlab{d}}){Lundquist}, {Sand}, {Fong},
  {Rastinejad}, {Paterson}, {Andrews}, {Wyatt}, {Christensen}, {Gibbs},
  {Shelly}, \& {Saguaro Collaboration}}]{gcn26473}
{Lundquist}, M.~J., {Sand}, D.~J., {Fong}, W.-F., {et~al.} 2019{\natexlab{d}},
  GRB Coordinates Network, 26473, 1

\bibitem[{{Lundquist} {et~al.}(2020{\natexlab{a}}){Lundquist}, {Sand},
  {Paterson}, {Fong}, {Rastinejad}, {Andrews}, {Wyatt}, {Christensen}, {Gibbs},
  {Shelly}, \& {Saguaro Collaboration}}]{gcn26750}
{Lundquist}, M.~J., {Sand}, D.~J., {Paterson}, K., {et~al.} 2020{\natexlab{a}},
  GRB Coordinates Network, 26750, 1

\bibitem[{{Lundquist} {et~al.}(2020{\natexlab{b}}){Lundquist}, {Sand},
  {Rastinejad}, {Paterson}, {Andrews}, {Fong}, {Wyatt}, {Christensen}, {Gibbs},
  {Shelly}, \& {Saguaro Collaboration}}]{gcn26753}
{Lundquist}, M.~J., {Sand}, D.~J., {Rastinejad}, J., {et~al.}
  2020{\natexlab{b}}, GRB Coordinates Network, 26753, 1

\bibitem[{{Ma{\'\i}z Apell{\'a}niz} \& {Weiler}(2018)}]{Maiz18}
{Ma{\'\i}z Apell{\'a}niz}, J., \& {Weiler}, M. 2018, \aap, 619, A180,
  \dodoi{10.1051/0004-6361/201834051}

\bibitem[{{Margalit} \& {Metzger}(2019)}]{MargalitMetzger2019}
{Margalit}, B., \& {Metzger}, B.~D. 2019, \apjl, 880, L15,
  \dodoi{10.3847/2041-8213/ab2ae2}

\bibitem[{{Margutti} {et~al.}(2017){Margutti}, {Berger}, {Fong}, {Guidorzi},
  {Alexander}, {Metzger}, {Blanchard}, {Cowperthwaite}, {Chornock},
  {Eftekhari}, {Nicholl}, {Villar}, {Williams}, {Annis}, {Brown}, {Chen},
  {Doctor}, {Frieman}, {Holz}, {Sako}, \& {Soares-Santos}}]{margutti17}
{Margutti}, R., {Berger}, E., {Fong}, W., {et~al.} 2017, \apjl, 848, L20,
  \dodoi{10.3847/2041-8213/aa9057}

\bibitem[{{Margutti} {et~al.}(2018){Margutti}, {Alexander}, {Xie}, {Sironi},
  {Metzger}, {Kathirgamaraju}, {Fong}, {Blanchard}, {Berger}, {MacFadyen},
  {Giannios}, {Guidorzi}, {Hajela}, {Chornock}, {Cowperthwaite}, {Eftekhari},
  {Nicholl}, {Villar}, {Williams}, \& {Zrake}}]{margutti2018}
{Margutti}, R., {Alexander}, K.~D., {Xie}, X., {et~al.} 2018, \apjl, 856, L18,
  \dodoi{10.3847/2041-8213/aab2ad}

\bibitem[{{Masci} {et~al.}(2019){Masci}, {Laher}, {Rusholme}, {Shupe}, {Groom},
  {Surace}, {Jackson}, {Monkewitz}, {Beck}, {Flynn}, {Terek}, {Landry},
  {Hacopians}, {Desai}, {Howell}, {Brooke}, {Imel}, {Wachter}, {Ye}, {Lin},
  {Cenko}, {Cunningham}, {Rebbapragada}, {Bue}, {Miller}, {Mahabal}, {Bellm},
  {Patterson}, {Juri{\'c}}, {Golkhou}, {Ofek}, {Walters}, {Graham}, {Kasliwal},
  {Dekany}, {Kupfer}, {Burdge}, {Cannella}, {Barlow}, {Van Sistine}, {Giomi},
  {Fremling}, {Blagorodnova}, {Levitan}, {Riddle}, {Smith}, {Helou}, {Prince},
  \& {Kulkarni}}]{ztfarchive}
{Masci}, F.~J., {Laher}, R.~R., {Rusholme}, B., {et~al.} 2019, \pasp, 131,
  018003, \dodoi{10.1088/1538-3873/aae8ac}

\bibitem[{{Mazaeva} \& {IKI FuN}(2020)}]{gcn26819}
{Mazaeva}, E., \& {IKI FuN}, G. 2020, GRB Coordinates Network, 26819, 1

\bibitem[{{McBrien} {et~al.}(2019{\natexlab{a}}){McBrien}, {Smartt}, {Smith},
  {Young}, {Denneau}, {Flewelling}, {Heinze}, {Tonry}, {Weiland}, {Gillanders},
  {Srivastav}, {O'Neil}, {Clark}, {Sim}, {Rest}, {Stalder}, {Stubbs},
  {Magnier}, {Schultz}, {Huber}, \& {Chambers}}]{gcn24197}
{McBrien}, O., {Smartt}, S., {Smith}, K.~W., {et~al.} 2019{\natexlab{a}}, GRB
  Coordinates Network, 24197, 1

\bibitem[{{McBrien} {et~al.}(2019{\natexlab{b}}){McBrien}, {Smith}, {Smartt},
  {Young}, {Srivastav}, {Gillanders}, {Clark}, {Fulton}, {Huber}, {Chambers},
  {Flewelling}, {Willman}, {Schultz}, {Magnier}, {Waters}, {Bulger},
  {Wainscoat}, \& {Wright}}]{gcn26364}
{McBrien}, O., {Smith}, K.~W., {Smartt}, S.~J., {et~al.} 2019{\natexlab{b}},
  GRB Coordinates Network, 26364, 1

\bibitem[{{McBrien} {et~al.}(2019{\natexlab{c}}){McBrien}, {Smartt}, {Smith},
  {Young}, {Srivastav}, {Clark}, {O'Neill}, {Fulton}, {McLaughlin}, {Chambers},
  {Huber}, {Schultz}, {de Boer}, {Bulger}, {Fairlamb}, {Lin}, {Lowe},
  {Magnier}, {Wainscoat}, {Willman}, {Rest}, {Stubbs}, {Chen}, \& {Pan-Starrs
  Collaboration}}]{mcbrien19gcn}
{McBrien}, O., {Smartt}, S.~J., {Smith}, K.~W., {et~al.} 2019{\natexlab{c}},
  GRB Coordinates Network, 26485, 1

\bibitem[{{McCully} {et~al.}(2017){McCully}, {Hiramatsu}, {Howell},
  {Hosseinzadeh}, {Arcavi}, {Kasen}, {Barnes}, {Shara}, {Williams},
  {V{\"a}is{\"a}nen}, {Potter}, {Romero-Colmenero}, {Crawford}, {Buckley},
  {Cooke}, {Andreoni}, {Pritchard}, {Mao}, {Gromadzki}, \&
  {Burke}}]{McCully+17}
{McCully}, C., {Hiramatsu}, D., {Howell}, D.~A., {et~al.} 2017, \apjl, 848,
  L32, \dodoi{10.3847/2041-8213/aa9111}

\bibitem[{{McCully} {et~al.}(2020){McCully}, {Hiramatsu}, {Burke}, {Andrews},
  {Peligrino}, {Howell}, {Arcavi}, {De Carvalho}, {F{\"o}rster}, {Foley},
  {Coulter}, {Kilpatrick}, {Sand}, {Valenti}, {Soares-Santos}, {Rembold},
  {Rest}, {Kasen}, {Metzger}, {Piro}, {Quataert}, {Ramirez-Ruiz}, {Wheeler},
  {Bauer}, {Bloom}, {Brink}, {Cooke}, {Clocchiatti}, {Filippenko}, {Freedman},
  {Garnavich}, {Horvath}, {Jha}, {Kirshner}, {Krisciunas}, {Lin}, {Madore},
  {Makler}, {Narayan}, {Prochaska}, {Riess}, {Sturani}, {Suntzeff}, {Tanaka},
  {Tucker}, {Vinko}, {Wang}, {Contreras}, {D'Andrea}, {Dimitriadis}, {Jones},
  {Lundquist}, {Olivares}, {Palmese}, {Pan}, {Scolnic}, {Zheng}, {Bernardo},
  {Bostroem}, {Murguia Berthier}, {Rodr{\'\i}guez}, {Rojas-Bravo}, {Siebert},
  \& {Souza}}]{gcn27073}
{McCully}, C., {Hiramatsu}, D., {Burke}, J., {et~al.} 2020, GRB Coordinates
  Network, 27073, 1

\bibitem[{{McKernan} {et~al.}(2018){McKernan}, {Ford}, {Bellovary}, {Leigh},
  {Haiman}, {Kocsis}, {Lyra}, {Mac Low}, {Metzger}, {O'Dowd}, {Endlich}, \&
  {Rosen}}]{McKernan2018}
{McKernan}, B., {Ford}, K.~E.~S., {Bellovary}, J., {et~al.} 2018, \apj, 866,
  66, \dodoi{10.3847/1538-4357/aadae5}

\bibitem[{{McMahon} {et~al.}(2013){McMahon}, {Banerji}, {Gonzalez}, {Koposov},
  {Bejar}, {Lodieu}, {Rebolo}, \& {VHS Collaboration}}]{McMahon2013}
{McMahon}, R.~G., {Banerji}, M., {Gonzalez}, E., {et~al.} 2013, The Messenger,
  154, 35

\bibitem[{{Metzger}(2017)}]{Metzger17}
{Metzger}, B.~D. 2017, Living Reviews in Relativity, 20, 3,
  \dodoi{10.1007/s41114-017-0006-z}

\bibitem[{{Metzger} \& {Fern{\'a}ndez}(2014)}]{metzger14}
{Metzger}, B.~D., \& {Fern{\'a}ndez}, R. 2014, \mnras, 441, 3444,
  \dodoi{10.1093/mnras/stu802}

\bibitem[{{Mooley} {et~al.}(2018){Mooley}, {Nakar}, {Hotokezaka}, {Hallinan},
  {Corsi}, {Frail}, {Horesh}, {Murphy}, {Lenc}, {Kaplan}, {de}, {Dobie}, {Chand
  ra}, {Deller}, {Gottlieb}, {Kasliwal}, {Kulkarni}, {Myers}, {Nissanke},
  {Piran}, {Lynch}, {Bhalerao}, {Bourke}, {Bannister}, \& {Singer}}]{mooley18a}
{Mooley}, K.~P., {Nakar}, E., {Hotokezaka}, K., {et~al.} 2018, \nat, 554, 207,
  \dodoi{10.1038/nature25452}

\bibitem[{{Morgan} {et~al.}(2020{\natexlab{a}}){Morgan}, {Palmese}, {Garcia},
  {Soares-Santos}, {Herner}, {Bom}, {Nicolaou}, {Vivas}, {Zenteno},
  {Garcia-Bellido}, \& {Desgw Collaboration}}]{GCN27366}
{Morgan}, R., {Palmese}, A., {Garcia}, A., {et~al.} 2020{\natexlab{a}}, GRB
  Coordinates Network, 27366, 1

\bibitem[{{Morgan} {et~al.}(2020{\natexlab{b}}){Morgan}, {Soares-Santos},
  {Annis}, {Herner}, {Garcia}, {Palmese}, {Drlica-Wagner}, {Kessler},
  {Garcia-Bellido}, {Sherman}, {Allam}, {Bechtol}, {Bom}, {Brout}, {Butler},
  {Butner}, {Cartier}, {Chen}, {Conselice}, {Cook}, {Davis}, {Doctor}, {Farr},
  {Figueiredo}, {Finley}, {Foley}, {Galarza}, {Gill}, {Gruendl}, {Holz},
  {Kuropatkin}, {Lidman}, {Lin}, {Malik}, {Mann}, {Marriner}, {Marshall},
  {Martinez-Vazquez}, {Meza}, {Neilsen}, {Nicolaou}, {Olivares E.},
  {Paz-Chinchon}, {Points}, {Quirola}, {Rodriguez}, {Sako}, {Scolnic}, {Smith},
  {Sobreira}, {Tucker}, {Vivas}, {Wiesner}, {Wood}, {Yanny}, {Zenteno},
  {Abbott}, {Aguena}, {Avila}, {Bertin}, {Bhargava}, {Brooks}, {Burke},
  {Carnero Rosell}, {Carrasco Kind}, {Carretero}, {da Costa}, {Costanzi}, {De
  Vicente}, {Desai}, {Diehl}, {Doel}, {Eifler}, {Everett}, {Flaugher},
  {Frieman}, {Gaztanaga}, {Gerdes}, {Gruen}, {Gschwend}, {Gutierrez},
  {Hartley}, {Hinton}, {Hollowood}, {Honscheid}, {James}, {Kuehn}, {Lahav},
  {Lima}, {Maia}, {March}, {Miquel}, {Ogando}, {Plazas}, {Roodman}, {Sanchez},
  {Scarpine}, {Schubnell}, {Serrano}, {Sevilla-Noarbe}, {Suchyta}, \&
  {Tarle}}]{Morgan20}
{Morgan}, R., {Soares-Santos}, M., {Annis}, J., {et~al.} 2020{\natexlab{b}},
  arXiv e-prints, arXiv:2006.07385.
\newblock \doarXiv{2006.07385}

\bibitem[{{Morgan} {et~al.}(2020{\natexlab{c}}){Morgan}, {Garcia},
  {Soares-Santos}, {Herner}, {Bom}, {Palmese}, {Tabbutt}, {Stringer}, {Vivas},
  {Zenteno}, {Puzia}, {Peng}, \& {Desgw Collaboration}}]{gcn27227}
{Morgan}, R., {Garcia}, A., {Soares-Santos}, M., {et~al.} 2020{\natexlab{c}},
  GRB Coordinates Network, 27227, 1

\bibitem[{{Morokuma} {et~al.}(2016){Morokuma}, {Tanaka}, {Asakura}, {Abe},
  {Tristram}, {Utsumi}, {Doi}, {Fujisawa}, {Itoh}, {Itoh}, {Kawabata}, {Kawai},
  {Kuroda}, {Matsubayashi}, {Motohara}, {Murata}, {Nagayama}, {Ohta}, {Saito},
  {Tamura}, {Tominaga}, {Uemura}, {Yanagisawa}, {Yatsu}, \& {Yoshida}}]{jgem}
{Morokuma}, T., {Tanaka}, M., {Asakura}, Y., {et~al.} 2016, \pasj, 68, L9,
  \dodoi{10.1093/pasj/psw061}

\bibitem[{{Nascimbeni} {et~al.}(2019){Nascimbeni}, {Salmaso}, {Tomasella},
  {Benetti}, {D'Avanzo}, {Cappellaro}, {Brocato}, \& {Grawita
  Collaboration}}]{gcn25661}
{Nascimbeni}, V., {Salmaso}, I., {Tomasella}, L., {et~al.} 2019, GRB
  Coordinates Network, 25661, 1

\bibitem[{{Nicholl} {et~al.}(2017){Nicholl}, {Berger}, {Kasen}, {Metzger},
  {Elias}, {Brice{\~n}o}, {Alexander}, {Blanchard}, {Chornock},
  {Cowperthwaite}, {Eftekhari}, {Fong}, {Margutti}, {Villar}, {Williams},
  {Brown}, {Annis}, {Bahramian}, {Brout}, {Brown}, {Chen}, {Clemens},
  {Dennihy}, {Dunlap}, {Holz}, {Marchesini}, {Massaro}, {Moskowitz},
  {Pelisoli}, {Rest}, {Ricci}, {Sako}, {Soares-Santos}, \&
  {Strader}}]{Nicholl17}
{Nicholl}, M., {Berger}, E., {Kasen}, D., {et~al.} 2017, \apjl, 848, L18,
  \dodoi{10.3847/2041-8213/aa9029}

\bibitem[{{Nicholl} {et~al.}(2019){Nicholl}, {Cartier}, {Pelisoli}, {Berger},
  {Blanchard}, {Eftekhari}, {Gomez}, {Hosseinzadeh}, {Villar}, {Williams},
  {Cowperthwaite}, {Alexander}, {Coppejans}, {Fong}, {Margutti}, {Terreran},
  {Chornock}, {Braga}, {Chomiuk}, {Strader}, {Clemens}, {Reichart}, {Drout},
  {Sand}, {Smith}, {Kasen}, \& {Metzger}}]{gcn24321}
{Nicholl}, M., {Cartier}, R., {Pelisoli}, I., {et~al.} 2019, GRB Coordinates
  Network, 24321, 1

\bibitem[{{Nordin} {et~al.}(2019){Nordin}, {Brinnel}, {van Santen}, {Bulla},
  {Feindt}, {Franckowiak}, {Fremling}, {Gal-Yam}, {Giomi}, {Kowalski},
  {Mahabal}, {Miranda}, {Rauch}, {Reusch}, {Rigault}, {Schulze}, {Sollerman},
  {Stein}, {Yaron}, {van Velzen}, \& {Ward}}]{Nordin2019}
{Nordin}, J., {Brinnel}, V., {van Santen}, J., {et~al.} 2019, \aap, 631, A147,
  \dodoi{10.1051/0004-6361/201935634}

\bibitem[{{Oates} {et~al.}(2019){Oates}, {Breeveld}, {Kuin}, {Marshall},
  {Brown}, {Gronwall}, {Page}, {De Pasquale}, {Siegel}, {D'Ai}, {D'Avanzo},
  {Barthelmy}, {Beardmore}, {Burrows}, {Campana}, {Cenko}, {Cusumano},
  {D'Elia}, {Evans}, {Giommi}, {Hartmann}, {Kennea}, {Klingler}, {Krimm},
  {Lien}, {Malesani}, {Meland ri}, {Nousek}, {O'Brien}, {Osborne}, {Pagani},
  {Page}, {Palmer}, {Perri}, {Racusin}, {Sakamoto}, {Sbarufatti},
  {Tagliaferri}, {Tohuvavohu}, {Troja}, \& {Swift Team}}]{gcn25901}
{Oates}, S.~R., {Breeveld}, A.~A., {Kuin}, N.~P.~M., {et~al.} 2019, GRB
  Coordinates Network, 25901, 1

\bibitem[{{Onori} {et~al.}(2019){Onori}, {Turatto}, {Benetti}, {Tomasella},
  {Nicastro}, {Palazzi}, {Melandri}, {Piranomonte}, {Yang}, {D'Avanzo},
  {Cecconi}, {Ghedina}, {Stoev}, {Brosio}, \& {Savaglio}}]{gcn24090}
{Onori}, F., {Turatto}, M., {Benetti}, S., {et~al.} 2019, GRB Coordinates
  Network, 24090, 1

\bibitem[{{Paterson} {et~al.}(2020){Paterson}, {Lundquist}, {Rastinejad},
  {Fong}, \& {Sand}}]{tnscands}
{Paterson}, K., {Lundquist}, M., {Rastinejad}, J., {Fong}, W., \& {Sand}, D.
  2020, Transient Name Server Discovery Report, 2020-3619, 1

\bibitem[{{Paterson} {et~al.}(2019{\natexlab{a}}){Paterson}, {Lundquist},
  {Sand}, {Fong}, {Rastinejad}, {Andrews}, {Wyatt}, {Christensen}, {Gibbs},
  {Shelly}, \& {Saguaro Collaboration}}]{gcn25915}
{Paterson}, K., {Lundquist}, M.~J., {Sand}, D.~J., {et~al.} 2019{\natexlab{a}},
  GRB Coordinates Network, 25915, 1

\bibitem[{{Paterson} {et~al.}(2019{\natexlab{b}}){Paterson}, {Lundquist},
  {Sand}, {Rastinejad}, {Fong}, {Andrews}, {Wyatt}, {Christensen}, {Gibbs},
  {Shelly}, {Leonard}, \& {Saguaro Collaboration}}]{gcn26360}
---. 2019{\natexlab{b}}, GRB Coordinates Network, 26360, 1

\bibitem[{{Pavana} {et~al.}(2019){Pavana}, {Anupama}, {Kiran}, \&
  {Bhalerao}}]{gcn24200}
{Pavana}, M., {Anupama}, G.~C., {Kiran}, B.~S., \& {Bhalerao}, V. 2019, GRB
  Coordinates Network, 24200, 1

\bibitem[{{Perego} {et~al.}(2014){Perego}, {Rosswog}, {Cabez{\'o}n},
  {Korobkin}, {K{\"a}ppeli}, {Arcones}, \& {Liebend{\"o}rfer}}]{perego14}
{Perego}, A., {Rosswog}, S., {Cabez{\'o}n}, R.~M., {et~al.} 2014, \mnras, 443,
  3134, \dodoi{10.1093/mnras/stu1352}

\bibitem[{{Perley} \& {Copperwheat}(2019)}]{gcn24202}
{Perley}, D.~A., \& {Copperwheat}, C.~M. 2019, GRB Coordinates Network, 24202,
  1

\bibitem[{{Perley} {et~al.}(2019{\natexlab{a}}){Perley}, {Copperwheat}, \&
  {Growth Collaboration}}]{PerleyCopperwheat19}
{Perley}, D.~A., {Copperwheat}, C.~M., \& {Growth Collaboration}.
  2019{\natexlab{a}}, GRB Coordinates Network, 26426, 1

\bibitem[{{Perley} {et~al.}(2019{\natexlab{b}}){Perley}, {Copperwheat}, \&
  {Taggart}}]{gcn24204}
{Perley}, D.~A., {Copperwheat}, C.~M., \& {Taggart}, K.~L. 2019{\natexlab{b}},
  GRB Coordinates Network, 24204, 1

\bibitem[{{Perley} {et~al.}(2019{\natexlab{c}}){Perley}, {Goobar}, {Kasliwal},
  {Coughlin}, {Miller}, {Ahumada}, {Jencson}, {Kumar}, {Kaplan}, {Anand},
  {Singer}, {Andreoni}, {Bhalerao}, {Goldstein}, {Duev}, {Cenko}, {Bellm},
  {de}, {Biswas}, {de}, \& {Bloom}}]{gcn24331}
{Perley}, D.~A., {Goobar}, A., {Kasliwal}, M.~M., {et~al.} 2019{\natexlab{c}},
  GRB Coordinates Network, 24331, 1

\bibitem[{{Perna} {et~al.}(2018){Perna}, {Chruslinska}, {Corsi}, \&
  {Belczynski}}]{Perna2018}
{Perna}, R., {Chruslinska}, M., {Corsi}, A., \& {Belczynski}, K. 2018, \mnras,
  477, 4228, \dodoi{10.1093/mnras/sty814}

\bibitem[{{Perna} {et~al.}(2019){Perna}, {Lazzati}, \& {Farr}}]{Perna2019}
{Perna}, R., {Lazzati}, D., \& {Farr}, W. 2019, \apj, 875, 49,
  \dodoi{10.3847/1538-4357/ab107b}

\bibitem[{{Pian} {et~al.}(2017){Pian}, {D'Avanzo}, {Benetti}, {Branchesi},
  {Brocato}, {Campana}, {Cappellaro}, {Covino}, {D'Elia}, {Fynbo}, {Getman},
  {Ghirlanda}, {Ghisellini}, {Grado}, {Greco}, {Hjorth}, {Kouveliotou},
  {Levan}, {Limatola}, {Malesani}, {Mazzali}, {Melandri}, {M{\o}ller},
  {Nicastro}, {Palazzi}, {Piranomonte}, {Rossi}, {Salafia}, {Selsing},
  {Stratta}, {Tanaka}, {Tanvir}, {Tomasella}, {Watson}, {Yang}, {Amati},
  {Antonelli}, {Ascenzi}, {Bernardini}, {Bo{\"e}r}, {Bufano}, {Bulgarelli},
  {Capaccioli}, {Casella}, {Castro-Tirado}, {Chassande-Mottin}, {Ciolfi},
  {Copperwheat}, {Dadina}, {De Cesare}, {di Paola}, {Fan}, {Gendre},
  {Giuffrida}, {Giunta}, {Hunt}, {Israel}, {Jin}, {Kasliwal}, {Klose}, {Lisi},
  {Longo}, {Maiorano}, {Mapelli}, {Masetti}, {Nava}, {Patricelli}, {Perley},
  {Pescalli}, {Piran}, {Possenti}, {Pulone}, {Razzano}, {Salvaterra},
  {Schipani}, {Spera}, {Stamerra}, {Stella}, {Tagliaferri}, {Testa}, {Troja},
  {Turatto}, {Vergani}, \& {Vergani}}]{Pian+17}
{Pian}, E., {D'Avanzo}, P., {Benetti}, S., {et~al.} 2017, \nat, 551, 67,
  \dodoi{10.1038/nature24298}

\bibitem[{{Podlesnyi} \& {Dzhatdoev}(2020)}]{Podlesnyi2020}
{Podlesnyi}, E., \& {Dzhatdoev}, T. 2020, arXiv e-prints, arXiv:2007.03086.
\newblock \doarXiv{2007.03086}

\bibitem[{{Pozanenko} {et~al.}(2018){Pozanenko}, {Barkov}, {Minaev}, {Volnova},
  {Mazaeva}, {Moskvitin}, {Krugov}, {Samodurov}, {Loznikov}, \&
  {Lyutikov}}]{Pozanenko+17}
{Pozanenko}, A.~S., {Barkov}, M.~V., {Minaev}, P.~Y., {et~al.} 2018, \apjl,
  852, L30, \dodoi{10.3847/2041-8213/aaa2f6}

\bibitem[{{Prentice} {et~al.}(2019){Prentice}, {Maguire}, {Skillen}, {Magee},
  \& {Clark}}]{AT2019dma}
{Prentice}, S.~J., {Maguire}, K., {Skillen}, K., {Magee}, M.~R., \& {Clark}, P.
  2019, Transient Name Server Classification Report, 2019-617, 1

\bibitem[{{Rosell} {et~al.}(2019){Rosell}, {Gebhardt}, {Zimmerman}, {Shetrone},
  {Fryer}, {Wheeler}, {Odewahn}, {McReynolds}, \& {LIGO Telescope Hobby-Eberly
  Response Team}}]{gcn25622}
{Rosell}, M.~J.~B., {Gebhardt}, K., {Zimmerman}, A., {et~al.} 2019, GRB
  Coordinates Network, 25622, 1

\bibitem[{{Safarzadeh} {et~al.}(2020){Safarzadeh}, {Ramirez-Ruiz}, \&
  {Berger}}]{Safarzadeh2020}
{Safarzadeh}, M., {Ramirez-Ruiz}, E., \& {Berger}, E. 2020, \apj, 900, 13,
  \dodoi{10.3847/1538-4357/aba596}

\bibitem[{{Salmaso} {et~al.}(2019){Salmaso}, {Tomasella}, {Benetti},
  {D'Avanzo}, {Cappellaro}, {Botticella}, {Martone}, {Rossi}, {Brocato}, \&
  {Grawita Collaboration}}]{gcn25619}
{Salmaso}, I., {Tomasella}, L., {Benetti}, S., {et~al.} 2019, GRB Coordinates
  Network, 25619, 1

\bibitem[{{Shappee} {et~al.}(2017){Shappee}, {Simon}, {Drout}, {Piro},
  {Morrell}, {Prieto}, {Kasen}, {Holoien}, {Kollmeier}, {Kelson}, {Coulter},
  {Foley}, {Kilpatrick}, {Siebert}, {Madore}, {Murguia-Berthier}, {Pan},
  {Prochaska}, {Ramirez-Ruiz}, {Rest}, {Adams}, {Alatalo}, {Ba{\~n}ados},
  {Baughman}, {Bernstein}, {Bitsakis}, {Boutsia}, {Bravo}, {Di Mille}, {Higgs},
  {Ji}, {Maravelias}, {Marshall}, {Placco}, {Prieto}, \& {Wan}}]{Shappee+17}
{Shappee}, B.~J., {Simon}, J.~D., {Drout}, M.~R., {et~al.} 2017, Science, 358,
  1574, \dodoi{10.1126/science.aaq0186}

\bibitem[{{Short} {et~al.}(2019{\natexlab{a}}){Short}, {Nicholl}, {Smartt},
  {Smith}, {Chambers}, {Huber}, {Anderson}, {Chen}, {Inserra}, {Yaron},
  {Young}, {Angus}, {Pursiainen}, {Wiseman}, {Taubenberger}, \&
  {Gromadzki}}]{gcn24215}
{Short}, P., {Nicholl}, M., {Smartt}, S.~J., {et~al.} 2019{\natexlab{a}}, GRB
  Coordinates Network, 24215, 1

\bibitem[{{Short} {et~al.}(2019{\natexlab{b}}){Short}, {Nicholl}, {Anderson},
  {Chen}, {Inserra}, {Yaron}, {Young}, {Angus}, {Pursiainen}, {Wiseman},
  {Taubenberger}, \& {Gromadzki}}]{gcn24269}
{Short}, P., {Nicholl}, M., {Anderson}, J., {et~al.} 2019{\natexlab{b}}, GRB
  Coordinates Network, 24269, 1

\bibitem[{{Smartt} {et~al.}(2017){Smartt}, {Chen}, {Jerkstrand}, {Coughlin},
  {Kankare}, {Sim}, {Fraser}, {Inserra}, {Maguire}, {Chambers}, {Huber},
  {Kr{\"u}hler}, {Leloudas}, {Magee}, {Shingles}, {Smith}, {Young}, {Tonry},
  {Kotak}, {Gal-Yam}, {Lyman}, {Homan}, {Agliozzo}, {Anderson}, {Angus},
  {Ashall}, {Barbarino}, {Bauer}, {Berton}, {Botticella}, {Bulla}, {Bulger},
  {Cannizzaro}, {Cano}, {Cartier}, {Cikota}, {Clark}, {De Cia}, {Della Valle},
  {Denneau}, {Dennefeld}, {Dessart}, {Dimitriadis}, {Elias-Rosa}, {Firth},
  {Flewelling}, {Fl{\"o}rs}, {Franckowiak}, {Frohmaier}, {Galbany},
  {Gonz{\'a}lez-Gait{\'a}n}, {Greiner}, {Gromadzki}, {Guelbenzu},
  {Guti{\'e}rrez}, {Hamanowicz}, {Hanlon}, {Harmanen}, {Heintz}, {Heinze},
  {Hernandez}, {Hodgkin}, {Hook}, {Izzo}, {James}, {Jonker}, {Kerzendorf},
  {Klose}, {Kostrzewa-Rutkowska}, {Kowalski}, {Kromer}, {Kuncarayakti},
  {Lawrence}, {Lowe}, {Magnier}, {Manulis}, {Martin-Carrillo}, {Mattila},
  {McBrien}, {M{\"u}ller}, {Nordin}, {O'Neill}, {Onori}, {Palmerio},
  {Pastorello}, {Patat}, {Pignata}, {Podsiadlowski}, {Pumo}, {Prentice}, {Rau},
  {Razza}, {Rest}, {Reynolds}, {Roy}, {Ruiter}, {Rybicki}, {Salmon}, {Schady},
  {Schultz}, {Schweyer}, {Seitenzahl}, {Smith}, {Sollerman}, {Stalder},
  {Stubbs}, {Sullivan}, {Szegedi}, {Taddia}, {Taubenberger}, {Terreran}, {van
  Soelen}, {Vos}, {Wainscoat}, {Walton}, {Waters}, {Weiland}, {Willman},
  {Wiseman}, {Wright}, {Wyrzykowski}, \& {Yaron}}]{Smartt+17}
{Smartt}, S.~J., {Chen}, T.~W., {Jerkstrand}, A., {et~al.} 2017, \nat, 551, 75,
  \dodoi{10.1038/nature24303}

\bibitem[{{Smartt} {et~al.}(2019{\natexlab{a}}){Smartt}, {Smith}, {Young},
  {McBrien}, {Gillanders}, {Srivastav}, {O'Neil}, {Clark}, {Sim}, {Denneau},
  {Flewelling}, {Heinze}, {Tonry}, {Weiland}, {Chambers}, {Huber}, {Magnier},
  {Schultz}, {Rest}, {Stalder}, \& {Stubbs}}]{gcn24078}
{Smartt}, S.~J., {Smith}, K.~W., {Young}, D.~R., {et~al.} 2019{\natexlab{a}},
  GRB Coordinates Network, 24078, 1

\bibitem[{{Smartt} {et~al.}(2019{\natexlab{b}}){Smartt}, {Srivastav}, W.,
  {Chen}, {Young}, {Fulton}, {Denneau}, {Flewelling}, {Heinze}, {Tonry},
  {Weiland}, {Rest}, {Stalder}, {Stubbs}, {McBrien}, {Dobson}, {Gillanders},
  {O'Neil}, {Clark}, \& {Sim}}]{gcn25922}
{Smartt}, S.~J., {Srivastav}, S., W., S.~K., {et~al.} 2019{\natexlab{b}}, GRB
  Coordinates Network, Circular Service, No.~25922, \#1 (2019/April-0), 25922

\bibitem[{{Smith} {et~al.}(2019{\natexlab{a}}){Smith}, {Young}, {Huber},
  {Chambers}, {Smartt}, {McBrien}, {Gillanders}, {Srivastav}, {O'Neil},
  {Clark}, {Sim}, {Magnier}, {Schultz}, {Denneau}, {Flewelling}, {Heinze},
  {Tonry}, {Weiland}, {Rest}, {Stalder}, \& {Stubbs}}]{gcn24096}
{Smith}, K.~W., {Young}, D.~R., {Huber}, M., {et~al.} 2019{\natexlab{a}}, GRB
  Coordinates Network, 24096, 1

\bibitem[{{Smith} {et~al.}(2019{\natexlab{b}}){Smith}, {Young}, {McBrien},
  {Gilland ers}, {Srivastav}, {Smartt}, {O'Neil}, {Clark}, {Sim}, {Chambers},
  {Huber}, {Magnier}, {Schultz}, {Denneau}, {Flewelling}, {Heinze}, {Tonry},
  {Weiland}, {Rest}, {Stalder}, \& {Stubbs}}]{gcn24210}
{Smith}, K.~W., {Young}, D.~R., {McBrien}, O., {et~al.} 2019{\natexlab{b}}, GRB
  Coordinates Network, 24210, 1

\bibitem[{{Smith} {et~al.}(2019{\natexlab{c}}){Smith}, {Young}, {McBrien},
  {Gilland ers}, {Srivastav}, {Smartt}, {O'Neil}, {Clark}, {Sim}, {Chambers},
  {Huber}, {Magnier}, {Schultz}, {Denneau}, {Flewelling}, {Heinze}, {Tonry},
  {Weiland}, {Rest}, {Stalder}, \& {Stubbs}}]{gcn24262}
---. 2019{\natexlab{c}}, GRB Coordinates Network, 24262, 1

\bibitem[{{Soares-Santos} {et~al.}(2017){Soares-Santos}, {Holz}, {Annis},
  {Chornock}, {Herner}, {Berger}, {Brout}, {Chen}, {Kessler}, {Sako}, {Allam},
  {Tucker}, {Butler}, {Palmese}, {Doctor}, {Diehl}, {Frieman}, {Yanny}, {Lin},
  {Scolnic}, {Cowperthwaite}, {Neilsen}, {Marriner}, {Kuropatkin}, {Hartley},
  {Paz-Chinch{\'o}n}, {Alexander}, {Balbinot}, {Blanchard}, {Brown}, {Carlin},
  {Conselice}, {Cook}, {Drlica-Wagner}, {Drout}, {Durret}, {Eftekhari}, {Farr},
  {Finley}, {Foley}, {Fong}, {Fryer}, {Garc{\'\i}a-Bellido}, {Gill}, {Gruendl},
  {Hanna}, {Kasen}, {Li}, {Lopes}, {Louren{\c{c}}o}, {Margutti}, {Marshall},
  {Matheson}, {Medina}, {Metzger}, {Mu{\~n}oz}, {Muir}, {Nicholl}, {Quataert},
  {Rest}, {Sauseda}, {Schlegel}, {Secco}, {Sobreira}, {Stebbins}, {Villar},
  {Vivas}, {Walker}, {Wester}, {Williams}, {Zenteno}, {Zhang}, {Abbott},
  {Abdalla}, {Banerji}, {Bechtol}, {Benoit-L{\'e}vy}, {Bertin}, {Brooks},
  {Buckley-Geer}, {Burke}, {Carnero Rosell}, {Carrasco Kind}, {Carretero},
  {Castander}, {Crocce}, {Cunha}, {D'Andrea}, {da Costa}, {Davis}, {Desai},
  {Dietrich}, {Doel}, {Eifler}, {Fernand ez}, {Flaugher}, {Fosalba},
  {Gaztanaga}, {Gerdes}, {Giannantonio}, {Goldstein}, {Gruen}, {Gschwend},
  {Gutierrez}, {Honscheid}, {Jain}, {James}, {Jeltema}, {Johnson}, {Johnson},
  {Kent}, {Krause}, {Kron}, {Kuehn}, {Kuhlmann}, {Lahav}, {Lima}, {Maia},
  {March}, {McMahon}, {Menanteau}, {Miquel}, {Mohr}, {Nichol}, {Nord}, {Ogand
  o}, {Petravick}, {Plazas}, {Romer}, {Roodman}, {Rykoff}, {Sanchez},
  {Scarpine}, {Schubnell}, {Sevilla-Noarbe}, {Smith}, {Smith}, {Suchyta},
  {Swanson}, {Tarle}, {Thomas}, {Thomas}, {Troxel}, {Vikram}, {Wechsler},
  {Weller}, {Dark Energy Survey}, \& {Dark Energy Camera GW-EM
  Collaboration}}]{Soares-Santos+17}
{Soares-Santos}, M., {Holz}, D.~E., {Annis}, J., {et~al.} 2017, \apjl, 848,
  L16, \dodoi{10.3847/2041-8213/aa9059}

\bibitem[{{Steeghs} {et~al.}(2019){Steeghs}, {Dyer}, {Galloway}, {Dhillon},
  {O'Brien}, {Ramsay}, {Pollacco}, {Thrane}, {Poshyachinda}, {Palle},
  {Ulaczyk}, {Cutter}, {Stanway}, {Ackley}, {Obradovic}, {Mong}, {Casey},
  {Brown}, {Rol}, {Mullaney}, {Littlefair}, {Makrygianni}, {Daw}, {Maund},
  {Starling}, {Eyles}, {Sawangwit}, {Mkrtichian}, {Awiphan},
  {Aukkaravittayapun}, {Irawati}, {Kennedy}, {Breton}, {Mata-Sanchez},
  {Heikkila}, \& {Kotak}}]{gcn24224}
{Steeghs}, D., {Dyer}, M., {Galloway}, D., {et~al.} 2019, GRB Coordinates
  Network, 24224, 1

\bibitem[{{Stein} {et~al.}(2019{\natexlab{a}}){Stein}, {Reusch}, {Perley},
  {Andreoni}, {Coughlin}, {Zwicky Transient Facility Collaboration}, \& {Global
  Relay of Transients Watching Observatories Happen
  Collaboration}}]{Stein19gcn}
{Stein}, R., {Reusch}, S., {Perley}, D., {et~al.} 2019{\natexlab{a}}, GRB
  Coordinates Network, 26437, 1

\bibitem[{{Stein} {et~al.}(2020){Stein}, {Ztf Collaboration}, \& {Growth
  Collaboration}}]{GCN26673}
{Stein}, R., {Ztf Collaboration}, \& {Growth Collaboration}. 2020, GRB
  Coordinates Network, 26673, 1

\bibitem[{{Stein} {et~al.}(2019{\natexlab{b}}){Stein}, {Kool}, {Karambelkar},
  {Kasliwal}, {Perley}, {Anand}, {Coughlin}, {Sharma}, {Singer}, {Andreoni},
  {Waratkar}, {Kumar}, {Khand agale}, {Deshmukh}, {Bhalerao}, {Anupama},
  {Dobie}, {Cenko}, {Ahmuda}, {Bellm}, {Kong}, {Franckowiak}, {Gatkine}, {Ztf
  Collaboration}, \& {Growth Collaboration}}]{gcn25634}
{Stein}, R., {Kool}, E., {Karambelkar}, V., {et~al.} 2019{\natexlab{b}}, GRB
  Coordinates Network, 25634, 1

\bibitem[{{Stein} {et~al.}(2019{\natexlab{c}}){Stein}, {Kool}, {Kumar},
  {Coughlin}, {Kasliwal}, {Perley}, {Anand }, {Sharma}, {Karambelkar},
  {Singer}, {Andreoni}, {Waratkar}, {Khand agale}, {Deshmukh}, {Bhalerao},
  {Anupama}, {Dobie}, {Cenko}, {Ahmuda}, {Bellm}, {Kong}, {Franckowiak},
  {Gatkine}, {Ztf Collaboration}, \& {Growth Collaboration}}]{gcn25656}
{Stein}, R., {Kool}, E., {Kumar}, H., {et~al.} 2019{\natexlab{c}}, GRB
  Coordinates Network, 25656, 1

\bibitem[{{Stein} {et~al.}(2019{\natexlab{d}}){Stein}, {Kasliwal}, {Kool},
  {Bellm}, {Andreoni}, {Ahumada}, {Coughlin}, {Anand}, {Singer}, {Cenko},
  {Kong}, {Deshmukh}, {Khand agale}, {Waratkar}, {Bhalerao}, {Anupama}, {Ztf
  Collaboration}, \& {Growth Collaboration}}]{gcn25899}
{Stein}, R., {Kasliwal}, M.~M., {Kool}, E., {et~al.} 2019{\natexlab{d}}, GRB
  Coordinates Network, 25899, 1

\bibitem[{{Stone} {et~al.}(2017){Stone}, {Metzger}, \& {Haiman}}]{Stone2017}
{Stone}, N.~C., {Metzger}, B.~D., \& {Haiman}, Z. 2017, \mnras, 464, 946,
  \dodoi{10.1093/mnras/stw2260}

\bibitem[{{Tanvir} {et~al.}(2017){Tanvir}, {Levan},
  {Gonz{\'a}lez-Fern{\'a}ndez}, {Korobkin}, {Mandel}, {Rosswog}, {Hjorth},
  {D'Avanzo}, {Fruchter}, {Fryer}, {Kangas}, {Milvang-Jensen}, {Rosetti},
  {Steeghs}, {Wollaeger}, {Cano}, {Copperwheat}, {Covino}, {D'Elia}, {de Ugarte
  Postigo}, {Evans}, {Even}, {Fairhurst}, {Figuera Jaimes}, {Fontes}, {Fujii},
  {Fynbo}, {Gompertz}, {Greiner}, {Hodosan}, {Irwin}, {Jakobsson},
  {J{\o}rgensen}, {Kann}, {Lyman}, {Malesani}, {McMahon}, {Melandri},
  {O'Brien}, {Osborne}, {Palazzi}, {Perley}, {Pian}, {Piranomonte}, {Rabus},
  {Rol}, {Rowlinson}, {Schulze}, {Sutton}, {Th{\"o}ne}, {Ulaczyk}, {Watson},
  {Wiersema}, \& {Wijers}}]{Tanvir+17}
{Tanvir}, N.~R., {Levan}, A.~J., {Gonz{\'a}lez-Fern{\'a}ndez}, C., {et~al.}
  2017, \apjl, 848, L27, \dodoi{10.3847/2041-8213/aa90b6}

\bibitem[{{Thakur} {et~al.}(2020){Thakur}, {Dichiara}, {Troja}, {Chase},
  {S{\'a}nchez-Ram{\'\i}rez}, {Piro}, {Fryer}, {Butler}, {Watson}, {Wollaeger},
  {Ambrosi}, {Becerra Gonz{\'a}lez}, {Becerra}, {Bruni}, {Cenko}, {Cusumano},
  {D'A{\`\i}}, {Durbak}, {Fontes}, {Gatkine}, {Hungerford}, {Korobkin},
  {Kutyrev}, {Lee}, {Lotti}, {Minervini}, {Novara}, {Parola}, {Pereyra},
  {Ricci}, {Tiengo}, \& {Veilleux}}]{Thakur2020}
{Thakur}, A.~L., {Dichiara}, S., {Troja}, E., {et~al.} 2020, \mnras, 499, 3868,
  \dodoi{10.1093/mnras/staa2798}

\bibitem[{{The Astropy Collaboration} {et~al.}(2018){The Astropy
  Collaboration}, {Price-Whelan}, {Sip{\H o}cz}, {G{\"u}nther}, {Lim},
  {Crawford}, {Conseil}, {Shupe}, {Craig}, {Dencheva}, {Ginsburg},
  {VanderPlas}, {Bradley}, {P{\'e}rez-Su{\'a}rez}, {de Val-Borro}, {Aldcroft},
  {Cruz}, {Robitaille}, {Tollerud}, {Ardelean}, {Babej}, {Bachetti}, {Bakanov},
  {Bamford}, {Barentsen}, {Barmby}, {Baumbach}, {Berry}, {Biscani}, {Boquien},
  {Bostroem}, {Bouma}, {Brammer}, {Bray}, {Breytenbach}, {Buddelmeijer},
  {Burke}, {Calderone}, {Cano Rodr{\'{\i}}guez}, {Cara}, {Cardoso},
  {Cheedella}, {Copin}, {Crichton}, {D{\'A}vella}, {Deil}, {Depagne},
  {Dietrich}, {Donath}, {Droettboom}, {Earl}, {Erben}, {Fabbro}, {Ferreira},
  {Finethy}, {Fox}, {Garrison}, {Gibbons}, {Goldstein}, {Gommers}, {Greco},
  {Greenfield}, {Groener}, {Grollier}, {Hagen}, {Hirst}, {Homeier}, {Horton},
  {Hosseinzadeh}, {Hu}, {Hunkeler}, {Ivezi{\'c}}, {Jain}, {Jenness}, {Kanarek},
  {Kendrew}, {Kern}, {Kerzendorf}, {Khvalko}, {King}, {Kirkby}, {Kulkarni},
  {Kumar}, {Lee}, {Lenz}, {Littlefair}, {Ma}, {Macleod}, {Mastropietro},
  {McCully}, {Montagnac}, {Morris}, {Mueller}, {Mumford}, {Muna}, {Murphy},
  {Nelson}, {Nguyen}, {Ninan}, {N{\"o}the}, {Ogaz}, {Oh}, {Parejko}, {Parley},
  {Pascual}, {Patil}, {Patil}, {Plunkett}, {Prochaska}, {Rastogi}, {Reddy
  Janga}, {Sabater}, {Sakurikar}, {Seifert}, {Sherbert}, {Sherwood-Taylor},
  {Shih}, {Sick}, {Silbiger}, {Singanamalla}, {Singer}, {Sladen}, {Sooley},
  {Sornarajah}, {Streicher}, {Teuben}, {Thomas}, {Tremblay}, {Turner},
  {Terr{\'o}n}, {van Kerkwijk}, {de la Vega}, {Watkins}, {Weaver}, {Whitmore},
  {Woillez}, \& {Zabalza}}]{astropy}
{The Astropy Collaboration}, {Price-Whelan}, A.~M., {Sip{\H o}cz}, B.~M.,
  {et~al.} 2018, ArXiv e-prints.
\newblock \doarXiv{1801.02634}

\bibitem[{{The LIGO Scientific Collaboration} {et~al.}(2020){The LIGO
  Scientific Collaboration}, {the Virgo Collaboration}, {Abbott}, {Abbott},
  {Abraham}, {Acernese}, {Ackley}, {Adams}, {Adhikari}, {Adya}, {Affeldt},
  {Agathos}, {Agatsuma}, \& {Aggarwal}}]{lvc_gw190521}
{The LIGO Scientific Collaboration}, {the Virgo Collaboration}, {Abbott}, R.,
  {et~al.} 2020, arXiv e-prints, arXiv:2009.01075.
\newblock \doarXiv{2009.01075}

\bibitem[{{Tody}(1986)}]{iraf1}
{Tody}, D. 1986, in \procspie, Vol. 627, Instrumentation in astronomy VI, ed.
  D.~L. {Crawford}, 733, \dodoi{10.1117/12.968154}

\bibitem[{{Tody}(1993)}]{iraf2}
{Tody}, D. 1993, in Astronomical Society of the Pacific Conference Series,
  Vol.~52, Astronomical Data Analysis Software and Systems II, ed. R.~J.
  {Hanisch}, R.~J.~V. {Brissenden}, \& J.~{Barnes}, 173

\bibitem[{{Tohuvavohu} {et~al.}(2019){Tohuvavohu}, {Kuin}, {Oates}, {Breeveld},
  {Marshall}, {Brown}, {Gronwall}, {Page}, {De Pasquale}, {Siegel}, {D'Ai},
  {D'Avanzo}, {Barthelmy}, {Beardmore}, {Burrows}, {Campana}, {Cenko},
  {Cusumano}, {D'Elia}, {Evans}, {Giommi}, {Hartmann}, {Kennea}, {Klingler},
  {Krimm}, {Lien}, {Malesani}, {Melandri}, {Nousek}, {O'Brien}, {Osborne},
  {Pagani}, {Page}, {Palmer}, {Perri}, {Racusin}, {Sakamoto}, {Sbarufatti},
  {Tagliaferri}, {Troja}, \& {Swift Team}}]{gcn25964}
{Tohuvavohu}, A., {Kuin}, N.~P.~M., {Oates}, S.~R., {et~al.} 2019, GRB
  Coordinates Network, 25964, 1

\bibitem[{{Tonry} {et~al.}(2018){Tonry}, {Denneau}, {Heinze}, {Stalder},
  {Smith}, {Smartt}, {Stubbs}, {Weiland }, \& {Rest}}]{tonry18}
{Tonry}, J.~L., {Denneau}, L., {Heinze}, A.~N., {et~al.} 2018, \pasp, 130,
  064505, \dodoi{10.1088/1538-3873/aabadf}

\bibitem[{{Troja} {et~al.}(2019){Troja}, {van Eerten}, {Ryan}, {Ricci},
  {Burgess}, {Wieringa}, {Piro}, {Cenko}, \& {Sakamoto}}]{troja19}
{Troja}, E., {van Eerten}, H., {Ryan}, G., {et~al.} 2019, \mnras, 489, 1919,
  \dodoi{10.1093/mnras/stz2248}

\bibitem[{{Utsumi} {et~al.}(2017){Utsumi}, {Tanaka}, {Tominaga}, {Yoshida},
  {Barway}, {Nagayama}, {Zenko}, {Aoki}, {Fujiyoshi}, {Furusawa}, {Kawabata},
  {Koshida}, {Lee}, {Morokuma}, {Motohara}, {Nakata}, {Ohsawa}, {Ohta},
  {Okita}, {Tajitsu}, {Tanaka}, {Terai}, {Yasuda}, {Abe}, {Asakura}, {Bond},
  {Miyazaki}, {Sumi}, {Tristram}, {Honda}, {Itoh}, {Itoh}, {Kawabata},
  {Morihana}, {Nagashima}, {Nakaoka}, {Ohshima}, {Takahashi}, {Takayama},
  {Aoki}, {Baar}, {Doi}, {Finet}, {Kanda}, {Kawai}, {Kim}, {Kuroda}, {Liu},
  {Matsubayashi}, {Murata}, {Nagai}, {Saito}, {Saito}, {Sako}, {Sekiguchi},
  {Tamura}, {Tanaka}, {Uemura}, \& {Yamaguchi}}]{Utsumi17}
{Utsumi}, Y., {Tanaka}, M., {Tominaga}, N., {et~al.} 2017, \pasj, 69, 101,
  \dodoi{10.1093/pasj/psx118}

\bibitem[{{Valeev} \& {Castro-Rodriguez}(2020)}]{gcn26764}
{Valeev}, A.~F., \& {Castro-Rodriguez}, N. 2020, GRB Coordinates Network,
  26764, 1

\bibitem[{{Valeev} \& {Font}(2020)}]{GCN26702}
{Valeev}, A.~F., \& {Font}, J. 2020, GRB Coordinates Network, 26702, 1

\bibitem[{{Valeev} {et~al.}(2019{\natexlab{a}}){Valeev}, {Hu}, {Castro-Tirado},
  {Fernand ez-Garcia}, {Sokolov}, {Carrasco}, {Castellon}, \&
  {Reverte-Paya}}]{gcn26421}
{Valeev}, A.~F., {Hu}, Y.~D., {Castro-Tirado}, A.~J., {et~al.}
  2019{\natexlab{a}}, GRB Coordinates Network, 26421, 1

\bibitem[{{Valeev} {et~al.}(2019{\natexlab{b}}){Valeev}, {Sokolov}, {Sokolov},
  {Hu}, {Li}, {Castro-Tirado}, {Carrasco}, {Sanchez-Ramirez},
  {Caballero-Garcia}, \& {Peseev}}]{gcn24092}
{Valeev}, A.~F., {Sokolov}, I.~V., {Sokolov}, V.~V., {et~al.}
  2019{\natexlab{b}}, GRB Coordinates Network, 24092, 1

\bibitem[{{Valeev} {et~al.}(2019{\natexlab{c}}){Valeev}, {Sokolov},
  {Castro-Tirado}, {Hu}, {Li}, {Ayala}, {Fernand ez-Garcia}, {Aceituno},
  {Carrasco}, {Castellon}, {Perez}, {Caballero-Garcia}, {Pand ey}, \&
  {Castro-Rodriguez}}]{gcn24317}
{Valeev}, A.~F., {Sokolov}, V.~V., {Castro-Tirado}, A.~J., {et~al.}
  2019{\natexlab{c}}, GRB Coordinates Network, 24317, 1

\bibitem[{{Valeev} {et~al.}(2019{\natexlab{d}}){Valeev}, {Hu}, {Castro-Tirado},
  {Aceituno}, {Fernandez-Garcia}, {Sokolov}, {Carrasco}, {Castellon}, \&
  {Castro-Rodriguez}}]{gcn26382}
{Valeev}, A.~F., {Hu}, Y.~D., {Castro-Tirado}, A.~J., {et~al.}
  2019{\natexlab{d}}, GRB Coordinates Network, 26382, 1

\bibitem[{{Valenti} {et~al.}(2017){Valenti}, {Sand}, {Yang}, {Cappellaro},
  {Tartaglia}, {Corsi}, {Jha}, {Reichart}, {Haislip}, \&
  {Kouprianov}}]{Valenti2017}
{Valenti}, S., {Sand}, D.~J., {Yang}, S., {et~al.} 2017, \apjl, 848, L24,
  \dodoi{10.3847/2041-8213/aa8edf}

\bibitem[{{Vieira} {et~al.}(2020){Vieira}, {Ruan}, {Haggard}, {Drout}, {Nynka},
  {Boyce}, {Spekkens}, {Safi-Harb}, {Carlberg}, {Fern{\'a}ndez}, {Piro},
  {Afsariardchi}, \& {Moon}}]{vieira_cfht}
{Vieira}, N., {Ruan}, J.~J., {Haggard}, D., {et~al.} 2020, arXiv e-prints,
  arXiv:2003.09437.
\newblock \doarXiv{2003.09437}

\bibitem[{{Villar} {et~al.}(2017){Villar}, {Guillochon}, {Berger}, {Metzger},
  {Cowperthwaite}, {Nicholl}, {Alexander}, {Blanchard}, {Chornock},
  {Eftekhari}, {Fong}, {Margutti}, \& {Williams}}]{villar17}
{Villar}, V.~A., {Guillochon}, J., {Berger}, E., {et~al.} 2017, \apjl, 851,
  L21, \dodoi{10.3847/2041-8213/aa9c84}

\bibitem[{{Vogl} {et~al.}(2019){Vogl}, {Floers}, {Taubenberger}, {Hillebrand
  t}, \& {Suyu}}]{GCN26504}
{Vogl}, C., {Floers}, A., {Taubenberger}, S., {Hillebrand t}, W., \& {Suyu}, S.
  2019, GRB Coordinates Network, 26504, 1

\bibitem[{{Wei} {et~al.}(2019){Wei}, {Xin}, {Antier}, {Wang}, {Leroy},
  {Turpin}, {SVOM Multi Messenger Astronomy Team}, \& {GWAC Team}}]{gcn25640}
{Wei}, J.~Y., {Xin}, L.~P., {Antier}, S., {et~al.} 2019, GRB Coordinates
  Network, 25640, 1

\bibitem[{{Wright} {et~al.}(2015){Wright}, {Smartt}, {Smith}, {Miller},
  {Kotak}, {Rest}, {Burgett}, {Chambers}, {Flewelling}, {Hodapp}, {Huber},
  {Jedicke}, {Kaiser}, {Metcalfe}, {Price}, {Tonry}, {Wainscoat}, \&
  {Waters}}]{wright15}
{Wright}, D.~E., {Smartt}, S.~J., {Smith}, K.~W., {et~al.} 2015, \mnras, 449,
  451, \dodoi{10.1093/mnras/stv292}

\bibitem[{{Wyatt} {et~al.}(2020){Wyatt}, {Tohuvavohu}, {Arcavi}, {Lundquist},
  {Howell}, \& {Sand}}]{Wyatt20}
{Wyatt}, S.~D., {Tohuvavohu}, A., {Arcavi}, I., {et~al.} 2020, \apj, 894, 127,
  \dodoi{10.3847/1538-4357/ab855e}

\bibitem[{{Yan}(2020)}]{yan20tns}
{Yan}, L. 2020, Transient Name Server Classification Report, 2020-2662, 1

\bibitem[{{Zackay} {et~al.}(2016){Zackay}, {Ofek}, \& {Gal-Yam}}]{ZOGY}
{Zackay}, B., {Ofek}, E.~O., \& {Gal-Yam}, A. 2016, \apj, 830, 27,
  \dodoi{10.3847/0004-637X/830/1/27}

\end{thebibliography}
\bibliographystyle{aasjournal}

\begin{appendices}
\section{Appendix: Localization figures and final candidates} \label{appendix:A}

\begin{deluxetable*}{ccccccccccl}[h!]
\tabletypesize{\scriptsize}
\tablecolumns{11}
\tablewidth{0pc}
\tablecaption{Viable SAGUARO candidates from O3.
\label{tab:obs}}
\tablehead{
\colhead {Event}	 &
\colhead {Candidate}	 &
\colhead {RA} &
\colhead {Dec}  &		
\colhead {m$_{can}$}  &
\colhead {MJD} &
\colhead {$\delta t_{rest}$}  &
\colhead {Host$\dagger$}   &
\colhead {m$_{host}$}   &
\colhead {$z_{host}$}   &
\colhead {$\delta r$}\\
\colhead {(Type)}	 &
\colhead {}	 &
\colhead {} &
\colhead {}  &		
\colhead {(\emph{Gaia} $G$)}  &
\colhead {(+58000)} &
\colhead {(days)}  &
\colhead {}   &
\colhead {}   &
\colhead {}   &
\colhead {}\\
}
\startdata
S190901ap & SAGUARO19k$^\ast$ & 
04:06:26.2 & -12:01:13 & 20.9 & 728.4880 & 0.48 & No$^c$ & $g > 25.6$ & \nodata & \\
(BNS) & /AT2019aaid & & & & & & & $r > 24.6$ & & \\
& & & & & & & & $z > 24.5$ & & .\\
\hline
GW190930\_133541 & SAGUARO19n & 21:53:14.0 & -06:43:38 & 20.6$^{[20]}$ & 759.1745 & 2.17 & Yes$^{a,h}$ &  $u$ = 20.7 & 0.208 $\pm$ 0.030$^d$ & \\
(BBH) & /AT2019aaig & & & & & & & $g$ = 20.0 & 0.168 $\pm$ 0.041$^f$ & \\
& & & & & & & & $r$ = 19.3 & & \\
& & & & & & & & $i$ = 18.9 & & \\
& & & & & & & & $z$ = 18.8  & & \\
\hline
S190930t & SAGUARO19l$^\ast$ & 20:40:05.2 & -01:39:09 & 20.9 & 758.1318 & 1.49 & Yes$^c$ & $g$ = 22.5 & \nodata & $\sim 0.3''$\\
(NSBH) & /AT2019aaie & & & & & & & $r$ = 21.5 & & \\
& & & & & & & & $z$ = 20.9 & & \\
& SAGUARO19m$^\ast$ & 20:15:45.8 & -07:55:44 & 18.6 & 759.1088 & 2.44 & No$^h$ & \nodata & \nodata & \\
& /AT2019aaif & & & & & & & & & \\
& SAGUARO19o$^\ast$ & 23:22:03.4 & -23:15:11 & 20.3 & 761.2123 & 4.49 & Yes$^b$ & $J$ = 15.1 & 0.07 $\pm$ 0.03$^e$ & $7.9''$\\
& /AT2019aaih & & & & & & & $H$ = 14.3 & 0.113 $\pm$ 0.008$^{f,g}$ & \\
& & & & & & & & $K$ = 14.1 & & \\
\hline
S200105ae & SAGUARO20h & 07:41:24.0 & 09:27:59 & 18.9$^{[20]}$ & 854.2061 & 0.49 & No$^h$ & \nodata & \nodata & \\
(NSBH) & /AT2020abgt & & & & & & & & & \\
\hline
S200128d & SAGUARO20i & 14:16:09.0 & -06:22:43 & 20.3$^{[240]}$ & 876.5531 & 0.28 & Yes$^c$ & $g$ = 19.5 & 0.065 $\pm$ 0.039$^f$ & $1.7''$\\
(BBH) & /AT2020abgu & & & & & & & $r$ = 19.1 & & \\
& & & & & & & & $z$ = 18.7 & & \\
\hline
\hline
\enddata
\tablecomments{$\dagger$Here we define the host as an underlying galaxy coincident with the candidate within $r < 4''$.
$^\ast$ Within the 50\% sky localization probability map.
$^{[X]}$ X times more luminous than models discussed here.
$^a$ Host magnitudes from \cite{SDSSDR12}.
$^b$ Host magnitudes from \cite{2MPZ}.
$^c$ Host magnitudes/limits from \cite{Dey2019}
$^d$ $z_{photo}$ from \cite{SDSSDR12}.
$^e$ $z_{photo}$ from \cite{2MPZ}.
$^f$ $z_{photo}$ from \cite{PSDR2}.
$^g$ $z_{photo} > 3\sigma$ outside LIGO/Virgo $d$.
$^h$ Nearby galaxy within 40$''$.}
\end{deluxetable*}

\begin{figure*}[h!]
\centering
\Large{\textbf{GW190408\_181802}}\par\medskip
\includegraphics[width=8.4cm]{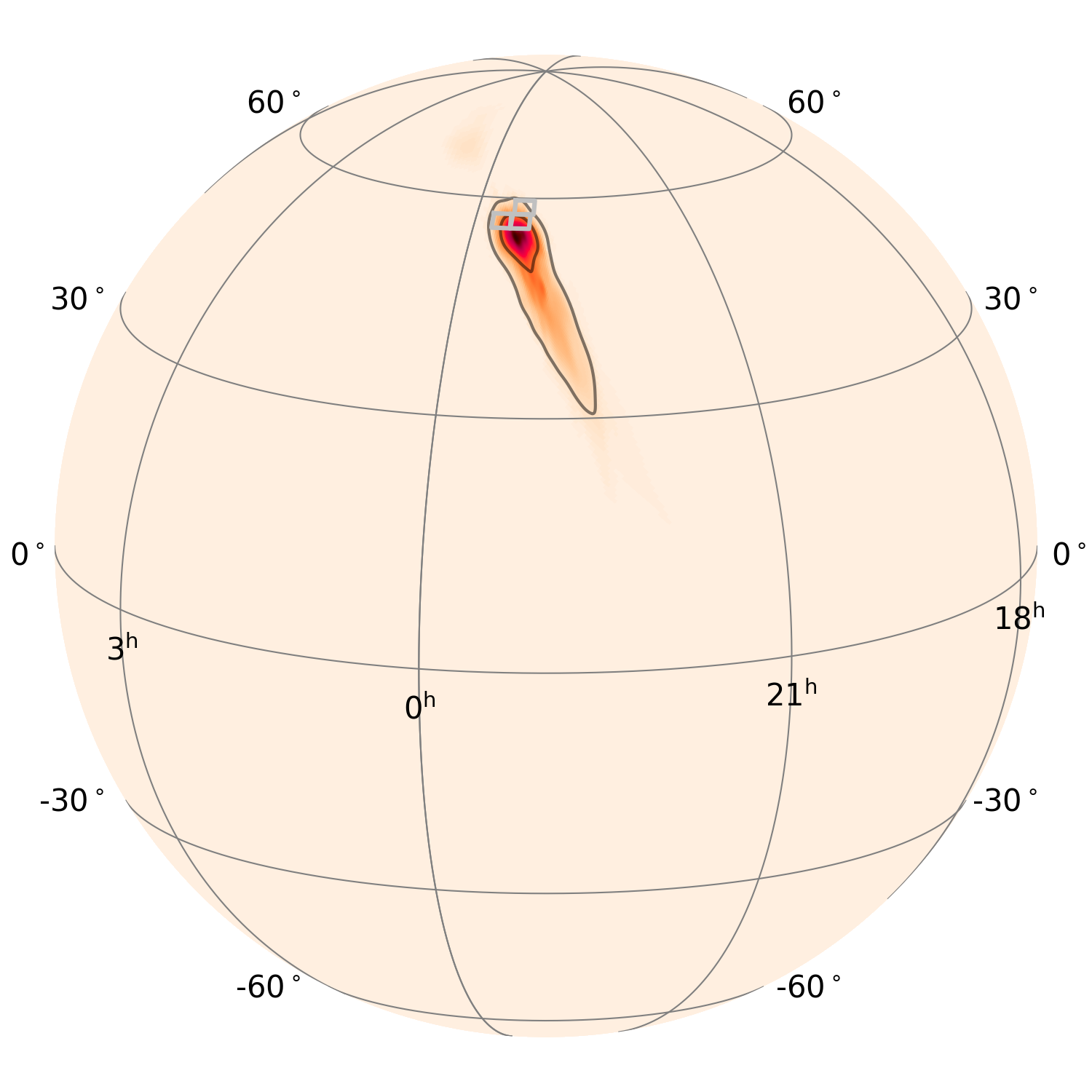}
\includegraphics[width=8.4cm]{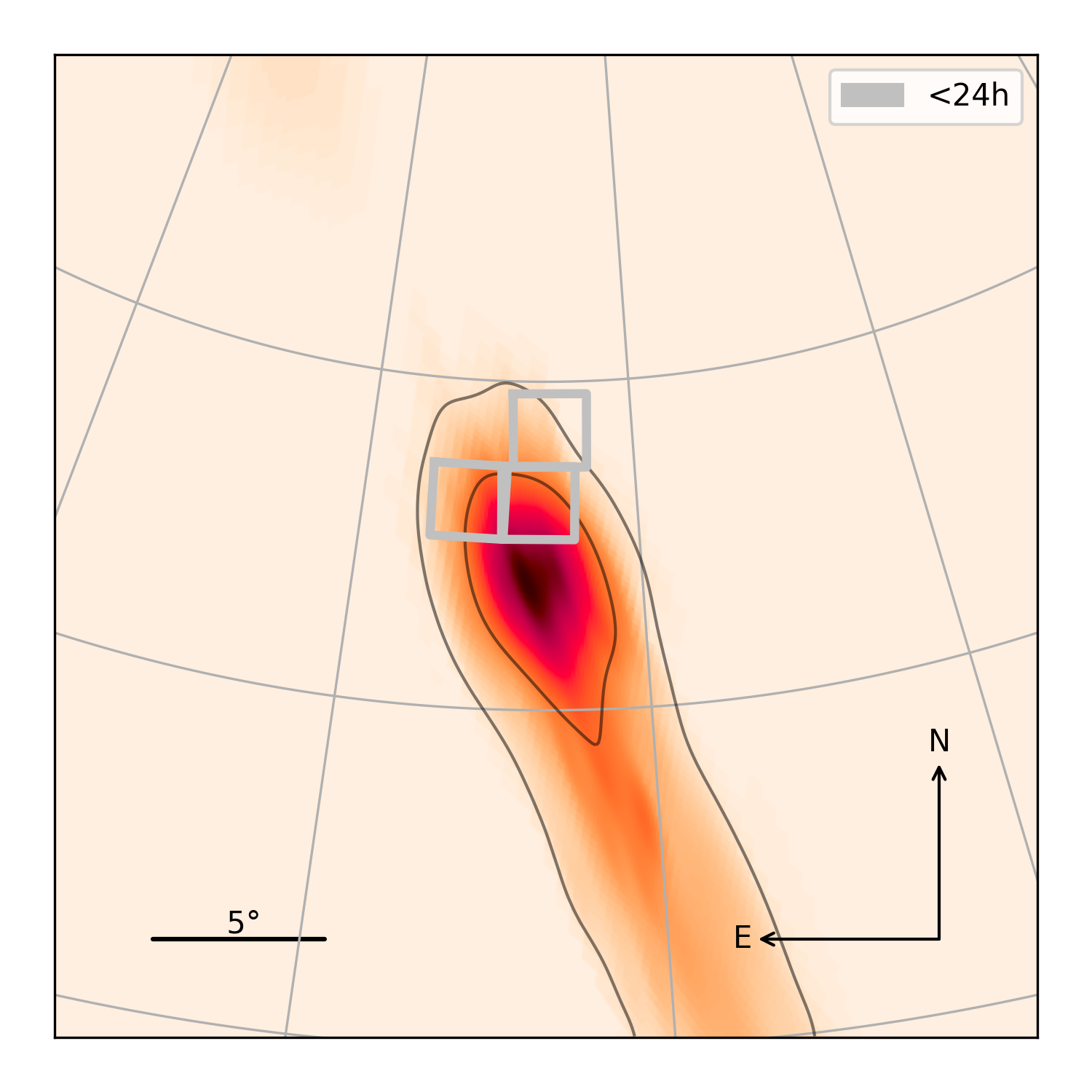}
\caption{GW localization of GW190408\_181802 overlaid with the 3 CSS fields that were triggered for this event.  These fields were observed within 24 hours of the GW detection and covered 13.73\% of the total localization probability.  A globe projection is shown on the left panel with a zoom in on the region of higher probability on the right panel.  These panels show the localization as a probability density map where darker colors indicate higher probability of containing the GW source. Contours indicate the 50\% and 90\% confidence levels for containing the GW event. \label{fig:190408an_loc}}
\end{figure*}

\begin{figure*}
\centering
\Large{\textbf{GW190425}}\par\medskip
\includegraphics[width=8.5cm]{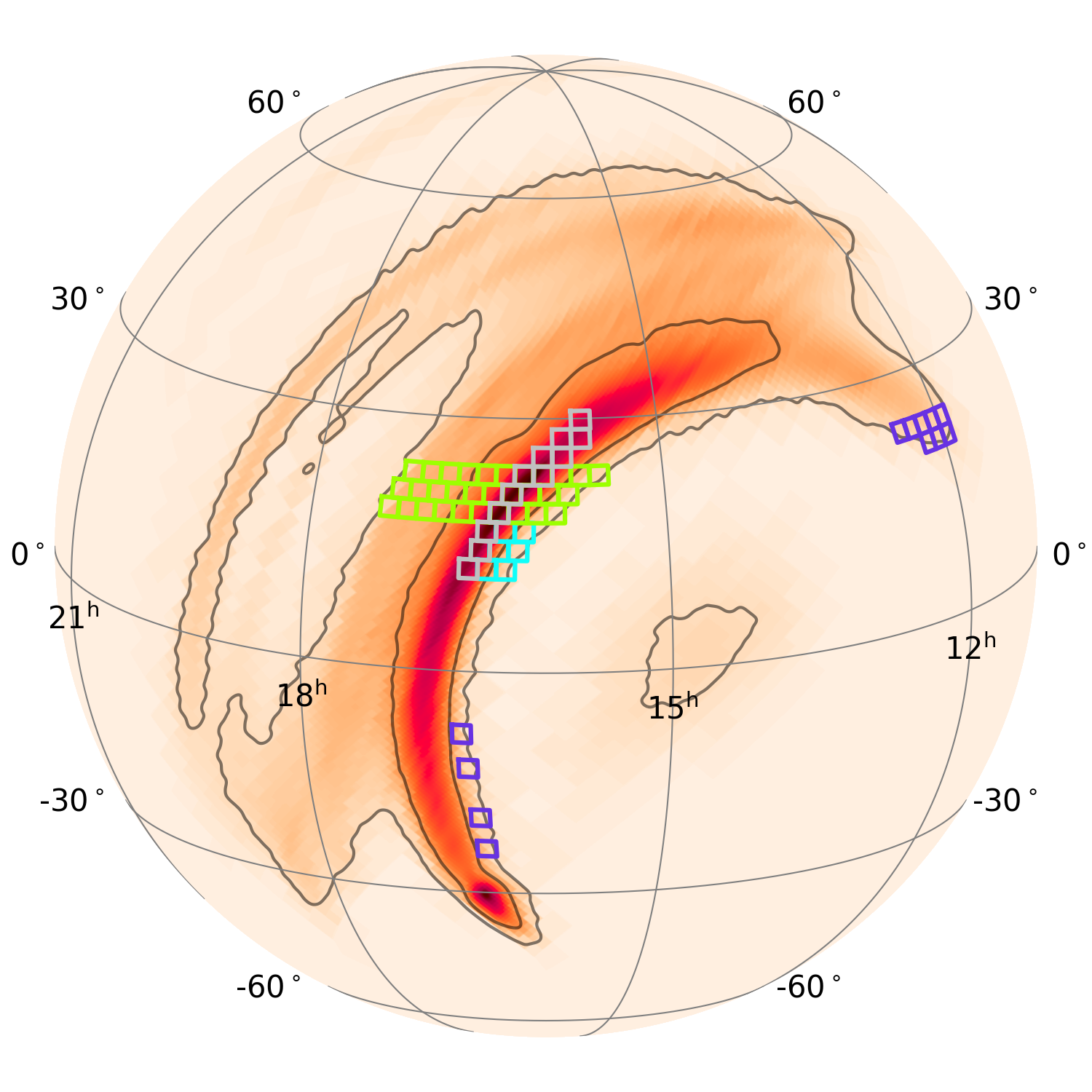}
\includegraphics[width=8.5cm]{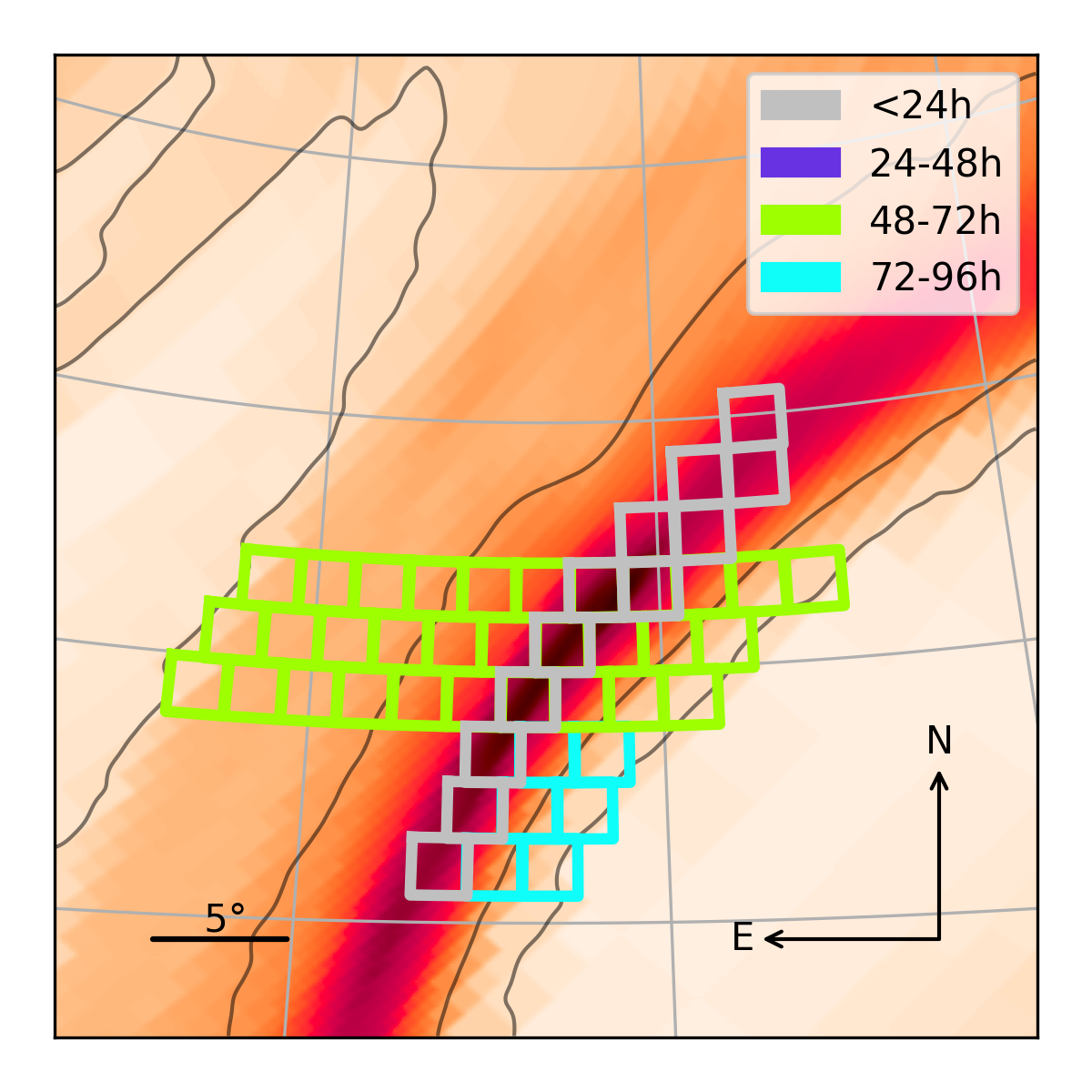}
\includegraphics[width=\linewidth]{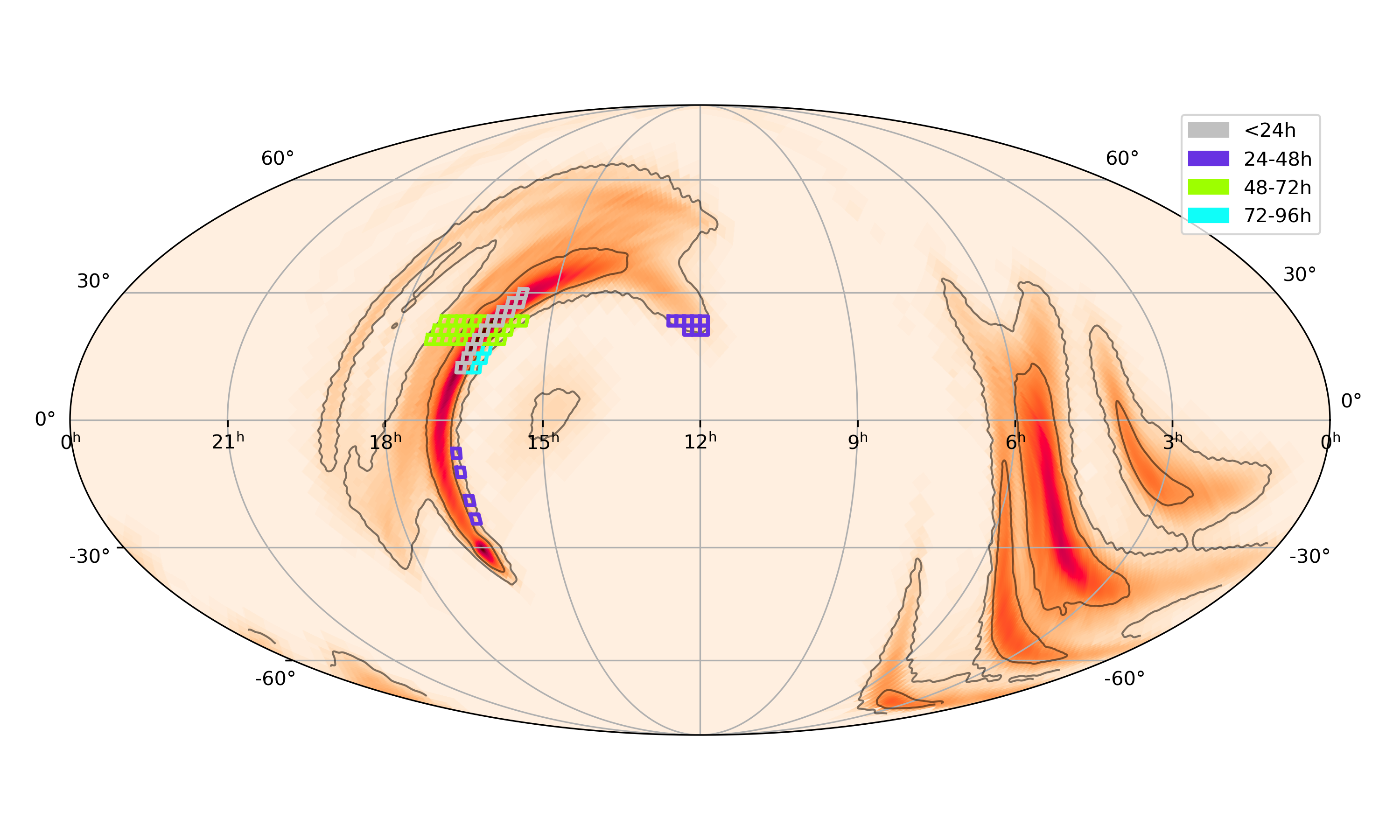}
\caption{Same as Figure~\ref{fig:190408an_loc} for GW190425 overlaid with the CSS fields that were observed within 96 hours of the GW detection. Twelve fields were triggered within 24 hours with four of these fields being revisited later.  In total, these fields covered 6.55\% of the total localization probability.  A Mollweide projection is included in the bottom panel to illustrate the full sky localization which had two large components.\label{fig:190425z_loc}}
\end{figure*}

\begin{figure*}
\centering
\Large{\textbf{GW190426\_152155}}\par\medskip
\includegraphics[width=8.5cm]{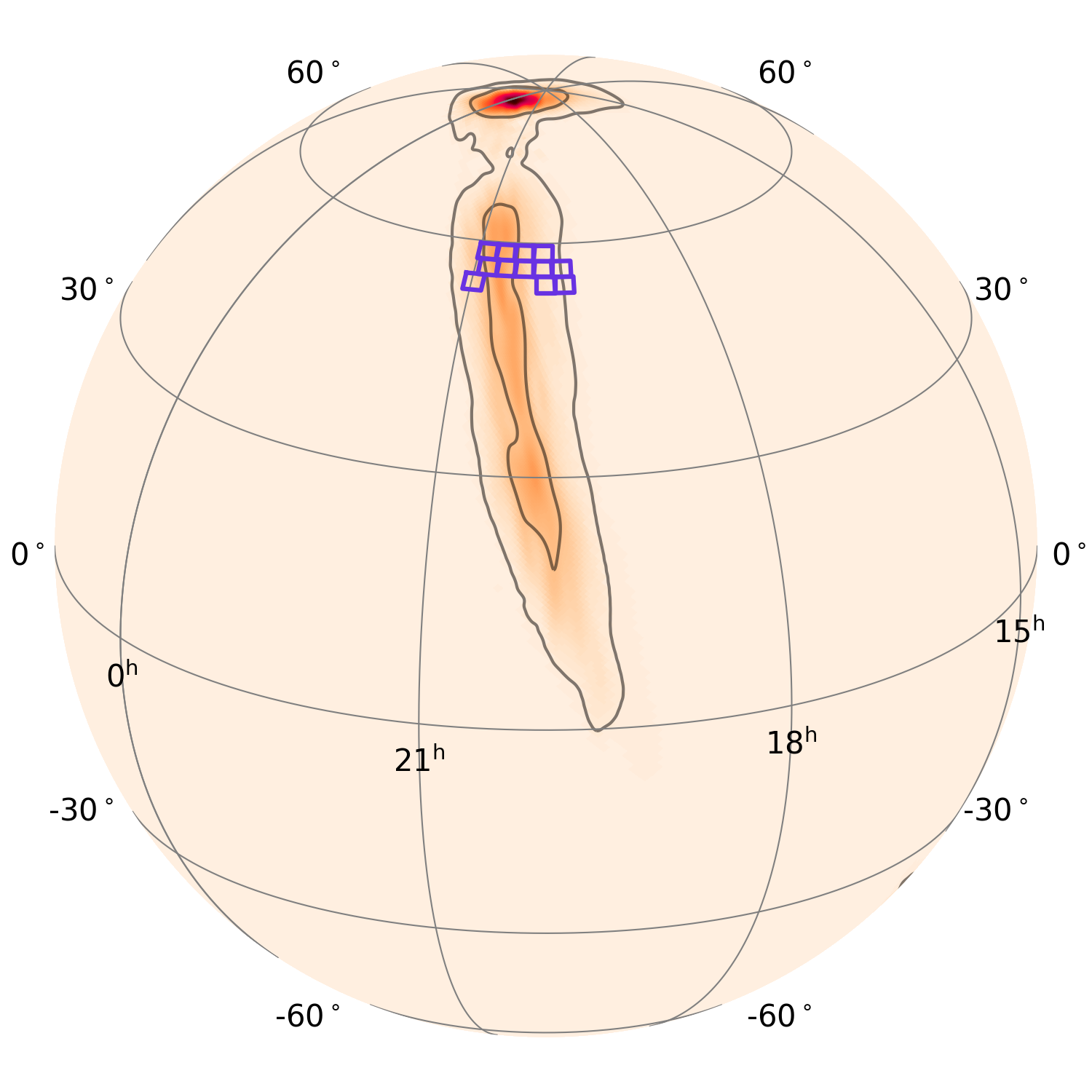}
\includegraphics[width=8.5cm]{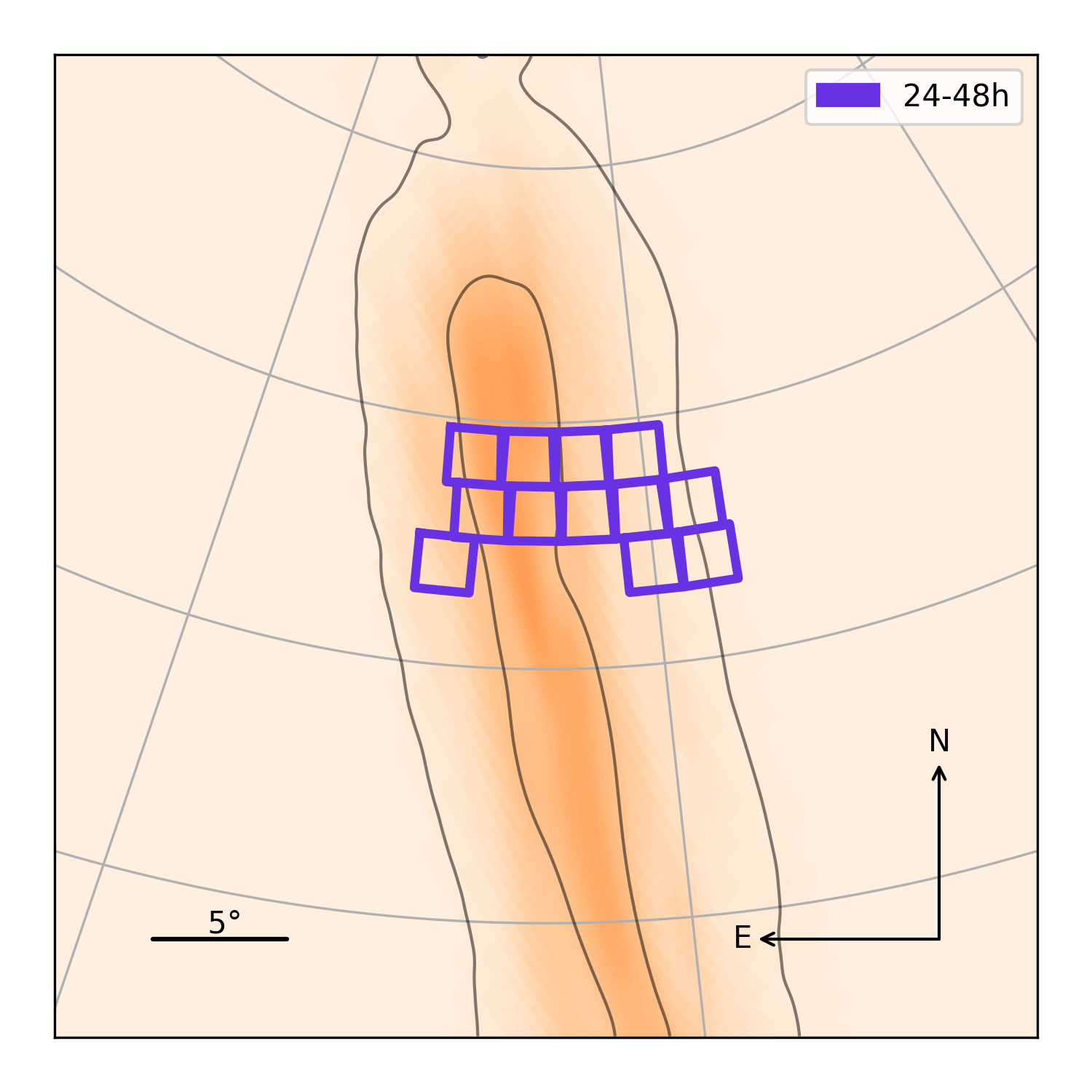}
\caption{Same as Figure~\ref{fig:190408an_loc} for GW190426\_152155 overlaid with the CSS fields that were triggered.  These fields were observed between 24 and 48 hours after the GW detection and covered 4.48\% of the total localization probability.  \label{fig:190426c_loc}}
\end{figure*}

\begin{figure*}
\centering
\Large{\textbf{GW190521}}\par\medskip
\includegraphics[width=\linewidth]{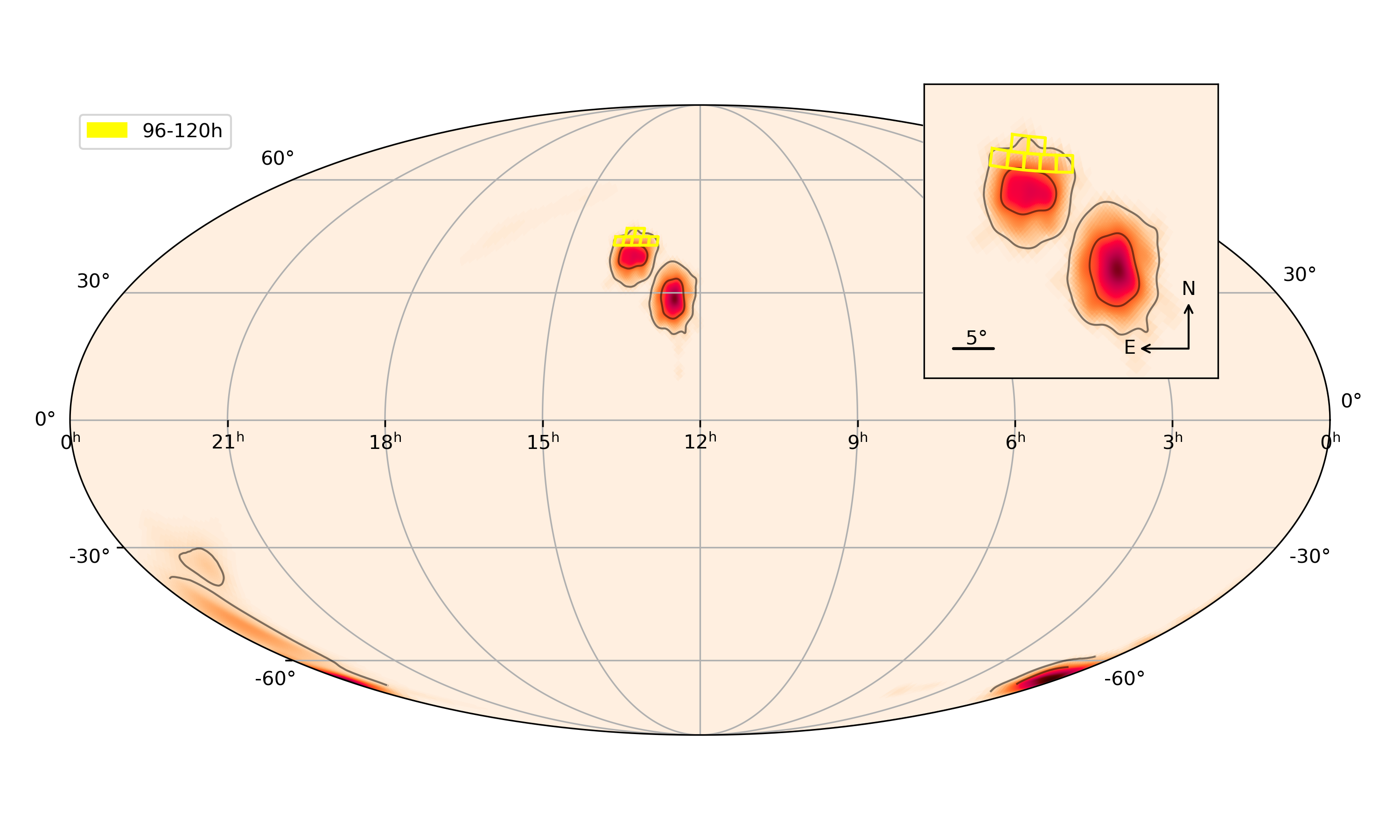}
\caption{Same as Figure~\ref{fig:190408an_loc} for GW190521 overlaid with the CSS fields that were observed within 120 hours of the GW detection.  These fields covered 2.27\% of the total localization probability.  A Mollweide projection is shown to illustrate the full sky localization, which also had a southern component. \label{fig:S190521g}}
\end{figure*}

\begin{figure*}
\centering
\Large{\textbf{GW190630\_185205}}\par\medskip
\includegraphics[width=\linewidth]{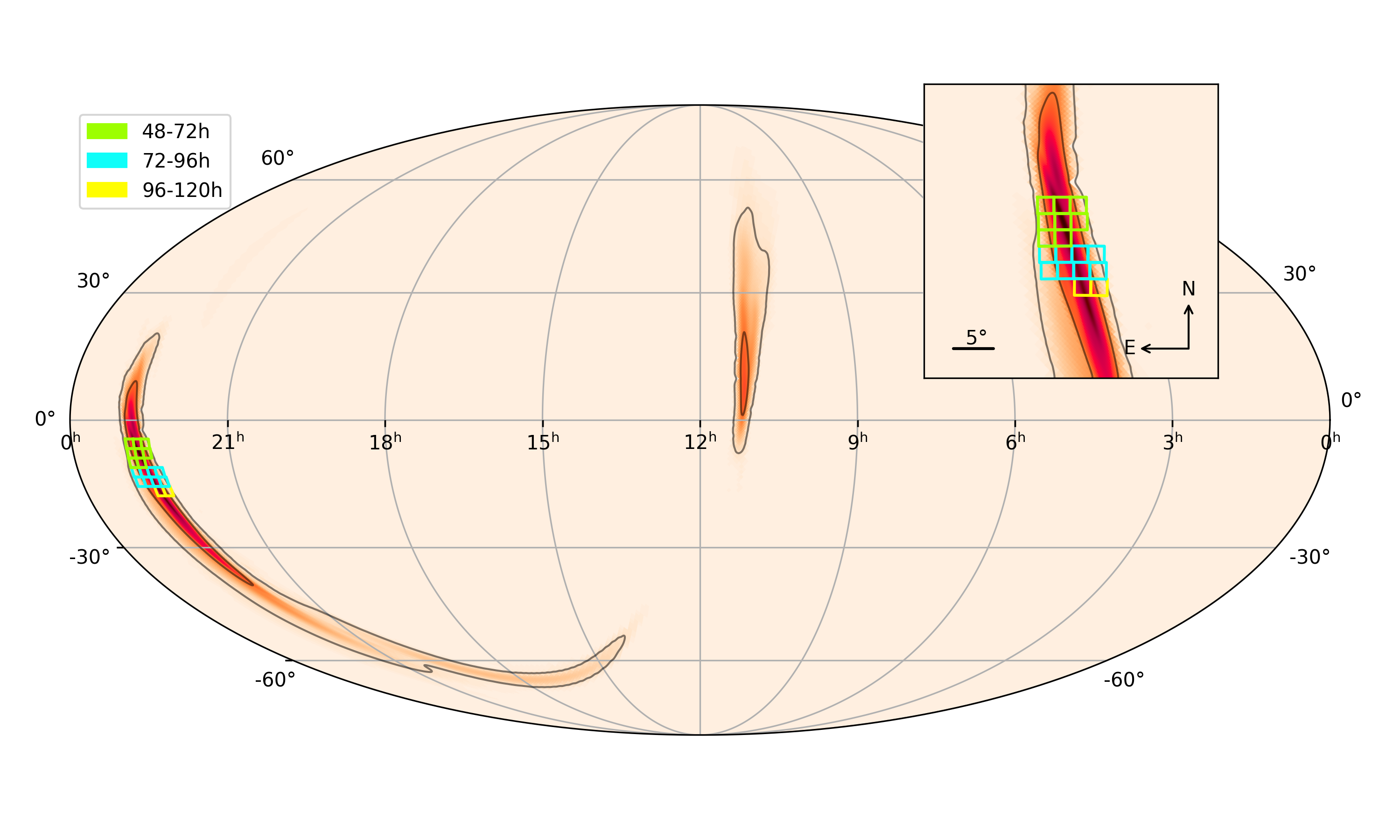}
\caption{Same as Figure~\ref{fig:190408an_loc} for GW190630\_185205 overlaid with the CSS fields that were observed within 120 hours of the GW detection.  These fields covered 25.29\% of the total localization probability.  A Mollweide projection is shown to illustrate the full sky localization, which also had a second component. \label{fig:S190630ag}}
\end{figure*}

\begin{figure*}
\centering
\Large{\textbf{S190901ap}}\par\medskip
\includegraphics[width=\linewidth]{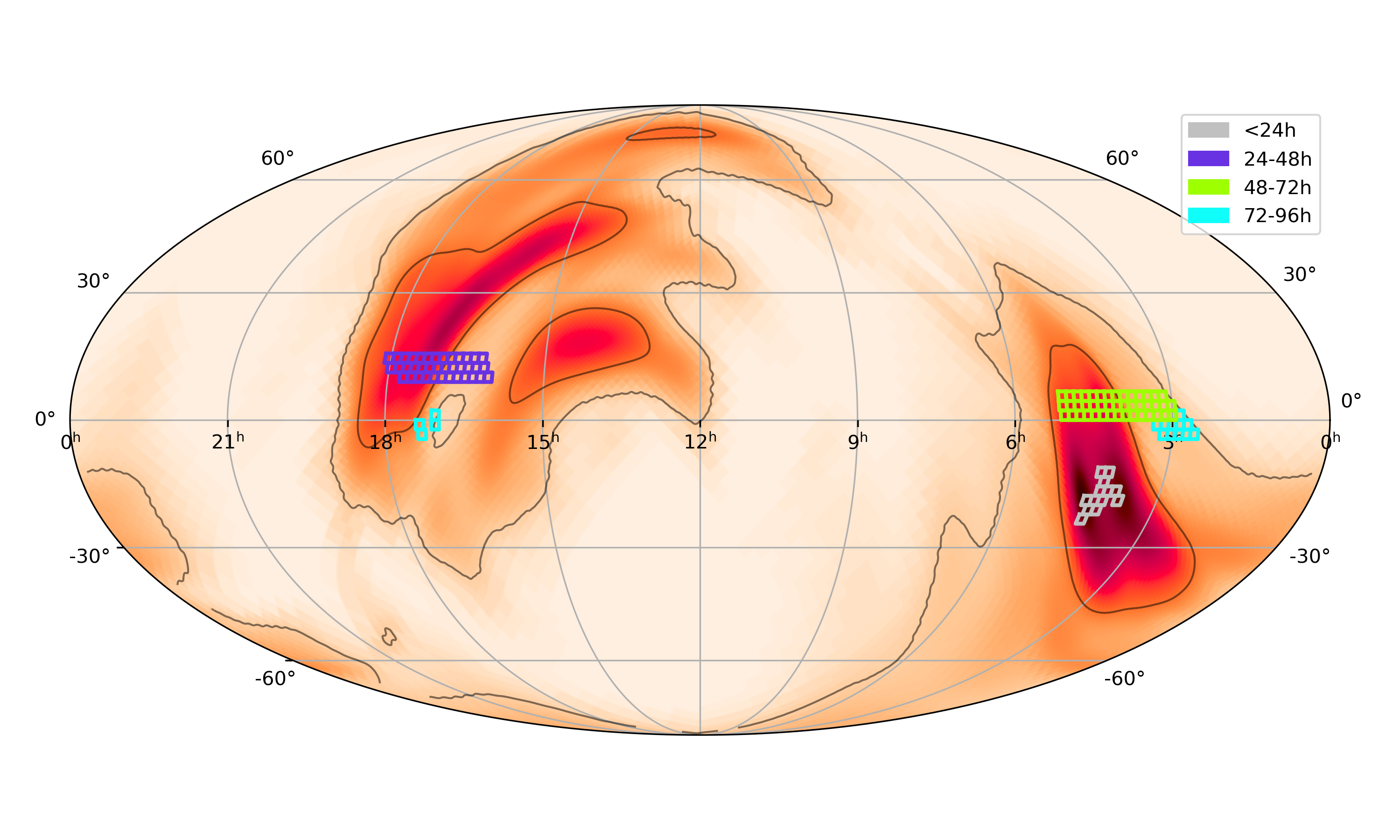}
\caption{Same as Figure~\ref{fig:190408an_loc} for S190901ap overlaid with the CSS fields that were observed within 96 hours of the GW detection.  These fields covered 5.79\% of the total localization probability.  A Mollweide projection is shown to illustrate the full sky localization, which had two large components. \label{fig:S190901ap}}
\end{figure*}

\begin{figure*}
\centering
\Large{\textbf{S190923y}}\par\medskip
\includegraphics[width=8.5cm]{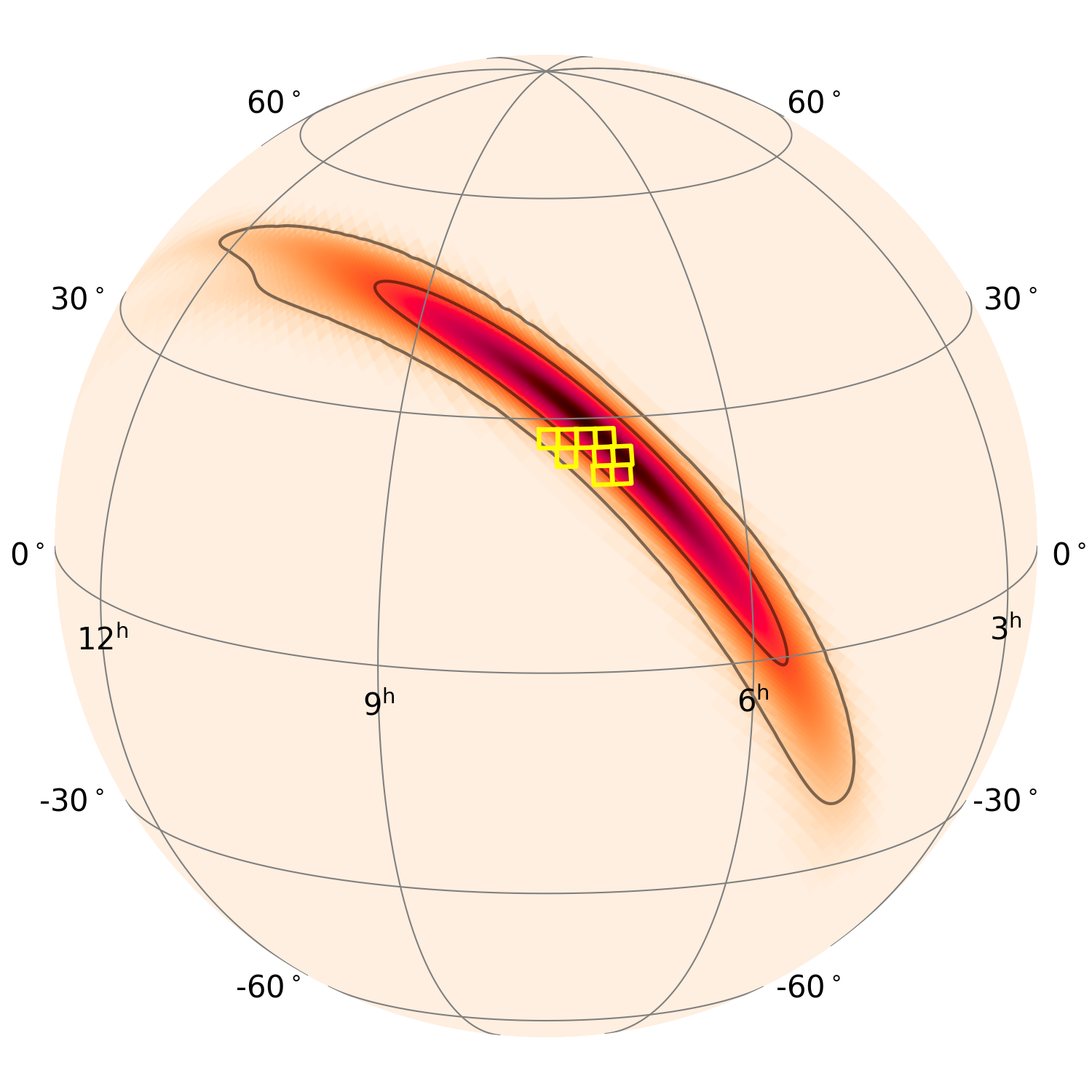}
\includegraphics[width=8.5cm]{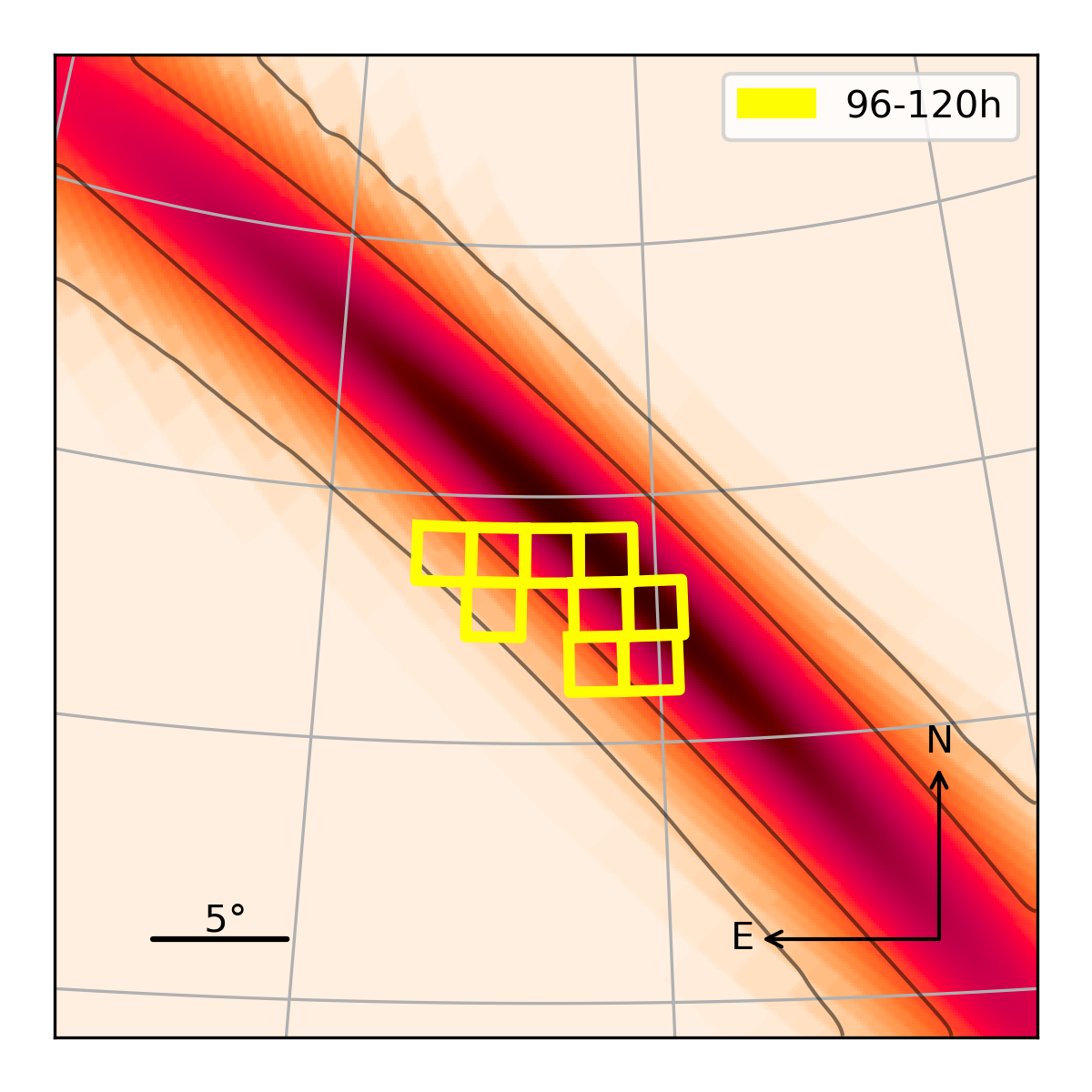}
\caption{Same as Figure~\ref{fig:190408an_loc} for S190923y overlaid with the CSS fields that were observed within 120 hours of the GW detection.  These fields covered 4.05\% of the total localization probability. \label{fig:S190923y}}
\end{figure*}

\begin{figure*}
\centering
\Large{\textbf{GW190930\_133541}}\par\medskip
\includegraphics[width=8.5cm]{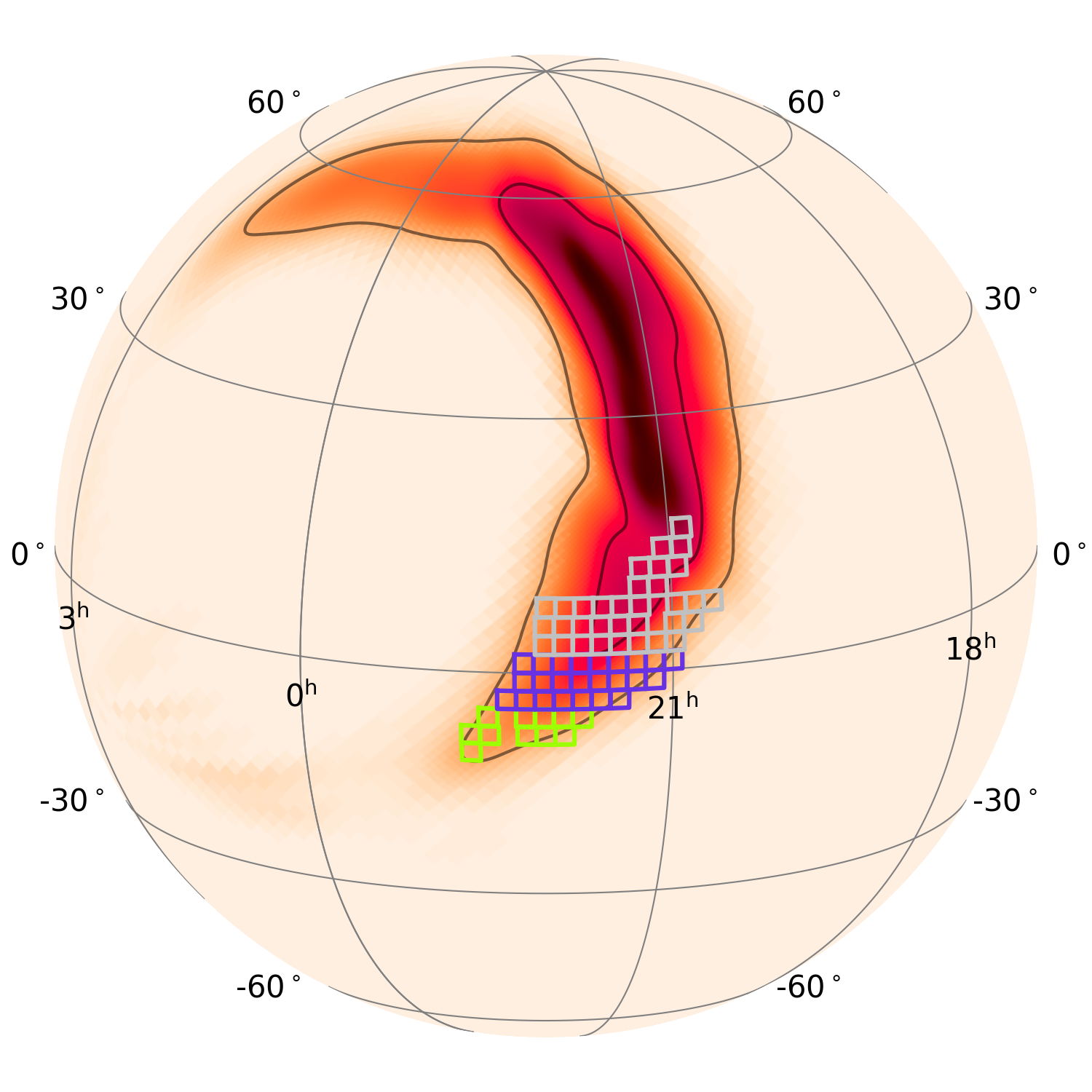}
\includegraphics[width=8.5cm]{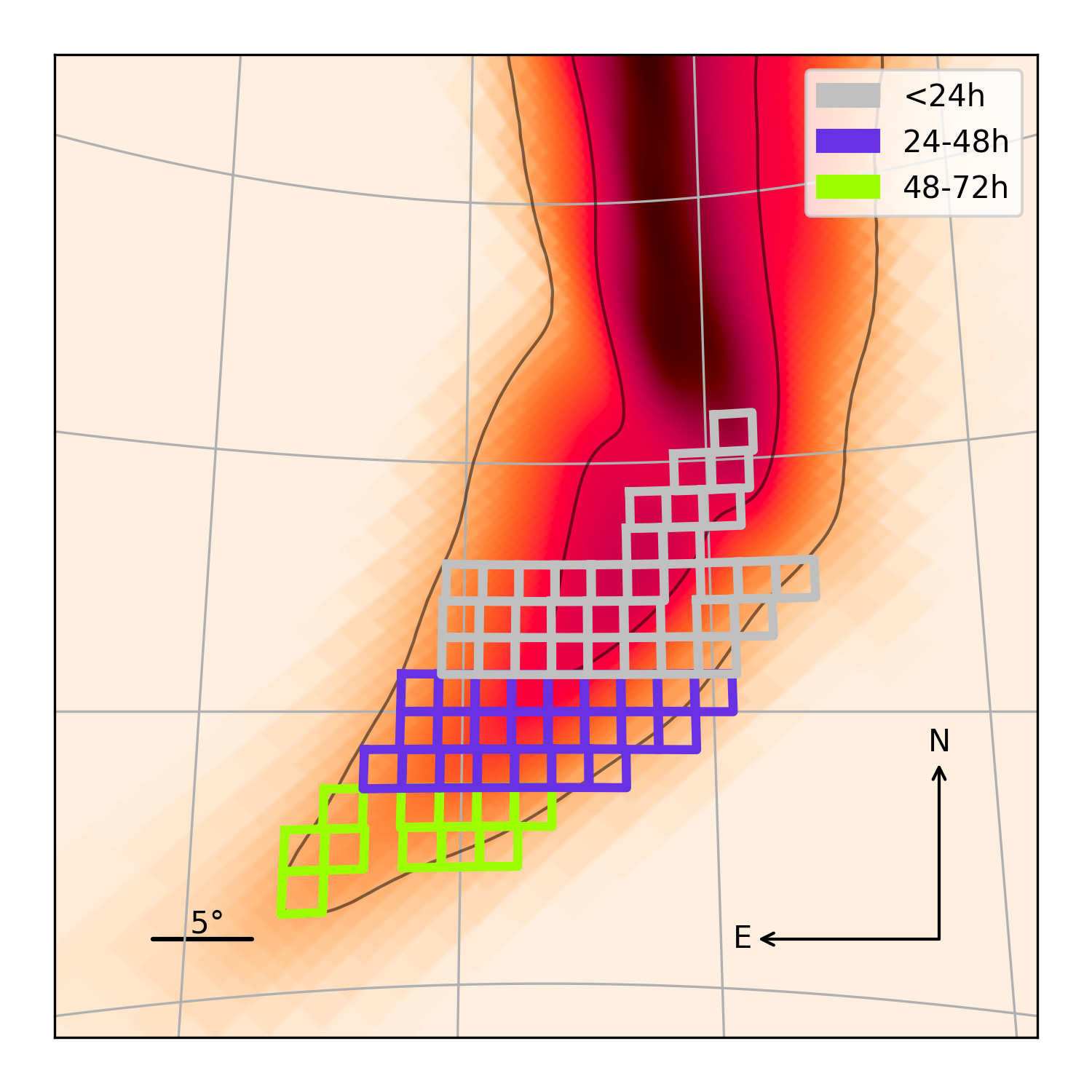}
\caption{Same as Figure~\ref{fig:190408an_loc} for GW190930\_133541 overlaid with the CSS fields that were observed within 72 hours of the GW detection.  These fields covered 13.13\% of the total localization probability.  Only the southern part of the localization could be observed due to the proximity to the Galactic Plane. \label{fig:S190930s}}
\end{figure*}

\begin{figure*}
\centering
\Large{\textbf{S190930t}}\par\medskip
\includegraphics[width=\linewidth]{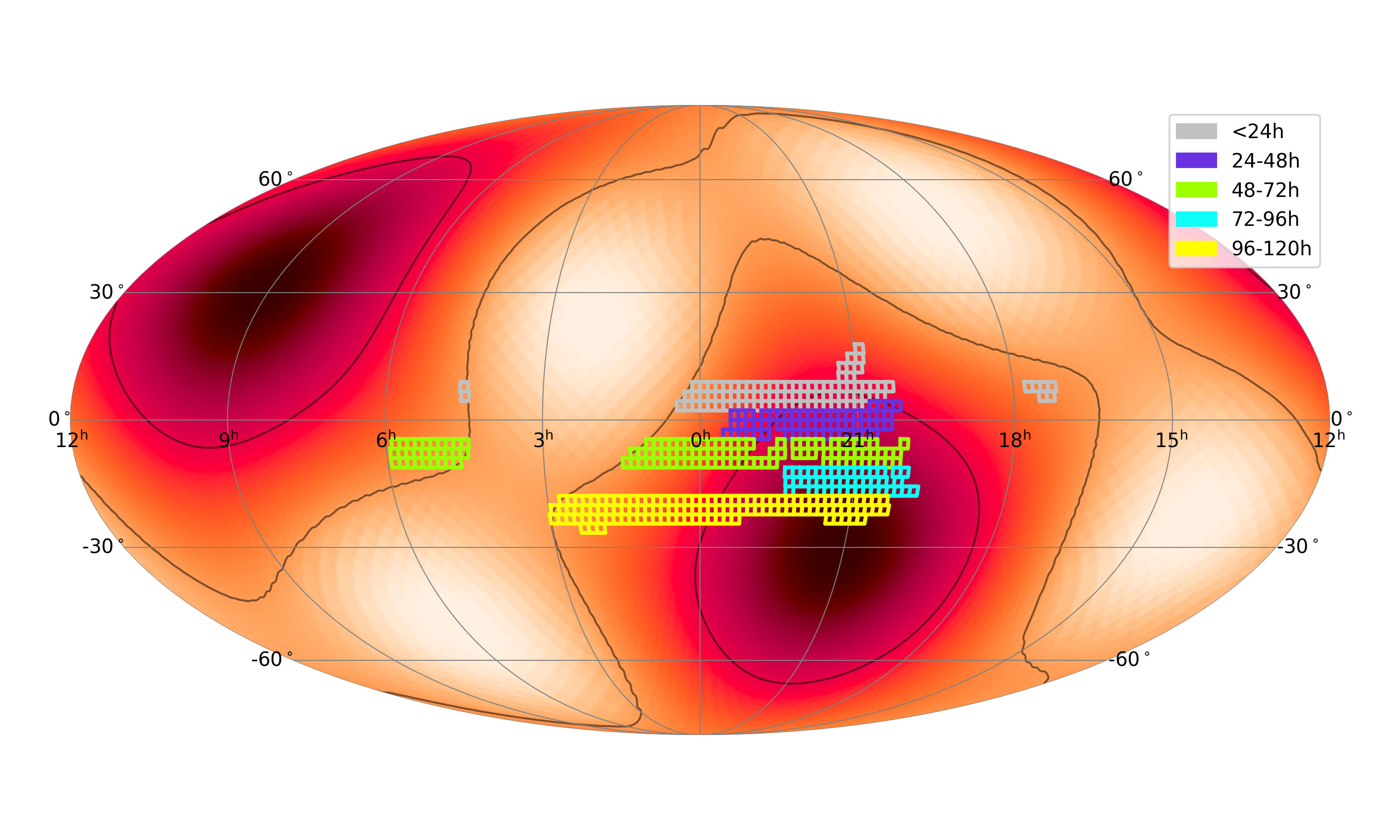}
\caption{Same as Figure~\ref{fig:190408an_loc} for S190930t overlaid with the CSS fields that were observed within 120 hours of the GW detection.  These fields covered 9.13\% of the total localization probability.  A Mollweide projection is included to illustrate the full sky localization which covers a majority of the sky. \label{fig:S190930t}}
\end{figure*}

\begin{figure*}
\centering
\Large{\textbf{S191205ah}}\par\medskip
\includegraphics[width=8.5cm]{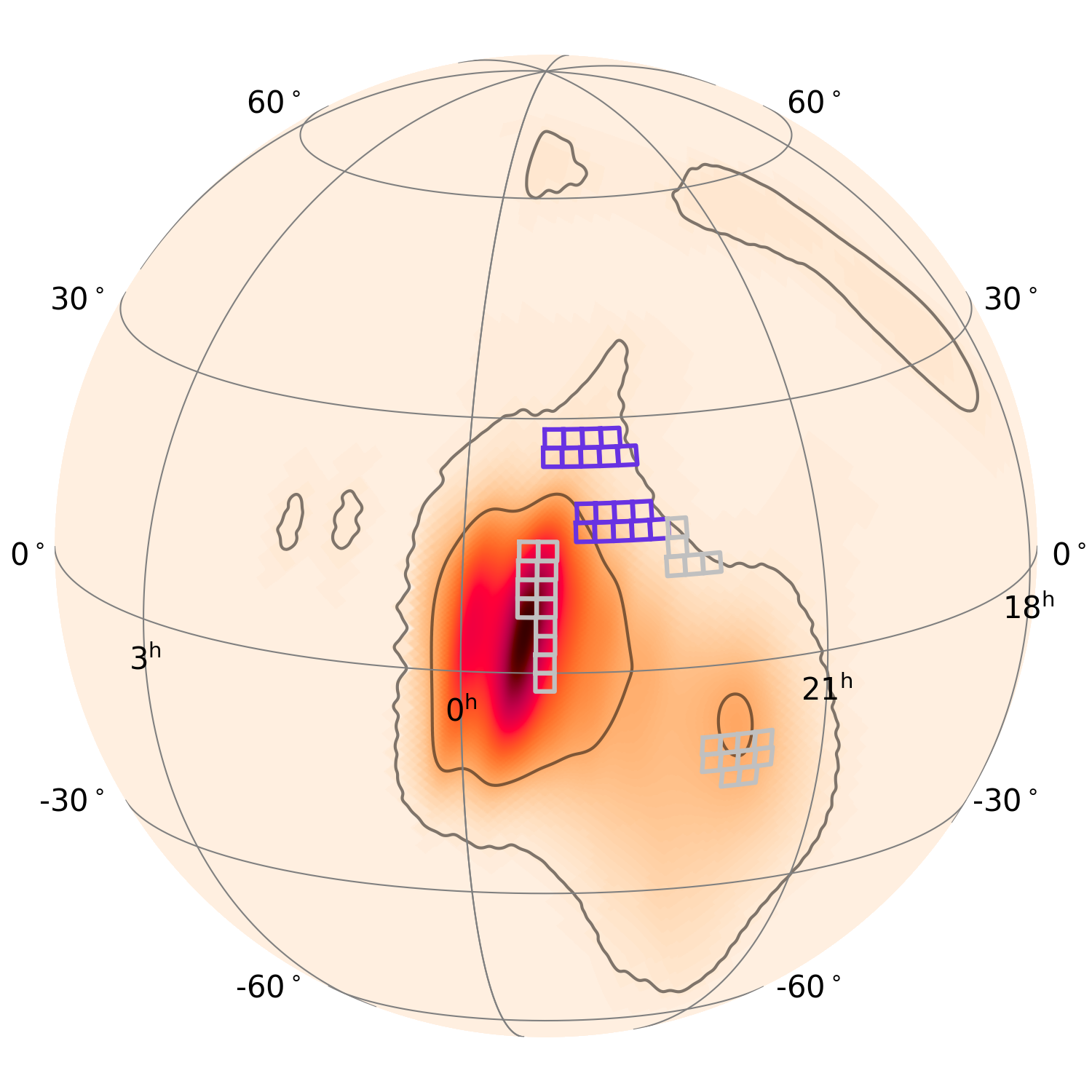}
\includegraphics[width=8.5cm]{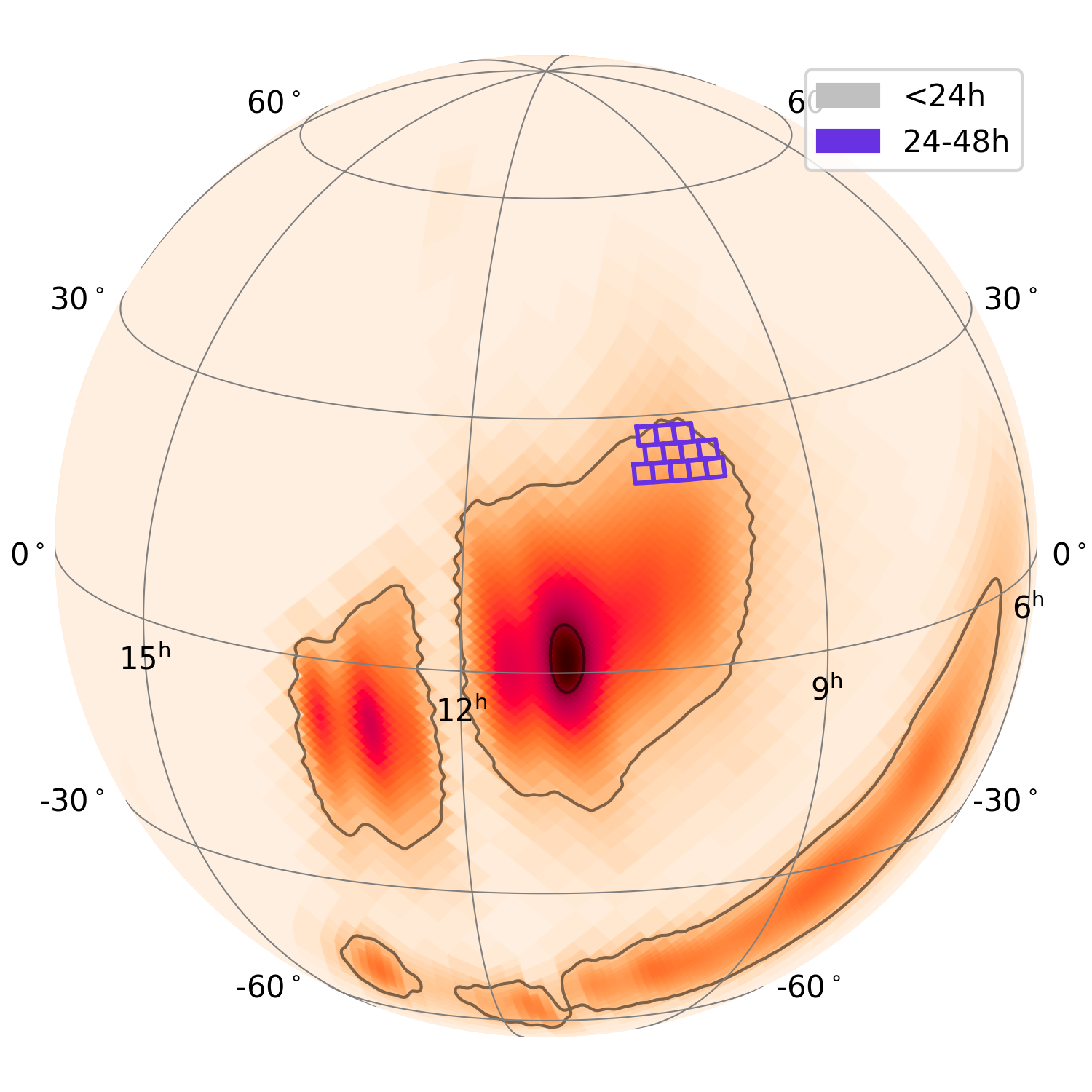}
\caption{Same as Figure~\ref{fig:190408an_loc} for S191205ah overlaid with the CSS fields that were observed within 48 hours of the GW detection.  These fields covered 10.91\% of the total localization probability.  Two globe projections were used to illustrate the multiple components of the localization. \label{fig:S191205ah}}
\end{figure*}

\begin{figure*}
\centering
\Large{\textbf{S191213g}}\par\medskip
\includegraphics[width=\linewidth]{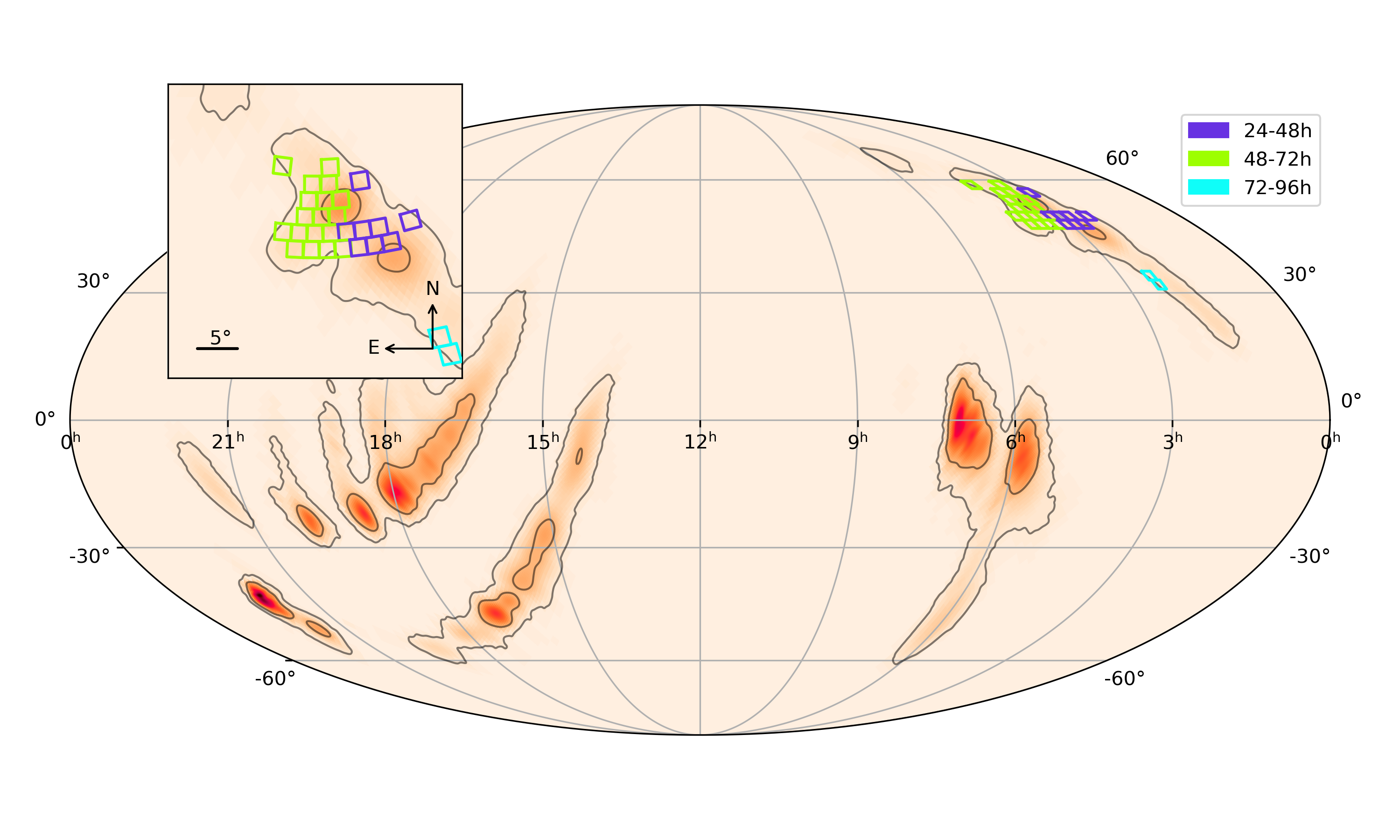}
\caption{Same as Figure~\ref{fig:190408an_loc} for S191213g overlaid with the CSS fields that were observed within 96 hours of the GW detection.  These fields covered 2.10\% of the total localization probability.  A Mollweide projection is shown to illustrate the full sky localization, which had multiple components.  The inset panel zooms in on the fields that were observed.  \label{fig:S191213g}}
\end{figure*}

\begin{figure*}
\centering
\Large{\textbf{S200105ae}}\par\medskip
\includegraphics[width=\linewidth]{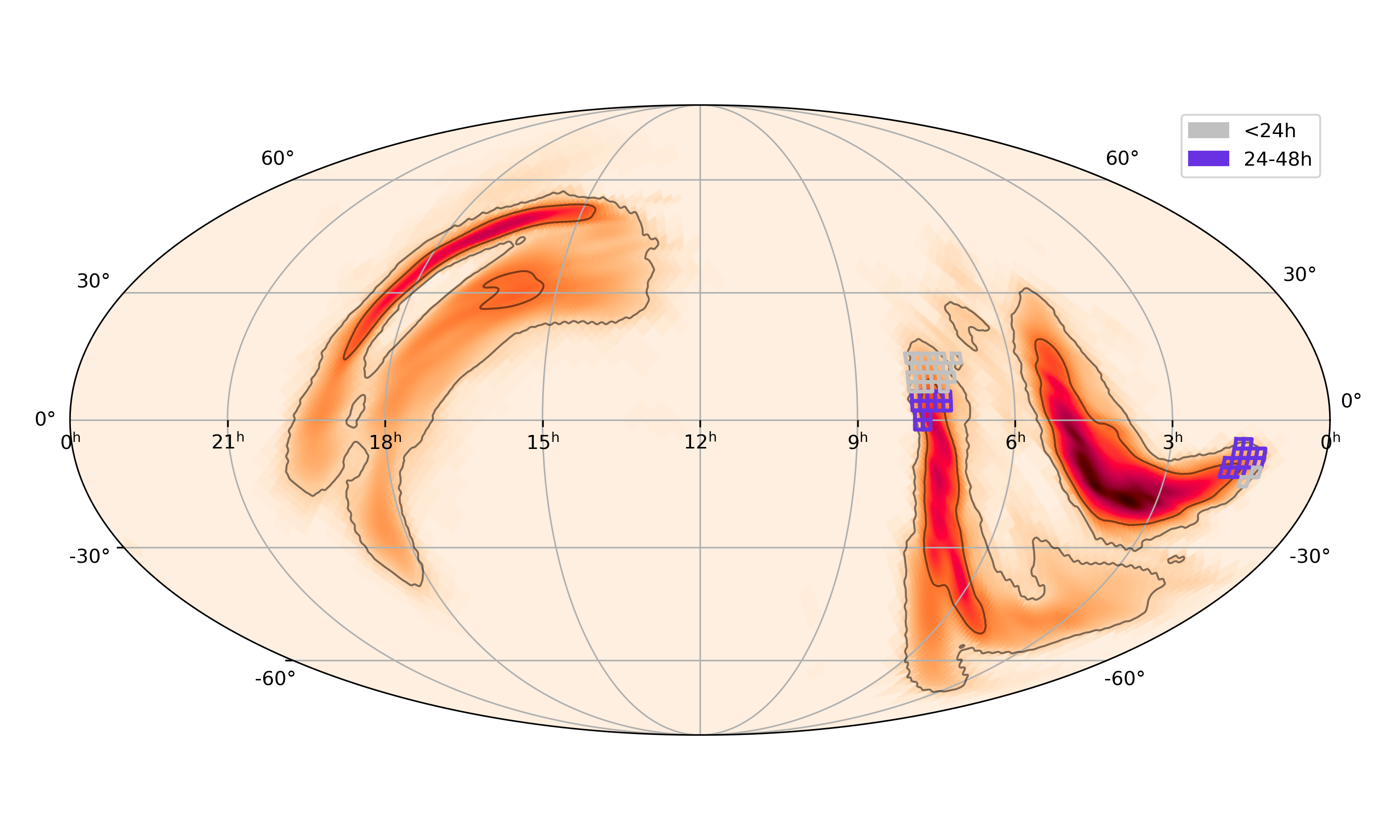}
\caption{Same as Figure~\ref{fig:190408an_loc} for S200105ae overlaid with the CSS fields that were observed within 48 hours of the GW detection.  These fields covered 2.84\% of the total localization probability.  A Mollweide projection is included to illustrate the full sky localization, which had multiple components.  \label{fig:S200105ae}}
\end{figure*}

\begin{figure*}
\centering
\Large{\textbf{S200114f}}\par\medskip
\includegraphics[width=8.5cm]{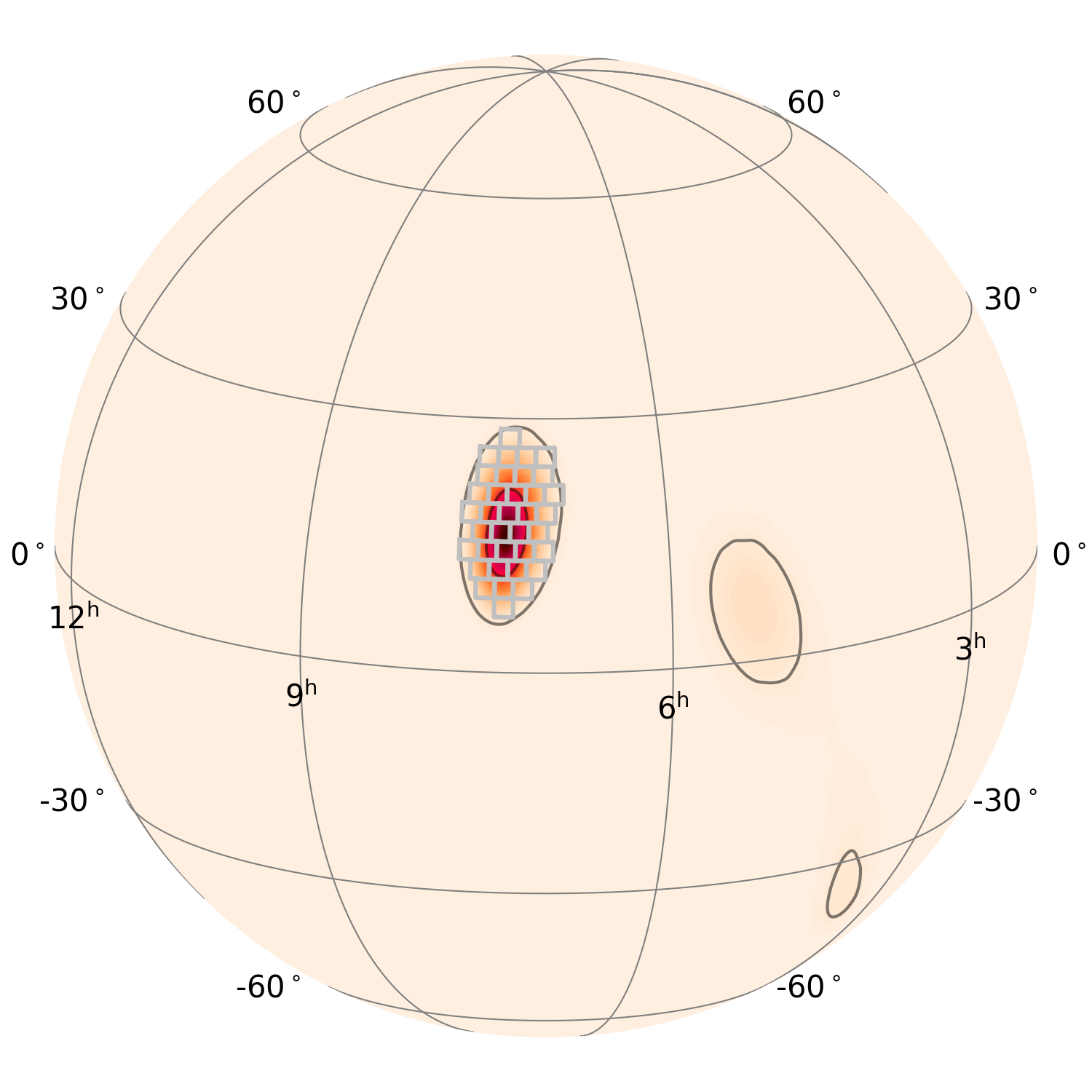}
\includegraphics[width=8.5cm]{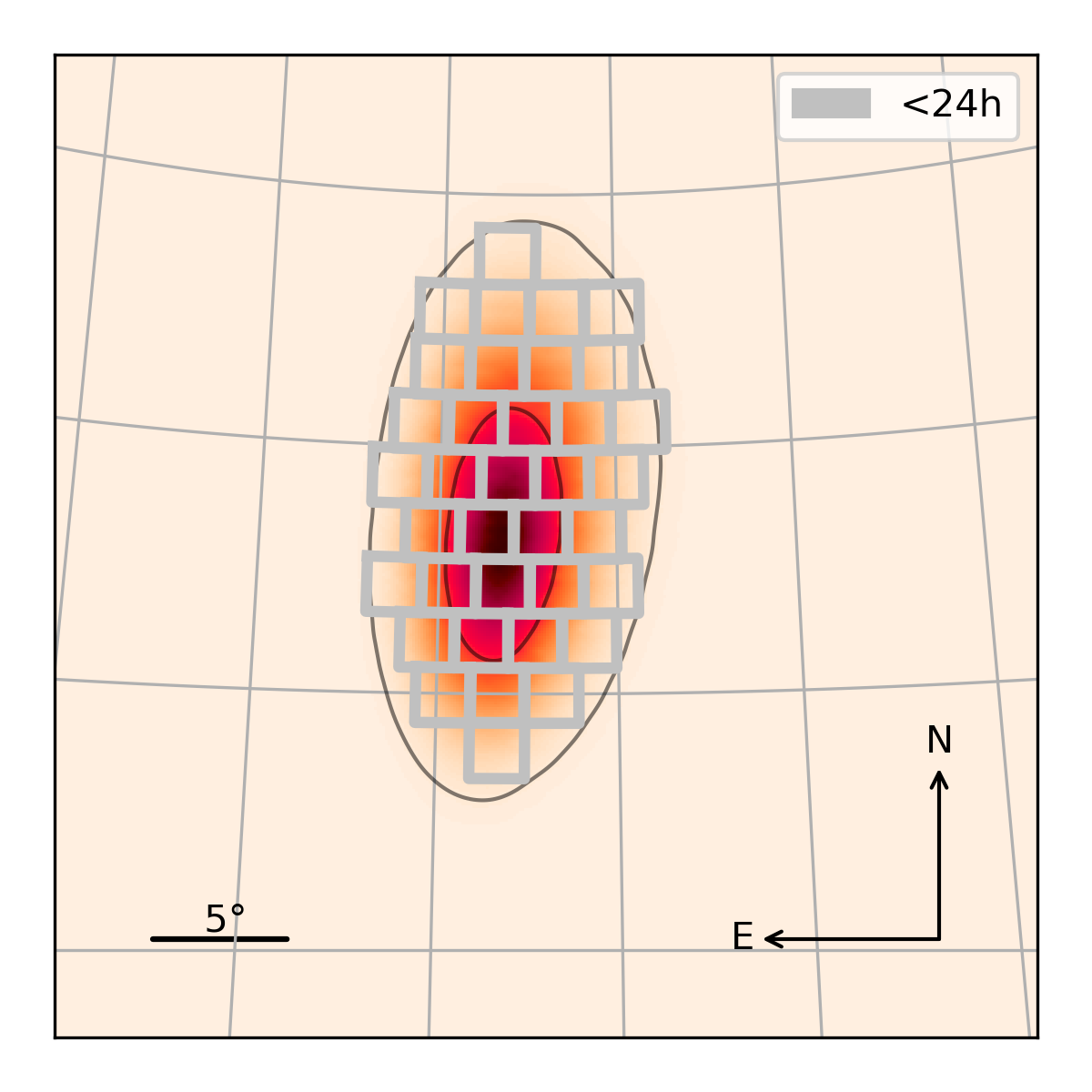}
\caption{Same as Figure~\ref{fig:190408an_loc} for S200114f overlaid with the 36 CSS fields that were observed within 24 hours of the GW detection.  These fields covered 86.30\% of the total localization probability. \label{fig:200114f_globe_loc}}
\end{figure*}

\begin{figure*}
\centering
\Large{\textbf{S200115j}}\par\medskip
\includegraphics[width=\linewidth]{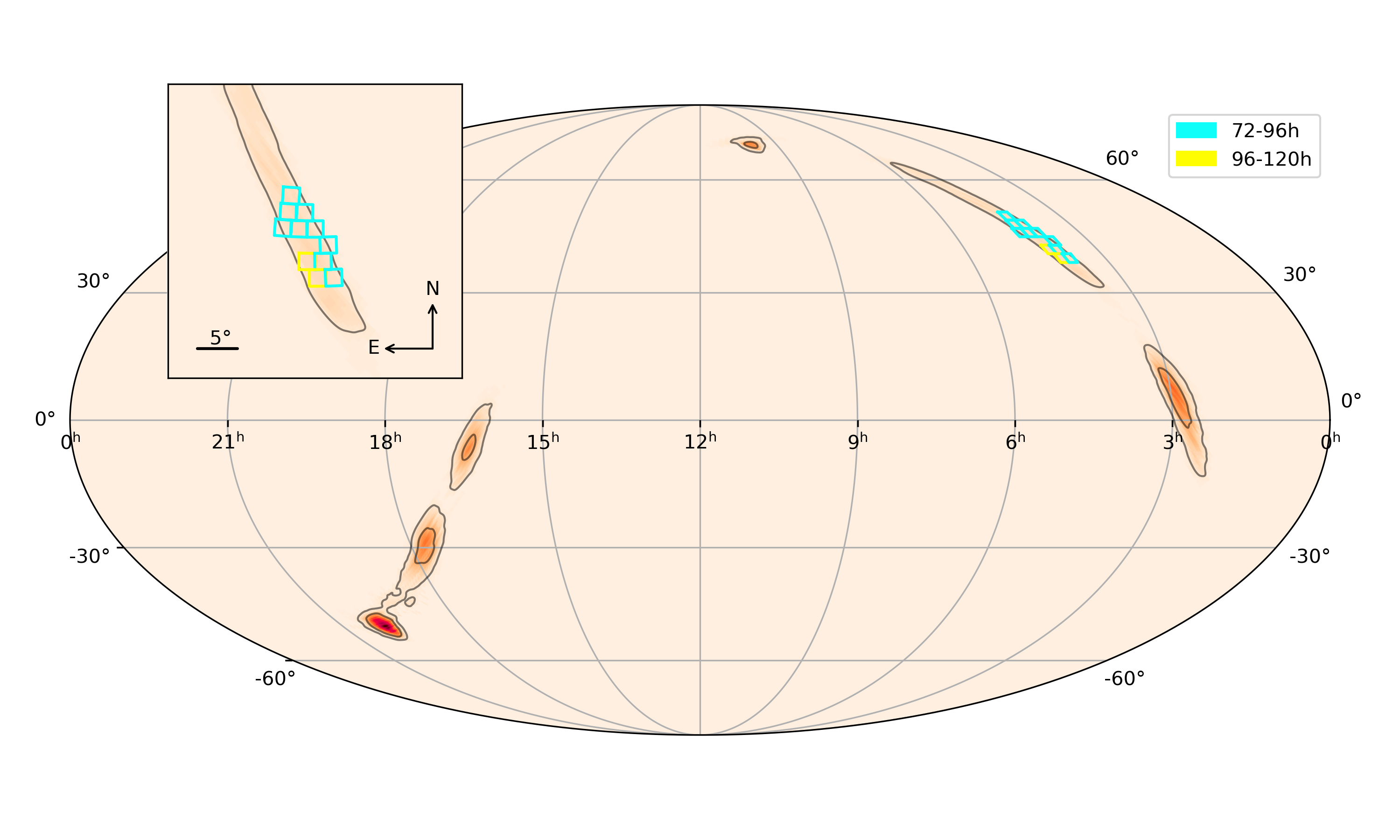}
\caption{Same as Figure~\ref{fig:190408an_loc} for S200115j overlaid with the CSS fields that were observed within 120 hour of the GW detection.  These fields covered 2.64\% of the total localization probability.  A Mollweide projection is shown to illustrate the full sky localization, which had multiple components.  The inset panel zooms in on the fields that were observed.  \label{fig:S200115j}}
\end{figure*}

\begin{figure*}
\centering
\Large{\textbf{S200128d}}\par\medskip

\includegraphics[width=\linewidth]{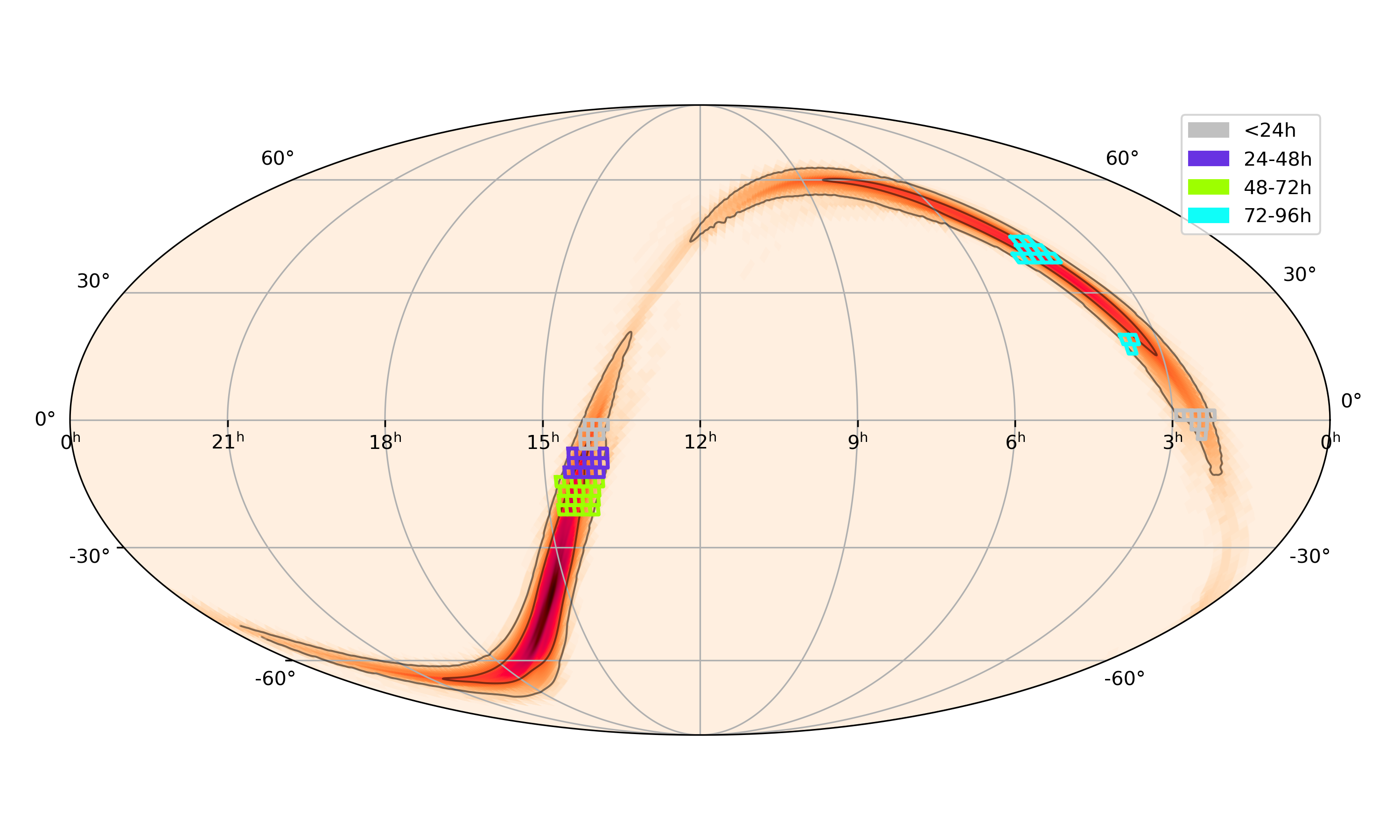}
\caption{Same as Figure~\ref{fig:190408an_loc} for S200128d overlaid with the CSS fields that were observed within 96 hours of the GW detection.  These fields covered 11.80\% of the total localization probability.  A Mollweide projection is shown to illustrate the full sky localization, which had both northern and southern components.  \label{fig:S200128d}}
\end{figure*}

\begin{figure*}
\centering
\Large{\textbf{S200224ca}}\par\medskip
\includegraphics[width=8.5cm]{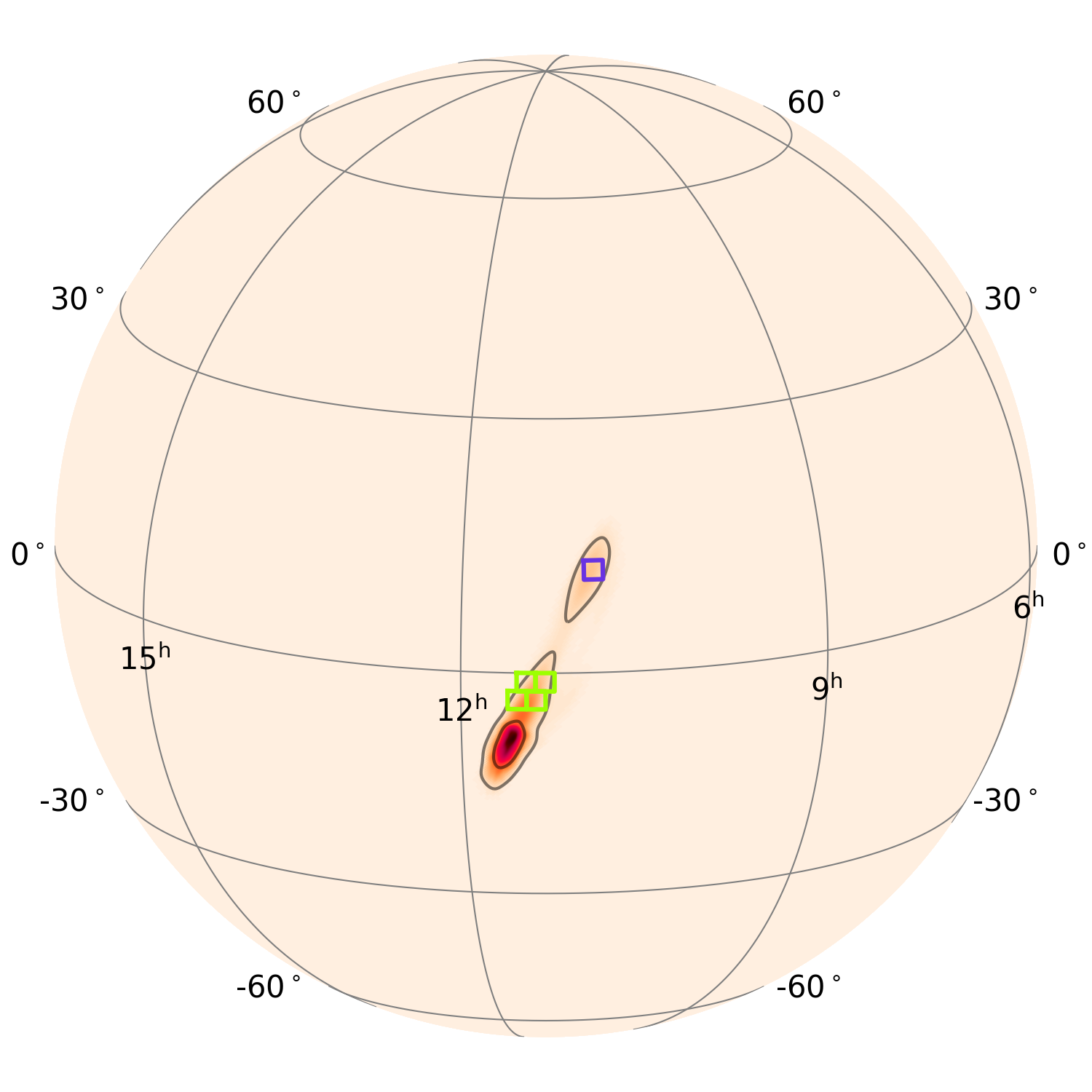}
\includegraphics[width=8.5cm]{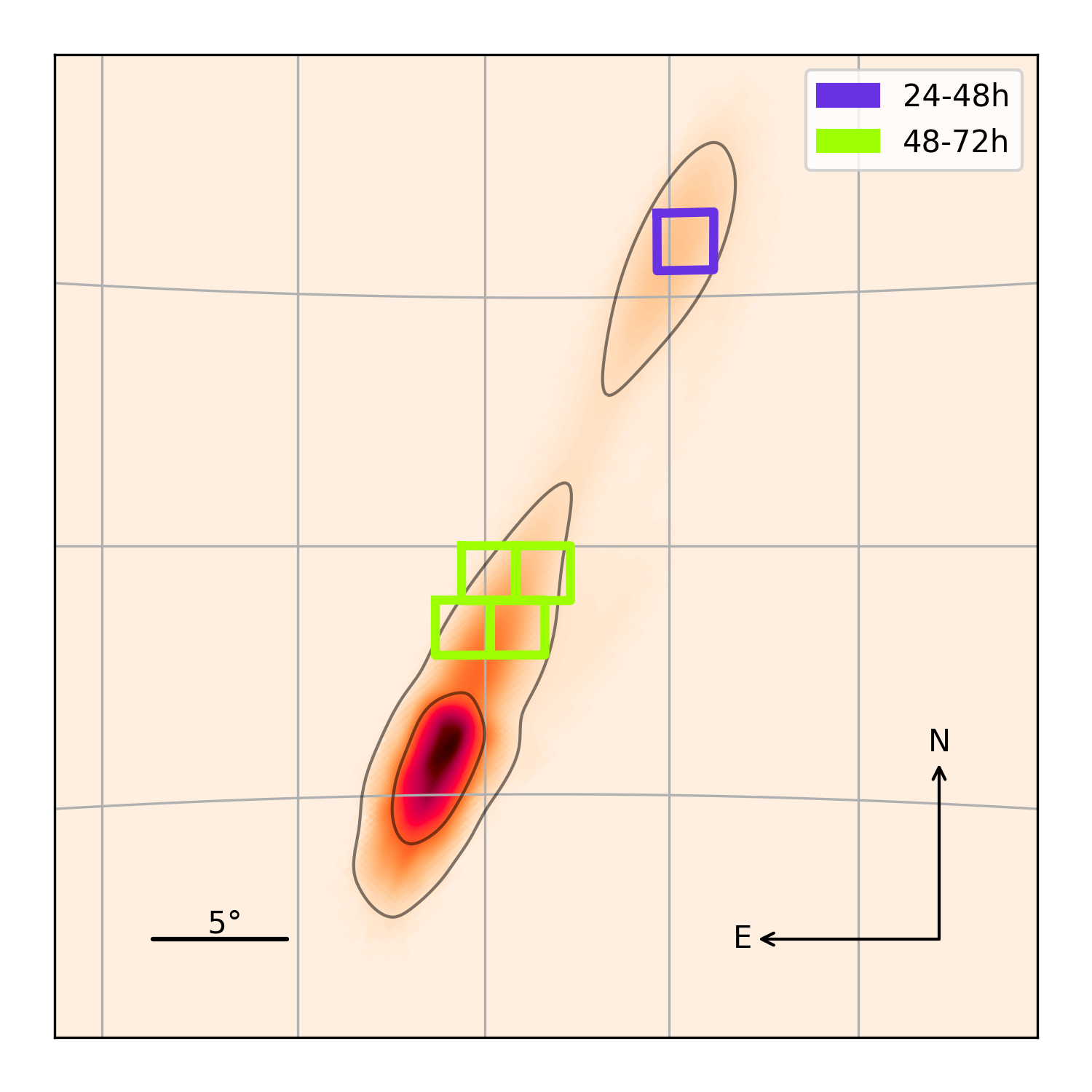}
\caption{Same as Figure~\ref{fig:190408an_loc} for S200224ca overlaid with the CSS fields that were observed within 72 hours of the GW detection.  These fields covered 11.42\% of the total localization probability.\label{fig:S200224ca}}
\end{figure*}

\begin{figure*}
\centering
\Large{\textbf{S200316bj}}\par\medskip
\includegraphics[width=\linewidth]{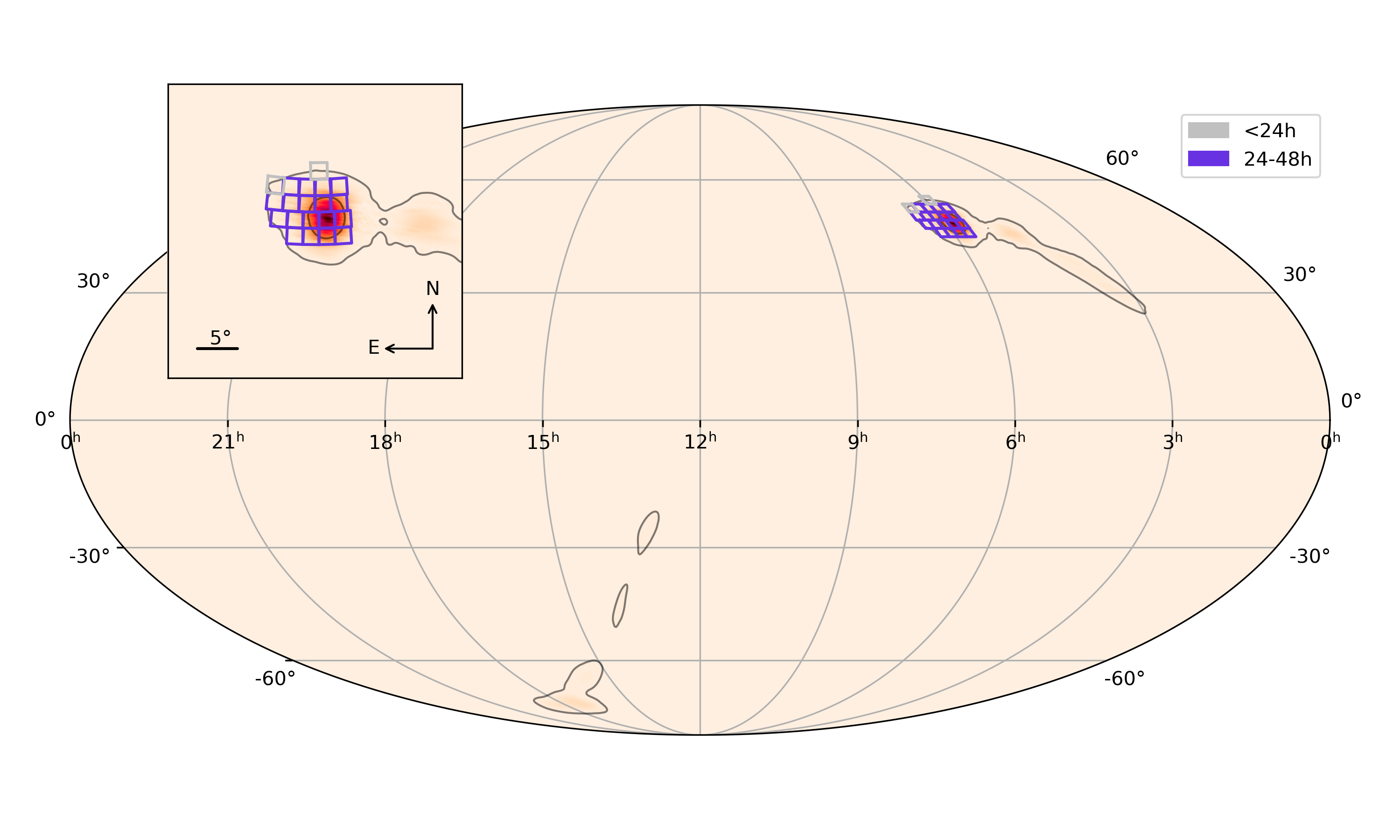}
\caption{Same as Figure~\ref{fig:190408an_loc} for S200316bj overlaid with the CSS fields that were observed within 48 hours of the GW detection.  These fields covered 73.24\% of the total localization probability.  A Mollweide projection is shown to illustrate the full sky localization, which also had southern components.  The inset panel zooms in on the fields that were observed.  \label{fig:S200316bj}}
\end{figure*}

\end{appendices}

\end{document}